\documentclass{uiucthesis2021}
\usepackage[utf8]{inputenc}
\usepackage[english]{babel}
\usepackage{csquotes}
\usepackage{microtype}
\usepackage{amsmath,amsthm,amssymb}
\usepackage[bookmarksdepth=3,linktoc=all,colorlinks=true,urlcolor=blue,linkcolor=blue,citecolor=blue]{hyperref}
\usepackage[capitalize]{cleveref}
\usepackage[style=ieee]{biblatex}
\usepackage{upgreek}
\usepackage{tikz}
\usepackage{cancel}
\usepackage{setspace}
\usepackage{bm}
\usepackage{graphicx}
\usepackage{color}
\usepackage{soul}
\usepackage{float}
\usepackage{hyperref}
\usepackage{epsfig}
\usepackage{epstopdf}
\usepackage[utf8]{inputenc}
\usepackage{epigraph}
\usepackage{mathtools}

\usepackage{physics}
\usepackage{tensor}
\usepackage[T1]{fontenc}

\newcommand{\barq}{{\bar q}}

\newcommand{\bars}{{\bar s}}

\newcommand{\tvec}[1]{\vec{#1}_\bot}

\newcommand{\ord}[1]{\mathcal{O} \left( #1 \right)}

\newcommand{\kompost}{K{\o}MP{\o}ST }

\newcommand{\G}[2]{\tensor*{\tilde{G}}{*^{#1}_{#2}}}
\newcommand{\F}[2]{\tensor*{\tilde{F}}{*^{#1}_{#2}}}
\newcommand{\GCS}[2]{\tensor*{G}{*^{#1}_{#2}}}
\newcommand{\FCS}[2]{\tensor*{F}{*^{#1}_{#2}}}
\newcommand{\tens}[3]{\tensor*{#1}{*^{#2}_{#3}}}

\newcommand{\xt}{\mathbf{x}}

\newcommand{\kt}{\mathbf{k}}
\newcommand{\rt}{\mathbf{r}}
\newcommand{\vkt}{\vqty{\mathbf{k}}}
\newcommand{\vrt}{\vqty{\mathbf{r}}}

\newcommand{\thydro}{\tau_\text{hydro}}

\newcommand{\BG}{\text{BG}}

\DeclareUnicodeCharacter{0301}{*************************************}
\makeatletter
\newsavebox\myboxA
\newsavebox\myboxB
\newlength\mylenA

\newcommand*\xoverline[2][0.75]{%
    \sbox{\myboxA}{$\m@th#2$}%
    \setbox\myboxB\null% Phantom box
    \ht\myboxB=\ht\myboxA%
    \dp\myboxB=\dp\myboxA%
    \wd\myboxB=#1\wd\myboxA% Scale phantom
    \sbox\myboxB{$\m@th\overline{\copy\myboxB}$}%  Overlined phantom
    \setlength\mylenA{\the\wd\myboxA}%   calc width diff
    \addtolength\mylenA{-\the\wd\myboxB}%
    \ifdim\wd\myboxB<\wd\myboxA%
       \rlap{\hskip 0.5\mylenA\usebox\myboxB}{\usebox\myboxA}%
    \else
        \hskip -0.5\mylenA\rlap{\usebox\myboxA}{\hskip 0.5\mylenA\usebox\myboxB}%
    \fi}
\makeatother
\newcommand{\code}[1]{\textsc{#1}}
\def\TRENTo{{\sc t\kern-.05em \lower.5ex\hbox{r}\kern-.025em e\kern-.05em n\kern-.05em t\kern-.09em}o}

\begin{document}

\title{Toward Initial Conditions of Conserved Charges}
\author{Patrick Carzon}
\department{Physics}
\phdthesis
\degreeyear{2023}
\committee{
    Associate Professor Jorge Noronha, Chair\\
    Assistant Professor Jacquelyn Noronha-Hostler, Director of Research\\
    Associate Professor Anne Sickles\\
    Assistant Professor Charles Gammie}
\maketitle
\frontmatter

\begin{abstract}

At top collider energies where baryon stopping is negligible, the initial state 
of heavy ion collisions is overall charge neutral and predominantly composed of gluons.  Nevertheless, there can also be significant local fluctuations of the baryon number, strangeness, and electric charge densities about zero, perturbatively corresponding to the production of quark/antiquark pairs.  These previously ignored local charge fluctuations can permit the study of charge diffusion in the quark-gluon plasma (QGP), even at top collider energies.  In this paper we present a new model denoted \code{iccing} (Initial Conserved Charges in Nuclear Geometry) which can reconstruct the initial conditions of conserved charges in the QGP by sampling a ($g \rightarrow q\bar{q}$) splitting probability over the initial energy density.  We find that the new charge distributions generally differ from the bulk energy density; in particular, the strangeness distribution is significantly more eccentric than standard bulk observables and appears to be associated with the geometry of hot spots in the initial state.  The new information provided by these conserved charges opens the door to studying a wealth of new charge- and flavor-dependent correlations in the initial state and ultimately the charge transport parameters of the QGP.
\end{abstract}

\begin{acknowledgments}

\epigraph{I want creation to penetrate you with so much admiration that wherever you go, the least plant may bring you clear remembrance of the Creator. A single plant, a blade of grass, or one speck of dust is sufficient to occupy all your intelligence in beholding the art with which it has been made.}{St. Basil the Great}

It has been a long path toward the completion of my PhD in physics and there are many who require recognition. With the death of my maternal grandparents, as well as my father, during the formative years of high school and undergrad, I was inspired to make a decision as to the priority of the many facets of life. Seeing the lives that were touched by my grandparents and father, through the attendance at their funerals, motivated me to put an inordinate emphasis on family and friendships above all else. As such, my life priorities have been organized in the following order: God, family, friends, work, and finally all else. I will address my appreciation according to this philosophy.

Above all, I am supremely appreciative of my Catholic faith, which has been the most important part of my journey through life. While being my primary motivation for studying physics, my relationship with Christ has grown tremendously during my PhD, and I do not know where I would be without it. From a purely utilitarian view, being Catholic has been of great value to my mental health, and without it I would not have been able to complete my degree. Everywhere I go, whether it be Rutgers in New Jersey, University of Illinois in Champaign-Urbana, or Copenhagen in Denmark, there are those that share similar perspectives on life that I made fast friends with. There have been several times when the stress has been so much that I considered stepping away from the pursuit of physics, but those friendships through Christ have been a great source of consolation and motivation to continue. From a spiritual point of view, these past five years have drawn me closer to God and the goal, to which all are called, of sainthood. My Latin Mass community in Jersey City fostered a deep appreciation for the ancient rite and the beauty of Sunday convivium! With much of my time spent in Champaign-Urbana, I have been blessed to be part of the Catholic Young Professionals (CYP) group, through which I obtained a deep appreciation for the study of the Bible, theology, and philosophy. The friendship and leadership of Ian Ludden and Laura Suttenfield have been unrivaled! Of particular grace was the spiritual direction I received through Msgr. Stanley Deptula, who helped me through some of the hardest times in my life so far. Lastly, I would be remiss to mention the Schola Choir at St. John Newman Center led by the wise Heath Morber. I started singing in the Schola back in 2019 and have refined that talent greatly since then! 

The love and support from my family has been extremely moving! While it was sorrowful to say goodbye to my grandparents, father, and others who have passed, it is reassuring to know that they are on the other side supporting me through their intercession. None of my family is perfect, but they have always been there. My father provided a great example of what it is to be a provider both materially and spiritually. I would not be here without his guidance and can only hope to be as great a man, father, and husband as he was. My mother deserves all the praise in the world! Having been homeschooled through high school, it is a fact that anything present in myself that can be called intelligence comes from her! She impressed upon my siblings and me that learning for the sake of learning could be fun and was worthy of pursuit. Her emotional support is without description. Despite the weight it put on her, we talked almost every day when I lived in New Jersey and frequently thereafter. My parents raised us to be very close to each other, and I will address each of my siblings with a separate statement. John embodies the stereotypical problematic sibling, from his inordinate amount of physical ailments to his instigative behavior. Without him, I would not have developed as a cooking aficionado, so that I could have brownies regularly, or have the ever present motivation to improve myself. With his particular empathy and drive, he will make a phenomenal physician! James is an interesting person, to say the least! Talking about any of his passions with him, such as juggling, movies, music, bouldering, or mathematics, provides some of the most insufferable experiences of my life! These deep interests, however, create one of the greatest humors I have ever met and give him the motivation to excel at the driest of subjects. While she has the least interest in STEM subjects among the family, Maria-Teresa is the most artistically talented and expressive individual I know. While the rest of us will be inspiring young minds or treating physical ailments, she has the power to change the world through her passion!  Olivia is the most sensitive among all of us, which is both her strength and weakness. Her avid interest in the written word has put her siblings to shame! My only regret is that I have lost her to the madness of theorems and proofs, failing to draw her to the Fermi estimates and “practical applications”. Anthony is the typical baby of the family; he is everyone’s favorite and a bit spoiled from it. This makes him, however, the distillation of all the best parts of us and the second heart of the family. I have had more success with him, personally inspiring his interest in video games and programming, for which I take full responsibility! While I know the paths my other siblings will walk, Tony has the ability to walk many paths. It will be interesting to see which he chooses. If the reader cannot tell, my family is very close to my heart; the love I have for them overflowing, and I only see it growing, both in members and depth.
 
 I value friendship very highly, sometimes to my detriment. Though it takes time for me to consider someone my friend, once they are there that status will never change and it comes with many perks! As such, I cannot cover all who fall under this category and will keep the list to those made during graduate school and address them in chronological order. My first year of my PhD was at Rutgers in New Jersey, where I met some great physicists in the making: Steven Clark, Daniel Seleznev, Atithi Acharya, and many others. I also met a great community in Jersey City of Latin Mass Catholics, of particular importance in that group, are Carlo de la Rama and Terry, who are great friends and my fun first roommates (although that was just for a transitional month). I greatly miss the post-Mass hangouts at the various cafes and restaurants in Jersey City. Among these east coast acquaintances, there is Travis Dore, one who I would refer to as a sleeper friend. As my physics big brother, he has a very odd drive and personality that is quite infectious. Despite staying with his family for a week, moving halfway across the country with him, and many interactions, it was not until much too late that I realized he had become a close friend who had shared in many of the important moments of my past five years. I am frightened to see where his passion for physics will take him! 

I have previously mentioned the CYP but more is required. It started with a little graduate student bible study led by the wise Ian Ludden and attended by the whimsical David Bianchi and Ryan Ash. These three people would become some of my greatest friends and the best roommates! As Bros of Burlison, we travelled through the pandemic together aided by many a beer-mosa, smash bros-ening, falling of towers, and “poorman” lattes. Ian is one of the wisest men I know, greatly gifted in leadership and teaching. Dave and Ryan are best friends who are welcoming and easy to hang out with. I do consider these men to be true brothers from significantly different mothers! From there, I became friends with the whole multitude of Chambana Catholic young (and not so young) adults. There were many potlucks, Holy Hours, movie nights, and fellowship opportunities shared with hobbits of all sizes and shapes (some more local to the Shire than others!). During this time, I ran a consecration to St. Joseph group which nosedived into the chaos that is, was, and shall be Inklings! United by a shared interest in literature, libations, and the impassioned conveyance of thoughts, this fellowship wandered many distant and ridiculous lands in search of the honey pot that is enlightened thought. It goes without saying that this land of enlightenment has yet to be reached. The last CYP related group to mention is the Brohim of Brohan. During my year of habitation with Micah Ventura, Lewis McAdow, and Matt Williams, I have been refined for that which lies ahead. 

I will end this section on friendships with the recognition that is required for fellow UIUC graduate students. I can be quite the introvert at times, but the self-proclaimed “Department of Vibes” was able to draw me out. It is with a heavy heart that I say goodbye to those storied conversations and arguments that ranged from the philosophical foundations of mathematics to intricacies of “bones/no bones”. While I have had interchange with other groups, such as the “Penthouse Pad”, it is the Vibes that I will miss the most. This is primarily due to the two people who have become very close friends and a major source of my interest in physics: Debora Mroczek and Jordi Salinas San Martin. One possible characterization of our friendship is that of Harry, Ron, and Hermione. Debora is clearly the one with all the smarts, or if not all an in proportionate amount! Jordi is the Ron to my Harry, that is to say, I am clearly the main character, but Jordi is an essential sidekick. One could also say, Jordi is to me as Sam is to Frodo (I don’t know where Debora factors into this analogy). While in my life I may view myself as the focal point, I would be nowhere without the friendship and wisdom of my best friend and our third wheel! {Honestly, a fairer analogy would be to Don Quixote, where I am the character of title and Jordi is to me my Sancho. He keeps my insanity in control on our journey toward the Dulcinea of Physics (or actually Travis might be more like Don Quixote…).} Analogies aside, I am excited for what the future holds for these two physicists and the possible collaborations and shenanigans we might get up to.

Having covered family and friends it is now time to acknowledge those most central to work. This, however, makes my advisors no less close friends! The person who started me down the dismal path of Heavy-Ion Collisions was George Moschelli. In undergraduate, George was my academic advisor and after some time became the advisor of my research. That transition was easy to make since I find his thought process to be extremely intuitive; there is no one I understand more clearly! It also helped that we share a lot of the same interests outside physics. Through him, I worked with Sean Gavin who is complementary to George in every way. Their relationship is like that of Jedi Master to Padawan although I am not sure who is who! In my senior year, Jacquelyn Noronha-Hostler came to give a presentation on her research. Her talk was extremely impressive and spoke to her communication skills. Along with teaching, another core passion of mine is to be an extremely good speaker and communicator. This was inspired by Fulton Sheen, developed by George Moschelli, and further pursued with Jacquelyn Noronha-Hostler. Jaki has one of the greatest minds I know! Somehow she is not only a great person and physicist, but she also has the bandwidth and energy to take on the largest projects in the community and effectively organize them. I don’t know of anyone else who could advise several post-docs, many graduate students, and a few undergrads while running an infinite number of collaborations. Oh, I almost forgot Jorge Noronha! Just like Jaki, Jorge has such an amazing capacity for activity, which, when paired with his sense of humor, makes him a behemoth of a personality! My last primary mentor is Matthew Sievert, a post-doc when I started and now in charge of his own research group. Like all of my other mentors, Matt thinks like a physicist, which made it easy to intellectually gel with him immediately. I would consider myself of moderate intelligence, but through the mentorship of those mentioned here, I have been able to accomplish things I would never have thought possible.

During my graduate career, I had the opportunity to work with many amazing people! There was a plethora of fellow graduate students in Heavy-Ion theory, collected by Jaki and Jorge, who are amazing physicists and great friends. A particular blessing were the many post-docs who inspired dedication to the craft and provided reassurance that the heights were not too lofty. Of course, there are those external collaborators that much of the work herein was produced with. It was a distinct joy to work with Mauricio Martinez, Matt Luzum, Soeren Schlichting, and Philip Plaschke. While the group of people at UIUC in Heavy-Ion theory are a special bunch, it has been great to meet others who share in the passion, particularly the Houston and Jyväskylä groups.

To the reader, I apologize for the length of this acknowledgements section! While the rest of this document is dedicated to the scientific pursuit of my time as a graduate student, it is not possible to convey the full reality of the work through just the sterile environment below. The work done here would have been impossible without every person mentioned above, so while it would have been acceptable to pare this section down, it would also have been an injustice and discredit to those truly responsible.

\end{acknowledgments}

\tableofcontents

\chapter{List of Publications}
\section*{Relevant Work (not including proceedings)}
The following is a list of published and unpublished work, relevant to this thesis, not including proceedings.
\begin{itemize}
    \item \fullcite{Carzon:2019qja}
    \item \fullcite{Carzon:2020xwp}
    \item \fullcite{Carzon:2021tif}
    \item \fullcite{Carzon:2023zfp}
    \item \fullcite{Plumberg:inPrep}
\end{itemize}

\section*{Other work (not including proceedings)}
The following is a list of other published and unpublished work, that will not be mentioned in this thesis
\begin{itemize}
  \item \fullcite{Carzon:InPrep}
\end{itemize}

\section*{Proceedings}
The following is a list of published conference proceedings
\begin{itemize}
    \item \fullcite{Carzon:2020ohg}
    \item \fullcite{Carzon:2022zpa}
\end{itemize}

\mainmatter

%%%%%%%%%%%%%%%%%%%%%%%%%%%%%%%%%%%%%%%%%%%%%%%%%%%%%%%%%%%%%%%%%%%%%%%%%%%
%
\chapter{Introduction}
%
%%%%%%%%%%%%%%%%%%%%%%%%%%%%%%%%%%%%%%%%%%%%%%%%%%%%%%%%%%%%%%%%%%%%%%%%%%%

\epigraph{It is dangerous to go alone! Take this!}{Old Man}

In the nucleus, there is an abundance of positively charged protons, which should repel each other according to the electromagnetic force. However, nuclei are generally quite stable pointing to the existence of the strong force. This fundamental force of nature is described by Quantum Chromodynamics (QCD), whose degrees of freedom are those of quarks and gluons, the fundamental particles from which nucleons are comprised of. A consequence of QCD is color confinement, i.e., quarks and gluons cannot be observed individually. To break color confinement, extremely high temperatures and pressures are required, which may only occur naturally in neutron stars. It is possible, however, to satisfy these conditions with heavy-ion collisions performed at particle accelerators, such as the Large Hadron Collider (LHC) and Relativistic Heavy Ion Collider (RHIC), which produces a quark-gluon plasma (QGP) that undergoes hydrodynamic evolution and coalesces into hadronic bound states. Studying these nuclear collisions provides insight into the properties of quarks, gluons, and the strong force.

\section{Quantum Chromodynamics}

\begin{figure}[ht]
\begin{centering}
   \includegraphics[width=0.8\textwidth]{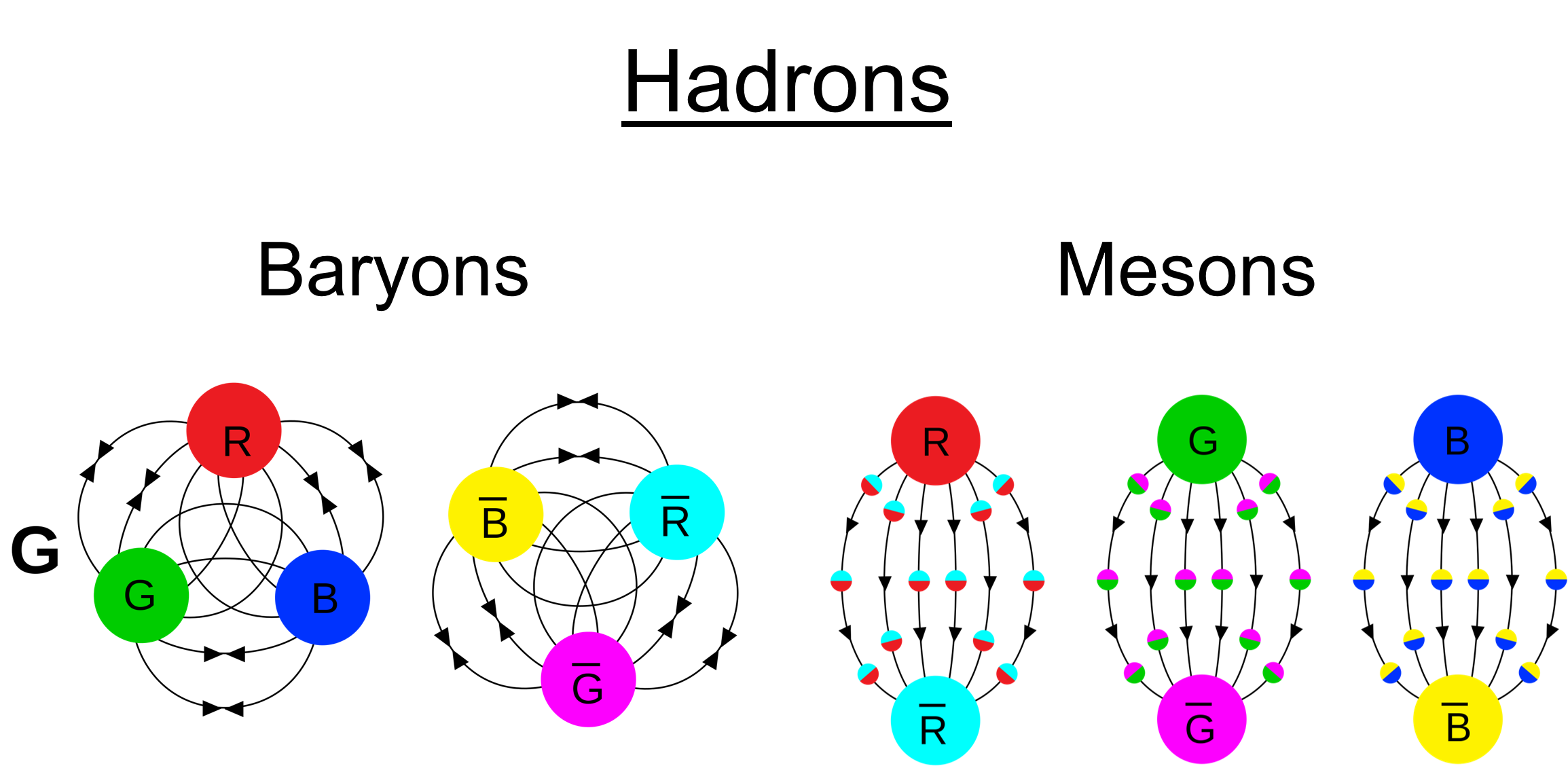}
   \caption{Cartoon of possible color confined states in QCD theory, where the connecting arrows represent gluon interactions.}
   \label{fig:ColorConfinementStates}
\end{centering}
\end{figure}

Quantum Chromodynamics is an SU(3) quantum field theory that describes the strong force using color charge. The fundamental particles of the theory are quarks and the interaction particle is the gluon (collectively referred to as partons), all of which carry color charge. Quarks are characterized by their mass and several charges: baryon, strangness, electric (BSQ), and spin. There are several features of QCD which are particularly important for heavy-ion collisions, namely: color confinement and gluon self-interaction. The strength of the strong force is such that quarks are not individually observable and always confined in hadrons. Hadrons come in two types: baryons and mesons\footnote{More recently, the existence of pentaquark \cite{LHCb:2015yax, LHCb:2019kea} and tetraquark \cite{Belle:2007hrb, Belle:2013shl, LHCb:2014wvs} states has been shown.}, composites of 3 and 2 quarks (referred to as valence quarks), respectively. Protons and neutrons are the most stable baryons and are composed of triplets of quarks that all carry color charge, such that the baryon is color neutral ($Red+Green+Blue=White$). Mesons are unstable states composed of a quark/anti-quark pair, such that the color charge is neutral, with the anti-quark having the anti-color of the quark. The possible color confined states are illustrated in Fig.~\ref{fig:ColorConfinementStates}. 

%__________________________________________________________________________
%
%__________________________________________________________________________
%

A unique property of QCD theory is that the interaction particle, the gluon, carries the charge of the theory and thus is allowed to undergo self interactions. One consequence of the gluon self-interaction, is that a gluon can split and increase the number of gluons. One can describe the composition of a QCD system with a parton distribution function (PDF). For nuclei at high energies, the gluon self-interaction leads to an exponential increase in gluon occupancy implying there is saturation of the PDF at high enough energies \cite{Aaron:2009aa}. Another consequence of color charged gluons is that they are free to split into quark/anti-quark pairs, also referred to as sea quarks. 

The baryon and electric charges of quarks are fractional quantum numbers and conserved quantities. Only strange quarks contain an integer strange charge, while all quarks have a non-zero baryon and electric charge. There are two terms that are used for the mass of the quarks: the "current" and constituent quark mass. In this work, I only use the current quark mass, which is the mass of individual quarks and can be found in the Particle Data Group (PDG). The constituent quark mass is the current mass plus a cloud of sea quarks and gluons, such that the mass of a given hadron is sum of these constituent masses. While this is fine in certain regimes, it is only an effective mass. There are 6 flavors of quark (up, down, strange, charm, bottom, and top), only the three lightest are relevant to this work (the heavier quarks are produced by hard scatterings which are not discussed here) and their properties are listed in Table \ref{t:charges}.

%__________________________________________________________________________
%
\begin{table}
\centering
\begin{tabular}{ | c || c | c | c | c | } \hline
Flavor & Mass & B & S & Q \\ \hline
u & 2.16 MeV & $\tfrac{1}{3}$ & 0 & $\tfrac{2}{3}$ \\ \hline
d & 4.67 MeV & $\tfrac{1}{3}$ & 0 & $-\tfrac{1}{3}$ \\ \hline
s & 93.4 MeV & $\tfrac{1}{3}$ & $-1$ & $-\tfrac{1}{3}$ \\ \hline
c & 1.27 GeV & $\tfrac{1}{3}$ & 0 & $\tfrac{2}{3}$\\ \hline
\end{tabular}
\caption{The conserved charges for the relevant quark flavors: baryon number (B), strangeness (S), electric charge (Q), charm (C). Masses of the light quarks here are the so-called "current quark masses" and the charm mass is the "running mass".}
\label{t:charges}
\end{table}
%
%__________________________________________________________________________

\begin{figure}[ht]
\begin{centering}
   \includegraphics[width=0.5\textwidth]{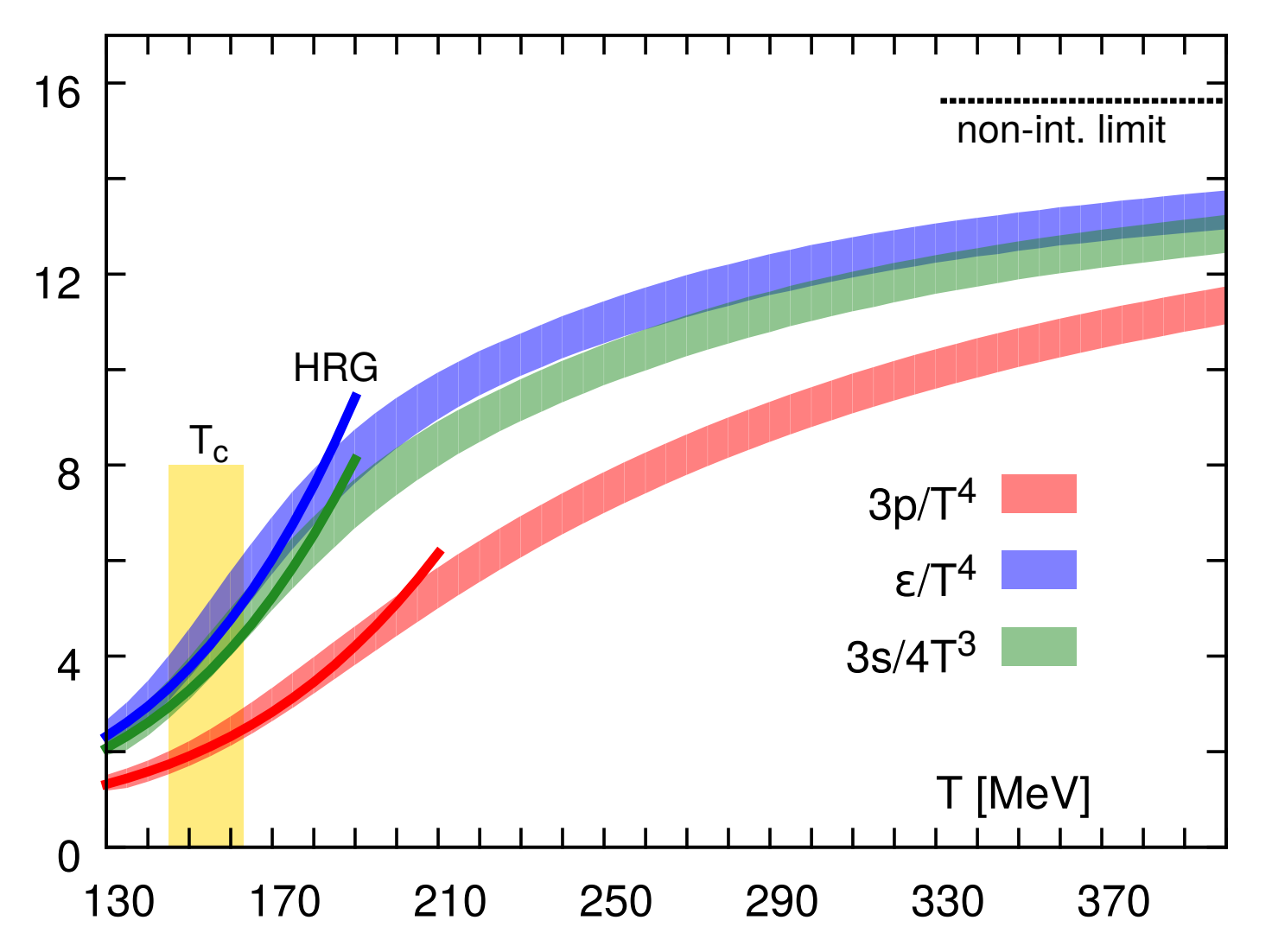}
   \caption{The bands represent lQCD calculations of the unit-less pressure, energy density, and entropy density versus temperature. The dark lines are results from HRG. The matching of lQCD to the hadron degrees of freedom indicates that there is a crossover phase transition around $T = (154 \pm 9) MeV$. Figure from Ref.~\cite{HotQCD:2014kol}.}
   \label{fig:Crossover}
\end{centering}
\end{figure}

A non-perturbative approach to solving QCD is lattice QCD (lQCD), which discreteizes the QCD Lagrangian and provides an numerical solution to the theory at $\mu_B = 0$. From lQCD, we see there is a crossover phase transition between quark and hadronic degrees of freedom \cite{Aoki:2006br}, which is illustrated by the matching of lQCD to the hadron resonance gas (HRG) in Fig.~\ref{fig:Crossover}. For heavy-ion collisions at high energies, QCD at $\mu_B = 0$ is a good description, since the nuclei can be described as highly gluon saturated objects and the resulting collision has a net baryon density of zero.

%__________________________________________________________________________
%
%__________________________________________________________________________
%

\section{Heavy-Ion Collisions}

\begin{figure}[ht]
   \includegraphics[width=\textwidth]{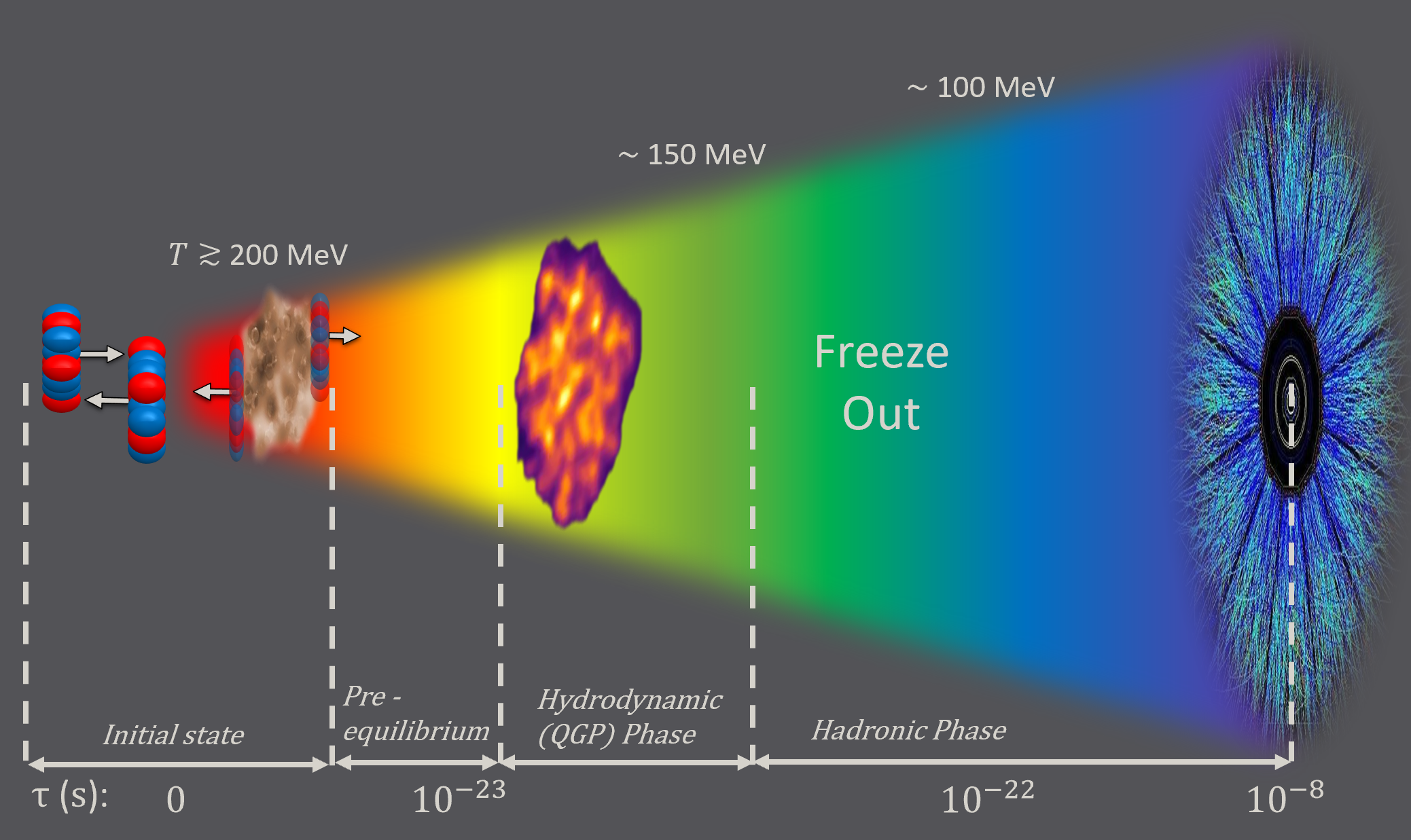}
   \caption{Illustration of the different steps in a nuclear collision from a modeling perspective. General information about approximate temperatures and times are included. }
   \label{fig:StepsOfNuclearCollision}
\end{figure}

Experimentally, we only have access to the final state particles produced by heavy-ion collisions and need to use them to extract properties of the system. This can be quite complicated due to the many stages of a collision, illustrated in Fig.~\ref{fig:StepsOfNuclearCollision}. The individual stages of a nuclear collision contain different dynamics and require the stitching together of simulations that cover each stage. In addition to the complexity introduced at each stage, the connection of them together also poses its own problems. Below is an overview of the different stages of heavy-ion collisions with a focus on the physics involved and the difficulties of switching between them.

%__________________________________________________________________________
%
%__________________________________________________________________________
%

\begin{figure}[ht]
   \includegraphics[width=\textwidth]{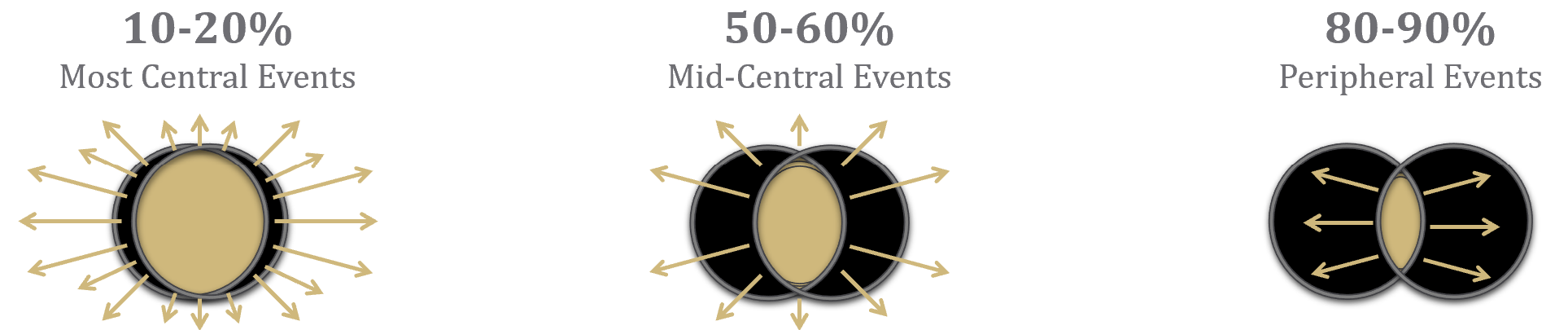}
   \caption{Illustration of nuclear collisions with respect to the centrality of the collision. The average geometry is represented by the overlap and the final multiplicity and momentum anisotropy is characterized by the arrows.}
   \label{fig:CentralityCartoon}
\end{figure}

Simulations of heavy-ion collisions begin with the determination of the initial state energy or entropy density profile that must be calculated from the cross section of the colliding nuclei. Being the farthest stage from experimental observation, it is quite difficult to determine the exact physical mechanisms involved in the initial collision. Luckily, the final state momentum anisotropy of the produced particles is highly dependent on the geometry of the initial state and we can use that to make some inferences about the required ingredients in the initial state. The geometry of the initial state is highly dependent on the impact parameter, or distance between the centers of the nuclei in the collision plane, which produces an ellipse dominated geometry. We cannot measure the impact parameter directly from experiment and so a proxy, called percent centrality, is defined with respect to the final state particle multiplicity. The relationship between percent centrality and the final multiplicity is illustrated in Fig.~\ref{fig:CentralityCartoon}, where the relationship is inversely proportional with the percent centrality clearly labeled and the multiplicity represented by the number of arrows in each collision. This illustration of the colliding nuclei as two overlapping smooth densities is quite simple and was the actual description of the system earlier in the development of the field. While this picture is able to describe the elliptic momentum anisotropy of the final state generally well, it fails at capturing higher order momentum anisotropies measured by experiment. This disagreement necessitated the inclusion of event-by-event fluctuations of the initial state geometry \cite{Osada:2001hw, Andrade:2006yh,Takahashi:2009na}, the largest contribution coming from nucleon fluctuations. This proved instrumental to the description of the collision and directly responsible for triangular flow \cite{Alver:2010gr}. The geometry of the initial state is well constrained and generally understood, except in the most extreme collisions (See Chap.~\ref{chap:ExplorationOfMultiplicityFluctuations} and Chap.~\ref{chap:V2toV3Puzzle}). What is less well constrained, is the characterization of the deposition of energy or entropy density by the collision, which has seen many different approaches that generally agree with experimental data \cite{Parkkila:2020mgo}. 

%__________________________________________________________________________
%
%__________________________________________________________________________
%

The result of the initial state simulation is some profile of energy or entropy density that describes the initial collision, which must then be used to initialize a hydrodynamic simulation that describes the QGP evolution. This connection is the first which introduces some issues. For a system to be described by hydrodynamics, it must have a separation of scales between the system size and interaction length. This correlates with the system being near thermal equilibrium. Unfortunately, the initial state can be quite far from equilibrium and is free to contain large gradients in the energy/entropy density. Plugging these far-from-equilibrium initial conditions directly into hydrodynamic simulations, leads to much of the system violating nonlinear causality constraints \cite{Bemfica:2020xym}. This can be partially alleviated by first using some pre-equilibrium phase to smooth out the gradients. This issue was, recently, quantified by Ref.~\cite{Plumberg:2021bme}, which found that without a pre-equilibrium stage approximately 30\% of the system was acausal (the rest being uncertain as to its status) and this was greatly reduced, but not fully eliminated, by this bridging stage.

The next main stage is the hydrodynamic evolution of the QGP created by the initial state. The system is described by the time evolution of the energy-momentum tensor $T^{\mu\nu}$ and charge currents $J^\mu$ using some equations of motion. The evolution is constrained by energy-momentum and charge conservation
\begin{subequations}
\begin{align}
    \partial_\mu T^{\mu\nu} & = 0\label{idealEOM1}
     \\
    \partial_\mu J^\mu & = 0 \label{idealEOM2} .
\end{align}
\end{subequations}
For a non-ideal system, these conservation equations do not close the system and so equations of motion (EOM) are required for a hydrodynamic formulation. One particular set of EOM comes from the Israel-Stewart description \cite{Israel:1979wp,Denicol:2012cn,Shen:2014vra,Ryu:2015vwa}
\begin{subequations}
    \begin{align}
        \tau_{\Pi} \dot{\Pi}+\Pi &= -\zeta \theta-\delta_{\Pi \Pi} \Pi \theta+\lambda_{\Pi \pi} \pi^{\mu \nu} \sigma_{\mu \nu} 
        \\
        \tau_\pi \dot{\pi}^{\langle\mu \nu\rangle}+\pi^{\mu \nu} &= 2 \eta \sigma^{\mu \nu}-\delta_{\pi \pi} \pi^{\mu \nu} \theta+\varphi_7 \pi_\alpha^{\langle\mu} \pi^{\nu\rangle \alpha} 
        \\
        -\tau_{\pi \pi} \pi_\alpha^{\langle\mu} \sigma^{\nu\rangle \alpha}+\lambda_{\pi \Pi} \Pi \sigma^{\mu \nu}
    \end{align} ,
    \label{eq:code}
\end{subequations}
where $\zeta$ is bulk viscosity, $\kappa$ is the diffusion coefficient, and $\eta$ is shear viscosity. The Israel-Stewart formulation is a second-order hydrodynamic theory and was the first to solve issues of acausality and instability present in first-order formulations. Alternative methods of solving hydrodynamic systems have been used, such as the Bemfica-Disconzi-Noronha-Kovtun (BDNK) formulation \cite{Bemfica:2017wps,Kovtun:2019hdm,Hoult:2020eho,Bemfica:2019knx,Bemfica:2020zjp} which is a causal first-order scheme.

The viscosity of the QGP can be extracted from simulations of event-by-event viscous relativistic hydrodynamics using realistic equations of state which match experimental measurements from LHC and RHIC \cite{Gardim:2012yp, Gardim:2012im, Heinz:2013bua, Shen:2015qta, Niemi:2015qia, Niemi:2015voa, Noronha-Hostler:2015uye, Gardim:2016nrr, Gardim:2017ruc, Eskola:2017bup, Giacalone:2017dud, Gale:2012rq, Bernhard:2016tnd, Giacalone:2016afq, Zhu:2016puf, Zhao:2017yhj, Zhao:2017rgg, Alba:2017hhe, Noronha-Hostler:2019ytn, Sievert:2019zjr, Bozek:2011if, Bozek:2012gr, Bozek:2013ska, Bozek:2013uha, Kozlov:2014fqa, Zhou:2015iba, Mantysaari:2017cni, Weller:2017tsr, Pratt:2015zsa, Moreland:2015dvc, Auvinen:2018uej}. The field has converged on a minimum in the shear viscosity to entropy density ratio of $\eta/s\approx 0.1\pm 0.1$ with a maximum in the bulk viscosity to entropy density ratio at $\zeta/s\approx 0.15\pm0.15$. This makes the QGP a near perfect fluid with the lowest viscosity ever measured. 

The hydrodynamic evolution is continued until such time as the temperature of the fluid has reached the freeze-out temperature, where a description of the system in quark and gluon degrees of freedom is no longer applicable. This freeze-out temperature is often assumed to be a value around 155 MeV, although recent work has supported the use of two freeze-out temperatures \cite{Bellwied:2018tkc,Bluhm:2018aei,Noronha-Hostler:2016rpd}. 

With the frozen out hydrodynamic system, a hadronization process must be executed, since the degrees of freedom appropriate at this temperature are hadronic. The hadronization of the fluid is another major switching of models which is required to describe the change in physics. This process is highly dependent on the particle list that is used, since more particles means more opportunities for direct matching. With advancements in experimental design and calculation techniques, there is an ever growing number of resonances and particles, which are tracked by the PDG. The system is now described by a hadron resonance gas and is simulated using some hadronic transport code, such as SMASH \cite{SMASH:2016zqf}. This HRG phase continues until the system reaches kinetic freeze-out, at which point theory and experiment can be directly compared \cite{Petersen:2018jag, Bass:1998ca, Bleicher:1999xi}.

This multi-step method of simulating heavy-ion collisions has been quite successful in describing experimental measurements, with a majority of collaborations being able to fit and predict standard observables, such as charged particle spectra and two-particle azimuthal anisotropies \cite{Song:2010mg, Bozek:2012qs, Gardim:2012yp, Bozek:2013uha, Niemi:2015qia, Ryu:2015vwa, McDonald:2016vlt, Bernhard:2016tnd, Gardim:2016nrr, Alba:2017hhe, Giacalone:2017dud, Eskola:2017bup, Weller:2017tsr, Schenke:2019ruo}. This has led to general consensus that event-by-event relativistic viscous hydrodynamics are successful in describing large AA collisions.

\section{Signatures of the QGP}

If the above description of heavy-ion collisions is correct, we must be able to see signatures of the QGP in experimental observables. There are, indeed, many of these signals with the most important being collective flow, jet quenching, and strangeness enhancement, which are discussed in detail below.

\subsection{Collective Flow}

\begin{figure}[ht]
    \centering
   \includegraphics[width=0.6\textwidth]{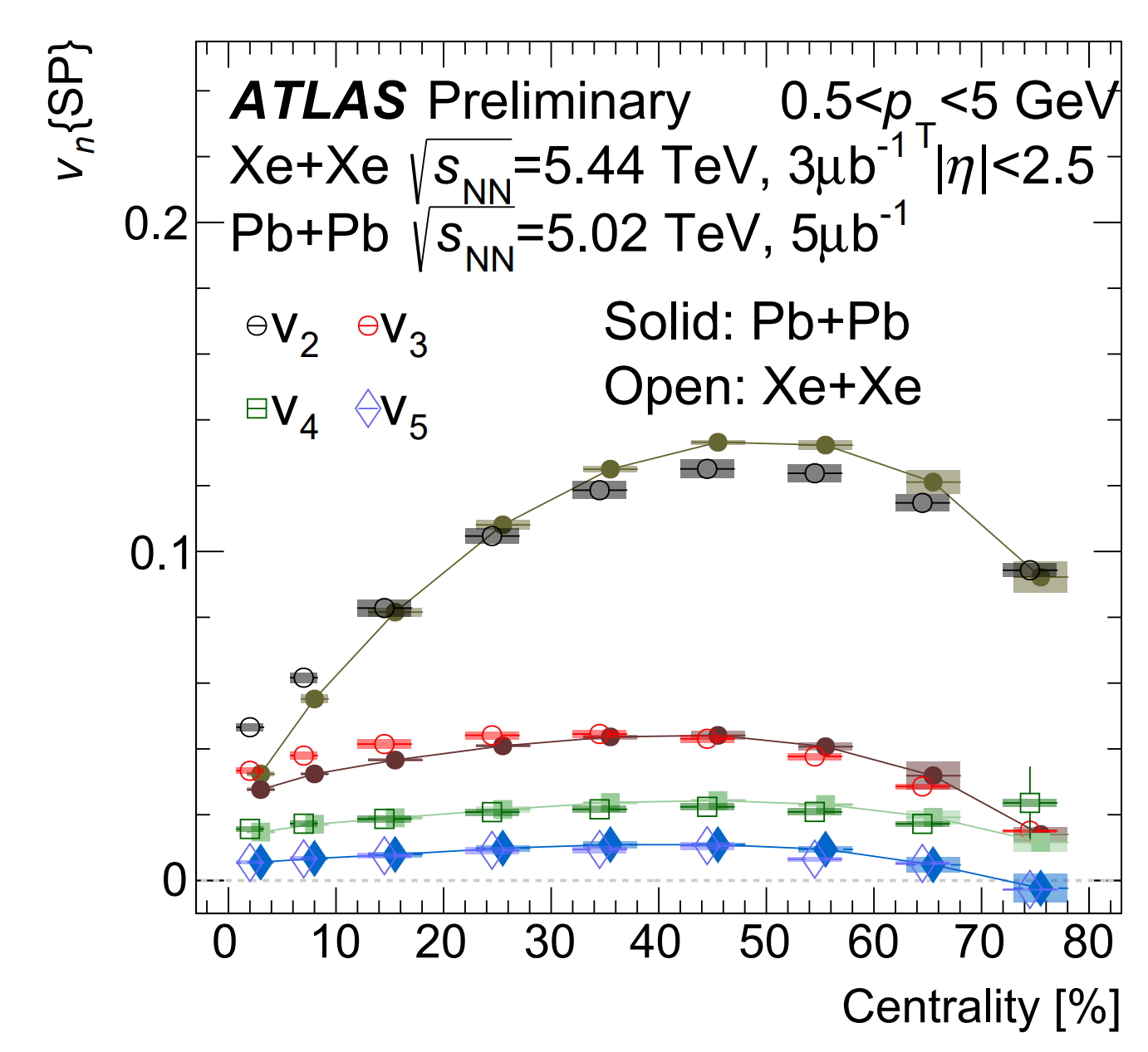}
   \caption{Measurement of integrated $v_n \{SP\}$ for Pb+Pb and Xe+Xe collisions. Figure from Ref.~\cite{ATLAS:2018iom}.}
   \label{fig:VnExample}
\end{figure}

If nuclear collisions create a QGP medium, then there must be a signature of collective flow that rules out the description of the system as solely that of a gaseous medium. This collective flow is observable from the momentum anisotropy of the final state particles, which is calculated for a single event, using the distribution of measured particles with respect to their transverse momentum ($p_T$), rapidity ($\eta$), and relative angle ($\phi$), and described by the complex vector 
\begin{equation}
    v_n e^{in\psi_n} (p_T,\eta) = \frac{\int^{2\pi}_0 d\phi \frac{dN}{d\phi dp_T d\eta} e^{in\phi}}
    {\int^{2\pi}_0 d\phi \frac{dN}{d\phi dp_T d\eta}} .
\end{equation}
Momentum anisotropy vectors of different orders represent flow of different order, with $v_2$ and $v_3$ being the most important and corresponding to elliptic and triangular flow, respectively. While smooth initial conditions are capable of describing the elliptic flow, triangular flow is driven by fluctuations in the initial state geometry and necessitates the description of quantum fluctuations in heavy-ion simulations. Since these fluctuations are inherent to the system, we use event-averaged flow using multiparticle cumulants:
\begin{equation}
    v_n\{2\} = \sqrt{\langle v_n^2 \rangle},
\end{equation}
where the RMS of the momentum anisotropy vector magnitude is the two-particle cumulant. A recent experimental measurement of $v_n \{SP\}$ (equivalent to $v_n \{2\}$) for Pb+Pb and Xe+Xe collisions with respect to centrality is presented in Fig.~\ref{fig:VnExample}.

%__________________________________________________________________________
%
%__________________________________________________________________________
%

Due to the low viscosity of the QGP, final state observables are highly sensitive to the geometry of the initial state, making it extremely important to have a correct description of the initial condition when reproducing the measured anisotropic flow and its multiparticle cumulants \cite{Renk:2014jja, Giacalone:2017uqx}. We can quantify the extent to which the final state flow depends on the initial state by looking at the deterministic response of viscous hydrodynamics to the initial geometry of the collision \cite{Noronha-Hostler:2015uye, Niemi:2015voa, Adam:2015ptt,Song:2010mg, Bozek:2012qs, Gardim:2012yp, Bozek:2013uha, Niemi:2015qia, Ryu:2015vwa, McDonald:2016vlt, Bernhard:2016tnd, Gardim:2016nrr, Alba:2017hhe, Giacalone:2017dud, Eskola:2017bup, Weller:2017tsr, Schenke:2019ruo}. The initial geometry of an individual event can be described by a Fourier series, where the coefficients are the eccentricity weights, $\varepsilon_{n} = | \langle r^{n} e^{i n \phi} \rangle |$, while the flow harmonics come from a Fourier series of the momentum anisotropy, $v_n$. The relationship has been found to be well described by a quasi-linear response \cite{Gardim:2011xv, Gardim:2014tya}, $v_n = \kappa_n \varepsilon_n$, where the response coefficient $\kappa_n$ is sensitive to the transport parameters of the hydrodynamics. As such, different initial conditions, when matched to the same experimental flow data, will result in different extractions of the viscosity and other transport parameters \cite{Luzum:2009sb,Bernhard:2015hxa,Heinz:2011kt}. The precision of measurements is such that differences in collective flow can be predicted between beam energies with in a few percent \cite{Noronha-Hostler:2015uye, Niemi:2015voa, Adam:2015ptt}. This high level of precision means that when we see disagreement between experimental flow data and theoretical calculations, it is likely that there is missing physics that is beyond our current model.

\subsection{Jet Quenching}

\begin{figure}[ht]
    \centering
   \includegraphics[width=0.6\textwidth]{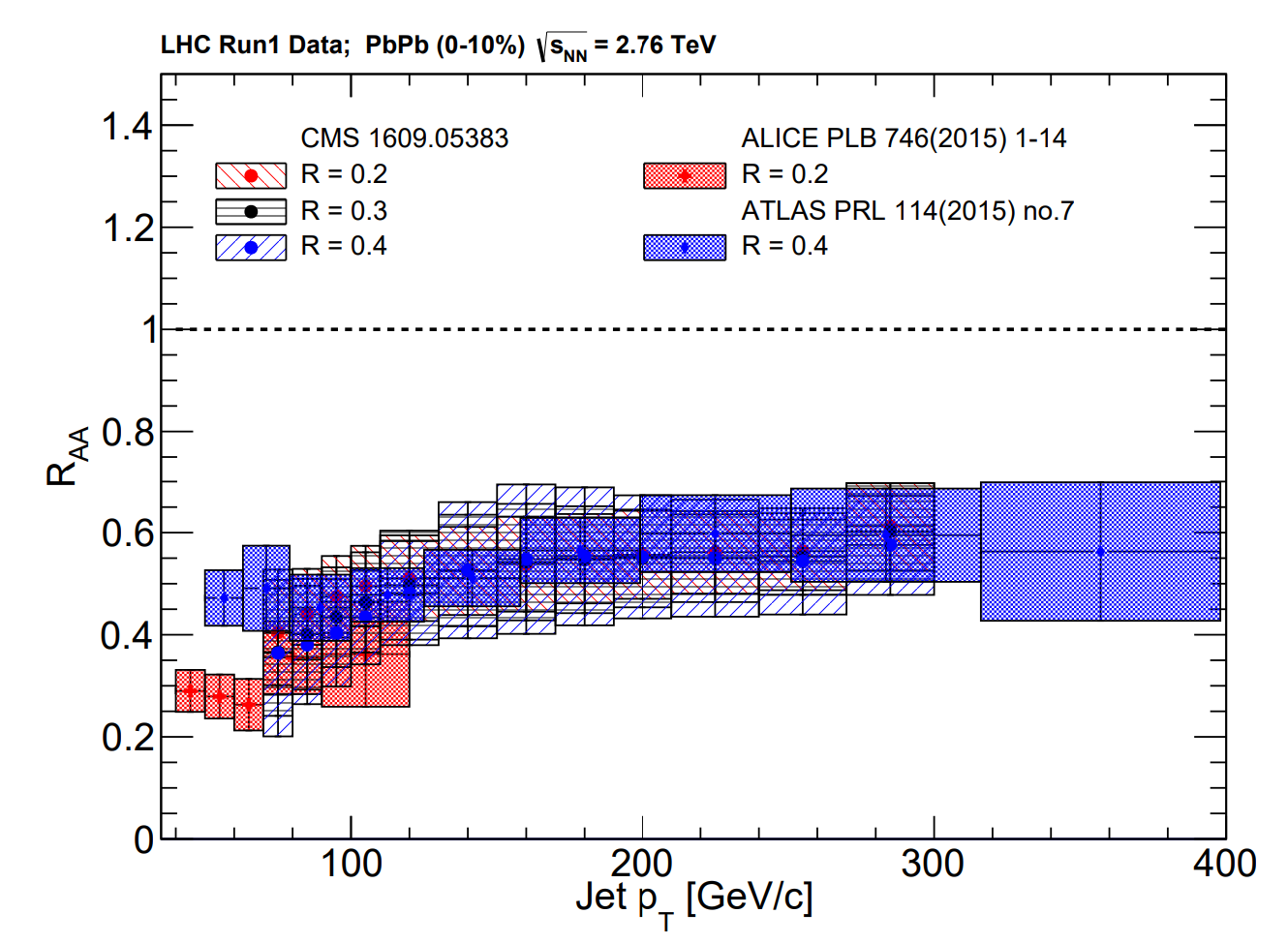}
   \caption{Comparison of $R_{AA}$ for several different jet radii, $R = 0.2,0.3,0.4$, across several experiments, illustrating that there is a consistent measurement of jet quenching in AA collisions. Figure from Ref.~\cite{Connors:2017ptx}.}
   \label{fig:JetQuenching}
\end{figure}

In nuclear collisions, not all processes are "soft" and lead to fluid-like behavior. Hard processes can occur that produce high momentum particles, called jets, that should interact with the hydrodynamic background as they traverse it. The jets can have two types of interaction with the fluid, generally, with one being deflection and the other energy loss, or quenching. We can measure jet production in proton-proton collisions, which are small enough that any jets should not have time to interact with any QGP medium that is created, if it is created. By comparing the jet production in nucleus-nucleus (AA) collisions to that seen in pp collisions, we can determine if jets are quenched when they have the opportunity to traverse the QGP medium. A useful way of quantifying this quenching is through the nuclear modification factor \cite{Connors:2017ptx} observable,
\begin{equation}
    R_{AA}(p_T) = \frac{1}{\langle T_{AA} \rangle} \frac{\frac{1}{N_{ev}} \frac{dN_{jet}}{dp_T}|_{AA}}{\frac{d\sigma_{jet}}{dp_T}|_{pp}},
\end{equation}
which is the ratio of jet production in AA collisions with respect to pp, and where $\langle T_{AA} \rangle$ is the nuclear thickness function and $\sigma_{pp}$ is the differential jet cross section for pp interactions. Experimental measurements of $R_{AA}$ show that this quenching is present in AA collisions, as illustrated by Fig.~\ref{fig:JetQuenching}, which compares data from several different experiments. 

%__________________________________________________________________________
%
%__________________________________________________________________________
%

\subsection{Strangeness Enhancement}

\begin{figure}[ht]
    \centering
   \includegraphics[width=0.5\textwidth]{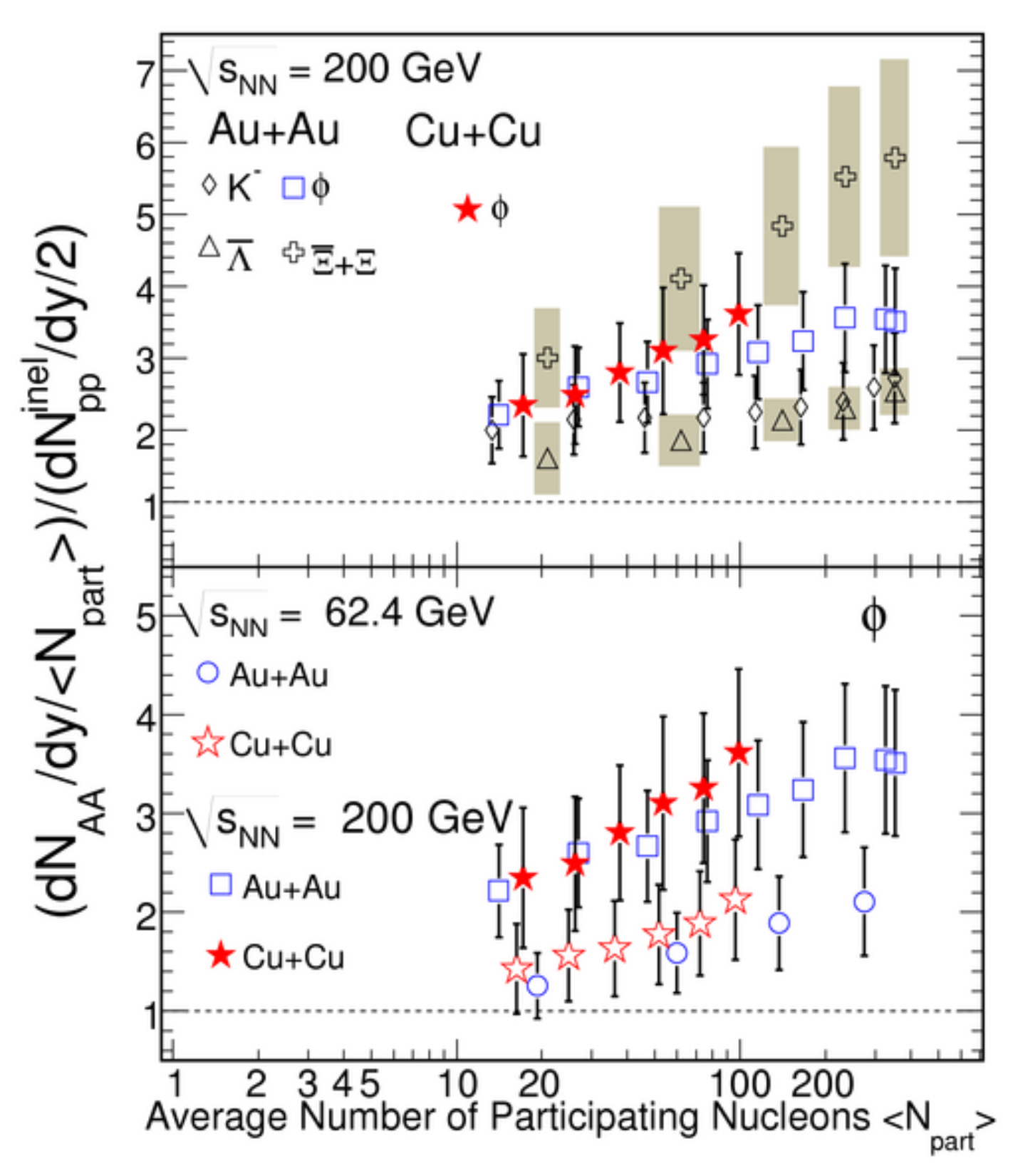}
   \caption{Experimental evidence of strangeness enhancement in Au+Au and Cu+Cu collisions at two different collision energies. Figure from Ref.~\cite{STAR:2008inc}.}
   \label{fig:StrangeEnahancement}
\end{figure}

In experimental measurements of heavy-ion collisions, we see a significant presence of strange hadrons, indicating production of $s\bars$ quark pairs in the system. In the QGP, a thermal source of these strange particles is through gluon-gluon interactions \cite{Rafelski:1982pu}. If there is a QGP medium, we should see an enhancement of strangeness in larger systems as compared to smaller systems \cite{STAR:2008inc}, where, if the QGP is even formed, strangeness production from thermal processes would be greatly reduced. Experimental observation of this strangeness enhancement is accomplished by taking the ratio of strange hadron multiplicity in AA collisions to pp collisions as shown in Fig.~\ref{fig:StrangeEnahancement}.

%__________________________________________________________________________
%
%__________________________________________________________________________
%

\section{Extension to Finite Chemical Potential}

\begin{figure}[ht]
\begin{centering}
   \includegraphics[width=0.8\textwidth]{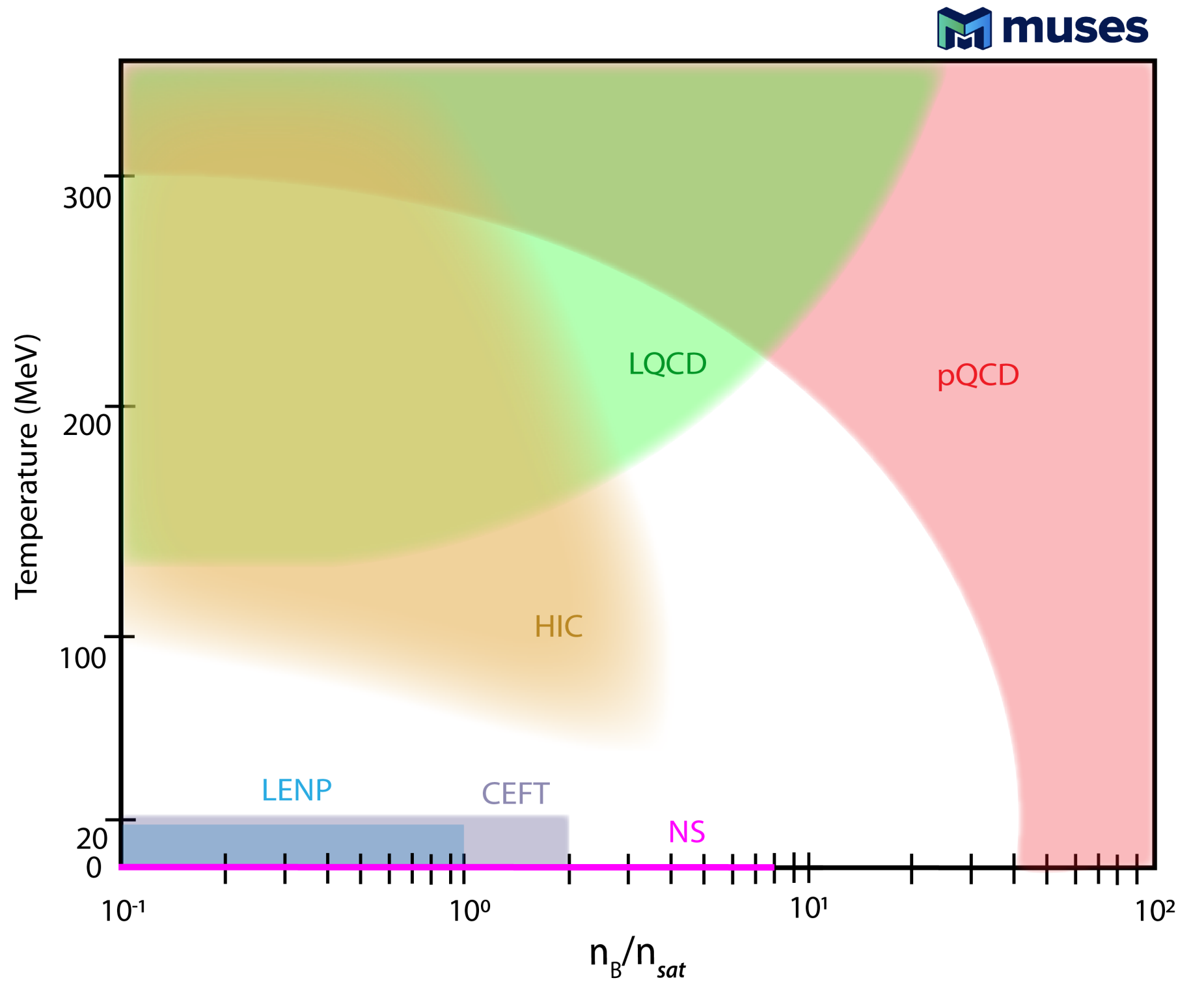}
   \caption{Illustration of the QCD phase diagram and regions described by different constraints and experiments, the important ones for this work being lattice QCD (LQCD) and heavy-ion collisions (HIC). Taken from Ref.~\cite{MUSES:2023hyz} }
   \label{fig:PhaseDiagram}
\end{centering}
\end{figure}

Theoretical calculations of the QCD equation of state are confined to a limited region of the phase diagram due to the complexity of the system. Although lQCD calculations of the QCD equation of state are numerically possible when the baryon chemical potential is zero, $\mu_B=0$, one can extend this out in a temperature dependent chemical potential by expanding $\mu_B$ around zero \cite{Borsanyi:2021sxv,Borsanyi:2022qlh}. This expansion is limited, though, due to the sign problem \cite{Dexheimer:2020zzs,Troyer:2004ge}, where the exponential of the QCD action becomes complex and creates an issue with a part of the calculation, called importance sampling. In Fig.~\ref{fig:PhaseDiagram}, an illustration of the QCD phase diagram shows the area described by lQCD and its overlap with the space probed by heavy-ion collisions, which is not fully covered by the lQCD equation of state. Work is being done to extend the description of the phase diagram, by stitching together equations of state in different regimes, and provide better constraints \cite{MUSES:2023hyz}.  

%__________________________________________________________________________
%
%__________________________________________________________________________
%

With the establishment of a crossover phase transition along the temperature axis by lQCD, the existence and location of a first-order phase transition and corresponding critical point in the QCD phase diagram becomes a topic of interest \cite{Athanasiou:2010kw, Athanasiou:2010vi}. The search for the QCD critical point has been a focus of the field, with the greatest effort coming from the Beam Energy Scan (BES) \cite{Lovato:2022vgq, Ratti:2020mzq, Achenbach:2023pba, Almaalol:2022xwv} that ran collisions at progressively lower energies to scan the phase diagram. 

Heavy-ion collisions can be described by a path through the QCD phase diagram, in terms of temperature and chemical potential. To a first order approximation, the QGP is assumed to be dominated by gluons and thus can be described by an energy density with net zero chemical potential. The only net source of charge comes from nucleons that are 'stopped' by the collision and trapped in the fluid. This process, called baryon stopping, occurs as the collision energy is decreased since the nuclei become less Lorentz contracted and are given time to interact and get caught in the QGP. As such, high energy collisions follow a path close to the temperature axis and as one goes to lower collision energies the path moves out in the baryon chemical potential axis. Initialization of the system can be represented by a point in the QGP regime corresponding to the average temperature and chemical potential, although a full description would consist of a distribution of initial points from each fluid cell. The properties of the system determine its path through the phase space, which is described by hydrodynamics. Eventually, the QGP undergoes a phase transition and freezes out into hadrons that are measured by detectors.

While the approximation of the initial collision at high energies, as one between two sheets of gluons, is consistent with a system with net-zero charge, there is the possibility of local fluctuations of charge in the initial state. These local fluctuations of charge would mean that a single collision sees a slice of the phase diagram, rather than a single path. Including charge fluctuations in heavy-ion simulations requires that the charge density is initialized in the initial condition. Additionally, a hydrodynamic code that incorporates conserved charge currents into the evolution of the medium is essential.

\section{Overview} \label{sec:Overview}

This thesis follows two related threads: the development of fluctuations in the initial state of heavy-ion collisions and the introduction of conserved charge fluctuations. To provide a common basis upon which the initial state is to be understood by the reader, Chap.~\ref{chap:InitialStatePhysics} details the historical development of initial state physics with a focus on sources of fluctuations and Chap.~\ref{Chap:Observables} establishes the relevant initial state observables while providing connection to final state experimental measurements and elucidating important observations about initial geometry. Development of fluctuations starts in Chap.~\ref{chap:ExplorationOfMultiplicityFluctuations} with an exploration of different choices for multiplicity fluctuations. Coupled with choices of multiplicity fluctuations, differences in the determination of the reduced thickness function are also investigated. In Chap.~\ref{chap:V2toV3Puzzle}, fluctuations arising from nuclear structure are applied to PbPb collisions in an attempt to solve the $v_2$-to-$v_3$ puzzle. 

Continuing the development of fluctuations but also introducing a new and novel source through conserved charges, Chap.~\ref{chap:ICCINGAlgorithm} details the importance of such fluctuations and their implementation through the \code{iccing} (Initial Conserved Charges In Nuclear Geometry) algorithm. An analysis of the model is performed in Chap.~\ref{chap:ICCINGResults}, which characterizes the new source of initial state geometry and explores the sensitivity of the model to its parameters. Finally, an extension of \code{iccing} is detailed through the application of pre-equilibrium dynamics of conserved charge perturbations in Chap.~\ref{chap:PreEquilibriumEvolution}. 

A summary of the \code{iccing} model and its various aspects is presented in Chap.~\ref{chap:Conclusion}, with special attention given to future directions of development and application of both conserved charges in the initial state and of other novel sources of fluctuations.

%%%%%%%%%%%%%%%%%%%%%%%%%%%%%%%%%%%%%%%%%%%%%%%%%%%%%%%%%%%%%%%%%%%%%%%%%%%
%
\chapter{Observables} \label{Chap:Observables}
%
%%%%%%%%%%%%%%%%%%%%%%%%%%%%%%%%%%%%%%%%%%%%%%%%%%%%%%%%%%%%%%%%%%%%%%%%%%%

\epigraph{A sharp knife is nothing without a sharp eye.}{Klingon Proverb}

It is useful to provide a common basis upon which to understand the analyses done in later chapters. In this chapter, I will start with a framing of relevant properties of nuclear collisions and then provide useful steps and observables for characterizing the initial state. Throughout, connections are made to experimental measurements and limitations that inform our description of the initial state.

To begin, we define the frame of reference to be the center of mass frame, since this greatly simplifies calculations thanks to symmetries of the system. For heavy-ion collisions which use two beams of nuclei accelerated to relativistic speeds, this definition simplifies the center of mass energy since, in this frame, the total 3-momentum is zero and is proportional to the energy of the nuclei:
\begin{equation}
    \sqrt{s} = 2 E .
\end{equation}
This relation between $\sqrt{s}$ and $E$ is a bit more complicated in fixed target collisions. The center of mass energy is often reported per nucleon and denoted as $\sqrt{s_{NN}}$. 

Now in the center of mass frame, one can use Cartesian coordinates, with which to measure distance and momentum, and orient the z-axis along the beam direction. However, using hyperbolic coordinates greatly simplifies calculations and the interpretation of physical processes. A consequence of special relativity, that supports this choice of hyperbolic coordinates, is that the nuclei undergo strong Lorentz contraction along the direction of motion. An intuition of this aspect can be obtained by calculating an approximation of the radius of nuclei in the beam direction. For top energies, the thickness of the colliding nuclei is approximately $R_{beam} = 0.004 - 0.1 fm$, while the radius in the x-y plane is on $O(1 fm)$. This disparity in lengths between the x-y plane and the beam axis focuses our interest on the x-y, or transverse, plane, where the structure will be easiest to characterize. In the transverse plane, we express quantities as a radial magnitude with some azimuthal angle.

\begin{figure}[ht]
    \centering
    \includegraphics[width=0.48\textwidth]{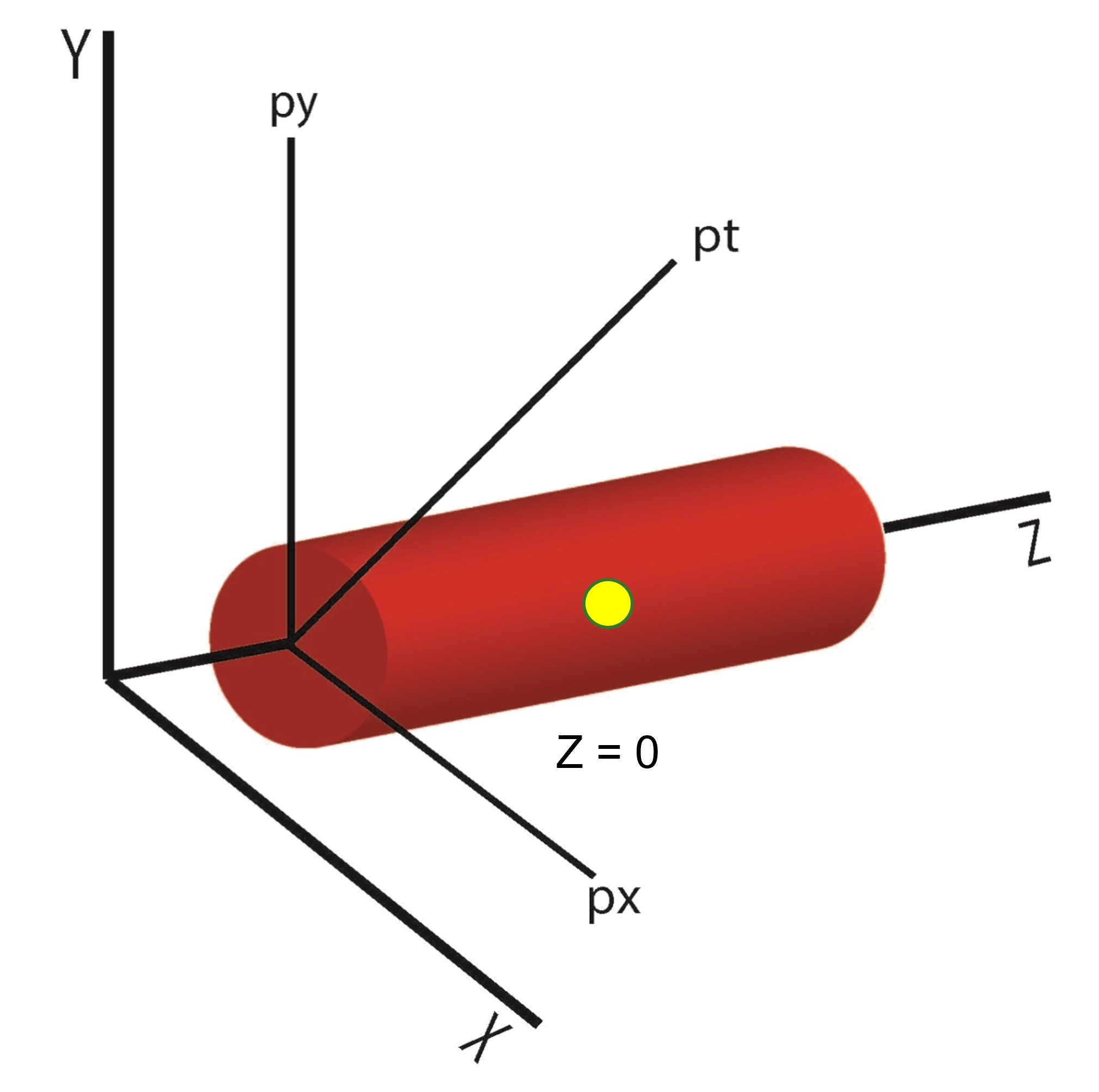}
    \includegraphics[width=0.48\textwidth]{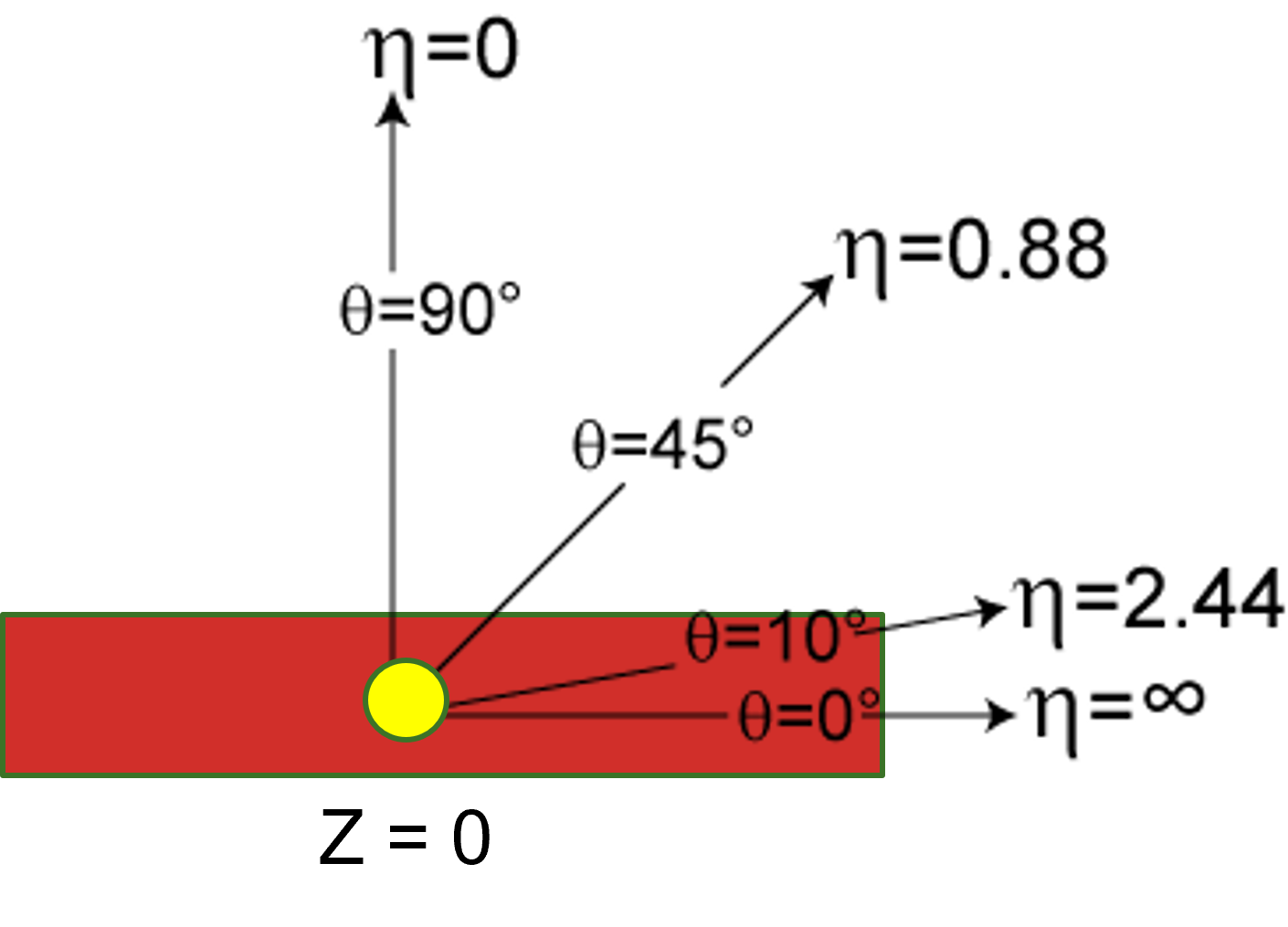}
    \caption{Cartoon of the coordinate system of heavy-ion collisions, where the beam direction is along the z-axis and the collision system is represented by the red tube with center of mass at the yellow point. On the left, is the Cartesian view and, on the right, is an illustration of pseudo-rapidity.}
    \label{fig:coordinatesystem}
\end{figure}

A consequence of the relativistic speeds of the colliding nuclei, is that velocity is not additive along the beam, or longitudinal, direction. By switching to a relativistic velocity in the longitudinal direction, we can restore the additive property. This replacement is done using rapidity coordinates defined as:
\begin{equation}
    cosh(y) = \gamma = \frac{1}{\sqrt{1-v_{z}^2}},
\end{equation}
where $\gamma$ is the conventional $\gamma$ factor from special relativity. What is actually measured by experiment is the pseudo-rapidity,
\begin{equation}
    \eta = \frac{1}{2} ln \left( \frac{|\vec{p}| + p_z}{|\vec{p}| + p_z} \right),
\end{equation}
where $p_z$ is the longitudinal momentum. The reason for using $\eta$ over $y$ is due to the fact that the momentum and not the velocity is measured by experimental detectors. For massless particles, these two coordinates are equal, $y = \eta$, but for massive particles their definitions diverge. At top LHC energies, the maximum range of rapidity is $|y| < 9$, with pseudo-rapidity bounded as $|\eta| < \infty$. These two definitions of longitudinal properties create a lot of confusion in the field, since the term 'rapidity' is often used, interchangeably, to refer to both $y$ and $\eta$. An illustration of the coordinate system utilized by heavy-ion collisions is presented in Fig.~\ref{fig:coordinatesystem}.

%__________________________________________________________________________
%
%__________________________________________________________________________
%

An important approximation, used in heavy-ion collisions, is that the system is Bjorken boost-invariant, which arises from the extreme Lorentz contraction of the colliding nuclei and the subsequent approximation of them as 2-D objects.  When boost-invariance holds, it implies that dynamics perpendicular to the beam direction, in the transverse plane, are independent of those along the beam direction, which can be described by identical transverse slices. The regime where this approximation is most applicable is at mid-rapidity, defined as $|\eta| < 1$ \footnote{Notice the sly shift of referring to pseudo-rapidity as rapidity due to the simplification it creates in the definition of mid-rapidity. It is important to keep in mind the distinctions between the two since this shift is done quite often in the literature.}, and starts to break down at large pseudo-rapidity (close to the beam axis), small systems, and low beam energies (where the longitudinal thickness of the nuclei is no longer trivial). As such, most emphasis is put on dynamics at mid-rapidity in the transverse plane, since this allows simplified simulations in 2 spatial dimensions rather than 3. Having developed a good understanding of physics at mid-rapidity, inclusion of longitudinal dynamics in simulations has seen much progress recently.

There are many properties of the collision that contribute to the initial state geometry with different weights. It is useful to separate out the largest contributor to the geometry, which is the "head-on-ness" of the collision, so that smaller signals may be discerned. The initial spatial and final momentum anisotropies are closely correlated to how head-on, or central, a collision is. Assuming collisions of spherical nuclei, a central collision will have a circular shape, producing many particles, and as one goes to more peripheral collisions the dominant geometry becomes more elliptic, producing fewer particles. 

This chapter serves as a central source for all the observables used throughout this work and is organized as follows. In Sec.~\ref{Sec:Centrality}, a description is made of centrality and the various ways it is determined and calculated. Next, the initial state eccentricities and final state flow harmonics are defined, in Sec.~\ref{sec:Eccentricities}, in the standard way. Sec.~\ref{sec:Eccentricities} also contains discussion of response theory, in Sec.~\ref{subsec:ResponseTheory}, and a description of sources and features of triangular structure, in Sec.~\ref{subsuc:Triangularity}. This is followed by Sec.~\ref{sec:EccentricityCalculation}, which is a description of the practical application of the initial state eccentricity calculation that is used in \code{iccing}. A description of the event averaged cumulants is presented in Sec.~\ref{sec:Cumulants}. Finally, some important comments, on the comparison of theoretical observables to experimental data and the plethora of definitions of the initial state spatial anisotropy, are given in Sec.~\ref{sec:ImportantComments}.

%---------------------------------------------------------------------------
%
\section{Centrality} \label{Sec:Centrality}
%
%---------------------------------------------------------------------------

The true characterization of the "head-on-ness" of nuclear collisions is by means of the impact parameter $b$, the distance between the centers of the colliding nuclei in the transverse plane: $\vec{b} = \vec{x}_A - \vec{x}_B$, where $\vec{x}_{A,B}$ are the locations of the centers of mass of the two colliding nuclei. Unfortunately, the impact parameter $b$ of the collision is inaccessible to experiment and so a proxy for the impact parameter must be used. Since the impact parameter describes how overlapped the two nuclei are, the proxy needs to reflect that relationship. Several properties of the collision that should be related to impact parameter are: number of nucleons that participated in the collision ($N_{part}$), number of nucleons that are spectators ($N_{spec}$), number of binary collisions ($N_{coll}$), number of charged particles produced ($N_{ch}$), and energy deposited in Zero-Degree Calorimeters ($E_{ZDC}$). Only the last two, $N_{ch}$ and $E_{ZDC}$, are directly accessible to the experiment, while the others must be determined through comparison to theoretical simulations. The measurements of centrality that are not directly from experiment are still important since they can be used to describe or classify different phenomena.

The average multiplicity of charged particles $N_{ch}$ should be directly correlated with impact parameter, since one can expect that a central event will produce many particles while a peripheral event will produce very few. The relationship is thus assumed to be monotonically decreasing in $N_{ch}$ with increasing $b$. To confirm this assumption, one must compare the final state distribution of events with respect to $N_{ch}$ against theoretical models that have been tuned to match that distribution in some way. This is where one of the core assumptions of the initial state comes into play, namely that the initial state density is proportional to the final state multiplicity. In the particularization of the hydrodynamic system at its freeze-out, the entropy must be conserved, or equivalently, their should be a one-to-one correspondence to the amount of energy at the end of hydro and the number of particles frozen out. Since the viscosity of the QGP is very low, we can assume that there is near conservation of entropy throughout the evolution. This logic justifies the comparison of the initial state total number density to the final state average multiplicity. 

Now comparing the distributions of total initial density and final multiplicity, the shape of the distribution matches well between experiment and theory for a given system, but the magnitudes are different. This indicates that the correct geometry is being simulated in theory, but the initial density is not normalized correctly. To fix this, a proportionality factor between the theoretical initial density and the experimental final multiplicity is extracted for an event in a given class (often a central collision),
\begin{align} \label{e:EntropyScaling}
s_{tot}^{FS} &= a_{entropy} \, \rho_{tot}^{IS} ,
\end{align}
where $s_{tot}^{FS}$ is the final state entropy, $\rho_{tot}^{IS}$ is the initial state total density, and $a_{entropy}$ is the proportionality factor between them. An estimation of this factor is provided in Sec.~\ref{subsec:EstimationOfA}. Now scaling the distribution of total initial density by this factor, the theoretical and experimental distributions match well. 

An important note here, is that the standard choice for theoretical model is a Monte-Carlo initial state code, which samples the positions of nucleons for each nucleus randomly. The fact that a particular choice of Monte-Carlo initial state fits the shape of the final state multiplicity, confirms that the global geometry is compatible with reality. The ability to fit the magnitude of the final state distribution is more dependent on the particular physics implemented in the given initial state model.

From this matching of the distribution of events with respect to $N_{ch}$ and total initial density, we can extract mean values of $b$, $N_{part}$, $N_{spec}$, and $N_{coll}$ by means of a mapping procedure through the definition of percent centrality classes. Centrality classes are defined to be fractions of the total integral of the distribution of events with respect to $N_{ch}$ from large $N_{ch}$ to small. The boundaries of these fractions are represented by $n_{\%}$. Small percent centrality correlates to events with largest $N_{ch}$. An example of the determination of the 10-20\% centrality class is shown here:
\begin{equation}
    \frac{\int_{\infty}^{n_{10}} \frac{d\sigma}{dN_{ch}}dN_{ch}}{\int_{\infty}^{0} \frac{d\sigma}{dN_{ch}}dN_{ch}} = 0.1
    \, \mathrm{and} \,
    \frac{\int_{\infty}^{n_{20}} \frac{d\sigma}{dN_{ch}}dN_{ch}}{\int_{\infty}^{0} \frac{d\sigma}{dN_{ch}}dN_{ch}} = 0.2
\end{equation}
This method is quite robust, allowing the comparison of geometry across different system sizes. Centrality classes are commonly defined in 10\%, 5\%, and 1\% increments. Sub-percent centrality classes have also been used in ultra-central and ultra-peripheral collisions to further differentiate small structural differences that are observable in those extremes. As one goes to smaller centrality bins, the number of events per bin must increase in order to keep the error bars constrained. 

Centrality driven geometry is highly elliptical, such that, for near-spherical nuclei, a central collision will be very circular in the transverse plane and mid-central (50-60\% centrality) will be the most elliptical. Plotting observables with respect to centrality provides an easy way to investigate their dependence on the isotropy of the system. An extremely useful case is in ultra-central collisions where sub-percent centrality bins make one sensitive to sources of nuclear deformation and fluctuations in particle production \cite{Noronha-Hostler:2019ytn}, since the event is very isotropic and small deviations become observable. 

\begin{figure} [ht]
    \centering
	\includegraphics[width=0.65\textwidth]{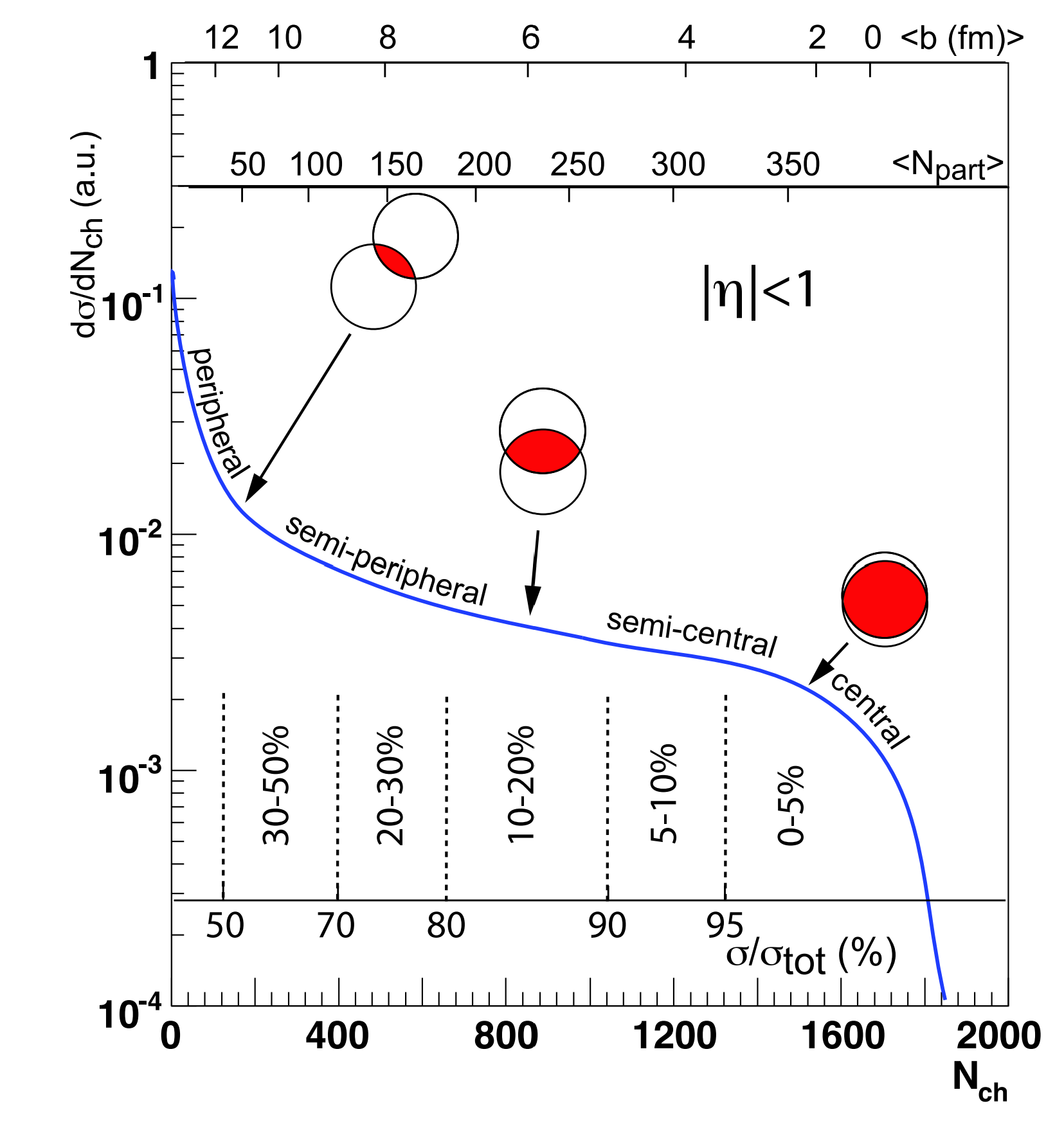}
	\caption{Illustration of the distribution of events with respect to impact parameter and proxies of impact parameter. The experimental axis is $N_{ch}$ and the relation to $\sigma\/\sigma_{tot}$, $\langle N_{part} \rangle$, and $\langle b \rangle$ coming from theoretical models. A general description of centrality classes is included with the dashed lines and illustrated by the cartoons of overlapping circles. Figure taken from \cite{Miller:2007ri}.}
	\label{fig:CentralityComparison}
\end{figure}

Also defining centrality classes with respect to the initial total density, now produces a mapping between final and initial state properties. From this mapping, we can extract mean values of $b$, $N_{part}$, $N_{spec}$, and $N_{coll}$ for each centrality class. Experimental observables are often reported in terms of these extracted properties and it is important to remember that these come from Monte-Carlo simulations. An illustration of the mapping between $N_{ch}$, $\langle N_{part} \rangle$, and $\langle b \rangle$ using centrality classes from the $d\sigma/dN_{ch}$ distribution is provided in Fig.~\ref{fig:CentralityComparison}. 

%__________________________________________________________________________
%
%
%__________________________________________________________________________

We can obtain $N_{spec}$ from experiment through the energy deposited in the ZDC, although there are severe limitations. In a collision, spectator nucleons see some deflection, pushing them away from the beam axis. The Zero-Degree Calorimeters are placed at far forward and backward rapidities in order to measure these deflected remnants. Spectating nucleons can form nuclear fragments with similar magnetic properties to the beam nuclei and thus escape detection by the ZDC. This happens more often in peripheral collisions, making $E_{ZDC}$ non-monotonically related to impact parameter. However, there is a monotonic relation between $E_{ZDC}$ and $b$ at small impact parameter. Thus, centrality determination can be made directly from $E_{ZDC}$ with supplementation from $N_{ch}$ in more peripheral events. Very fine resolution of ultra-central bins is possible using the ZDC \cite{STAR:2015mki}.

\begin{figure} [ht]
    \centering
	\includegraphics[width=0.65\textwidth]{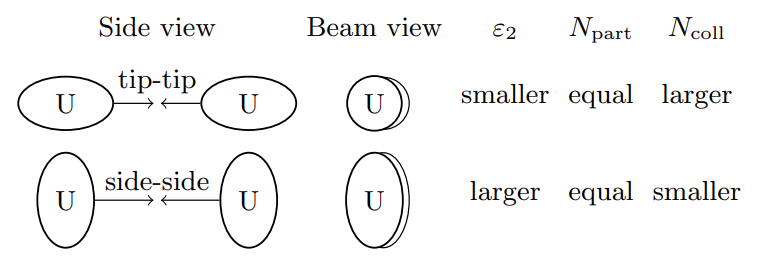}
	\caption{The different orientations of collisions of U+U with the relative elliptic geometry, $N_{part}$, and $N_{coll}$ provided. Figure taken from \cite{Moreland:2014oya}.
	}
	\label{fig:UUSchematic}
\end{figure}

Using extracted mean values from the initial state as proxies for centrality can provide useful information that is otherwise inaccessible. An important example of this was in the description of collisions of highly deformed nuclei, specifically $^{238}U$. This uranium isotope has a prolate shape that translates to a very elliptic profile, when the major axis is parallel to the transverse plane, and a very circular profile, when oriented along the beam axis. These two orientations lead to different types of collisions, tip-to-tip and side-to-side, that are illustrated in Fig.~\ref{fig:UUSchematic}. In these two situations, determination of $N_{part}$ and $N_{coll}$ for the same central centrality class becomes complicated. For a central collision, both tip-to-tip (TT) and side-to-side (SS) collisions would contain the same number of participating nucleons, but due to the "lining up" of nucleons along the beam axis in tip-to-tip, they would have different numbers of binary collisions $N_{coll}^{TT} > N_{coll}^{SS}$. Early Glauber models, deemed two-component Glauber, assumed that $N_{ch}$ scaled with binary collisions, but with the measurement of ultra-central U+U this was proven false \cite{Pandit:2013uiv,Wang:2014qxa}. This failing of two-component Glauber was one of the inciting incidents that led to the creation of \code{trento} \cite{Moreland:2014oya}, which was able to match U+U measurements \cite{Wertepny:2019yye}. The success of \code{trento} in this description can be traced back to their postulate that the entropy of the initial state must be scale invariant: N nucleon-nucleon collisions produces the same entropy of one N-N collision. Although initial state centrality proxies, such as $N_{part}$ and $N_{coll}$, are model dependent, they can be used to investigate phenomena that percent centrality is insensitive to.

%__________________________________________________________________________
%
%
%__________________________________________________________________________

\subsection{Estimation of Entropy Proportionality Factor} \label{subsec:EstimationOfA}

The typical value of the entropy proportionality factor, $a_{entropy}$ from Eq.~\ref{e:EntropyScaling}, is on $O(100)$. The motivation for this can be illustrated by a simple heuristic argument. Consider the final hadronic state of a heavy-ion collision to be composed of an ideal gas of $\pi^+, \pi^-, \pi^0$. By the equipartition theorem, the energy density for this ideal gas is given by $\epsilon = \tfrac{9}{2} n T$ in units where $k_B = 1$. In a similar way, the ideal gas law gives $p = n T$, which, when joined with the first law of thermodynamics $\epsilon = T s - p$, yields the relation $s = \tfrac{11}{2} n$ between the entropy density and number density. Then, assuming a matching of the final pion entropy at freezeout to the fluid entropy, together with a nearly ideal (isentropic) hydrodynamic state, the total entropy of the initial state is directly related to the number of pions produced in the final state by
\begin{align}
\int d^3 x \, s_0 = \frac{11}{2} N_\mathrm{final} .
\end{align}
Assuming by isospin symmetry that the charged $\pi^\pm$ states account for $2/3$ of $N_\mathrm{final}$, and changing to Milne coordinates $d^3 x = \tau_0 \, d^2 x_\bot \, d\eta$, we have
\begin{align}
\tau_0 \int d^2 x_\bot \, s_0 (\vec{x}_\bot) = \frac{33}{4} \frac{dN_{ch}}{d\eta}
\end{align}
Then assuming the model Eq.~\ref{e:EntropyScaling} and taking $\frac{dN_{ch}}{d\eta} \approx \frac{dN_{ch}}{dy} \sim \ord{1000}$, $\tau_0 \approx 0.6 \, \mathrm{fm}$, and $\int d^2 x_\bot \, T_R (\vec{x}_\bot) \approx 140$ for a typical central event (here we use the \code{trento} model with $p=0$ \cite{Bernhard:2016tnd}), this gives
\begin{align}
a = \frac{
\tfrac{33}{4} \frac{dN_{ch}}{dy}
}{
\tau_0 \, \int d^2 x_\bot \, T_R (\vec{x}_\bot)
}
\approx
\frac{
\tfrac{33}{4} (1000)
}{
(0.6 \, \mathrm{fm}) \, (140)
}
\sim \ord{100 \, \mathrm{fm}^{-1}},
\end{align}
which is roughly consistent with the value from Ref.~\cite{Bernhard:2016tnd}.

%---------------------------------------------------------------------------
%
\section{Initial and Final State Anisotropies} \label{sec:Eccentricities}
%
%---------------------------------------------------------------------------

Experimentally, we only have access to the particles produced in the final state of the collisions. A description of the dynamics of the collision must be obtained from observables derived from the properties of the final state particles. A useful characterization of the final state is through a quantification of the momentum anisotropy, from which many properties of the evolution can be extracted, specifically the shear and bulk viscosities. The momentum anisotropy is a measurement of the azimuthal dependence of the final state particles. 

When constructing the initial state, an estimator of the final state flow would prove extremely useful in discerning correct descriptions of the involved physics. This estimator should come from the spatial anisotropy of the initial collision, since the final state flow is generated from the hydrodynamic response to the initial geometry. 

\begin{figure} [ht]
    \centering
	\includegraphics[width=\textwidth]{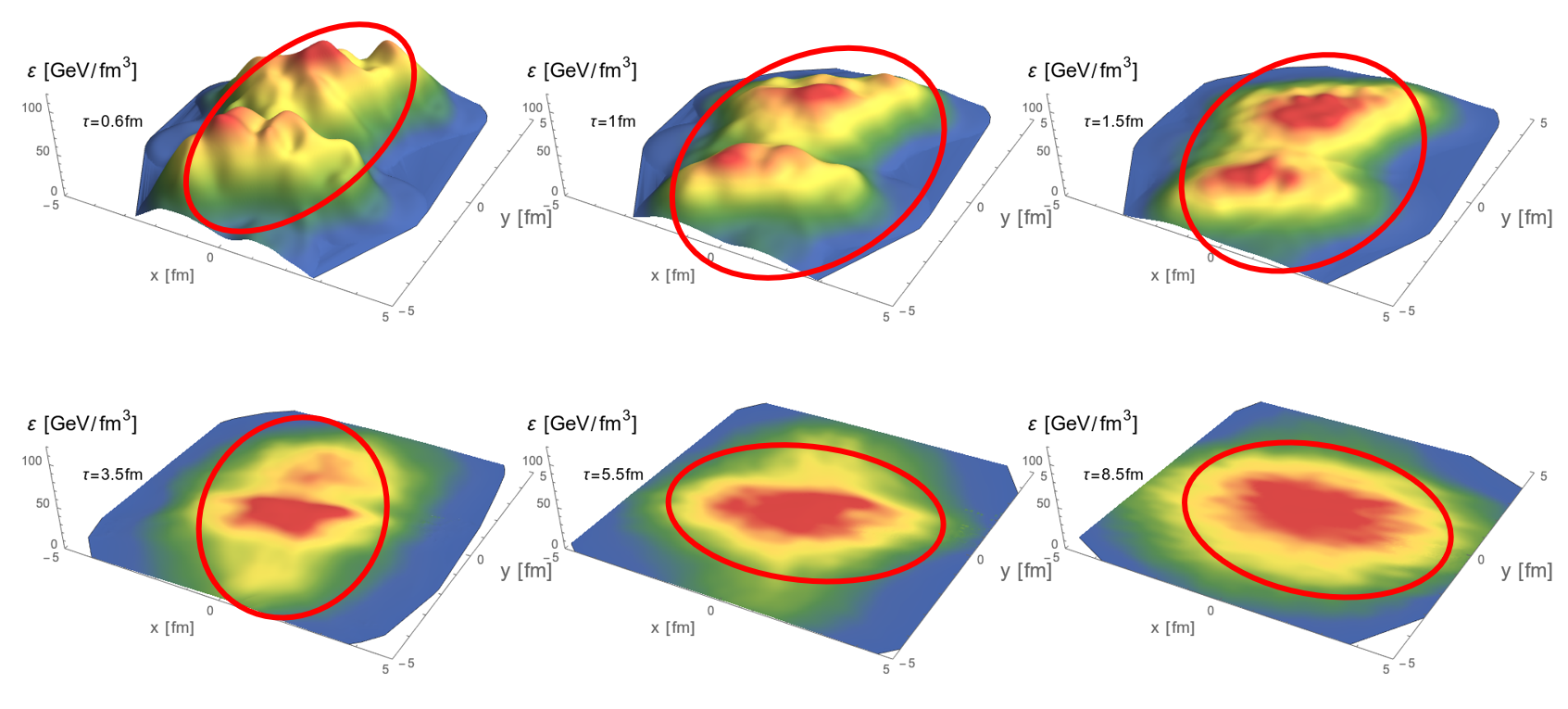}
	\caption{Profile of the energy density of a collision at different time steps in the hydrodynamic evolution. At the beginning (top left), all information is in the energy density distribution and as the evolution progresses it is transferred into the momentum anisotropy (bottom right). The dominant elliptic shape of the energy density and momentum anisotropy is illustrated by the red outline. Figure from Ref.~\cite{Noronha-Hostler:2014dqa}.}
	\label{fig:GeometryTranslation}
\end{figure}

The initial geometry of the collision can be described by gradients in the energy density, which translates to pressure gradients when switching to a hydrodynamic description of the system. These pressure gradients drive expansion of the system and translate initial geometric structure into anisotropic flow in the final state. An example of this translation process is presented in Fig.~\ref{fig:GeometryTranslation}, where, from the top left to bottom right, progressive snapshots in time of the system are plotted. In the top left corner of Fig.~\ref{fig:GeometryTranslation}, is the earliest time step of the hydrodynamic evolution, where no expansion has yet occurred and all structure is from the initial collision profile. Despite the significant geometrical fluctuations, there is an overall elliptic shape, with the major axis along the y-axis, present at this first stage which is highlighted in red. As the system expands (toward the right and down), the initial elliptic geometry is softened by viscous expansion and translated into an ellipsoidal anisotropy in momentum space. In the final state (bottom right), the initial elliptic geometry along the y-axis has been translated to elliptic momentum anisotropy in the x-axis. The rotation of orientation of the ellipse is because the largest spatial gradients in the initial state are along the minor axis of the ellipse and push out along that direction more than along the major axis.

%__________________________________________________________________________
%
%
%__________________________________________________________________________

A good intuition can be obtained by general observations of the evolution of the system, but a more quantitative description is required to extract properties of the system. The key to this extraction is that the initial geometry anisotropy generates the final momentum anisotropy and, consequently, the final state flow can be described as a response to the initial state geometry. From experimental measurements, we have access to the momentum distribution of the particles produced in the final state, which can be decomposed in to a series of flow harmonics using a Fourier series. Likewise, the initial state geometry can be described by a Fourier expansion of the spatial anisotropy into different contributing eccentricities. In this way, a specific flow harmonic should be a function of the initial eccentricity of the same order. This relationship between the initial and final state can be quantified by looking at the response of the final flow to varying initial geometries. Response theory is explored in more detail in Sec.~\ref{subsec:ResponseTheory}.

The example in Fig.~\ref{fig:GeometryTranslation}, focuses on the elliptic geometry, which is the dominant contribution driven mostly by the impact parameter. However, there is finer structure in the initial state, which has a significant impact on the initial geometry and leads to finite higher order geometry. This fine structure comes, primarily, from nucleon fluctuations and is known to be a direct source of triangular geometry in the initial state, as discussed in Sec.~\ref{subsuc:Triangularity}.

%..........................................................................
%
\subsection{Flow Harmonics}
%
%..........................................................................

Similar to centrality classification, we must first establish the observables accessible to experiment in order to inform our characterization of the initial state geometry. Experiment has access to the particles emitted from heavy-ion collisions and these can be described by an underlying probability distribution. Of particular interest is the azimuthal anisotropy of the produced particles, which can be expanded in a Fourier series of the azimuthal angle $\phi$:
\begin{align}
    E \frac{dN}{d^3p} \equiv \frac{1}{2\pi} \frac{\sqrt{m^2 + p_T^2 cosh(\eta)^2}}{p_T cosh(\eta)} \frac{dN}{p_T dp_T d\eta} \left[ 1 + \sum_{n=1}^{\infty} v_n cos \, n (\phi-\psi_n)\right], 
\end{align}
with $m$ and $p_T$ being the mass and transverse momentum of individual particles and $\eta$ is the pseudorapidity. The azimuthal anisotropy is represented by the $v_n$ coefficients, which are functions of $p_T$ and $\eta$. The orientation angles $\psi_n$ are, likewise, functions of $p_T$ and $\eta$ and defined so that all sine terms in the expansion vanish. 

The full flow vector can be written as a complex number \cite{Luzum:2013yya}, such that the explicit definition is:
\begin{align} \label{e:Vdef2}
	\bm{V_n} \equiv v_n \, e^{i n \phi_n} (p_T, \eta) = 
	\frac{\int_0^{2\pi} \frac{dN}{d\phi dp_T d\eta} \, e^{i n \phi} \, d\phi}
        {\int_0^{2\pi} \frac{dN}{d\phi dp_T d\eta} \, d\phi}.
\end{align}
Each harmonic of the the Fourier series describes a different type of flow, the first three being: directed flow $v_1$, elliptic flow $v_2$, and triangular flow $v_3$. The definition of the flow vectors also ensures that they are normalized.

For a symmetric collision, all of the orientation angles $\psi_n$ will be aligned with the impact parameter, such that all odd harmonics will be zero. This was the motivation behind the definition of the "event plane" angle and was assumed to be a good approximation for collisions of the same ion at mid-rapidity. This definition is what led to the assumption that there was no triangular flow in heavy-ion collisions \cite{Alver:2010gr}. This, however, is not the case, with event-by-event fluctuations in the initial state producing finite triangular flow. A correct description of these initial state fluctuations, thus, becomes very important for understanding the dynamics of the system. 

%..........................................................................
%
\subsection{Eccentricities}
%
%..........................................................................

We choose to quantify the initial state geometry using a measure of the spatial anisotropy, since this provides a useful analog to the momentum anisotropy measured by experiments. In order for the quantifiers of the initial geometry to be good estimators of the final state flow, they need to transform in the same way as the flow vectors $\bm{V_n}$ i.e., they need to be translationally and rotationally invariant functions as well as containing other discrete symmetries \cite{Gardim:2011xv,Gardim:2014tya}. The standard characterization of the spatial anisotropy, which contains all of the required properties, are the eccentricity vectors $\bm{\mathcal{E}_n}$ in the complex plane, 
\begin{align} \label{e:ecc1}
\bm{\mathcal{E}_n} \equiv \varepsilon_n \, e^{i n \psi_n} \equiv -
\frac{\int r dr d\phi \, r^n e^{i n \phi} \, f(r, \phi)}
{\int r dr d\phi \, r^n \, f(r, \phi)} ,
\end{align}
where $f(r,\phi)$ is the initial state density and $\varepsilon_n$ and $\psi_n$ are the magnitude and complex angle of the vector. 

\begin{figure} [ht]
    \centering
	\includegraphics[width=0.65\textwidth]{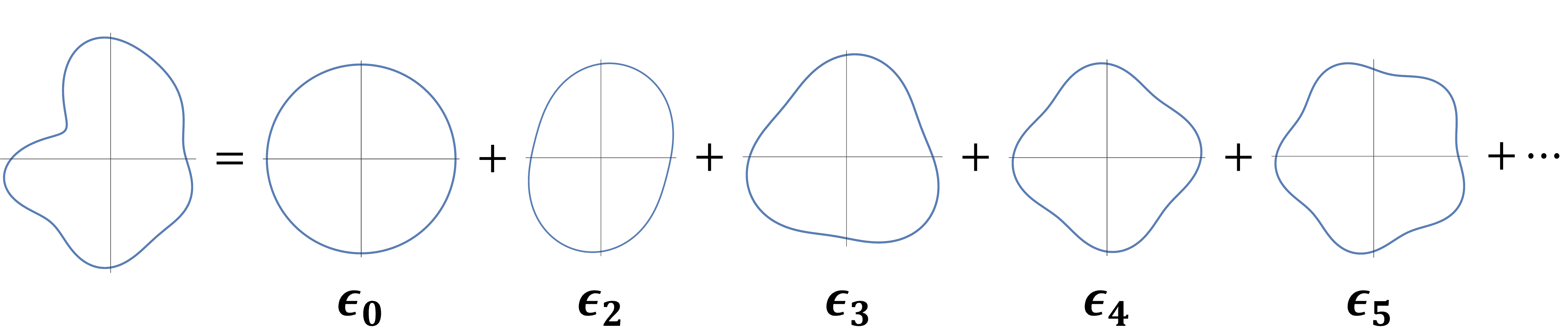}
	\caption{Illustration of the different geometric anisotropies described by $\varepsilon_n$.
	}
	\label{fig:EccentricityExpansion}
\end{figure}

The index $n$ indicates different weights of the vector and corresponds to terms in the Fourier expansion of the initial state density \cite{Coleman-Smith:2012kbb,Floerchinger:2013rya}. Additionally, $n$ correlates with different geometric anisotropies of the initial state, such that $n=2$ describes the elliptic geometry and $n=3$ describes the triangular. This is illustrated in Fig.~\ref{fig:EccentricityExpansion}. The definition of the eccentricity vectors includes a normalization (the denominantor of Eq.~\ref{e:ecc1}), such that the magnitude of the vector is normalized between 0 and 1, where a value of 0 indicates there is no presence of a given geometry. This normalization simplifies the expansion by removing the need for the $n=0$ term to indicate a circular initial state, since this case is covered by all other $\varepsilon_n=0$. The $n=1$ term corresponds to the dipole asymmetry and vanishes when calculated in the center-of-mass frame.
%__________________________________________________________________________
%
%
%__________________________________________________________________________

It is convenient to rewrite the eccentricity vector in terms of the complex position vector $\bm{r} \equiv x + i y$ through $r^n e^{i n \phi} = \bm{r}^n$:
\begin{align} \label{e:ecc2}
\bm{\mathcal{E}}_n \equiv -
\frac{\int d^2 \bm{r} \, \bm{r}^n \, f(\bm{r})}
{\int d^2 \bm{r} \, |\bm{r}|^n \, f(\bm{r})} \: ,
\end{align}
where boldface is used to denote a complex vector. These definitions of eccentricity, Eq.~\ref{e:ecc1} and Eq.~\ref{e:ecc2}, applies only in the center-of-mass frame or whatever is the central frame of the density $f$. For a general coordinate system, this is
\begin{align} \label{e:ecc3}
\bm{\mathcal{E}}_n \equiv -
\frac{\int d^2 \bm{r} \, (\bm{r} - \bm{r}_{CMS})^n \, f(\bm{r})}
{\int d^2 \bm{r} \, |\bm{r} - \bm{r}_{CMS}|^n \, f(\bm{r})}
\end{align}
with the center-of-mass vector
\begin{align}
\bm{r}_{CMS} \equiv
\frac{\int d^2 r \, \bm{r} \, f(\bm{r})}
{\int d^2 r \, f(\bm{r})}
=
\frac{1}{f_{tot}} \, \int d^2 r \, \bm{r} \, f(\bm{r}) .
\end{align}
An important consequence of this definition is that the directed eccentricity $\bm{\mathcal{E}}_1$ vanishes identically, since
\begin{align}
\bm{\mathcal{E}}_1 &\propto
\int d^2 r \, (\bm{r} - \bm{r}_{CMS}) \, f(\bm{r})
\notag \\ &=
\int d^2 r \, \bm{r} \, f(\bm{r})
-
\bm{r}_{CMS} \, \int d^2 r \, f(\bm{r})
\notag \\ &=
f_{tot} \: \bm{r}_{CMS} - f_{tot} \: \bm{r}_{CMS} = 0.
\end{align}

\subsection{Response Theory} \label{subsec:ResponseTheory}

With the eccentricity, $\bm{\mathcal{E}_n}$, and flow, $\bm{V_n}$, vectors, we can explore the dependence of the final state momentum anisotropy on the initial state geometry. The response of the flow to the initial geometry can be described as
\begin{equation}
    \bm{V_n} = f(\bm{\mathcal{E}_m}) + \delta_n ,
\end{equation}
where $f(\bm{\mathcal{E}_m})$ is the estimator of $\bm{V_n}$ as a function of the initial eccentricity and $\delta_n$ is the residual, or difference between the estimator and $\bm{V_n}$. The estimator, $f(\bm{\mathcal{E}_m})$, is any function of the initial eccentricities, but is chosen such that different contributions from $\bm{\mathcal{E}_m}$ can be parameterized using response coefficients. Furthermore, $f(\bm{\mathcal{E}_m})$ does not need to be a function of the same order vector as the flow, i.e., $m\neq n$. For a given estimator, the parameters are tuned to minimize the residual and provide the best possible mapping of initial geometry to final flow.

The estimator must be informed by the properties of the eccentricities and flow harmonics, specifically that $\bm{\mathcal{E}_n}$ transforms like $\bm{V_n}$ under azimuthal rotations, which constrains the function space. In Ref.~\cite{Teaney:2010vd}, a method was developed for characterizing the response of the flow to the initial state through a generating function that can be described by a series of terms. By looking at the leading order term, the simplest form for the estimator is that of linear response scaling:
\begin{equation} \label{eq:LinearResponse}
    f(\bm{\mathcal{E}_n}) = \kappa_n \bm{\mathcal{E}_n},
\end{equation}
where $\kappa_n$ is the linear response coefficient. The focus in this work is on $n=2,3$, which both have leading order linear response, though this is not the case for other harmonics which may include leading order non-linear response, as is the case with $\bm{V_4}$ \cite{Gardim:2014tya}
   
\begin{figure} [ht]
    \centering
	\includegraphics[width=0.65\textwidth]{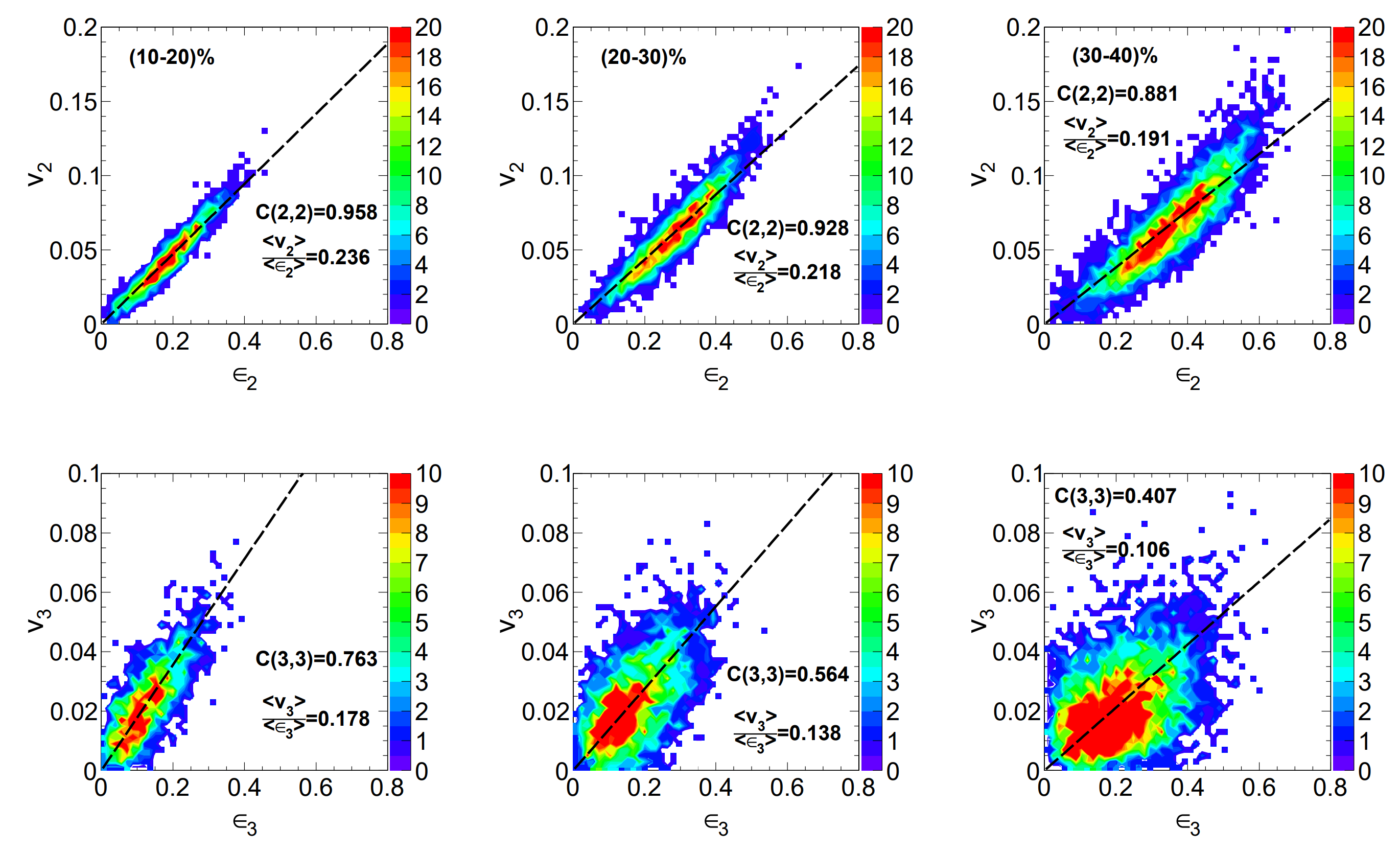}
	\caption{Relation of event-by-event initial state geometry, from an MC-Glauber model, to its final state flow, using a relativistic transport code. The elliptic and triangular geometry-flow relations, for three different centrality classes, are shown in the top and bottom rows, respectively. The linear relation for each geometry and centrality is represented by the dashed lines. Figure taken from \cite{Plumari:2015cfa}.
	}
	\label{fig:LinearResponseDeviation}
\end{figure}

The accuracy of this linear estimator can be gauged by plotting the event-by-event distribution of flow harmonics versus eccentricity, which was first done in Ref.~\cite{Niemi:2012aj}. In Ref.~\cite{Plumari:2015cfa}, the event-by-event distributions were analyzed across several centrality classes in AuAu at $\sqrt{s_{NN}} = 200 GeV$ for elliptic and triangular flow, which are shown in Fig.~\ref{fig:LinearResponseDeviation}. Along with the distribution plotted in Fig.~\ref{fig:LinearResponseDeviation}, the authors of Ref.~\cite{Plumari:2015cfa} included lines for each centrality and geometry to represent the linear response relationship, where the slope is the ratio $\langle v_n \rangle / \langle \varepsilon_n \rangle$ and equal to $\kappa_n$, which is the response coefficient for the magnitudes and not vector quantities. Qualitatively, we see that the linear response is a good approximation for the relationship between $\varepsilon_n$ and $v_n$ in central collisions, but deviations grow with an increase in centrality. The quality of linear response as an estimator is, here, described by a correlation coefficient, C(n,n), that describes the event-by-event deviation of $\varepsilon_n$ and $v_n$ from their average values. This correlation coefficient is commonly referred to as a Pearson Coefficient \footnote{More details included below.} and was first defined in Ref.~\cite{Gardim:2014tya}. As one goes to more peripheral centralities, the correlation coefficient, where $C(n,n) \approx 1$ means there is a strong correlation, deviates from unity reflecting a failure in the linear response approximation in this regime.

%__________________________________________________________________________
%
%
%__________________________________________________________________________

There are two features of the distributions in Fig.~\ref{fig:LinearResponseDeviation} that characterize the failure of linear response. The first is an enhancement of flow, with respect to linear response, in highly elliptic mid-central collisions, which indicates that non-linear corrections to the hydrodynamic response become significant \cite{Noronha-Hostler:2015dbi, Sievert:2019zjr, Rao:2019vgy}. The other feature is the smearing of the relationship between $\varepsilon_n$ and $v_n$, which increases with centrality. This smearing comes from fluctuations of the initial state geometry and non-linear response, which is larger for triangular flow since it is fluctuation driven (See Sec.~\ref{subsuc:Triangularity}).
    
We can improve our estimator by including non-linear correction terms, with the lowest term that still preserves rotational symmetry and analyticity being of a cubic form:
\begin{equation}
    f(\bm{\mathcal{E}_n}) = \kappa_n \bm{\mathcal{E}_n} + \kappa_n' |\bm{\mathcal{E}_n}|^2 \bm{\mathcal{E}_n} ,
\end{equation}
where $\kappa_n'$ is the cubic response coefficient.

\begin{figure}[ht]
    \centering
    \includegraphics[width=0.48\textwidth]{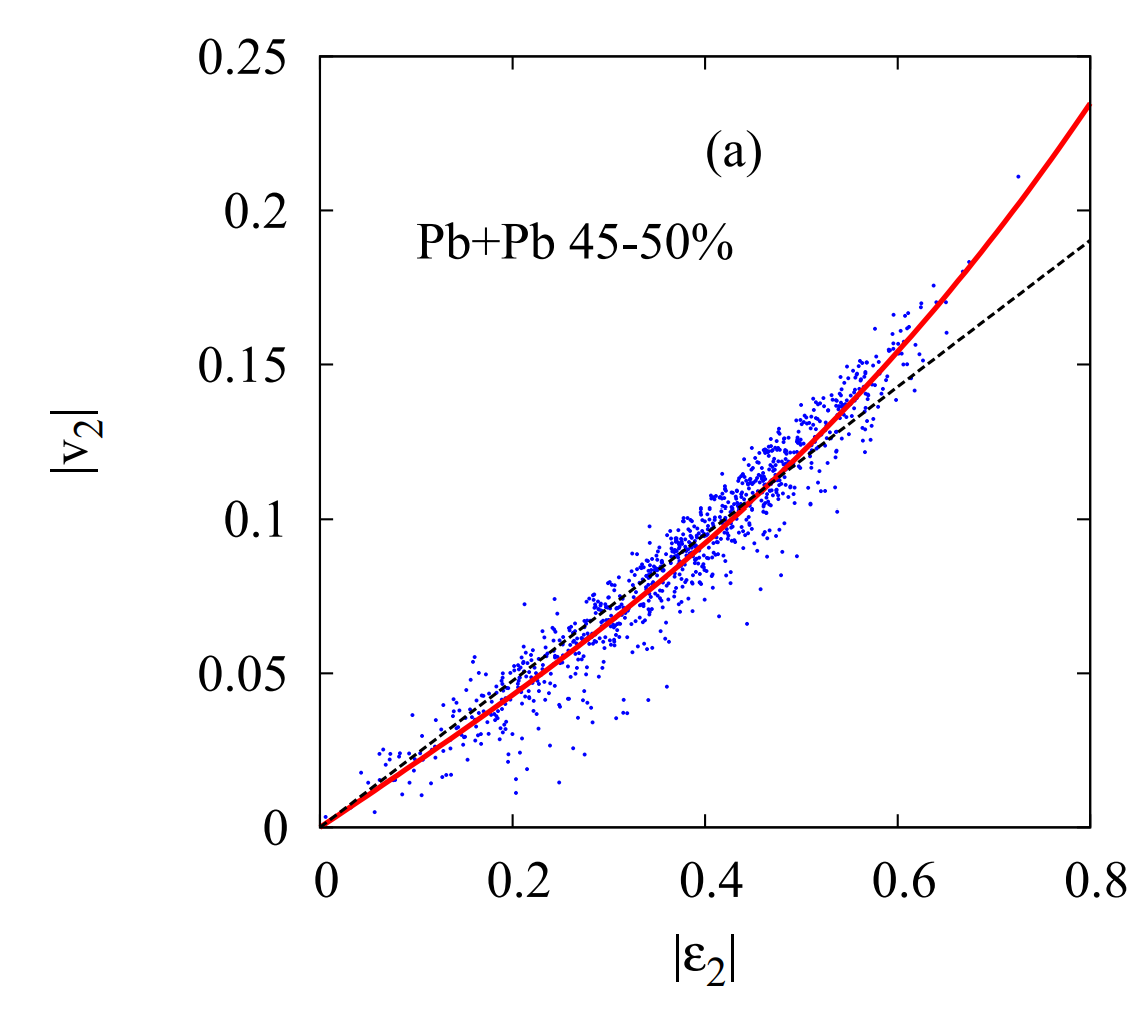}
    \includegraphics[width=0.48\textwidth]{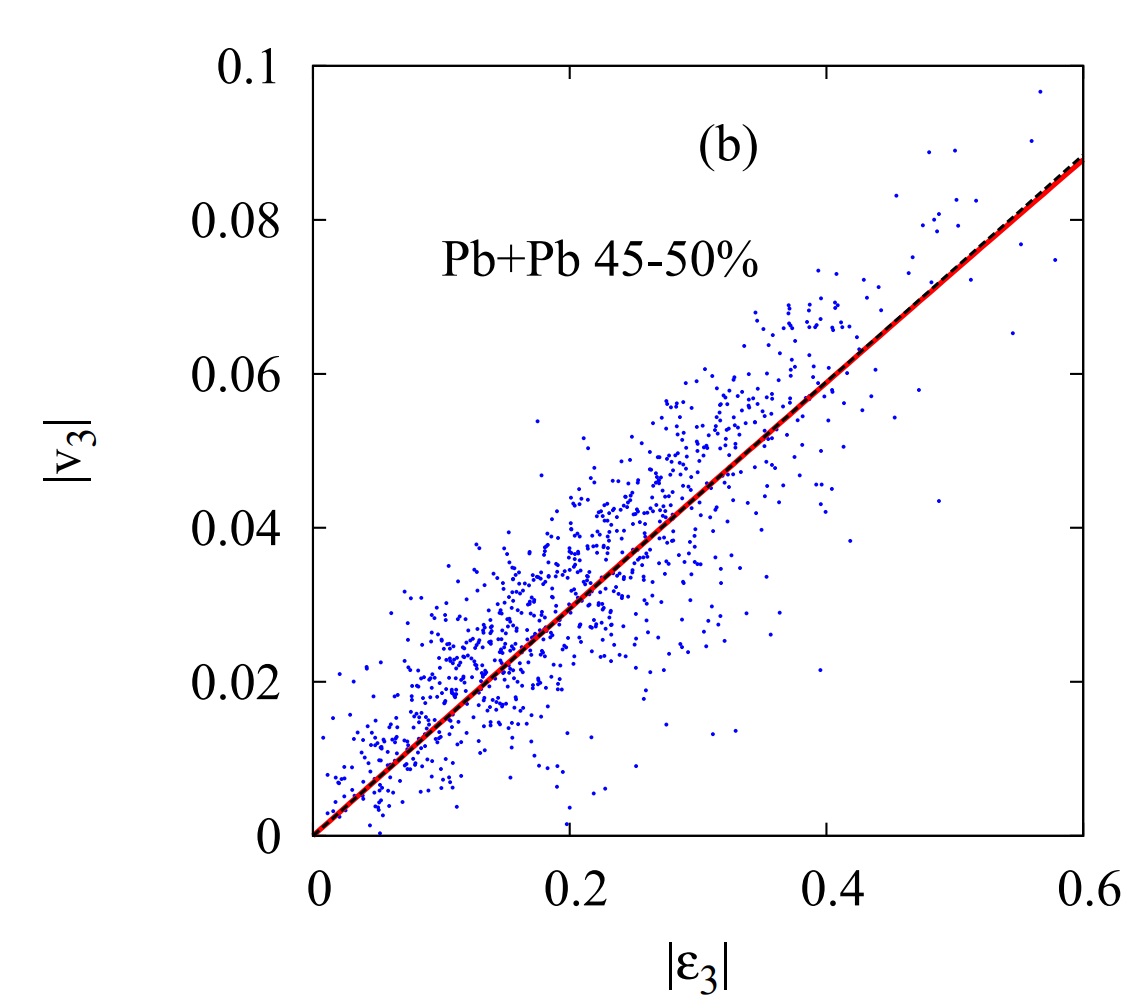}
    \caption{The correlation of initial geometry and flow magnitudes for Pb+Pb collisions at $2.76 TeV$ in 45-50\% centrality, where each point is a different initial condition. The linear and cubic estimators, dotted and solid lines respectively, are plotted for the elliptic (a) and triangular (b) geometry-flow relations. Taken from \cite{Noronha-Hostler:2015dbi}.}
    \label{fig:OrderedResponse}
\end{figure}

\begin{figure}[ht]
    \centering
     \includegraphics[width=0.48\textwidth]{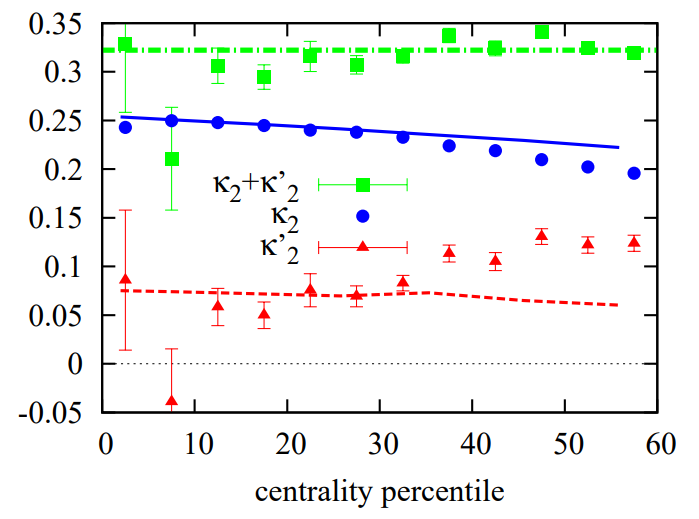}
    \caption{The response coefficients with respect to percent centrality for the linear (blue), cubic (red), and linear+cubic (green) estimators. The points represent an event-by-event calculation of the coefficients and the lines come from smooth initial conditions. Taken from \cite{Noronha-Hostler:2015dbi}.}
    \label{fig:ResponseCoefficentsMagnitude}
\end{figure}

In Fig.~\ref{fig:OrderedResponse}, the effect of including the cubic response term is illustrated for mid-central PbPb collisions in the 45-50\% centrality class, with the black dashed line representing the linear response and the solid red including the cubic term. We see that for large initial ellipticity, in this centrality window, there is a deviation away from the linear relationship and the inclusion of the cubic response term is able to more accurately describe the enhancement of flow. For triangularity, there is no difference between linear and cubic response, with the linear term sufficient for describing the relationship. While this is generally true, mixed harmonic terms play an important role in peripheral collisions \cite{Gardim:2011xv,Gardim:2014tya,Hippert:2020kde}. Investigating the magnitudes of the different response coefficients, provides insight into when the different response terms become important for a description of the system. The magnitudes of the linear, cubic, and linear+cubic response coefficients as a function of percent centrality are plotted in Fig.~\ref{fig:ResponseCoefficentsMagnitude}, with the lines reporting the magnitudes for smooth initial conditions and the points for fluctuating initial conditions. We see that the linear response, $\kappa_2$, is more pronounced than cubic, $\kappa_2'$, across all centralities. The cubic response is consistent with zero for 0-10\% centrality and becomes enhanced above 35\% centrality with a contribution on the order of 10\%. The linear response is a good predictor up until 35\% centrality when it starts to deviate from the distribution. The sum of both linear and cubic is flat across all centralities, indicating that when linear response starts to fail is when cubic response terms become important in describing the flow harmonics.

%__________________________________________________________________________
%
%__________________________________________________________________________
%

%__________________________________________________________________________
%
%__________________________________________________________________________
%

\begin{figure}[ht]
    \centering
    \includegraphics[width=0.48\textwidth]{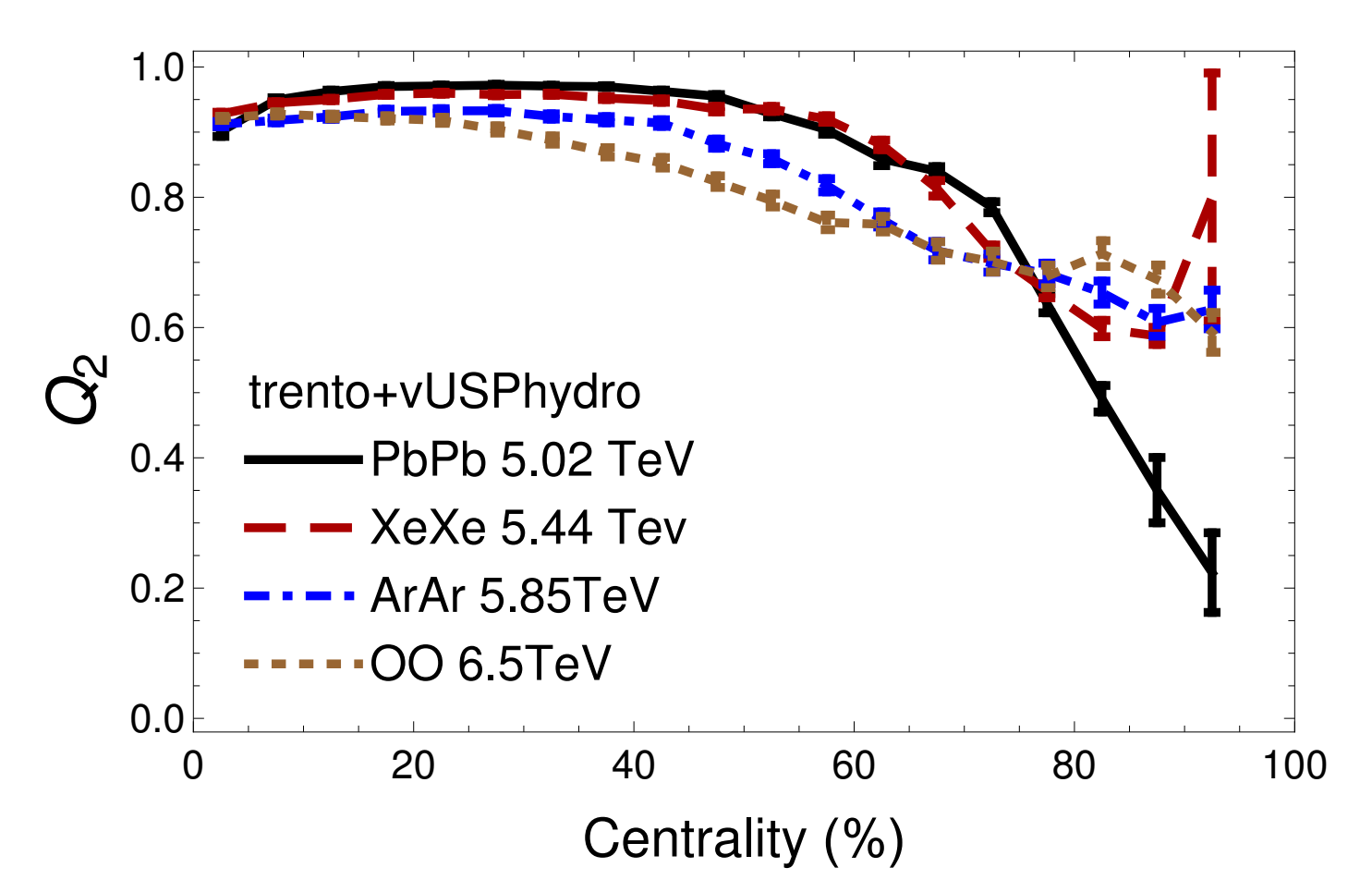}
    \includegraphics[width=0.48\textwidth]{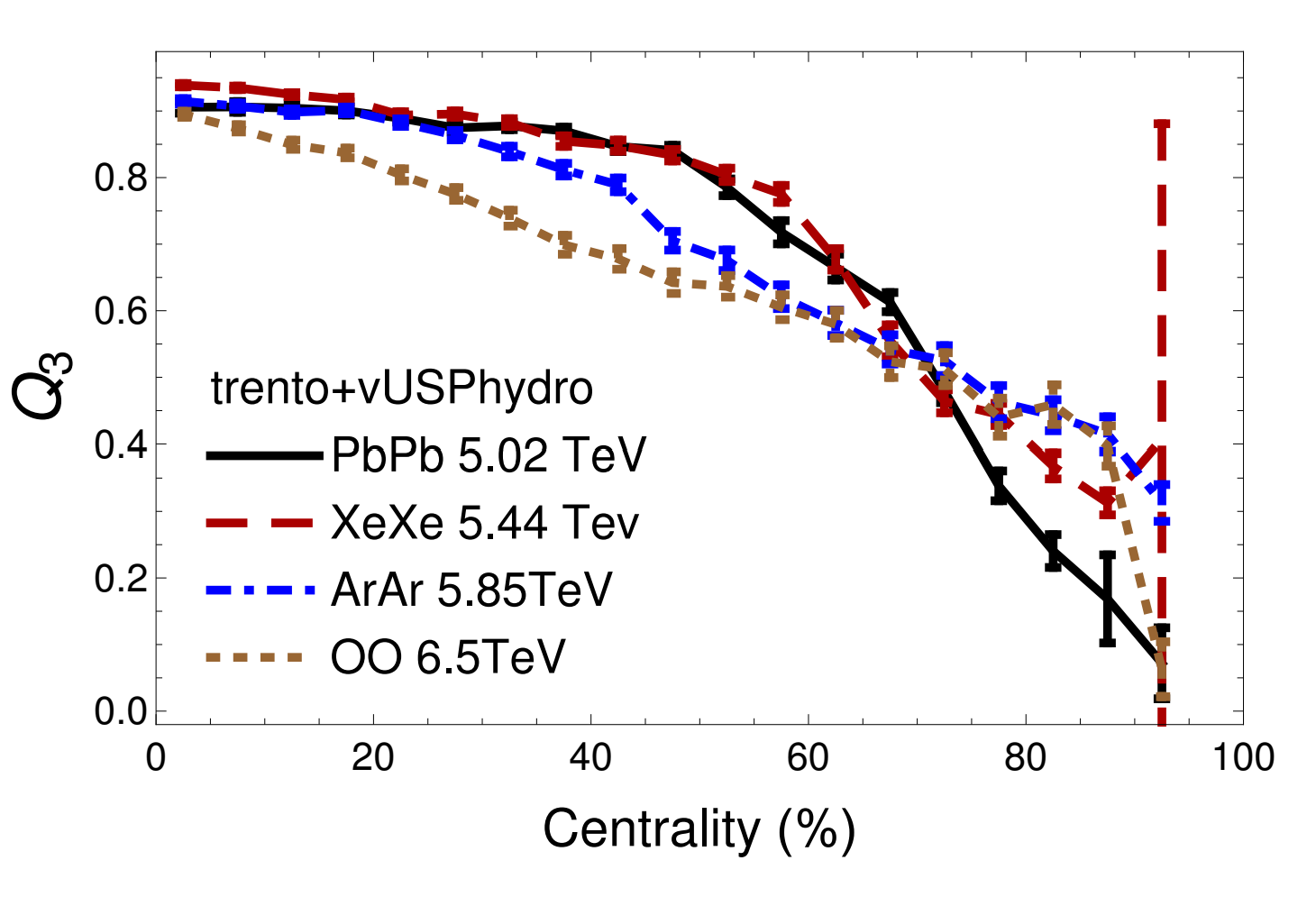}
    \caption{Plots of the Pearson Coefficients of elliptic and triangular geometry that show when the linear approximation fails. Taken from \cite{Sievert:2019zjr}.}
    \label{fig:pearsoncoefficients}
\end{figure}

The quantification of the effectiveness of linear response can be directly specified by using Pearson coefficients, first used in Ref.~\cite{Gardim:2014tya}, to compare the eccentricities ${\varepsilon_n, \phi_n}$ to the flow vectors ${v_n,\psi_n}$ \cite{Betz:2016ayq,Gardim:2011xv}. We can define the Pearson coefficient as:
\begin{equation}
    Q_n = \frac{\langle v_n \varepsilon_n cos(n[\psi_n - \phi_n]\rangle}{\sqrt{\langle |\varepsilon_n|^2 \rangle \langle |v_n|^2 \rangle}} ,
\end{equation}
where both the magnitudes and angles of each event are taken into account. For perfect linear response, $Q_n = 1$ and deviations indicate the increasing contribution of higher order terms. In Ref.~\cite{Sievert:2019zjr}, the authors calculated the Pearson coefficient for ellipticity and triangularity as a function of centrality for several different system sizes. These results are shown in Fig.~\ref{fig:pearsoncoefficients}, where one can see that linear response is a very good description of the elliptic flow up to 50\% centrality and triangular flow up to 20\% centrality. This gives us a good guide on where linear response should not be used: elliptic flow in peripheral collisions and mid-central to peripheral in triangular flow. Additionally, the predictive power of linear response has a dependence on the systems size, which is magnified for $v_3$.

%__________________________________________________________________________
%
%__________________________________________________________________________
%

Linear response provides an extremely important tool for studying the initial state since it alows us to make more direct comparisons to the final state observables. Description of this relations power is continued in Sec.~\ref{sec:Cumulants} and plays an important role throughout the rest of this work, specifically Chap.~\ref{chap:V2toV3Puzzle}.

\subsection{Triangularity} \label{subsuc:Triangularity}

At first, measurements of triangular flow were performed with respect to the event plane angle (often defined with respect to the angle of the elliptic flow vector, $\psi_2$) instead of the more consistent choice of $\psi_3$ \cite{Alver:2010gr}, and led to the conclusion that triangular flow in heavy-ion collisions was not present. However, using the correct observable allowed for the measurement of non-zero triangular flow in heavy-ion collisions, which set off a revolution in the description of initial state physics. Previously, the initial conditions for collisions were modeled using smooth profiles, but while this approximation was well motivated before the measurement of finite $v_3$, it produced no source from which triangular flow could be generated. To solve this problem fluctuations in the initial state geometry were introduced \cite{Alver:2010gr, Takahashi:2009na}.

The authors of Ref.~\cite{Petersen:2012qc} set out to show this relationship explicitly by taking initial conditions for two fixed impact parameters, corresponding to central and mid-central collisions, and progressively averaging over profiles for each. This produces initial conditions at different levels of smoothness and explicitly illustrates the dependence of geometry on the fluctuations of the initial state. I will summarize a few of the key findings of Ref.~\cite{Petersen:2012qc} to provide a quick reference for discussions of triangularity throughout this work.

%__________________________________________________________________________
%
\begin{figure}[ht]
    \centering
    \includegraphics[width=0.70\textwidth]{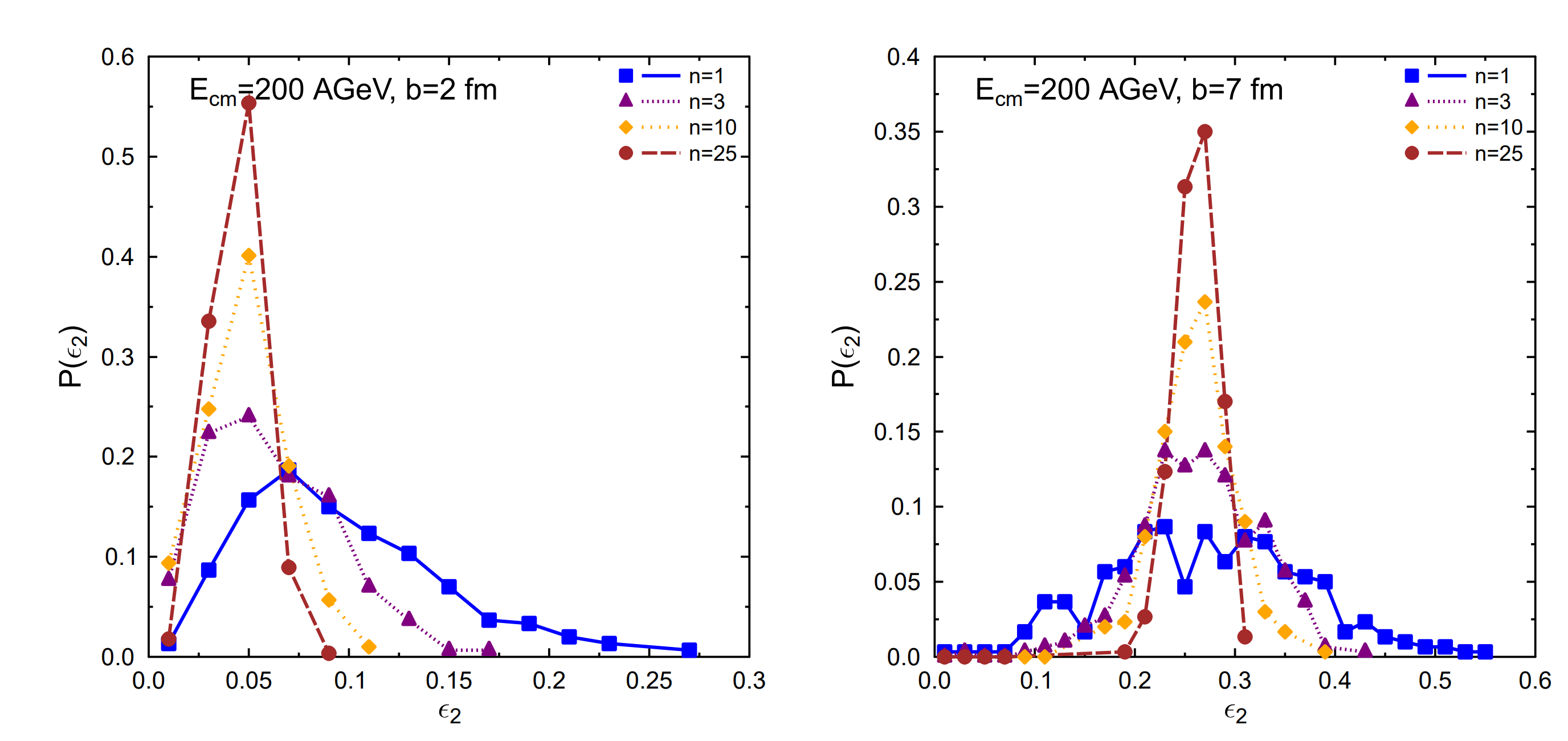}
    \\
    \includegraphics[width=0.68\textwidth]{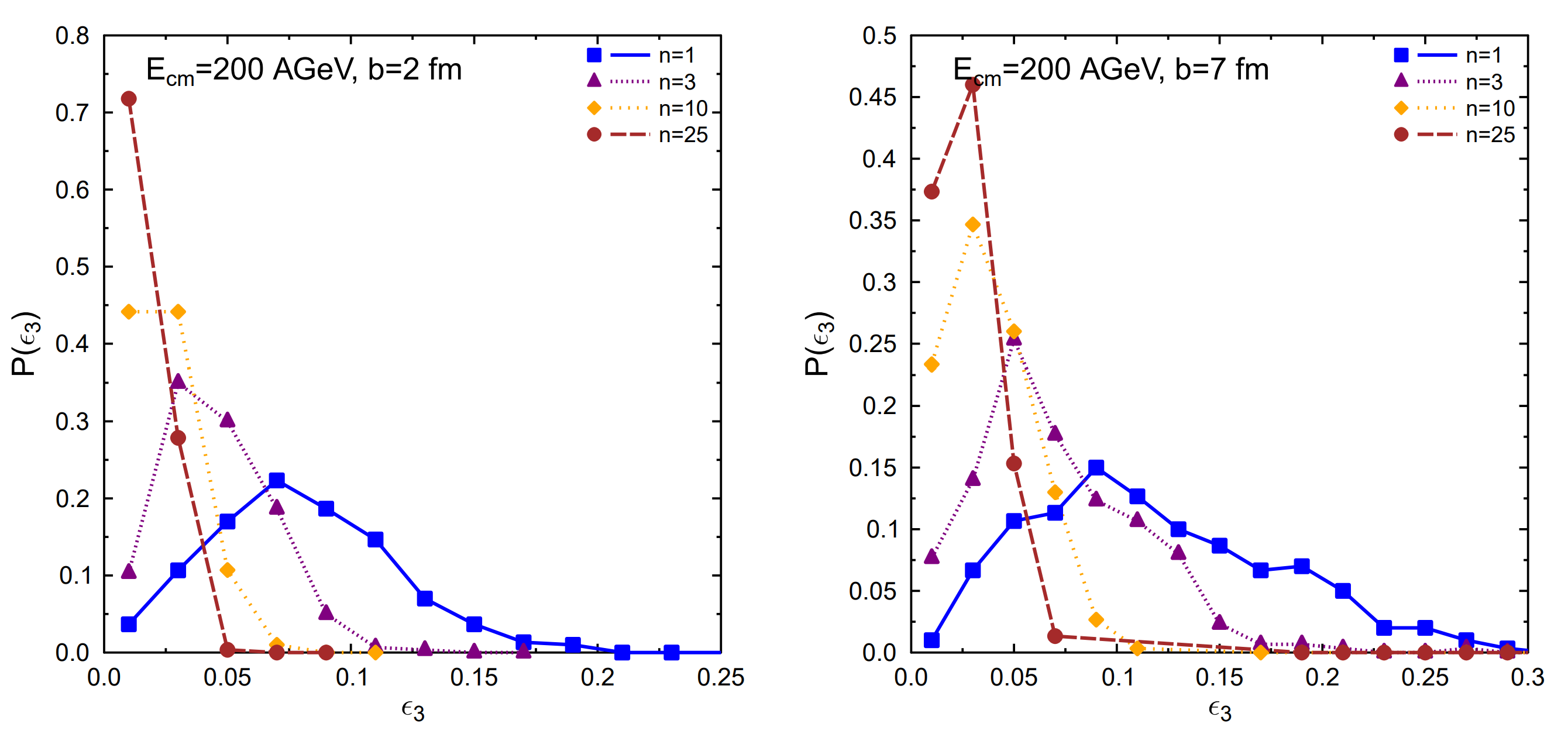}
    \caption{Probability distribution of elliptic (top) and triangular (bottom) initial state geometries for central (left) and mid-central (right) collisions. The blue distributions are those for event-by-event geometries and the other distributions correlate to increasing averages over initial state geometries, thus reducing the granularity of fluctuations. Figure taken from \cite{Petersen:2012qc}.}
    \label{fig:EccentricityDistributionGranularity}
\end{figure}
%__________________________________________________________________________
%

Using this progressive averaging scheme, we first look at the distribution of events by their eccentricity for both impact parameters in Fig.~\ref{fig:EccentricityDistributionGranularity}. For the elliptic geometry (Fig.~\ref{fig:EccentricityDistributionGranularity} top), there is a clear distribution in the event ellipticities centered on a finite value of $\varepsilon_2$. As one averages over events, the distribution becomes narrow and settles on a non-vanishing central value: $\approx 0.05$ for central and $\approx 0.25$ for mid-central. Looking at triangular geometry (Fig.~\ref{fig:EccentricityDistributionGranularity} bottom), we see that for the distribution of unaveraged events there is a clear distribution centered at finite $\varepsilon_3$, though as more events are averaged over the distribution becomes narrow and the peak shifts toward zero. This differentiates the elliptic geometry as being mainly sourced from impact parameter physics and triangular geometry from fluctuations of the initial state geometry. 

%__________________________________________________________________________
%
\begin{figure} [ht]
    \centering
	\includegraphics[width=0.75\textwidth]{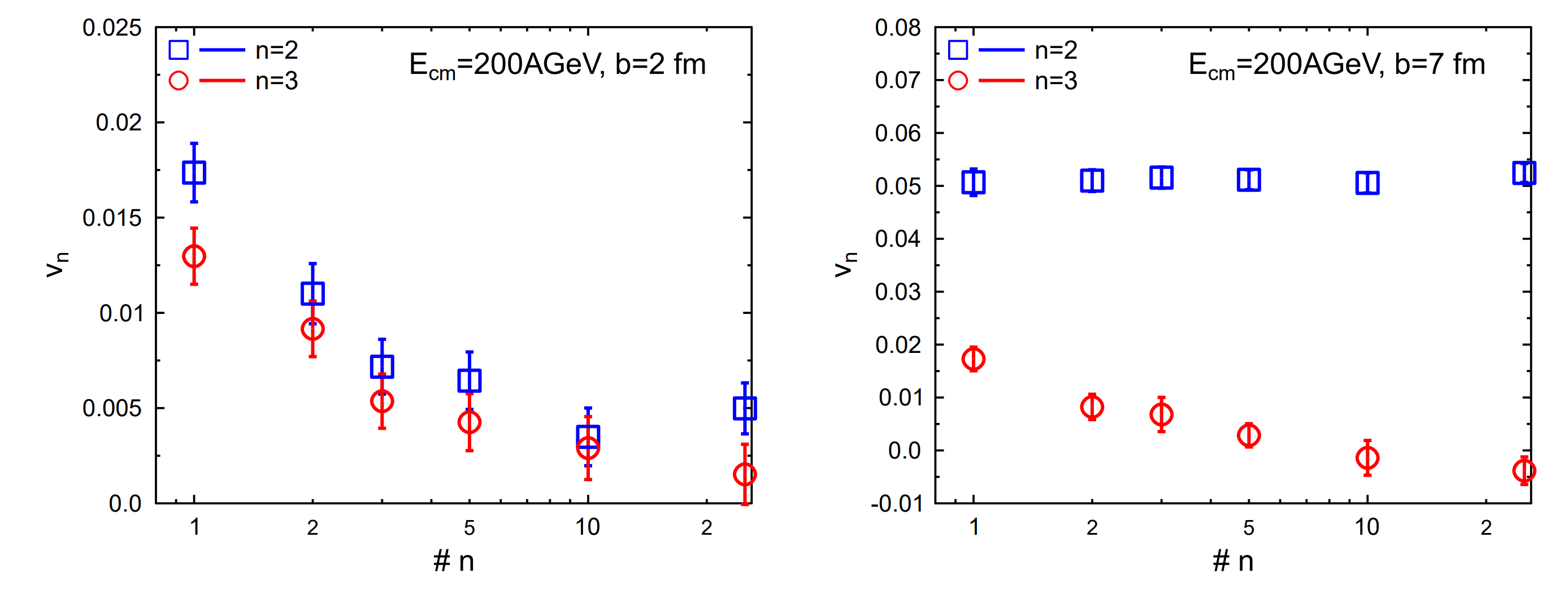}
	\caption{The elliptic and triangular flow magnitudes from central (left) and mid-central (right) events with respect to the number of initial conditions averaged over. Figure taken from \cite{Petersen:2012qc}.
	}
	\label{fig:FluctuationDrivenTriangularity}
\end{figure}
%
%__________________________________________________________________________

Now we can look at the elliptic and triangular flow that these progressively averaged initial state geometries generate in Fig.~\ref{fig:FluctuationDrivenTriangularity}. For central collisions (Fig.~\ref{fig:FluctuationDrivenTriangularity} left), we see that both $v_2$ and $v_3$ decrease in magnitude as the initial conditions are smoothed out. While the ellipticity stops decreasing and levels off to a small but non-zero value, the triangular flow continues decreasing until it is consistent with zero. This is seen with more clarity in mid-central collisions (Fig.~\ref{fig:FluctuationDrivenTriangularity} right), where elliptic flow remains the same despite the number of profiles averaged over, while the triangular flow is heavily suppressed and ultimately vanishes. From this, we can conclude that the elliptic flow is mainly impact parameter driven while triangular flow is fluctuation driven. Another interesting observation, is that elliptic flow in central collisions is much more sensitive to fluctuations.

%%%%%%%%%%%%%%%%%%%%%%%%%%%%%%%%%%%%%%%%%%%%%%%%%%%%%%%%%%%%%%%%%%%%%%%%%%%
%
\section{Implementation of Eccentricity Calculation}\label{sec:EccentricityCalculation}
%
%%%%%%%%%%%%%%%%%%%%%%%%%%%%%%%%%%%%%%%%%%%%%%%%%%%%%%%%%%%%%%%%%%%%%%%%%%%

To calculate the eccentricities of energy and different charge distributions, in this work, I use the eccentricity vector as defined in the imaginary plane and disentangle the magnitude from the angular information. Since this process is a bit involved, this section provides a derivation of this method and a description of its implementation in the \code{iccing} code.

\subsection{Initial Condition}

%__________________________________________________________________________
%
\begin{figure}[ht]
    \centering
	\includegraphics[width=0.5\textwidth]{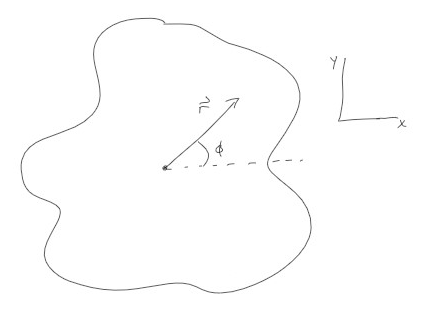}
	\caption{Cartoon of 2-D initial condition with specification of radial vector used to calculate eccentricities of the initial state.}
	\label{f:EccentricityInitialCondition}
\end{figure}
%
%__________________________________________________________________________

In order to describe an initial condition with some density geometry, there needs to be a consistent reference for coordinates which is taken to be the center of mass. From this reference point, a vector can be drawn to any point in the density and represented by a magnitude, $r^2=(\Delta x)^2 + (\Delta y)^2$, and angle, $\phi = arctan(\frac{\Delta y}{\Delta x})$. In order to calculate different orders of eccentricities, the radius vector needs to be rescalable:
\begin{equation}
    r^m = \rho*(r^2)^{m/2}
\end{equation}
where $\rho$ is the local density and $m$ is the radius weight.

\subsection{Definition of Eccentricity}

The eccentricity vector can be defined as an average over all points in the initial condition, here represented by the angle brackets $\langle ... \rangle$, where each point can be described with separately weighted magnitude and angular components \cite{Luzum:2013yya}. The convention is to use $m$ as the radial weight and $n$ as the angular weight, 
\begin{equation}
    \bm{\mathcal{E}_{mn}} = \langle r^m e^{in\phi} \rangle .
\end{equation}
Using Euler's formula, we can separate this into its real and imaginary components:
\begin{equation}
    \bm{\mathcal{E}_{mn}} = \langle r^m cos(n\phi) \rangle + i \langle r^m sin(n\phi) \rangle .
\end{equation}
Which allows the magnitude and angular components to be calculated separately:
\begin{equation}
   \bm{\mathcal{E}_{mn}} = 
   \begin{cases}
        \psi_n, & \frac{1}{N} arctan\frac{\langle r^m sin(n\phi) \rangle}{\langle r^m cos(n\phi) \rangle} 
        \\
        \epsilon_{mn}, & \frac{1}{r_{tot}^m} \sqrt{{\langle r^m cos(n\phi) \rangle}^2 + {\langle r^m sin(n\phi) \rangle}^2 }
    \end{cases}, 
\end{equation}
where $\psi_n$ is the orientation angle of the eccentricity vector, $N$ is the number of points in the initial state, and $r_{tot}^m$ is the normalization for the magnitude and is defined as 
\begin{equation}
r^m_{tot} = \sum\limits_{i}^{N} r^m_i.
\end{equation}
These equations can be rewritten for easier implementation in C++: 
\begin{equation}
\psi_n = \frac{1}{N} \sum\limits_{i}^{N} \tan^{-1}(\frac{r^m_i \sin(n\phi_i)}{r^m_i \cos(n\phi_i)}),
\end{equation}

\begin{equation}
\epsilon_{mn} = \frac{1}{r^m_{tot}} \sum\limits_{i}^{N} r^m_i \cos(n[\phi_i-\psi_n]).
\end{equation}

%---------------------------------------------------------------------------
%
\section{Cumulants}\label{sec:Cumulants}
%
%---------------------------------------------------------------------------

Due to event-by-event fluctuations, the underlying distribution of particles for each collision event is different and the reconstruction of the flow anisotropy $v_n$ from $N$ particles from the distribution carries a statistical uncertainty of approximatly $1\/\sqrt{2N}$. For a typical event, the anisotropy is of a few percent and the number of particles detected is on the order of a few thousand, which results in an uncertainty of 50\% for a single event. This requires us to use correlations between particles that can be averaged over many events and allow determination of moments of the event-by-event distribution of the anisotropy $v_n$ and orientation angle $\psi_n$.

The particles emitted by a collision are not, generally, independent of each other but are correlated. We can calculate the distribution of pairs of particles in the transverse plane as
\begin{align} \label{eq:2ParticleDistribution}
    \frac{dN_{2}}{d^2_{p_1}d^2_{p_2}} = \frac{dN}{d^2_{p_1}} \frac{dN}{d^2_{p_2}} + \delta_2 (p_1, p_2) , 
\end{align}
where the second term $\delta_2 (p_1, p_2)$ is the intrinsic correlation between pairs of particles. The first term of Eq.~\ref{eq:2ParticleDistribution} is the correlation of each particle with the global event and represents the collective behavior of the system, dubbed "flow", while the second term represents "non-flow" correlations. Hydrodynamic simulations do not include the intrinsic correlations between particles, hence the label of "non-flow", and so experimental calculations are designed to minimize non-flow correlations. 

Suppression of non-flow contributions to the anisotropic flow can be accomplished by using multi-particle cumulants \cite{Poskanzer:1998yz,Borghini:2000sa,Borghini:2001zr,STAR:2002hbo,Bilandzic:2010jr}. The two- and four-particle flow cumulants, in terms of multi-particle correlations, are defined as
\begin{subequations}
\begin{align}
v_n \{2\} &\equiv \sqrt{
\left\langle \frac{1}{N_2} \int_{p_1 p_2} e^{i n (\phi_1 - \phi_2)} \,
\frac{dN_2}{d^2 p_1 d^2 p_2}
\right\rangle
}
\\
v_n \{4\} &\equiv \sqrt[4]{
2 \left\langle \frac{1}{N_2} \int_{p_1 p_2} e^{i n (\phi_1 - \phi_2)} \,
\frac{dN_2}{d^2 p_1 d^2 p_2} \right\rangle^2 -
\left\langle \frac{1}{N_4} \int_{p_1 \cdots p_4} e^{i n (\phi_1 + \phi_2 - \phi_3 - \phi_4)} \,
\frac{dN_4}{d^2 p_1 \cdots d^2 p_4} \right\rangle
} ,
\end{align}
\end{subequations}
where $N_2$ and $N_4$ are the number of particle pairs and quadruplets, respectively \cite{Luzum:2013yya}. If the multi-particle correlations arise entirely from independent particle emission coupled to a collective flow (i.e., the non-flow contribution is negligible), then the multi-particle distributions factorize on an event-by-event basis, and the cumulants can be written entirely in terms of the statistical distribution of flow harmonics $v_n$:
\begin{subequations} \label{e:Vcumdefs}
\begin{align}
v_n \{2\} &= \sqrt{ \left\langle v_n^2 \right\rangle }
\\ \notag \\
v_n \{4\} &= \sqrt[4]{
2 \left\langle v_n^2 \right\rangle^2 -
\left\langle v_n^4 \right\rangle
} \label{e:Vcumdefs2}
\notag \\ &=
v_n \{2\} \,
\sqrt[4]{
1  - \frac{\mathrm{Var}(v_n^2)}{\langle v_n^2 \rangle^2}
}.
\end{align}
\end{subequations}
Thus, when all correlations come from flow, $v_n \{2\}$ measures the root-mean-square (RMS) of the $n^\mathrm{th}$ harmonic flow, and $v_n \{4\}$ measures the variance of the $n^\mathrm{th}$ harmonic. The 4-particle cumulant describes the fluctuations of the given harmonic flow, where the suppression from unity corresponds to an increase in the amount of fluctuations. While the 2-particle cumulant is always real, the 4-particle cumulant can become imaginary, due to the assumption that the underlying distribution for each centrality class is of a certain form and well behaved. In the case that $v_n \{4\}$ becomes imaginary, we plot the magnitude of the observable as negative.

In a similar way, we quantify the initial state geometry using cumulants analogous to Eq.~\ref{e:Vcumdefs} for the initial eccentricities $\varepsilon_n$:
\begin{subequations} \label{e:Ecumdefs}
\begin{align}
\varepsilon_n \{2\} &= \sqrt{ \left\langle \varepsilon_n^2 \right\rangle }
\\ \notag \\
\varepsilon_n \{4\} &= \sqrt[4]{
2 \left\langle \varepsilon_n^2 \right\rangle^2 -
\left\langle \varepsilon_n^4 \right\rangle
}
\notag \\ &=
\varepsilon_n \{2\} \,
\sqrt[4]{
1  - \frac{\mathrm{Var}(\varepsilon_n^2)}{\langle \varepsilon_n^2 \rangle^2}
} .
\end{align}
\end{subequations}

The approximately linear response of $v_n$ to $\varepsilon_n$ is extremely useful, since it allows us to separate out medium effects by taking a ratio of cumulants.
Ratios of multi-particle cumulants, specifically the ratio of the four-particle to two-particle cumulants, have been shown to provide important constraints on the initial state \cite{Giacalone:2017uqx}. Using the assumption of linear response (See Sec.~\ref{subsec:ResponseTheory}), we can take the ratio $\varepsilon_n\{4\}/\varepsilon_n\{2\}\/$ to cancel out the response coefficient $\kappa_n$ and provide a direct comparison to $v_n\{4\}/v_n\{2\}$, which can be measured experimentally. 
As seen in Eq.~\ref{e:Vcumdefs2}, this ratio also quantifies the amount of fluctuations in the quantity $v_n$ (or $\varepsilon_n$): the limit $\varepsilon_n \{4\} / \varepsilon_n \{2\} \rightarrow 1$ corresponds to no fluctuations ($\langle \varepsilon_n^2 \rangle^2 = \langle \varepsilon_n^4 \rangle$), with more fluctuations in $v_n$ or $\varepsilon_n$ the further below $1$ the cumulant ratio $\varepsilon_n \{4\} / \varepsilon_n \{2\}$ falls.

\section{Important Comments} \label{sec:ImportantComments}

Some important caveats must be mentioned here. When theoretical simulations are performed at mid-rapidity, they are truly done in the transverse plane centered on the center of mass, while experimental measurements are performed in some pseudo-rapidity window around $\eta = 0$. This is important in experimental calculations for net-particle fluctuations, like kurtosis or directed flow, due to momentum conservation issues. While this difference does not generally present an issue in making comparisons between theory and experiment, it is important to be aware of. Another difference that is important to mention is that, due to detector limitations, experimental measurements are done using only charged particles resulting from the collision, whereas theory simulations have access to both charged and neutral particles.  This requires theory to constrict analysis to only charged particles if comparison with experimental data is to be attempted. While largely unimportant for the work presented here, it is useful to be aware of the disconnect between theoretical simulations and experiment.

\begin{figure} [ht]
    \centering
	\includegraphics[width=0.75\textwidth]{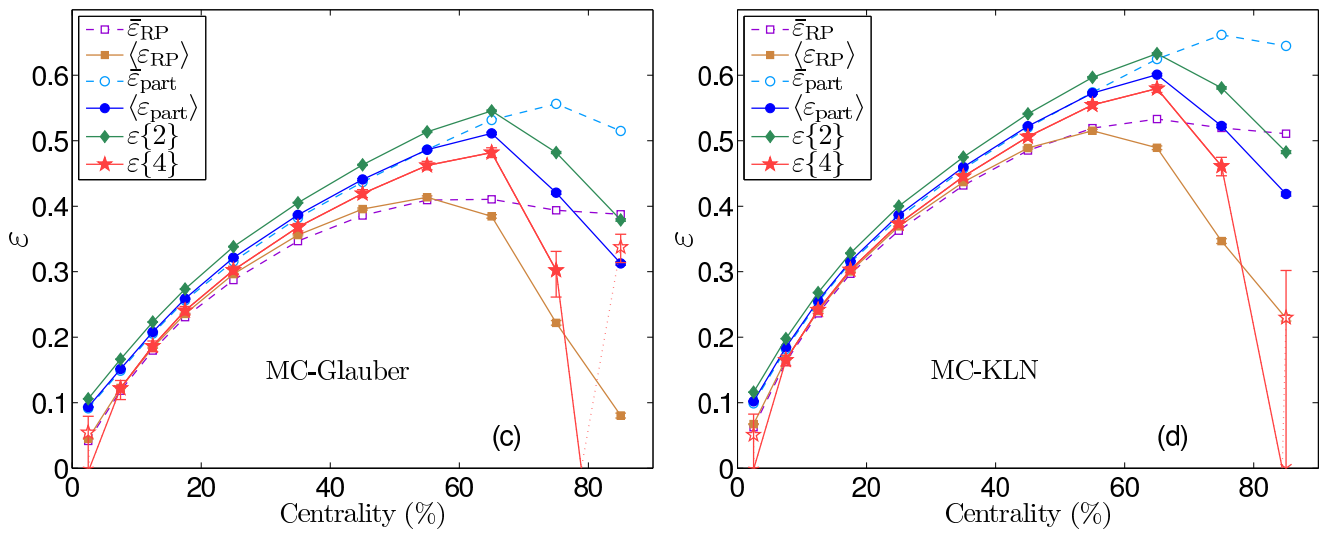}
	\caption{Comparison of various elliptic geometry estimators for MC-Glauber (left) and MC-KLN (right) initial state models. The estimator $\varepsilon\{2\}$ is synonymous with the standard cumulant definition. Figure taken from \cite{Qiu:2011iv}.}
	\label{fig:AlternateGeometryQuantifiers}
\end{figure}

Historically, the quantification of event averaged geometry has been done in many different ways with the cumulant description seeing wide adoption by the field as the best characterization \cite{Luzum:2012da}. Typically, when referring to work done before this consensus on cumulants one must be careful to consider the exact definition used by such works. The authors of Ref.~\cite{Qiu:2011iv} compare many of these different methods in Fig.~\ref{fig:AlternateGeometryQuantifiers}, which plots the elliptic geometry across centrality for two different initial state models. While there are significant differences between the different estimators, there is general consensus on the increase of elliptic geometry with respect to centrality and a suppression in peripheral events. The universal presence of these features allows some conjectures to be made about the initial state without needing to specify the estimator method for each result. 

%__________________________________________________________________________
%
%
%__________________________________________________________________________

%---------------------------------------------------------------------------
%
\chapter{Initial State Physics} \label{chap:InitialStatePhysics}
%---------------------------------------------------------------------------
%

\epigraph{The whole matter of the world must have been present at the beginning, but the story it has to tell may be written step by step.}{Fr. Georges Lemaitre}

\begin{figure}[ht]
    \centering
    \includegraphics[width=0.6\textwidth]{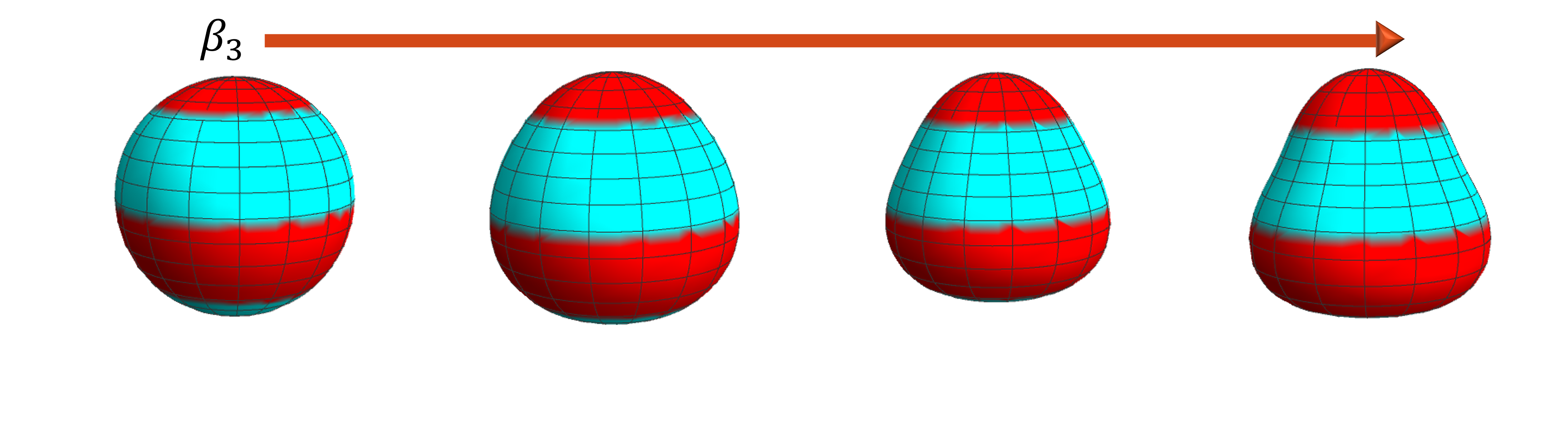}
    \caption{Cartoon of a nuclei that contains an octupole deformation that increases toward the right.}
    \label{fig:DeformedNuclei}
\end{figure}

Most initial state models essentially follow the same method to determine the geometry of the collision, and differ in their approach to energy/entropy deposition. We begin by describing the distribution of nucleons in the nucleus. The Woods-Saxon potential, introduced in Ref.~\cite{Woods:1954zz}, while agreeing well with experimental data across many systems, is the easiest way to describe the distribution of nucleons and is formulated as:
\begin{align} \label{eq:WoodsSaxon}
    \rho(r,\theta) = \rho_0 \left[ 1 + \exp\left(\frac{r - R}{a}\right)\right]^{-1},
\end{align}
where $\rho_0$ is the nucleon density in the center of the nucleus, $a$ is the "skin thickness", and $R$ is the nuclear radius. This distribution assumes a spherical nucleus, though, that assumption can be relaxed by allowing the nuclear radius to have angular dependence, $R(\theta,\phi)$, and expanding it in a series of spherical harmonics. This angular dependent radius can be simplified, due to Lorentz contraction of the nucleus, to only depend on $\phi$ and can be expressed as:
\begin{align} \label{e:Rdef}
R(\theta) = R \Big(1 + \beta_2 Y_{20} (\theta) + \beta_3 Y_{30} (\theta) 
+\beta_4 Y_{40} (\theta) 
+ \cdots \Big),
\end{align}
where $Y_{l m}$ are the spherical harmonics and the $\beta_l$ coefficients are the nuclear deformation parameters.
The effect of an octupole nuclear deformation is illustrated in Fig.~\ref{fig:DeformedNuclei}, where the octupole coefficient $\beta_3$ increases toward the right. 
%__________________________________________________________________________
%
%__________________________________________________________________________
%

Now we would like to know the probability for a nucleon from the projectile 
to collide with one from the target. First, we characterize the probability per unit area (in the transverse plane) of a nucleon being in a given position. We call this a thickness function:
\begin{align} \label{eq:ThicknessFunction}
    T_{A,B} (x,y) = \int \rho_{A,B} (x,y,z) dz,
\end{align}
where $\rho_{A,B}$ is the nucleon density of the specified nucleus, here taken to be the Woods-Saxon. By taking the product of the projectile and target thickness functions, we get the joint probability per unit area of nucleons being located in the same differential area $d^2s$. The reduced thickness function, so called due to its reduction of two thickness functions into a third, is defined as the integral over some function of the thickness functions A and B:
\begin{align} \label{eq:ReducedThicknessFunction}
    T_R (x,y) = \int f(T_A (x,y), T_B (x,y)) d^2s,
\end{align}
The geometry of a  collision will be dependent on the overlap of the two colliding nuclei, which is described by the impact parameter $b$. The impact parameter for a given collision is determined by sampling the distribution
\begin{align}
    \frac{d\sigma}{db} = 2\pi b .
\end{align}
The thickness functions can be rewritten so that the impact parameter dependence is explicitly included
\begin{align} 
    T_R (\Vec{b}) = \int f(T_A (\vec{s}), T_B (\vec{s}-\vec{b})) d^2s,
\end{align}
where $\vec{s}$ is the center of nucleus A and $\vec{b}$ describes the center of nucleus B with relation to A.

Having established a characterization for the overlap of two nuclei in space, it is still left to determine if they indeed interact. This must be done by including the inelastic nucleon-nucleon (NN) cross section $\sigma_{NN}^{inel}$, such that the probability of an interaction to occur is 
\begin{align} \label{eq:ProbOfInteraction}
   p(\vec{s}, \vec{b}) = \sigma_{NN}^{inel} T_R(\vec{s}, \vec{b}).
\end{align}
We can determine that the correct probability distribution for having $n$ interactions is the binomial distribution by using properties of the system, namely: number of nucleons is a discrete number, interactions are binary processes, probability of nucleons to interact is constant, and each interaction is independent. The probability mass function of the binomial distribution is:
\begin{align} 
   P(n) = \left[ {AB \atop n} \right] p^n (1-p)^{AB-n} ,
\end{align}
where the binomial coefficient is defined as
\begin{align} 
    \left[ {AB \atop n} \right] \equiv \frac{AB!}{n!(AB-n)!} .
\end{align}
Here, $n$ is the number of successful collisions, which come with a probability of $p$, and $AB$ is the number of potential collisions, where $AB-n$ is the number of failed collisions with probability $(1-p)$. 

So far, this has been the probability for a single NN collision, but what is needed is the probability of collisions of two heavy-ions. We can get the total probability of at least one NN collision to occur in the heavy-ion collision by summing over the probabilities for each individual nucleon to collide:
\begin{align} \label{eq:ProbCollNuclei}
   p_{inel}^{AB} \equiv \sum_{n=1}^{AB} P(n) = \sum_{n=1}^{AB} \left[ {AB \atop n} \right] p^n (1-p)^{AB-n} .
\end{align}
To fully normalize the collision probability, the outcome for 0 collisions to occur must be considered, which has the form of 
\begin{align} 
   P(0) = (1-p)^{AB} .
\end{align}
This allows us to rewrite the sum from Eq.~\ref{eq:ProbCollNuclei} as
\begin{align} 
   \sum_{n=1}^{AB} P(n) &= \sum_{n=0}^{AB} P(n) - P(0)
   \\
   &= 1 - (1-p)^{AB-n} .
\end{align}
Finally, with this we can determine the number of nucleon-nucleon collisions $N_{coll}$ that occur in a heavy-ion collision between nuclei A and B by weighting the summation by $n$ such that
\begin{align}
    N_{coll} &= \sum_{n=1}^{AB} nP(n)
    \\
    &= ABp .
\end{align}

Having determined the probability distribution of $N_{coll}$, there remains the relation of this to the reduced thickness function and impact parameter that describes the heavy-ion collision. Though the probability for any interaction to occur in the collision of A and B is just the inelastic cross-section $\sigma_{inel}$, this probability contains a strong dependence on impact parameter, which is important for experimental comparison, and so we use an impact parameter dependent cross-section:
\begin{align}
    \frac{d^2 \sigma_{inel}}{db^2} = P(n, \vec{b}) ,
\end{align}
which requires the second derivative with respect to $b$, since it is a vector quantity describing an area. Revisiting Eq.~\ref{eq:ProbOfInteraction}, we can integrate over the transverse area to obtain the probability in terms of the impact parameter dependent reduced thickness function:
\begin{align}
    p(\vec{b}) = \sigma_{NN}^{inel} \int T_R (\vec{s},\vec{b}) d^2\vec{s} = \sigma_{NN}^{inel} T_R (\vec{b}) .
\end{align}
It is now straightforward to express the cross-section in terms of the reduced thickness function, since the original probability of a nucleon-nucleon interaction can be written as an impact parameter dependent quantity:
\begin{align}
    \frac{d^2 \sigma_{inel}}{db^2} &= P(n, \vec{b}) 
    \\
    &= 1 - (1-p(\vec{b}))^{AB}
    \\
    &= 1 - (1-\sigma_{NN}^{inel} T_R(\vec{b}))^{AB} .
\end{align}
With this, we can finally calculate the number of collisions, or number of binary collisions, with respect to impact parameter as
\begin{align}
    N_{coll}(\vec{b}) = AB \sigma_{NN}^{inel} T_R (\vec{b}) .
\end{align}

Though $N_{coll}$ tells us how many individual nucleon-nucleon interactions occur in a heavy-ion collision, it does not provide information about how many nucleons participated in the collision, since a single nucleon could have multiple collisions. To determine the number of participants $N_{part}$, we can describe the individual cross-sections for each nucleus as 
\begin{align}
    \frac{d^4\sigma_{inel}^A}{db^2ds^2} &= 1 - (1 - \sigma_{NN}^{inel} T_A (\vec{s}) )^A
    \\
    \frac{d^4\sigma_{inel}^B}{db^2ds^2} &= 1 - (1 - \sigma_{NN}^{inel} T_B (\vec{s}-\vec{b}) )^B
\end{align}
and the total number of participants, sometimes referred to as "wounded nucleons", are  
\begin{align}
    N_{part}(\vec{b}) &= A \int T_A (\vec{s})\frac{d^4\sigma_{inel}^B}{db^2ds^2} d^2\vec{s} + B \int T_B (\vec{s}-\vec{b})\frac{d^4\sigma_{inel}^A}{db^2ds^2} d^2\vec{s}
    \\
    &= A \int T_A (\vec{s}) \left[ 1 - (1 - \sigma_{NN}^{inel} T_B (\vec{s}-\vec{b}) )^B \right] d^2\vec{s}
    \\
    &+ B \int T_B (\vec{s}-\vec{b}) \left[ 1 - (1 - \sigma_{NN}^{inel} T_A (\vec{s}) )^A \right] d^2\vec{s} .
\end{align}
Using $N_{part}$, we can also get the number of spectator nucleons (those that do not have any interactions) trivially:
\begin{align}
    N_{spec} = A + B - N_{part} .
\end{align}

Now we can describe the density profile, number of collisions, and number of participants of an initial state of a heavy-ion collision through the determination of the reduced thickness function. However, the exact functional combination of the projectile and target thickness functions, $T_A$ and $T_B$, to create the reduced thickness function has not been determined and is left as a general function. The determination of the correct function for $T_R$ is the topic of much of the historical development of initial state modeling, as well as more complex constructions of $T_A$ and $T_B$.

This chapter follows in three parts: the first, Sec.~\ref{sec:HistoryOfInitialState}, looks at the historical development of initial state modeling with a focus on fluctuations in the initial state, the second, Sec.~\ref{sec:Intro:ModernInitialStates}, details state-of-the-art initial state models, and the third, Sec.~\ref{sec:Intro:FurtherDevelopmentOfInitialState} addresses current developments.

%---------------------------------------------------------------------------
%
\section{Historical Evolution of Our Understanding of the Initial State} \label{sec:HistoryOfInitialState}
%---------------------------------------------------------------------------
%

Development of initial state modeling can be separated into two parallel paths: phenomenological and theoretical. The phenomenological approach consists of starting with the simplest assumptions and introducing further complexity in response to disagreement with experimental observations. This is different from more theoretical methods, which start from fundamental physics and build out complexity. Both of these methods tend to progress in tandem, despite their difference in perspective, due to the crux of development hinging on experimental measurements. However, they both provide useful insights into the source of new physics.

The root of the phenomenological initial state is the Glauber model \cite{Glauber:1987bb,Miller:2007ri}, which treated heavy-ion collisions as the combination of independent nucleon-nucleon collisions and defined the reduced thickness function to be the addition of the two participating thickness functions:
\begin{align}
    T_R^{Glauber} = T_A + T_B .
\end{align}
Another popular variation on this method was the two-component Glauber, which added a multiplicative term to the standard Glauber reduced thickness function and allowed for the contribution of each term to be balanced by a parameter $\alpha$:
\begin{align}
    T_R^{2-Comp-Glauber} = (1-\alpha)T_A T_B + \alpha (T_A + T_B) .
\end{align}
For these phenomonological models, the reduced thickness functions was assumed to be proportional to entropy.

There are several threads of theoretical initial state models, but here I will focus on ones based off Color Glass Condensate (CGC) theory \cite{McLerran:1993ni,Gelis:2010nm}. CGC initial state approaches use observations from the PDF as inspiration (Fig.~\ref{f:HERAPDF}). As one goes to the high energy limit, or small-x, the nucleon is dominated by gluons when compared to valence and sea quark distributions. 

This is due to the explosive growth of the gluon PDF at small $x$, which cannot continue indefinitely as it would lead to a violation of unitarity. Instead, at sufficiently high densities, nonlinear corrections become important and are predicted to lead to a slowing, or saturation, of the gluon density at very small $x$ \cite{McLerran:1993ni, McLerran:1993ka, McLerran:1994vd, Kovchegov:2012mbw,Balitsky:1995ub,Kovchegov:1999yj,JalilianMarian:1997jx,JalilianMarian:1997gr,Kovner:2000pt,Iancu:2000hn,Ferreiro:2001qy}. The saturated state is characterized by a coherent condensate of gluons whose dynamics is well-described by classical fields with high occupation number. 

While much of the initial state construction process, from the introduction to this chapter, is used by CGC models, the key difference comes from the specific construction of $T_A$ and $T_B$ and the functional combination of them to get the reduced thickness function. The CGC-based approach calculates thickness functions for the target and projectile nuclei using gluon distributions rather than the geometrical observations of Glauber. The reduced thickness function in CGC models is not directly calculated, but calculations of the initial-state energy density at early times show a proportionality to the binary collision density $T_A T_B$ which can be used to define $T_R$ in a CGC context:
\begin{align}
    \epsilon \propto T_R^{CGC} = T_A T_B ,
\end{align}
where $T_R$ is proportional to the energy density rather than the entropy density, assumed by Glauber models \cite{Nagle:2018ybc, Lappi:2006hq, Chen:2015wia, Romatschke:2017ejr}. 

The path I will follow is mainly that of the Kharzeev-Levin-Nardi (KLN) approach \cite{Kharzeev:2000ph,Kharzeev:2001gp,Kharzeev:2002ei,Kharzeev:2004if}, due to its similar development as Glauber and historically good comparisons. The KLN model is a $k_\perp$-factorization of the number distribution of produced gluons. The original KLN model had an issue coming from edge effects \cite{Drescher:2006pi,Drescher:2006ca}, where the nucleon density near the edge of nuclei (and peripheral collisions) was being over estimated and lead to a disagreement with experimental data. A new version \cite{Drescher:2006ca} of the approach was formulated as the factorized KLN (fKLN) model\footnote{This is a weird name since both versions are factorizations!}, which did the factorization at a different part in the calculation then the original KLN. For more information on these models, see Ref.~\cite{Drescher:2006ca}. 

\subsection{Smooth Initial Conditions} \label{sec:Intro:OpticalGlauber}

Fundamentally, the Glauber method assumes that a nucleus-nucleus collision can be understood by looking at the individual nucleon-nucleon collisions that make up the event. Since the nucleus is a quantum system, the location of each individual nucleon is free to fluctuate, which makes the calculation of the true cross section of a collision difficult. An approximation can be made here that simplifies the complexity of the initial state model by assuming that the nucleus can be described by a smooth density. This approximation is called the optical limit and assumes the incoming nucleons see the target nucleus as a smooth density. While the optical limit is able to describe many features of the heavy-ion collision, it is an insufficient description of the initial state. This assumption, of a smooth density for the colliding nuclei, was quite prevalent in initial state modeling, with similar approximations made by CGC models such as KLN.

\begin{figure}[ht]
    \centering
    \includegraphics[width=0.6\textwidth]{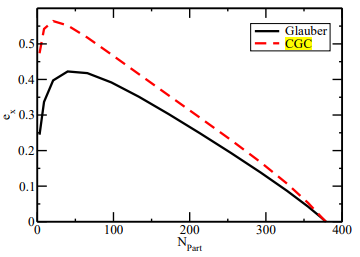}
    \caption{Comparison of elliptic initial state geometry as a function of $N_{part}$ between Optical Glauber and CGC (here it is a KLN variant) models for Au+Au at $\sqrt{s}=200 GeV$. Figure sourced from \cite{Luzum:2008cw}.}
    \label{fig:GlauberVsCGC}
\end{figure}

Using this smooth initial condition, we can start to investigate the structure of the initial state that comes directly from the impact parameter. Revisiting Fig.~\ref{fig:CentralityCartoon}, we see that the dominant geometry coming from collisions of smooth nuclei will be elliptic in nature and highly correlated to the percent centrality, with the ellipticity increasing with respect to centrality. Often, obseravbles are presented with respect to $N_{part}$, which shows a decreasing elliptic geometry with an increase in $N_{part}$ as the system becomes more central \cite{Drescher:2006ca}. This behavior should be consistent for all initial state models and is what we see Fig.~\ref{fig:GlauberVsCGC}, which compares the elliptic geometry from optical Glauber to KLN. Both models produce the same geometrical relationship with respect to $N_{part}$, but the CGC model produces an enhanced elliptic initial state. This difference in magnitude requires different viscosities, for each model, in order for both to fit experimental data, as evidenced by Ref.~\cite{Luzum:2008cw}. 

%__________________________________________________________________________
%
%__________________________________________________________________________
%

\begin{figure}[ht]
    \centering
    \includegraphics[width=0.6\textwidth]{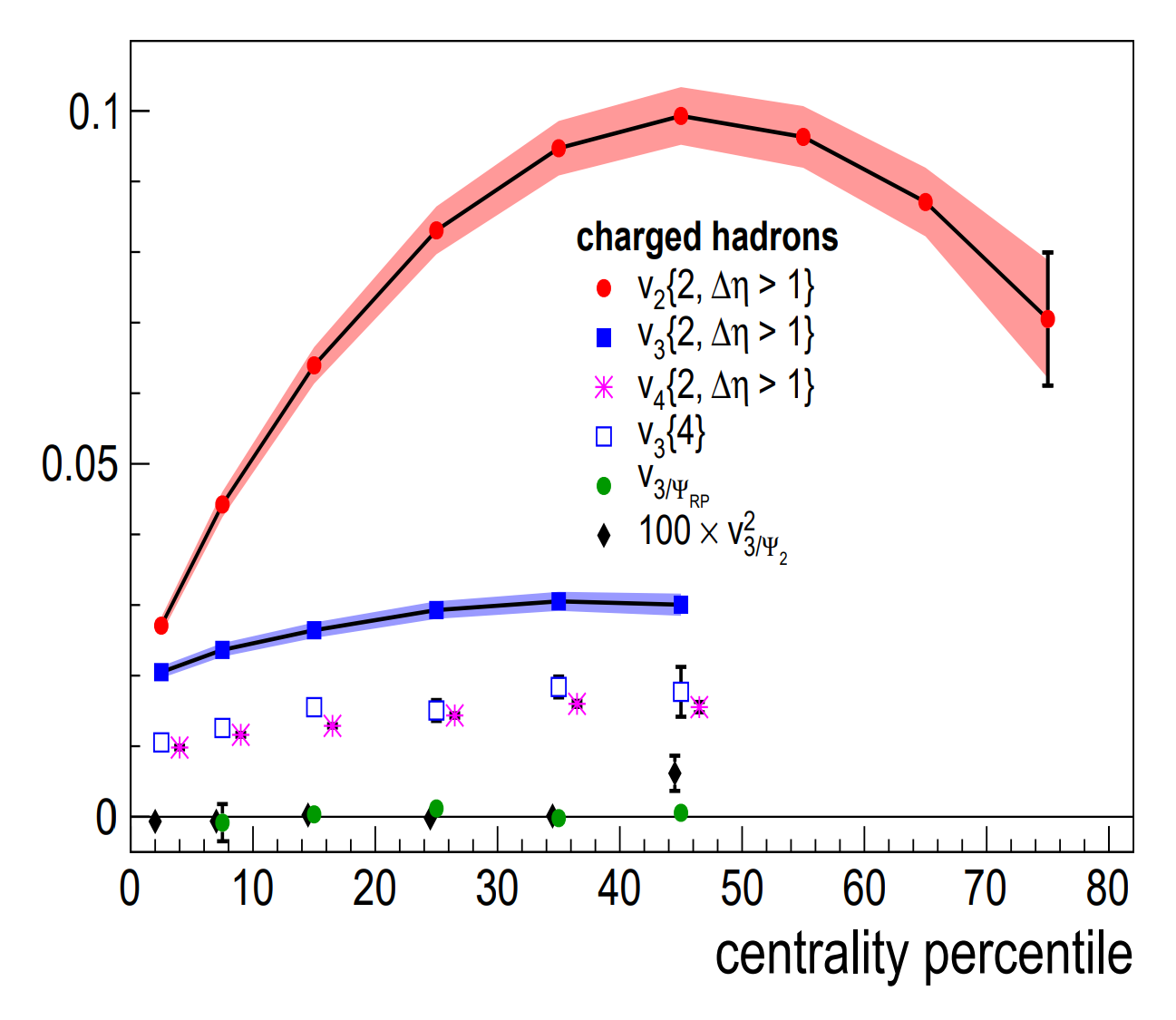}
    \caption{Experimental measurement of flow estimated by multi-particle cumulants for PbPb at $\sqrt{s}=2.76 TeV$. Figure sourced from \cite{Bilandzic:2011ww}.}
    \label{fig:ExperimentalFlow}
\end{figure}

Thanks to linear response, see Sec.~\ref{sec:Eccentricities}, we can compare this result to experimental measurements. In Fig.~\ref{fig:ExperimentalFlow}, the estimated flow using multi-particle cumulants is shown for the measurement of PbPb at $\sqrt{s}=2.76 TeV$ in ALICE. Of interest to this discussion, are the red circles and the solid blue squares, which correspond to the elliptic and triangular flow, respectively. Whereas the optical Glauber predicted a linear relationship between elliptic geometry and centrality, the experimental data shows that this is generally the case with final state flow until about 40-60\% centrality where the elliptic flow peaks and then is significantly suppressed. Some of this dampening of elliptic flow at high centrality can be described by the breakdown of linear response (Sec.~\ref{sec:Eccentricities}) and the increased influence of viscosity on the system. 

%__________________________________________________________________________
%
%__________________________________________________________________________
%

Up until 2010, triangular flow was largely ignored by the community due to several reasons, but the chief among them was that it had not been measured. This missing of triangular flow came, partially, from the fact that $v_3$ and all higher order $v_n$ had been calculated with respect to the "event plane angle", which was defined as the angle of the second flow harmonic $\psi_2$. Additionally, events were rotated so that all of the event plane angles were aligned. This led to the miscalculation of $v_3$ as consistent with zero. This error was corrected \cite{Alver:2010gr} by using the correct angle $\psi_3$, which is correlated with triangular flow, for the calculation of $v_3$. This resulted in a finite measurement of triangular flow by ALICE in heavy-ion collisions \cite{Bilandzic:2011ww}. This new measurement correlated with the, at the time, recent development in initial state modeling of the inclusion of event-by-event fluctuations, which arise due to the quantum nature of the system \cite{Alver:2010gr, Takahashi:2009na}. This became a necessary component of the initial state, since a smooth initial condition provides no geometrical source for triangular flow (See Sec.~\ref{subsuc:Triangularity}).

\subsection{Nucleon (Position) Fluctuations} \label{sec:Intro:NucleonFluctuations}

Nuclei are quantum systems and, as such, any individual nucleus will have a unique configuration of nucleon positions. The optical Glauber model assumes smooth nucleon densities for the colliding nuclei, which is consistent with an average over many nucleon configurations. Including quantum fluctuations in the initial state can be easily done by constructing configurations of nucleons for each nuclei on an event-by-event basis. This development was made possible due to advances in computing technology and randomized algorithms, specifically Monte-Carlo (MC) algorithms, which makes calculations by randomly sampling a relevant probability distribution. The dependence of sampling on a distribution, made MC algorithms a perfect candidate for relaxing the optical approximation since the Glauber model was built on the probability density of nucleons in the nucleus. 

The new Monte-Carlo Glauber \cite{Miller:2007ri} algorithms sampled nucleon positions, for each colliding nucleus, from the Woods-Saxon distribution and deposited their density in the transverse plane using a gaussian profile. The thickness functions, $T_{A,B}$, now become integrals over the individual nucleon densities. After constructing the projectile and target in this way, the reduced thickness function is calculated by sampling a random impact parameter. In this way, the density profile of each collision will fluctuate on an event-by-event basis. A useful side effect of the MC-Glauber approach, is that observables can be calculated with methods that are more similar to those used in experiment (See Sec.~\ref{Chap:Observables}).

\begin{figure}[ht]
    \centering
    \includegraphics[width=0.6\textwidth]{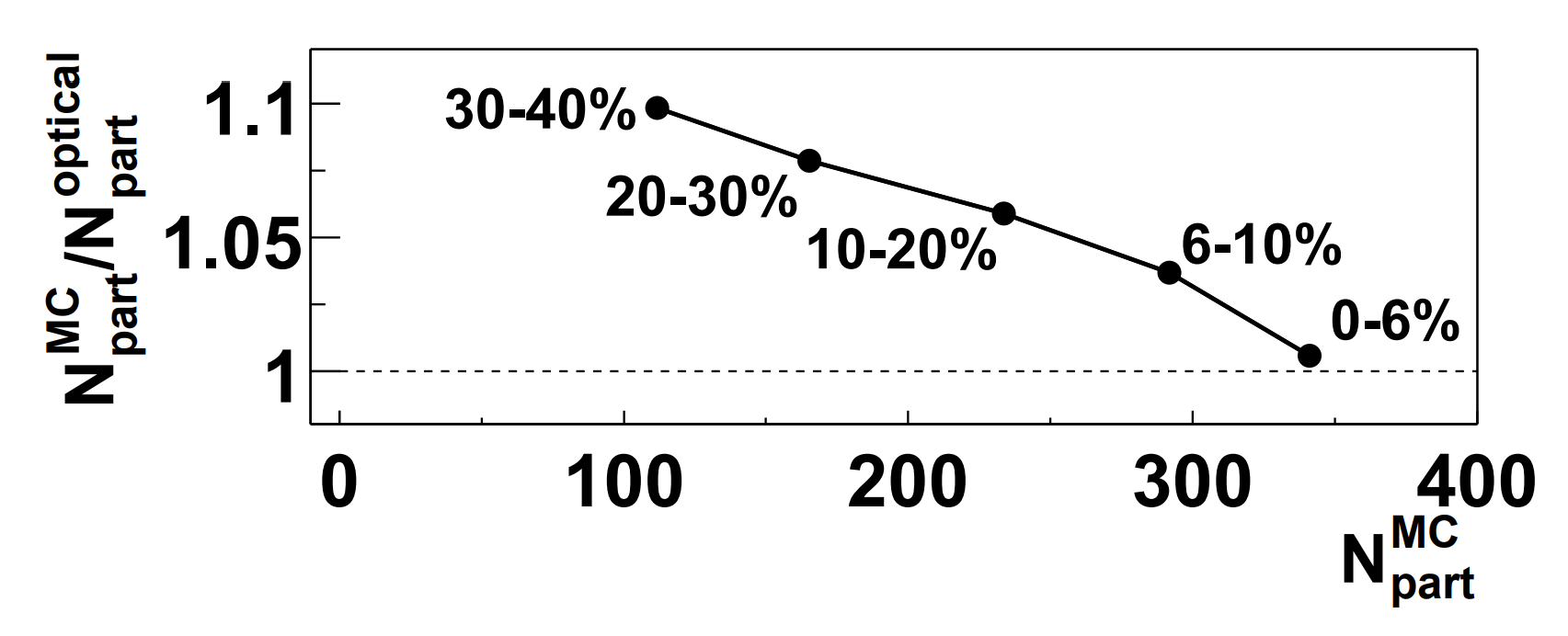}
    \caption{Percent effect that including nucleon fluctuations has on the determination of Number of Participants across centrality classes. System considered here is AuAu at $\sqrt{s_{NN}}=130 GeV$. Figure sourced from \cite{Miller:2007ri}.}
    \label{fig:NumberOfParticipantsNucleonFluctuations}
\end{figure}

We can quantify how good of an approximation the Optical Glauber was by comparing the number of participants, in that model, with those calculated in the MC-Glauber simulations. \footnote{For more details on Optical Glauber and the comparison to Monte-Carlo Glauber see Ref.~\cite{Bilandzic:2011ww}.} The ratio of $N_{part}$ from MC-Glauber to Optical Glauber is plotted in Fig.~\ref{fig:NumberOfParticipantsNucleonFluctuations} with respect to MC-Glauber number of participants. Each point in Fig.~\ref{fig:NumberOfParticipantsNucleonFluctuations} is labled with its corresponding centrality class. We see that for the most central collisions, the Optical Glauber approximation is very close to the number of participants predicted by MC-Glauber. However, as one goes to more peripheral events, we see that the Optical Glauber model under predicts the number of participants by as much as 10\%. 

%__________________________________________________________________________
%
%__________________________________________________________________________
%

\begin{figure}[ht]
   \begin{center}
   \includegraphics[width=0.48\textwidth]{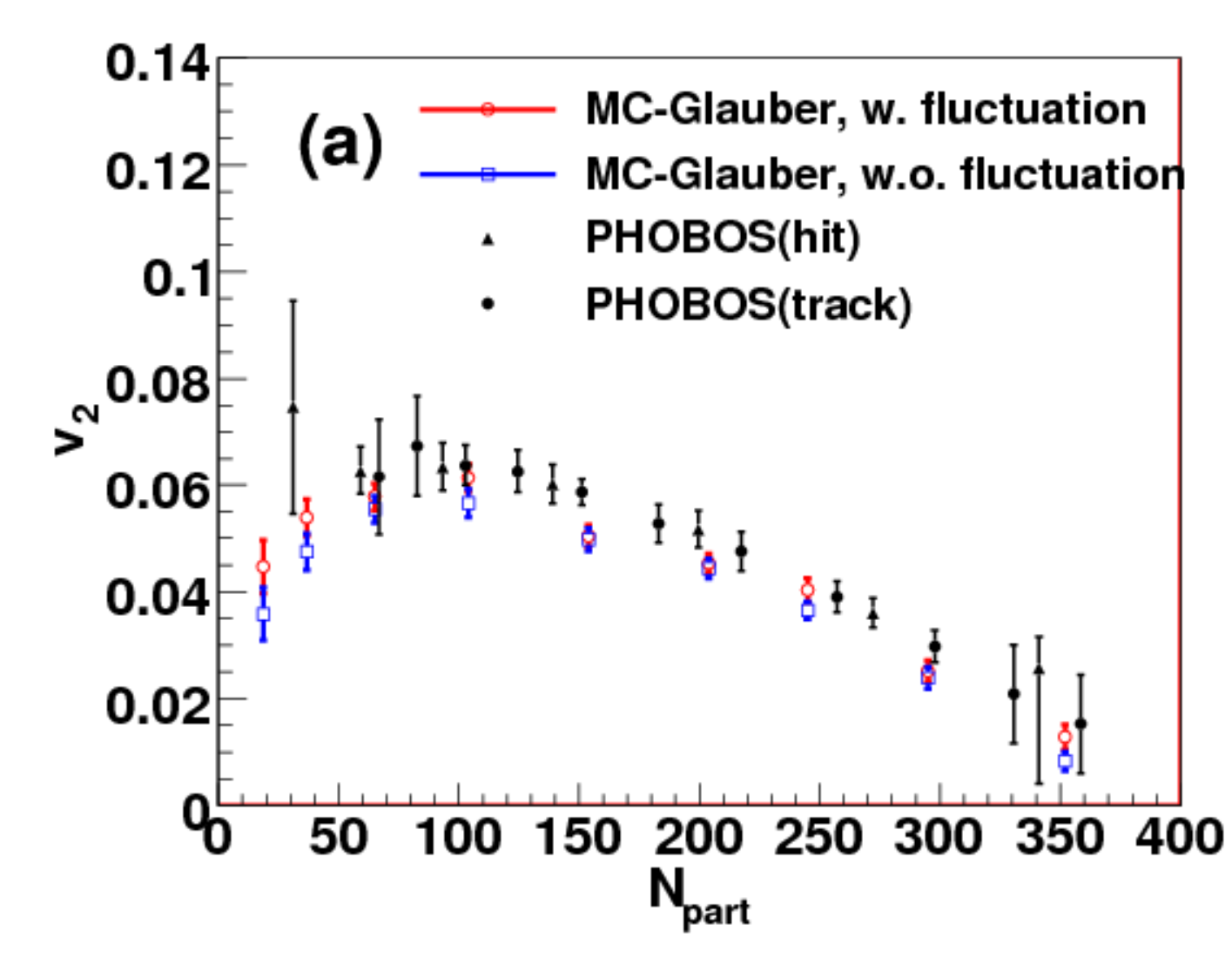}
   \includegraphics[width=0.48\textwidth]{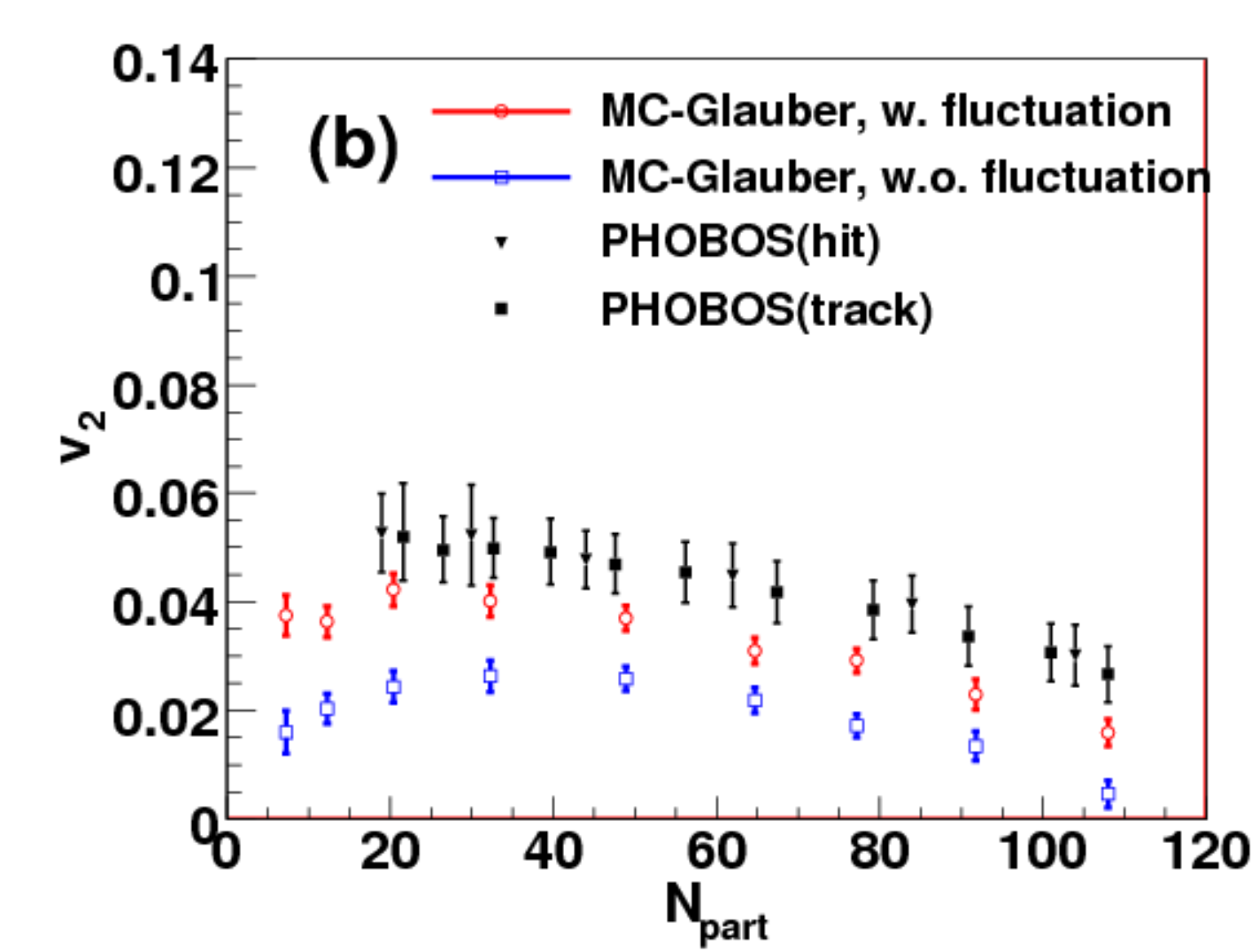}
   \caption{Comparison of $v_2$ for AuAu (A) and CuCu (B), at $\sqrt{s_{NN}}=200 GeV$ using the phenomenological MC-Glauber model. Blue points are generated without fluctuations of nucleon positions, while the red points include these fluctuations. Figures sourced from \cite{Hirano:2009ah} with PHOBOS data from \cite{PHOBOS:2004vcu, PHOBOS:2006dbo}. }
   \label{fig:MCGlauberNucleonFluctuations}
   \end{center}
\end{figure}

Having quantified the effect nucleon fluctuations have on the number of participants, we can look at how the MC-Glauber model affects the geometry of the initial state. In Fig.~\ref{fig:MCGlauberNucleonFluctuations}, the elliptic flow with respect $N_{part}$ is calculated using MC-Glauber with and without nucleon fluctuations and compared to measurements from PHOBOS for AuAu and CuCu at $\sqrt{s_{NN}}=200 GeV$ \cite{PHOBOS:2004vcu, PHOBOS:2006dbo}. Looking at the large system of AuAu, we see that for both models, there is a peak in $v_2$, around the same spot as experiment, and a suppression in more peripheral collisions. There is also a small enhancement of elliptic flow, with the inclusion of nucleon fluctuations, across all $N_{part}$, the most significant increase coming from peripheral collisions. Both models are able to match experimental data for Au+Au though Optical Glauber under predicts the elliptic flow in peripheral events more than MC-Glauber. The real importance of including nucleon fluctuations is seen in the elliptic flow of the smaller system of CuCu. We see a significant enhancement in elliptic flow for MC-Glauber with fluctuations, even changing the behavior of the most peripheral points by leveling out the slope. This behavior indicates that nucleon fluctuations are most important for describing peripheral events and small systems.

%__________________________________________________________________________
%
%__________________________________________________________________________
%

\begin{figure}[ht]
    \begin{center}
    \includegraphics[width=0.48\textwidth]{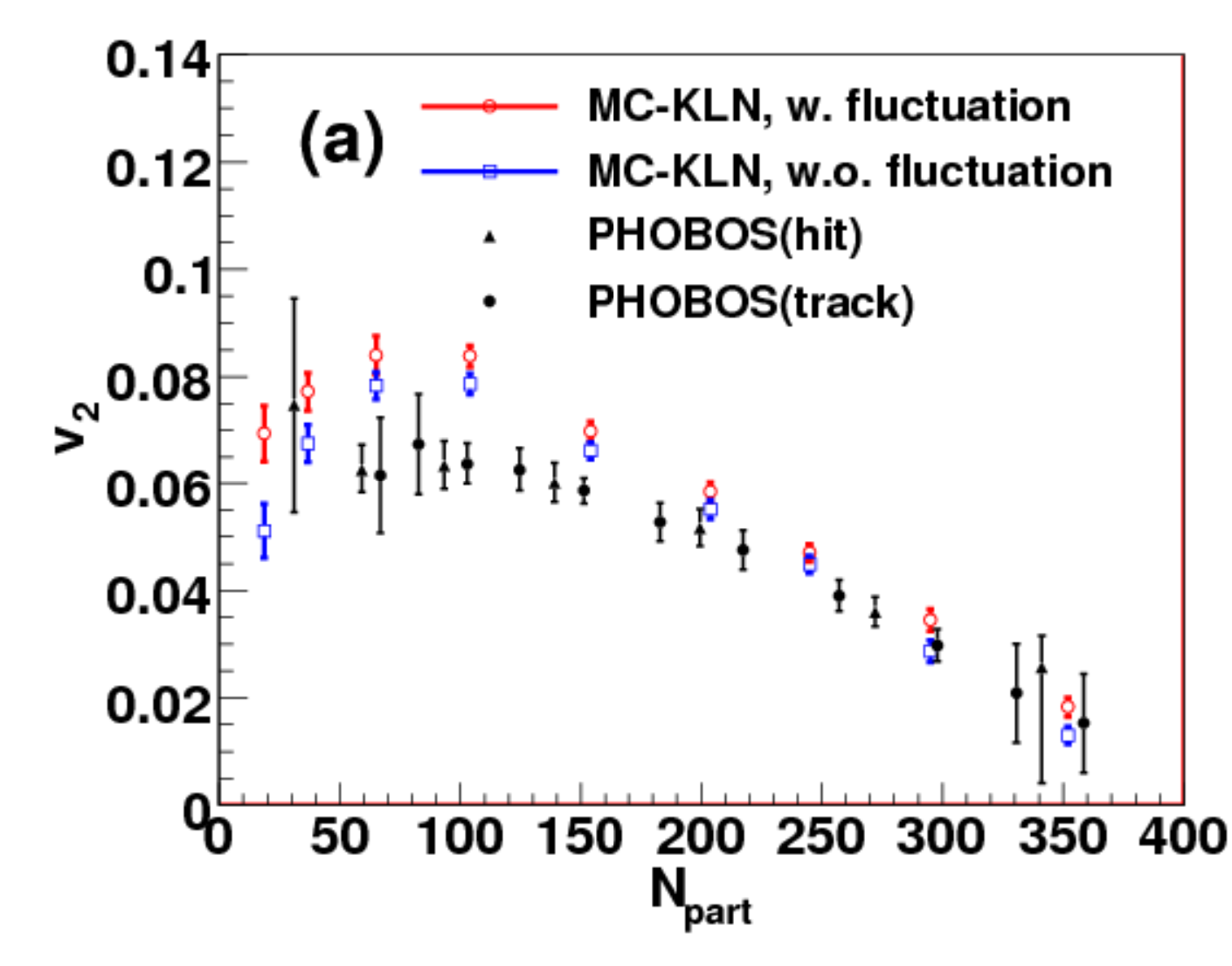}
    \includegraphics[width=0.48\textwidth]{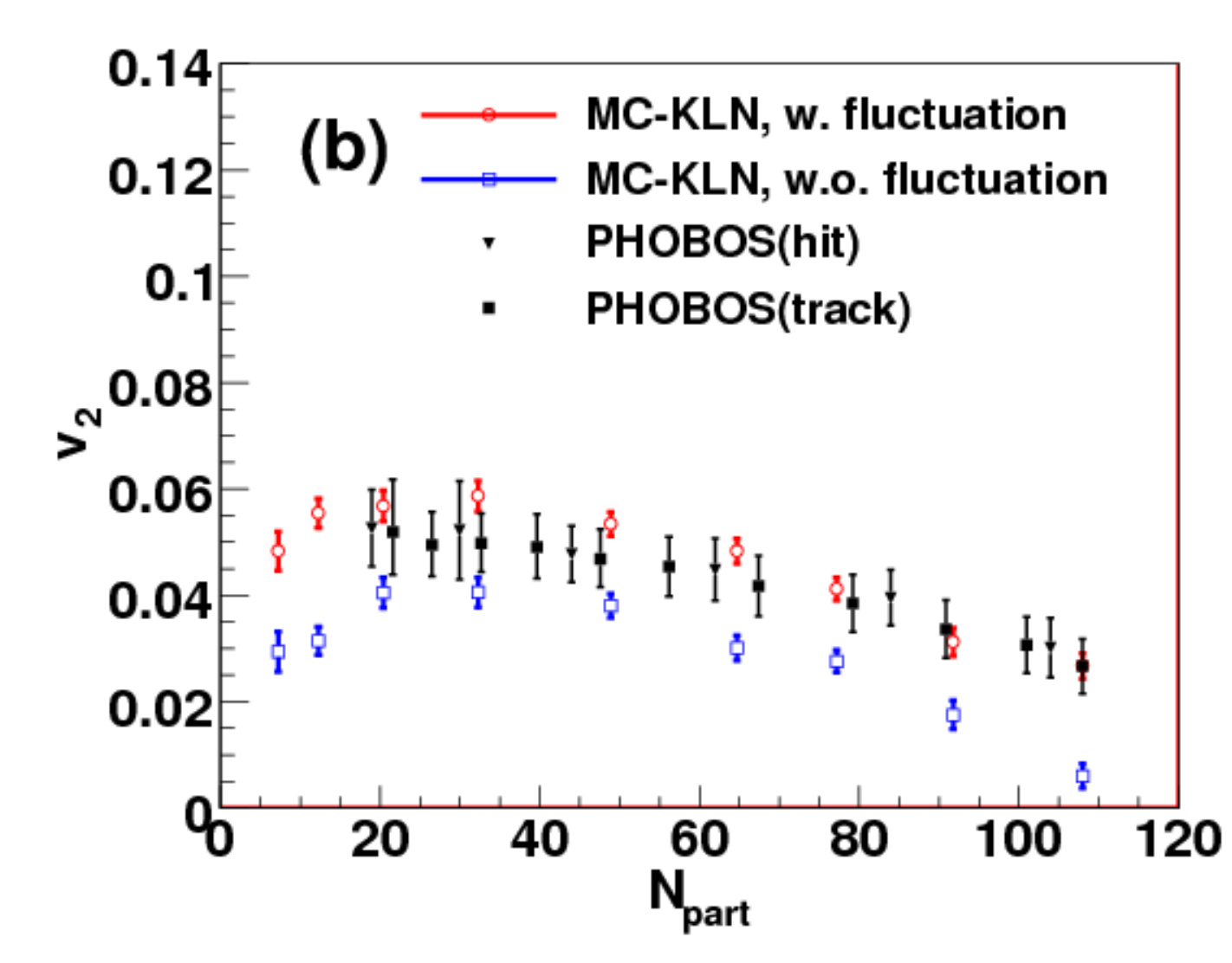}
    \caption{Comparison of $v_2$ for AuAu (A) and CuCu (B), at $\sqrt{s}=200 GeV$ using the CGC MC-KLN model. Blue points are generated without fluctuations of nucleon positions, while the red points include these fluctuations. Figures sourced from \cite{Hirano:2009ah} with PHOBOS data from \cite{PHOBOS:2004vcu, PHOBOS:2006dbo}.}
    \label{fig:MCKLNNucleonFluctuations}
    \end{center}
\end{figure}

We can also look at the effect of nucleon fluctuations in a CGC model with the Monte-Carlo version of fKLN \cite{Drescher:2006ca}. In Fig.~\ref{fig:MCKLNNucleonFluctuations}, we see the same behavior as MC-Glauber from Fig.~\ref{fig:MCGlauberNucleonFluctuations}, where including nucleon fluctuations increases elliptic flow with the most significant areas being peripheral collisions of AuAu and the smaller system of CuCu. Comparing both versions of MC-KLN with MC-Glauber, we see that the CGC models consistently predict larger elliptic flows. This means a larger shear viscosity is required for CGC-based initial states to reproduce the same elliptic flow as Glauber based approaches \cite{Song:2010mg}. 
%__________________________________________________________________________
%
%__________________________________________________________________________
%

\begin{figure}[ht]
    \centering
    \includegraphics[width=\textwidth]{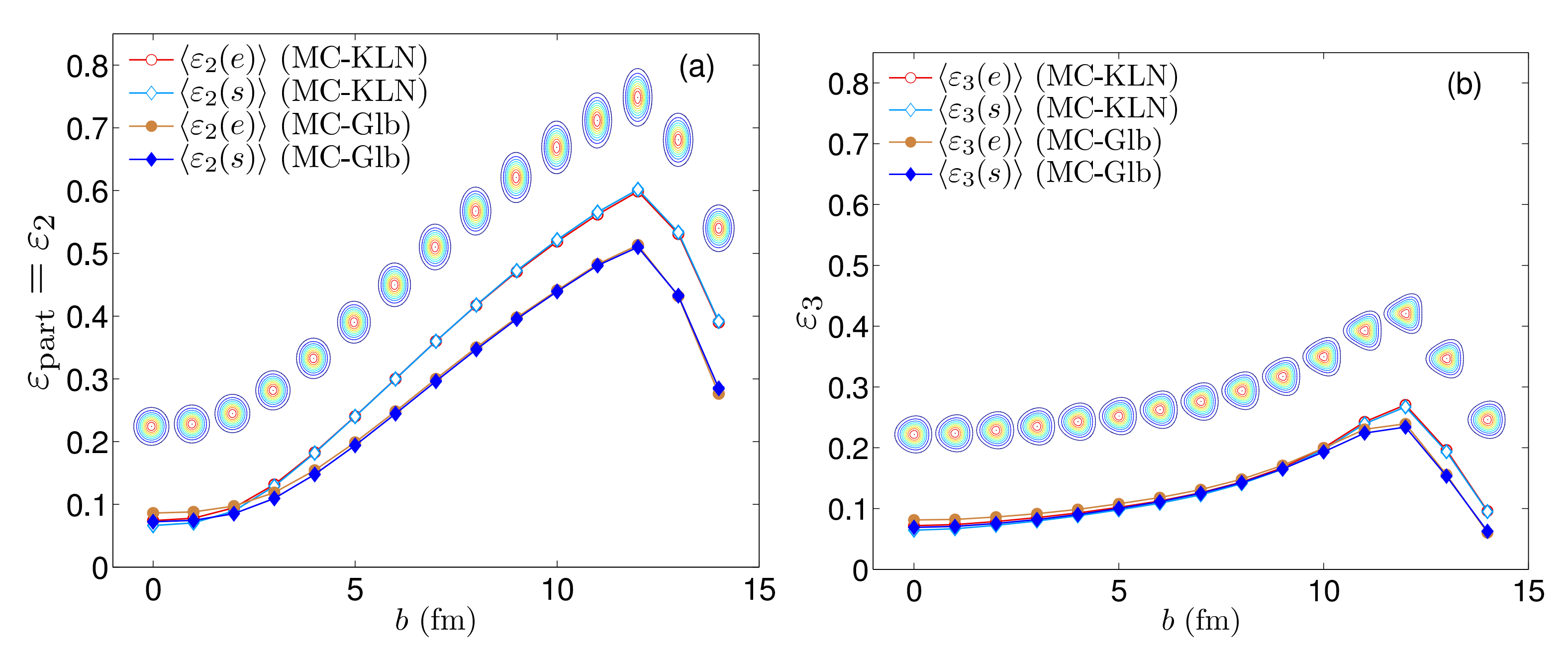}
    \caption{Comparison of elliptic and triangular geometry between MC-Glauber and MC-KLN initial state models with respect to impact parameter. These eccentricities are calculated using the energy (e) and entropy (s) densities. Density profiles displayed with the observables are simple deformed Gaussian profiles. System considered here is AuAu at $\sqrt{s_{NN}}=200 GeV$. Figure sourced from \cite{Qiu:2011iv}.}
    \label{fig:TriangularityFromNucleonFluctuations}
\end{figure}

The main benefit of Monte-Carlo initial state models is that the inclusion of nucleon fluctuations produces finite triangularity. This is well illustrated in Fig.~\ref{fig:TriangularityFromNucleonFluctuations} (b), where the third order eccentricity is plotted from both the MC-Glauber and MC-KLN models versus impact parameter. Comparing to the ellipticity in Fig.~\ref{fig:TriangularityFromNucleonFluctuations} (a), we see that $\varepsilon_3$ follows a similar trend to $\varepsilon_2$, with a peak and suppression in more peripheral collisions, but is much flatter and of an overall lower magnitude. The most central collisions are interesting since it seems like $\varepsilon_2 \approx \varepsilon_3$ (See Chap.~\ref{chap:V2toV3Puzzle}). The eccentricities plotted in Fig.~\ref{fig:TriangularityFromNucleonFluctuations} with respect to the energy (circles) and entropy (diamonds) densities of the initial state. We see that eccentricity is independent of the choice of distribution, energy or entropy, but different production mechanisms can have a significant effect.
%__________________________________________________________________________
%
%__________________________________________________________________________
%

The difference between Optical and Monte-Carlo Glauber models, boils down to when an average is taken. Due to the quantum nature of heavy-ion collisions and the limits of experimental accessibility, it is necessary to make calculations across many collision events and observe the average behavior. The Optical Glauber model takes this average individually over the colliding nuclei before the collision occurs, while the MC-Glauber model takes the average after the fluctuation nuclei collide. This translates to a sensitivity to different structures in the initial state, with MC-Glauber being sensitive to event-by-event fluctuations and Optical Glauber only capturing impact parameter dependent structure.

Despite the improvements from including nucleon fluctuations, there were still open problems concerning the initial state. One of these was an inability of models to get the multiplicity distributions of smaller systems, specifically dAu (See Sec.~\ref{sec:Intro:MultiplicityFluctuations}). Additionally, the fluctuations from these Monte-Carlo simulations were not enough to fully explain measurements of $v_n$ on an event-by-event basis \cite{Renk:2014jja,Giacalone:2017uqx}. This was a particular problem in ultra-central and peripheral collisions.

\subsection{Nucleon (Multiplicity) Fluctuations} \label{sec:Intro:MultiplicityFluctuations}

While the geometry of the initial state was influencing the addition of new physics processes to models, other issues were developing. The most important of these was the inability of models to reproduce the multiplicity distributions of experimental measurements, specifically in small deformed systems. Here, multiplicity distributions describe the distribution of events with respect to the number of charged particles produced by the collision. This distribution is used to define centrality classes, since the impact parameter of the collision is inherently tied to the number of charged particles produced. 

\begin{figure}[ht]
    \includegraphics[width=\textwidth]{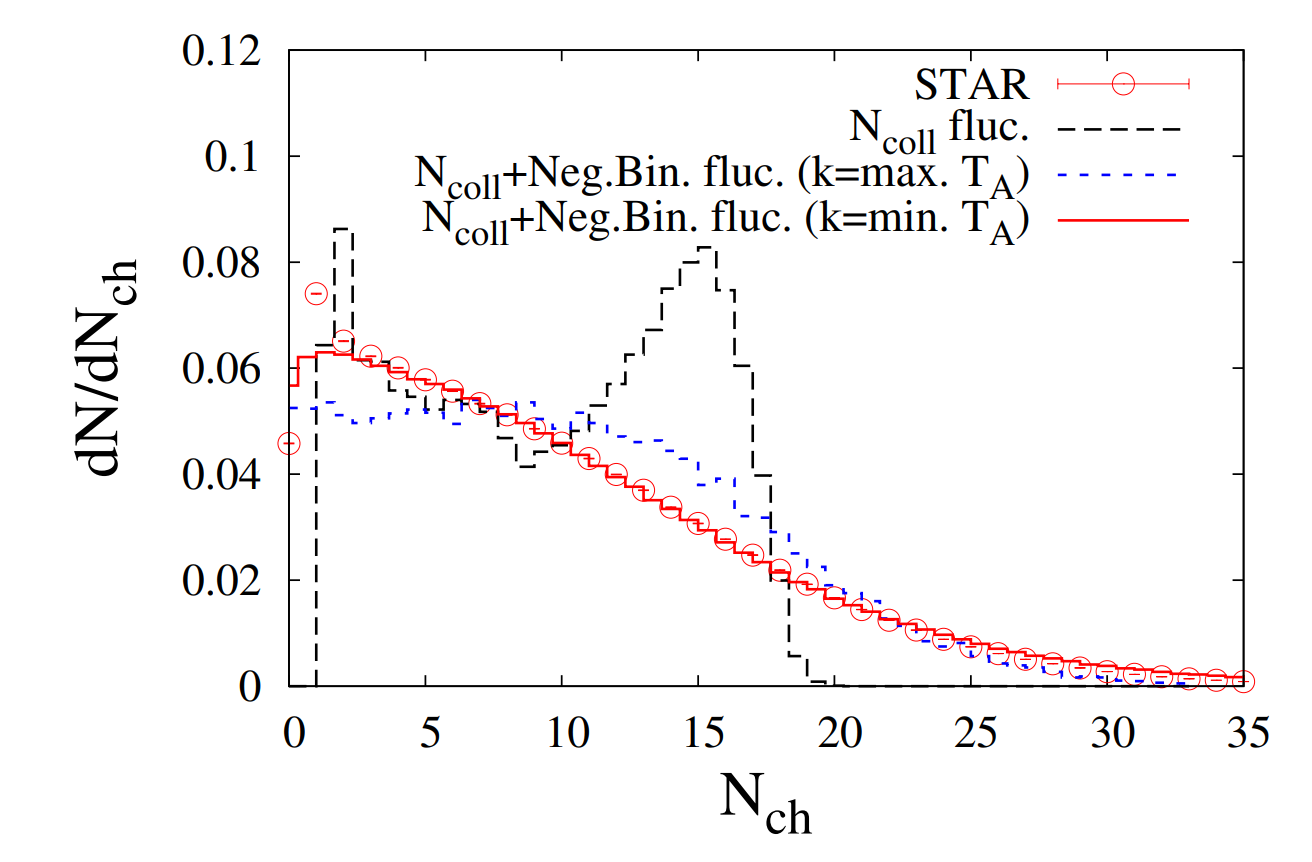}
    \caption{Including multiplicity fluctuations fixes disagreement between theory predictions for dAu, at $\sqrt{s_{NN}}=200 GeV$, multiplicity distributions. A Negative Binomial Distribution was used for the multiplicity fluctuations here. Figure sourced from \cite{Dumitru:2012yr}. See Sec.~\ref{chap:ExplorationOfMultiplicityFluctuations} for a further exploration.}
    \label{fig:MultiplicityResponseMultiplicityFluctuations}
\end{figure}

The most striking example of this multiplicity distribution failure is found in examining collisions of deuteron and gold, dAu. The gold ion acts as a mostly shperical background on which the highly elliptic deuteron impacts. In Monte-Carlo initial state models, the multiplicity distribution of this system, illustrated in Fig.~\ref{fig:MultiplicityResponseMultiplicityFluctuations}, contains two peaks, one at low and another at high multiplicity. The low-side peak corresponds to only one of the nucleons of deuteron colliding with the Au ion, while the high-side peak signifies participation from both nucleons. We can compare this distribution with experimental measurements from STAR, in Fig.~\ref{fig:MultiplicityResponseMultiplicityFluctuations} (red circles), and see that the Monte-Carlo approach with fluctuations of nucleons (black dashed line) is unable to match the smooth experimental data and does not capture the high multiplicity tail. Nucleon fluctuations introduced a granularity in the initial state that is inconsistent with reality.

%__________________________________________________________________________
%
%__________________________________________________________________________
%

Revisiting the process of the MC-Glauber model, the profile of nucleons is assumed to be Gaussian and be normalized to one. This introduced a granularity to the initial state, that wasn't present in Optical Glauber, and appears inconsistent with experimental measurements. By introducing fluctuations of the normalization of each nucleon, the peaks in the multiplicity distribution of dAu can be softened. We could also use these fluctuations to address the high multiplicity tail by allowing for large fluctuations in the nucleon weight. This new type of fluctuation is called multiplicity fluctuations. 

We can look to the CGC-based implementation of multiplicity fluctuations to provide some physical intuition. The core assumption of CGC models is that nucleons become saturated by gluons at high energies and can be described by a local gluon density with high occupation. We can relax this assumption by allowing the density of gluons to fluctuate. This casts fluctuations in multiplicity in terms of the fluctuating gluon density of individual nucleons. While this is a useful framing device, the source of multiplicity fluctuations is still unknown and is modeled with an assumed spectrum of fluctuations. 

The question now becomes how we implement these multiplicity fluctuations. The choice of distribution with which to sample these fluctuations is somewhat arbitrary (See Chap.~\ref{chap:ExplorationOfMultiplicityFluctuations}) as long as the maximum is centered on 1 and there are large positive fluctuations. In Ref.~\cite{Dumitru:2012yr}, they use a negative binomial distribution (NBD) to describe these fluctuations, defined as:
\begin{align} \label{eq:NBD}
    P_{NBD}(\omega,k) = \frac{\Gamma(k+\omega)}{\Gamma(k)\gamma(\omega+1)} \frac{\bar{\omega}^\omega k^k}{(\bar{\omega}+k)^{\bar{\omega}+k}} ,
\end{align}
where $\bar{\omega}$ is the mean multiplicity in a given region and k is the fluctuation parameter. They justified this choice citing the success of its application in describing the multiplicity distributions of proton anti-proton collisions \cite{UA5:1985hzd}. This, in turn, was applied to $p \bar{p}$ collisions due to its usefulness in particle physics. 

How multiplicity fluctuations, sampled from an NBD, affect the multiplicity distribution of dAu is illustrated in Fig.~\ref{fig:MultiplicityResponseMultiplicityFluctuations}. The $k$-parameter controls the amount of fluctuations, where a small value corresponds to many while a large value means fewer fluctuations. We see that when $k$ is minimized (red), or fluctuations are maximized, the model fits the experimental multiplicity distribution. When $k$ is maximized (blue), we see that fewer fluctuations are included and the multiplicity distribution is somewhere between the model with no multiplicity fluctuations and the one with them maximized.

\begin{figure}[ht]
    \includegraphics[width=\textwidth]{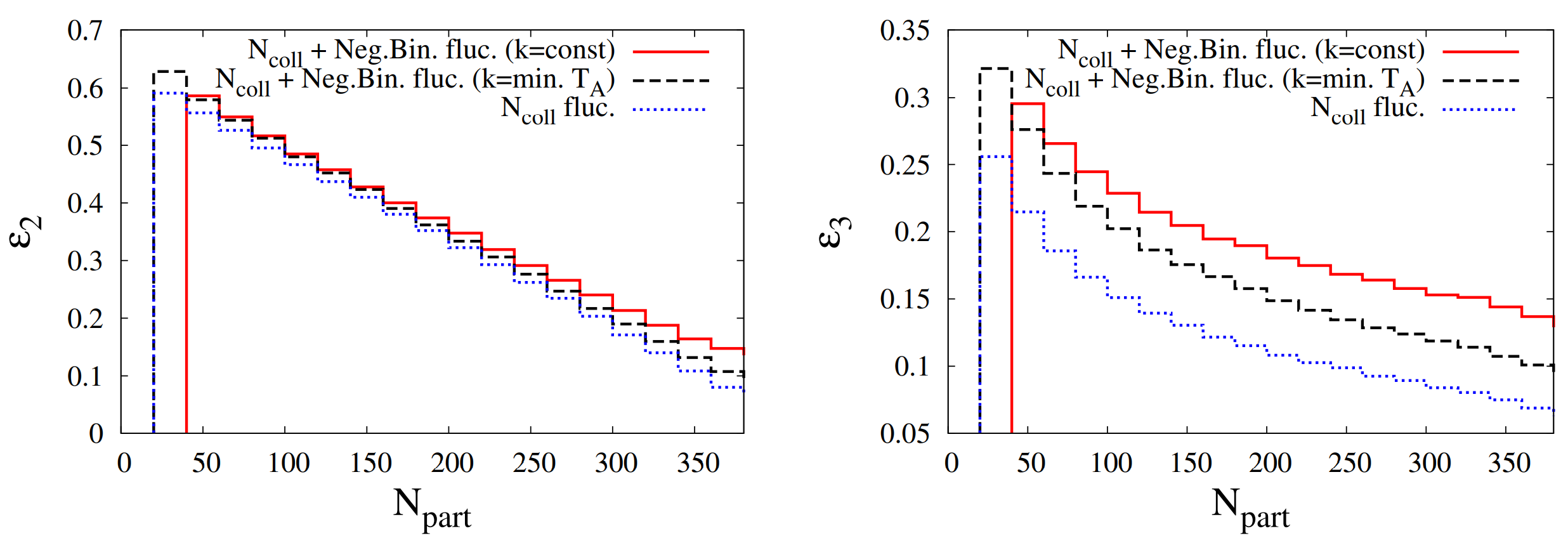}
    \caption{Response of initial state eccentricities to the inclusion of multiplicity fluctuations. System considered here is AuAu at $\sqrt{s_{NN}}=200 GeV$.  A Negative Binomial Distribution was used for the multiplicity fluctuations here. Figure sourced from \cite{Dumitru:2012yr}.}
    \label{fig:EccentricityResponseMultiplicityFluctuations}
\end{figure}

Now lets see how these new fluctuations change the initial state geometry. The elliptic and triangular geometry are plotted in Fig.~\ref{fig:EccentricityResponseMultiplicityFluctuations}, left and right respectively, using CGC-based Monte-Carlo including fluctuations in nucleon configurations (blue) and multiplicity fluctuations (black and red). For the ellipticity, we see that there is very little effect from the multiplicity fluctuations except in the most central collisions. Looking at the triangularity, we see a significant enhancement for all collisions when increasing multiplicity fluctuations.

%__________________________________________________________________________
%
%__________________________________________________________________________
%

\begin{figure}[ht]
    \includegraphics[width=\textwidth]{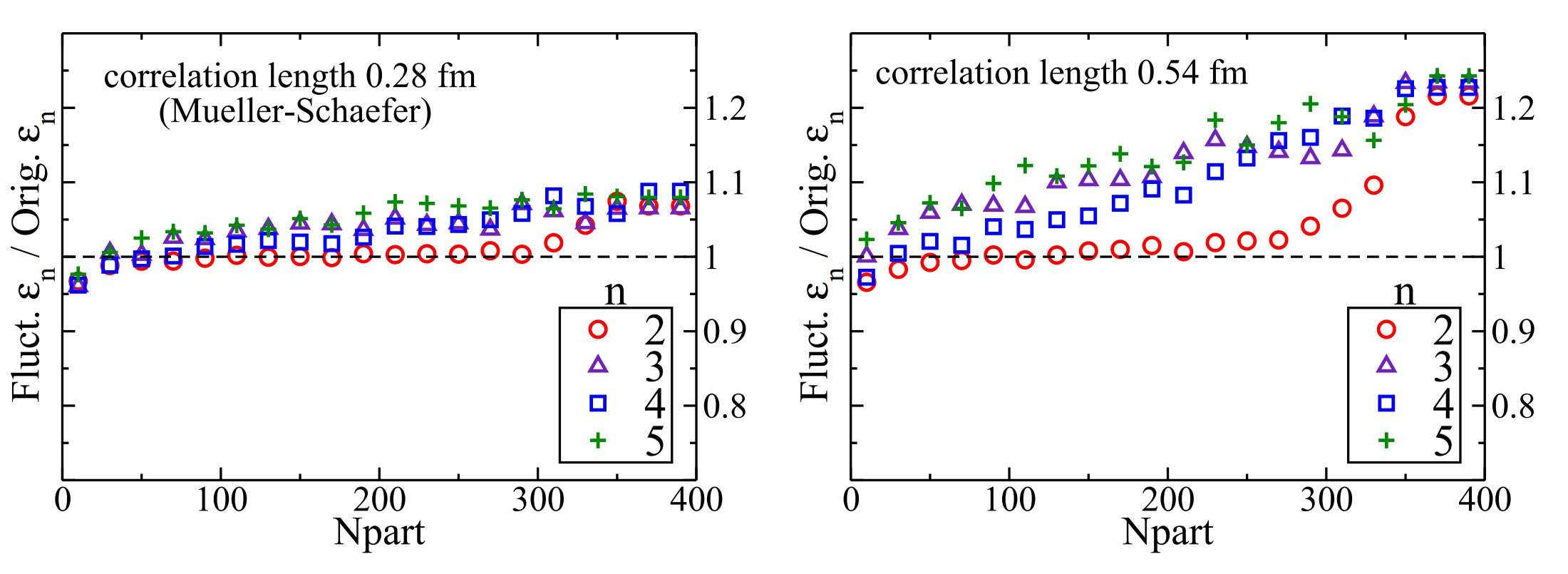}
    \caption{Percent response of initial state eccentricities to the inclusion of multiplicity fluctuations. System considered here is AuAu at $\sqrt{s_{NN}}=200 GeV$. This was generated using a modified version of MC-KLN with a Negative Binomial Distribution for the multiplicity fluctuations. Figure sourced from \cite{Moreland:2012qw}.}
    \label{fig:PercentEccentricityResponseMultiplicityFluctuations}
\end{figure}

In Ref.~\cite{Moreland:2012qw}, the authors better quantified the effect of multiplicity fluctuations on initial state geometry. Here they used the MC-KLN model and combined it with a Gaussian random field that fluctuated with an NBD. This fluctuating background is parameterized by a correlation length, dependent on the saturation scale, which controls the transverse energy density fluctuations of gluon fields. An approximate lower bound on the correlation length, 0.28, is derived from AuAu collisions at $\sqrt{s_{NN}}=200 GeV$. An upper bound is estimated by allowing the correlation length to be of the same order as the nucleon radius, 0.54. The ratio of fluctuated-to-original geometry is shown in Fig.~\ref{fig:PercentEccentricityResponseMultiplicityFluctuations}, with the lower bound on fluctuations on the left and the upper bound on the right. For the lower bound, we observe that the elliptic geometry is only affected in central collisions and sees an enhancement of $\approx10\%$, while the higher order geometries see modification across centralities with a maximum effect of the same order at central collisions. For the higher bound, we see the effect increase in central collisions to $\approx20\%$ with a significant enhancement for higher order eccentricities. 

%__________________________________________________________________________
%
%__________________________________________________________________________
%

There have been other implementations of multiplicity fluctuations across initial state models. An early adopter was the GLauber Initial-State Simulation AND mOre (GLISSANDO) model \cite{Broniowski:2007nz}, which provided options for fluctuations sampled from a Poisson,
\begin{align} \label{eq:Poisson}
    P_{Poisson}(\omega,k) = \frac{k^{\omega k}e^{-k}}{(\omega k)!},
\end{align}
or $\Gamma$,
\begin{align} \label{eq:Gamma}
    P_{\Gamma}(\omega,k) = \frac{k^k \omega^{k-1} e^{-k\omega}}{\Gamma(k)},
\end{align}
distribution. The Poisson distribution generates discrete weights 
\begin{align} 
    \omega = 0, \, \frac{1}{k} \, \frac{2}{k} \, \frac{3}{k} ... \,,
\end{align}
while the $\Gamma$ distribution generates continuous weights 
\begin{align}
    \omega \in [0,\infty) ,
\end{align}
both with $\langle \omega \rangle =1$ and $\sigma(\omega) = 1\/\sqrt{k}$.
In Ref.~\cite{Bozek:2013uha}, they looked at how well these two distributions were able to match the multiplicity distribution of pPb, as measured by CMS \cite{CMS:2012qk}, and found that the Poisson fluctuations were insufficient while the $\Gamma$ ones were adequate. They also provided some rationale to the jump from NBD to $\Gamma$ fluctuations: \textit{"the observed multiplicity distributions can be described as convolution of the number of participant nucleons and a negative binomial distribution. At the stage of the formation of the initial fireball it is equivalent to imposing fluctuations of the entropy deposited per participant nucleon following the $\Gamma$ distribution"} \cite{Bozek:2013uha}. In Ref.~\cite{Rogly:2018ddx}, the authors argue that the $\Gamma$ distribution is essentially a continuous version of the negative binomial distribution when the mean value is much larger that one.

The first use of multiplicity fluctuations, described as $Q_s$ fluctuations, in CGC models was in Ref.~\cite{ McLerran:2015lta} using a log-normal distribution:
\begin{equation} 
    \label{eq:lognormflucs}
    P_{log-normal}(\omega,k)=\frac{2}{\omega k\sqrt{2\pi}}e^{-\frac{\ln^{2}(\omega^{2})}{2k^{2}}}.
\end{equation}
The motivation for using the log-normal distribution is cited to be Ref.~\cite{Iancu:2004es}, which applied the distribution in a different context and explicitly state that the choice is an unproven conjecture. This justification is similar to that for the NBD used in Ref.~\cite{Dumitru:2012yr}, which underscores our lack of knowledge concerning the correct distribution\footnote{This is, indeed, one of the many coxambre of the field!}.

Multiplicity fluctuations are most important in central collisions, where they have the most effect on the geometry, and for higher order geometries. This brings up an interesting observation, which I will explore in more depth in Chap.~\ref{chap:V2toV3Puzzle}, that central collisions are the most sensitive to the smallest order fluctuations in the initial state.

\section{Modern Initial State Models} \label{sec:Intro:ModernInitialStates}

\begin{figure}[ht!]
    \includegraphics[width=\textwidth]{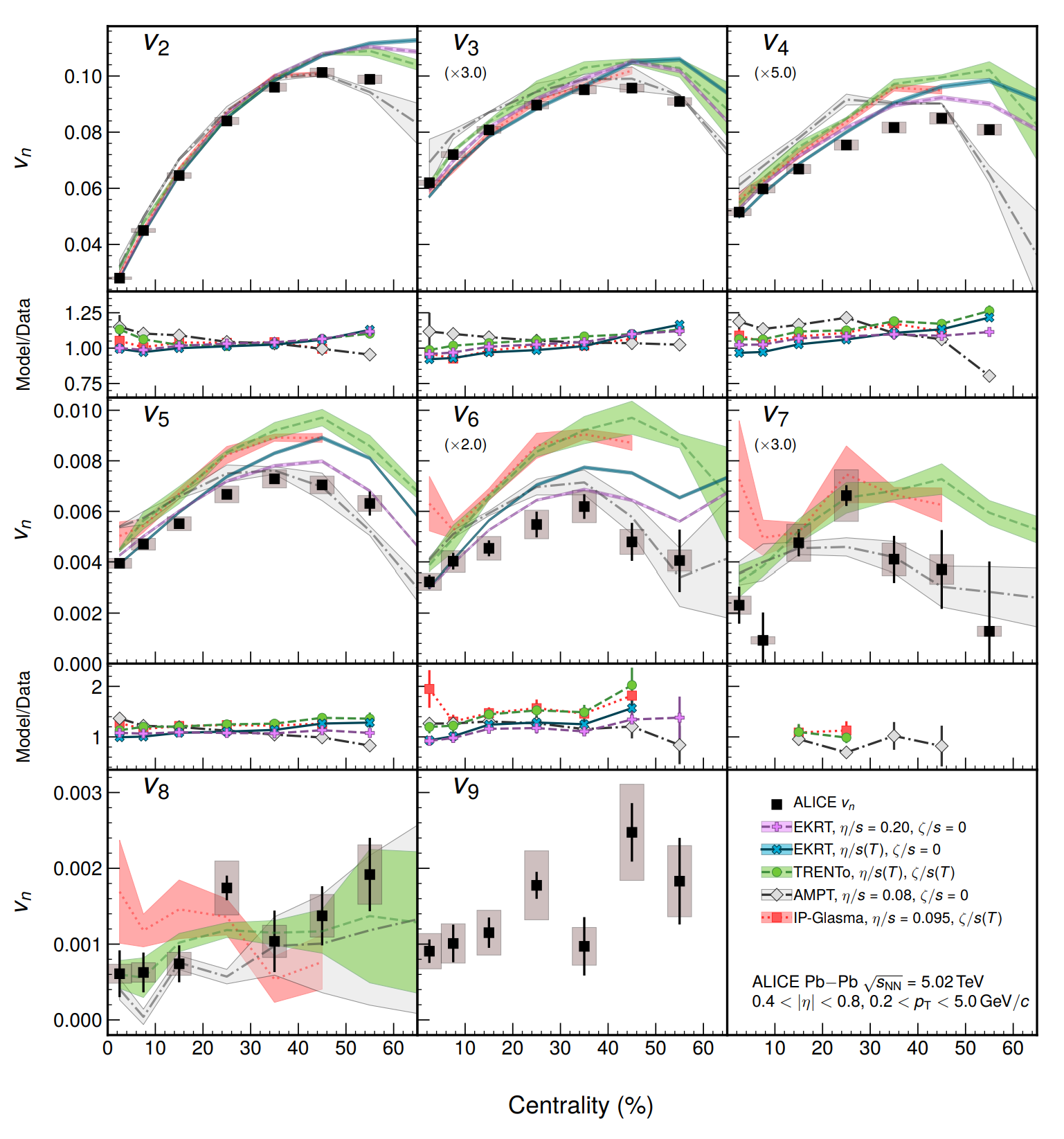}
    \caption{Modern comparison of many flow harmonics from experiment with state-of-the-art initial state models. Figure sourced from \cite{Parkkila:2020mgo}.}
    \label{fig:ExperimentalVsModelFlowHarmonics_Current}
\end{figure}

Having followed the development of initial state models, we now come to modern approaches in simulating the initial collision of heavy ion collisions. Parallel to the developments made in modeling, experimental methods and data collection has greatly advanced with measurements now spanning flow harmonics up to order $n=9$, as shown in Fig.~\ref{fig:ExperimentalVsModelFlowHarmonics_Current}. The final state flow, in Fig.~\ref{fig:ExperimentalVsModelFlowHarmonics_Current}, is compared to results from several of the current leaders in initial state modeling. Important to note, is that flow from models is dependent on the medium parameters, such as shear viscosity, which accounts for differences between different initial conditions (specifically the two EKRT curves; one with a fixed shear viscosity and the other with a temperature dependent shear viscosity). In the subsequent sections, I will focus on two of the models here, \code{trento} \cite{Moreland:2014oya} and IP-Glasma \cite{Schenke:2012wb}, that are the modern extensions of phenomenological and CGC-based approaches, respectively. A brief overview of other popular initial state models is also provided.

%__________________________________________________________________________
%
%__________________________________________________________________________
%

\subsection{\code{trento}} \label{sec:Trento}

The successor to MC-Glauber is the \code{trento} model introduced in Ref.~\cite{Moreland:2014oya}. This initial condition generator is built similarly to the MC-Glauber model, with all fluctuation sources detailed above, but with more general forms for the different physics components allowing for a more model agnostic construction. In addition to the detailed parameterization, \code{trento} is also highly optimized and computationally cheap compared to other initial state generators. I will detail the broad strokes of the process implemented in \code{trento}, how it diverges from the previous generations, and the additions I have made for use in this work. 

Like MC-Glauber, \code{trento} starts by specifying the thickness functions of the projectile and target by sampling nucleon positions from a parameterized Woods-Saxon profile, Eq.~\ref{eq:WoodsSaxon}, and depositing them as Gaussian blobs of entropy. This Woods-Saxon profile is generalized to allow for deformations (See Eq.~\ref{e:Rdef}).

Part of the motivation behind the creation of \code{trento}, were the, then, recent measurements from RHIC \cite{Pandit:2013uiv,Wang:2014qxa} of ultra-central U+U collisions, which showed that the multiplicity does not scale with number of binary collisions. This excluded the two-component Glauber method as a valid description of the initial state \cite{Goldschmidt:2015qya}. This inspired the creators of \code{trento} to require that initial state entropy should be scale-invariant,
\begin{align} \label{eq:ThicknessFunctionScaleInvariance}
    f(cT_A, cT_B) = cf(T_A,T_B),
\end{align}
which is mathematically equivalent to saying that N nucleon-nucleon collisions should produce the same entropy as one N-N collision.

The functional form of the reduced thickness is, thus, parameterized using the generalized mean,
\begin{align}   \label{e:genmean}
    s (T_A , T_B) \propto \left( \frac{T_A^p + T_B^p}{2} \right)^{1/p} ,
\end{align}
where $p$ is any real number, controlling the entropy production for the collision, and the resulting function is ensured to be scale-invariant. While $p$ is a continuous parameter, special values are: $p=-1$ corresponding to the arithmetic mean, $p=0$ or geometric mean, and $p=1$ being the harmonic mean. A Bayesian analysis was been done with \code{trento} \cite{Bernhard:2016tnd}, thanks to its low run time, and found that the preferred reduced thickness function is given by $p=0$,
\begin{align}
    T_R (\vec{x}_\bot) \overset{(p=0)}{=} \sqrt{T_A (\vec{x}_\bot) \, T_B (\vec{x}_\bot)} .
\end{align}
The reduced thickness profile is assumed to be proportional to the entropy density such that the relationship can be parameterized as,
\begin{align} \label{e:Trentomodel}
s(\vec{x}_\bot) &= a_{entropy} \, T_R (\vec{x}_\bot) ,
\end{align}
where the proportionality factor, with units of $fm^{-1}$, must be tuned for each collision system. For PbPb collisions at $5.02 TeV$, the value of $a$ was found to be $~120 fm^{-1}$ \cite{Bernhard:2016tnd}, which is also consistent with $119 fm^{-1}$ used in Ref.~\cite{Alba:2017hhe} for a more specific parameter set. In practice, this proportionality factor can change significantly with respect to beam energy, system size, and medium parameters.  

Multiplicity fluctuations are implemented by adding a weight factor to the definition of the projectile and target thickness functions,
\begin{align} \label{eq:FluctuatingThicknessFunction}
    T_{A,B} (x,y) = \omega_{A,B} \int \rho_{A,B} (x,y,z) dz,
\end{align}
where $\omega$ is randomly sampled from a $\Gamma$ distribution, Eq.~\ref{eq:Gamma}. The use of the $\Gamma$ distribution for multiplicity fluctuations is likely due to its popularization from the GLISSANDO model implementation, as mentioned above. The Bayesian analysis \cite{Bernhard:2016tnd} found $k=1.6$ to best fit data.

The Bayesian analysis, done in Ref.~\cite{Bernhard:2016tnd}, established \code{trento} as a leading initial state model. The parameterization coming from this analysis, has been successful in several predictions across system size, with one being Ref.~\cite{Giacalone:2017dud} which looked at XeXe collisions. Since, more Bayesian analyses have been done with \code{trento} due to its highly parametric construction. The JETSCAPE, in Ref.~\cite{Everett:2020xug}, and Trajectum collaborations, in Refs.~\cite{Nijs:2020roc,JETSCAPE:2020shq,JETSCAPE:2020mzn}, used \code{trento} as the initial state for full collision simulations, and extracted transport coefficients using Bayesian analyses. \code{trento} has been extensively used within the field to test various theories, reproduce averaged multiplicities and standard collective flow observables, and study nuclear structure \cite{Barbosa:2021ccw,Giacalone:2021uhj,Giacalone:2020awm,Plumberg:2021bme,Summerfield:2021oex,Carzon:2020xwp,Giacalone:2020lbm,Giacalone:2020dln,Martinez:2019jbu,Sievert:2019zjr,Giacalone:2017dud,Li:2021nas,Zhao:2020wcd,Liu:2019jxg}.

The creators of \code{trento} have also been very active in improving the model and including new features that keep it at the cutting edge. The first big evolution came with the inclusion of sub-nucleonic structure \cite{Moreland:2017kdx}, which allows for the specification of the number of hotspots in each nucleon, emulating valence quark sampling. Substructure in the initial state is most important in small systems and ultra-central collisions. A Bayesian analysis was done using this extension to the model in Ref.~\cite{Moreland:2018gsh,Moreland:2018jos}. The most recent development has come in the extension to a 3 dimensional framework \cite{Ke:2017xyi}.

While \code{trento} is extremely powerful and well tested, there are areas in which it is not complete. The functional form of the reduced thickness function used by \code{trento} is great for its parameterization, but it does not span the whole space of functions that have been shown to work. The IP-Glasma model \cite{Schenke:2012wb}, discussed in the next section, generates an energy scaling relationship that is linear in the constituent thickness functions, $\epsilon \propto T_A T_B$, and is not accessible by \code{trento}. The IP-Jazma model \cite{Nagle:2018ybc}, discussed below, uses the linear scaling along with multiplicity fluctuations sampled from a log-normal distribution. The \code{trento} code also requires internal definitions of the Woods-Saxon profile of each species, which is very easy to implement but makes it difficult to run analyses on many deformations of the same nuclei (See Sec.~\ref{chap:V2toV3Puzzle}). I have modified \code{trento} to explicitly include the linear scaling reduced thickness function, log-normal multiplicity fluctuations, and a way to fully specify deformed nuclei on run time. This version of \code{trento} was used in producing the work herein.

\subsection{CGC Approaches} \label{sec:CGCInitialState}

The phenomenological approach is not the only way of constructing the initial state, although aspects of it are present in all models. The most prolific theoretical approach is using CGC effective field theory to perform microscopic calculations to determine the initial energy/entropy density produced in the initial stages of the collision \cite{McLerran:1993ni, McLerran:1993ka, McLerran:1994vd}. This approach relies on the observation that the proton is saturated by gluons at high energies. There are three regimes in which this method is applied: dilute-dilute, dilute-dense, and dense-dense collisions. The signifiers of 'dilute' and 'dense' refer to the occupation number of gluons in the projectile and target. This translates to dilute-dilute corresponding to p-p collisions, dilute-dense to d-Au, and dense-dense to Au-Au. Each of these regimes provide assumptions that simplify certain calculations. Here I will focus on a particular thread of CGC initial state models that starts with IP-Sat \cite{Kowalski:2003hm} and continues with IP-Glasma \cite{Schenke:2012wb}, MSTV \cite{Mace:2018vwq}, and ends with IP-JAZMA \cite{Nagle:2018ybc}.

The motivation for creating the IP-Sat (Impact Parameter Saturated dipole) model was to better describe Deep Inelastic Scattering (DIS) observables by incorporating the distribution of proton impact parameter into the calculation of the cross section of relevant processes \cite{Kowalski:2003hm}. This was then extended to heavy-ion collisions and used to calculate the cross sections and saturation scales in those processes. 

The IP-Sat model \cite{Kowalski:2003hm}, for calculating the impact parameter dependent cross section, was then joined with the MC-Glauber nucleon sampling and the classical Yang-Mills description of color fields in the IP-Glasma model \cite{Schenke:2012wb}. The process applied in IP-Glasma starts with getting nucleon configurations from MC-Glauber, then calculating the density of those nucleons and the resulting reduced thickness density using the IP-Sat model and Glasma formulation. The resulting initial condition contains event-by-event fluctuations from the nucleon positions and in the color degrees of freedom of the gluon fields. These color field fluctuations are comparable to multiplicity fluctuations, as defined above, and follow a negative-binomial distribution. The original formulation of IP-Glasma was in the dense-dense regime and later extended to dilute-dilute in Ref.~\cite{McLerran:2015qxa}. Of those discussed here, IP-Glasma is the most popular of the CGC-like models and has been shown to reliably fit a wide range of experimental observables \cite{Gale:2012rq,Gale:2013da,Ryu:2015vwa,Bzdak:2013zma,Schenke:2012hg,Mantysaari:2017cni,Mantysaari:2016ykx,Dusling:2015gta,Schenke:2015aqa,Schenke:2020mbo,Schenke:2020uqq}. 

The scale of the color field fluctuations is on the order of the grid spacing in the simulation and creates a "spikey" initial condition as compared to the smoother distribution from \code{trento}. If the fluctuations in the color fields of CGC models are averaged over, then one finds that the small scale structure produced by these fluctuations has a minimal role when compared to experimental observables \cite{Noronha-Hostler:2015coa, Gardim:2017ruc, Mazeliauskas:2015vea}. In fact, if one averages over the color field fluctuations on a nucleon level, then the IP-Glasma initial condition should be comparable to that from phenomenological models. This leads to the energy density from IP-Glasma being proportional to the product of the two thickness functions $\epsilon \propto T_A T_B$ \cite{Nagle:2018ybc, Lappi:2006hq, Chen:2015wia, Romatschke:2017ejr}. The linear scaling from IP-Glasma can be included in \code{trento}, which I explore in Chap.~\ref{chap:ExplorationOfMultiplicityFluctuations}, and provides similar results to those from IP-Glasma itself. The fact that this reduction of the model agrees with the fully complex version and with experimental data, calls into question the usefulness of the small fluctuations of the color fields.

In 2013, the authors of Ref.~\cite{Nagle:2013lja} suggested the performance of a geometry scan in heavy-ion collisions stating that MC-Glauber simulations predict the survival of triangular flow in $He^3$. This observation also brought into question what the smallest drop of QGP might be. This was later followed by predictions made by CGC models \cite{Habich:2014jna,Shen:2016zpp,Welsh:2016siu}. With these predictions, RHIC and LHC made measurements of $pAu$, $dAu$, and $He^3 Au$ that were strikingly similar to those in A+A collisions which signified the presence of a hydrodynamic phase \cite{PHENIX:2018lia}. An alternative theory had also been developed that claimed the signals of flow in these small systems was entirely from correlations in the initial state and did not require the presence of a hydrodynamic phase \cite{Dusling:2012iga,Dusling:2013oia}. This approach was investigated in Ref.~\cite{Mace:2018vwq} using a new CGC initial state model, in the dilute-dense regime, that was later named MSTV, after the authors Mace, Skokov, Tribedy, and Venugopalan. The authors concluded that these hydrodynamic signatures are not discluded from being dominantly initial state effects.

Another initial state model, called IP-JAZMA \cite{Nagle:2018ybc}, was created to be a centralized implementation of some of the physics from both MSTV and IP-Glasma, while also allowing input from any MC-Glauber model. 
%The claims made by the authors of Ref.~\cite{Mace:2018vwq}, were called into question by the authors of Ref.~\cite{Nagle:2018ybc}, who took it upon themselves to test these claims by constructing another initial state model called IP-JAZMA \cite{Nagle:2018ybc}. IP-JAZMA was created to be an open-source implementation of some of the physics from both MSTV and IP-Glasma, while also allowing input from any MC-Glauber model.
An interesting feature of MSTV, is that the handling of color field fluctuations diverged from the rest of community by sampling from a log-normal distribution rather than the now standard NBD and $\Gamma$. With this new distribution, MSTV was able to reproduce the multiplicity distribution of $dAu$, which is supported by IP-JAZMA \cite{Nagle:2018ybc}. A detailed investigation of the role of the multiplicity fluctuation distribution and the parameterization of the initial entropy density is conducted in Chap.~\ref{chap:ExplorationOfMultiplicityFluctuations}, where I include the linear scaling and log-normal distributions in \code{trento} and look at both $dAu$ and $AuAu$.

\subsection{Other Approaches} \label{sec:OtherModernISModels}

The last main thread of initial state models are those that use a dynamic approach to calculating the initial energy/entropy density. These models are transport codes that are aimed at treating non-equilibrium dynamics and naturally include nucleon and multiplicity fluctuations. The main issue with these transport approaches, is that the systems do not thermalize well and are harder to connect to hydrodynamic codes which require local thermalization. Some progress has been made in resolving this issue \cite{vanderSchee:2013pia}. Several of these models are: NEXUS \cite{Gardim:2012yp,Gardim:2011xv}, EPOS \cite{Werner:2010aa,Werner:2012xh,Werner:2012ca}, UrQMD \cite{Petersen:2008dd,Steinheimer:2007iy,Petersen:2009vx,Petersen:2010md,Petersen:2010zt,Petersen:2011sb}, and AMPT \cite{Pang:2012he}. Another saturation model is EKRT, which generates the initial energy density from mini-jet production calculated from pQCD \cite{Niemi:2015qia}.

Another important aspect of the initial state is the use of pre-equilibrium evolution of the initial condition up to some local equilibrium is reached and the event can be run hydrodynamically. An early version was that of free-streaming \cite{Heinz:2002rs}. Another of these pre-equilibrium evolutions is that of the \kompost model which is explored more in Chap.~\ref{chap:PreEquilibriumEvolution}.

%---------------------------------------------------------------------------
%
\section{Further Development of the Initial State} \label{sec:Intro:FurtherDevelopmentOfInitialState}
%---------------------------------------------------------------------------
%

There has been much development in modeling of the initial state over a short period, and while most models are able to match experimental data well, there are many questions left to explored. 

While it is well established that the QGP exists in large systems, signals of it have been found in the small system of pPb (ATLAS \cite{Chatrchyan:2013nka, Aaboud:2017acw, Aaboud:2017blb, Aad:2013fja}, CMS \cite{Sirunyan:2018toe, Chatrchyan:2013nka, Khachatryan:2014jra, Khachatryan:2015waa, Khachatryan:2015oea, Sirunyan:2017uyl}, and ALICE \cite{ABELEV:2013wsa, Abelev:2014mda}) and have been matched quantitatively by hydrodynamics \cite{Bozek:2011if, Bozek:2012gr, Bozek:2013ska, Bozek:2013uha, Kozlov:2014fqa, Zhou:2015iba, Zhao:2017rgg, Mantysaari:2017cni, Weller:2017tsr, Zhao:2017rgg}, though other explanations have been formulated \cite{Greif:2017bnr, Schenke:2016lrs, Mantysaari:2016ykx, Albacete:2017ajt}. A recent beam energy scan of $^3$HeAu and dAu by experimentalists at the RHIC PHENIX detector \cite{Aidala:2018mcw, Adare:2018toe} has also shown these systems to contain signs of the QGP. Additionally, a recent analysis \cite{Schenke:2019pmk} looked at $v_2\{2\}$ in AuAu and dAu and found that precise measurements of $v_2$ in these systems would be helpful in identifying effects of the initial state momentum anisotropy. While the existence of the QGP in small systems is of great interest, there are causality issues with the use of hydrodynamics to describe these systems due to their size.

Incorporating knowledge from low-energy nuclear structure in initial state models is a promising direction of development, with recent interest at a high with the isobar runs at the LHC \cite{STAR:2021mii}. There is failure of heavy-ion simulations in correctly describing the dynamics of ultra-central (See Chap.~\ref{chap:V2toV3Puzzle}) and ultra-peripheral collisions. The failure at ultra-central collisions is particularly concerning, since it is the regime in which the assumptions of models are most valid. However, it opens the door to new physics, particularly nuclear structure from low-energy. Other areas of ongoing exploration, concerning nuclear structure are: nuclear deformations in both small \cite{Lim:2019cys,Citron:2018lsq,Rybczynski:2019adt,Noronha-Hostler:2019ytn} and large ions and the effect of a neutron skin on the structure of ions \cite{Abrahamyan:2012gp}.

The upcoming Electron Ion Collider (EIC) \cite{Accardi:2012qut} will be an important part of the future for the initial state, since it will allow us to probe small systems with high statistics and probe nuclear structure with much more precision than previously.

Modern initial state models have made developments in the description of sub-nucleonic fluctuations (though it is still unknown the exact scale that initial conditions probe \cite{Gardim:2017ruc,Noronha-Hostler:2015coa,Gardim:2017ruc,Nagle:2018ybc,Hippert:2020kde,NunesdaSilva:2020bfs}), rapidity structure \cite{Zhao:2022ugy,Ke:2017xyi,Schenke:2016ksl,Ke:2017xyi}, new sources of fluctuations, and the inclusion of new degrees of freedom. This last one, specifically for conserved charges in the initial state, is the focus of much of this work.

%%%%%%%%%%%%%%%%%%%%%%%%%%%%%%%%%%%%%%%%%%%%%%%%%%%%%%%%%%%%%%%%%%%%%%%%%%%
%
\chapter{Exploration of Multiplicity Fluctuations} \label{chap:ExplorationOfMultiplicityFluctuations}
%
%%%%%%%%%%%%%%%%%%%%%%%%%%%%%%%%%%%%%%%%%%%%%%%%%%%%%%%%%%%%%%%%%%%%%%%%%%%

Fluctuations in the initial state are important for a full description of experimental data. One of the more recent sources of fluctuations has been from multiplicity, see Sec.~\ref{sec:Intro:MultiplicityFluctuations} for more of the developmental history. The key reason to include these multiplicity fluctuations in the initial state is to match experimental data from central collisions of deformed nuclei. Glauber models without these fluctuations significantly over predict the multiplicity distributions. The linch-pin system that facilitated these fluctuations was deformed uranium, which two-component Glauber predicted would have a large difference in multiplicity between tip-on-tip (scaling like $N^{2}_{part}$) and side-on-side collisions (scaling like $N_{part}$). This came from the fact that for tip-on-tip collisions the nucleons would be "lined up" and, thus, produce many more binary collisions than side-on-side. The slope of the elliptic flow versus multiplicity, coming from this model, was much steeper than the experimental data from STAR and led to that model being ruled out as adequate to describe the initial state \cite{Pandit:2013uiv,Wang:2014qxa}. 

This inadequacy of the phenomenological Glauber model inspired many to introduce fluctuations that would fix this issue. The one I will focus on, in this chapter, is \code{trento}, which adopted a general scaling criterion, the generalized mean (\ref{e:genmean}), that enforced multiplicity of tip-on-tip collisions to scale comparably to side-on-side collisions. This forces a homogeneous scaling, described by Eq.~\ref{eq:ThicknessFunctionScaleInvariance}, and reduces the strong difference in multiplicity between tip-on-tip and side-on-side collisions of deformed nuclei, bringing the slope of $v_2$ versus $N_{ch}$ into agreement with the experimental data. Several Bayesian analyses of \code{trento} \cite{Moreland:2018gsh,Everett:2020xug,Nijs:2020roc} have shown a preference of $p=0$ within the functional form Eq.~\ref{e:genmean}. This corresponds to an initial entropy density $s \propto \sqrt{T_{A}T_{B}}$. An important assumption of these analyses was that the event-by-event multiplicity fluctuations would be described by a one-parameter $\Gamma$ distribution (see Eq.~Eq.~\ref{eq:Gamma}) \cite{Moreland:2018gsh}. 

This generalized mean is not the only scaling that is known to describe the ultra-central UU data, however. The authors of Ref.~\cite{Noronha-Hostler:2019ytn} were able to fit experimental data using a modified version of \code{trento} that contained a type of binary collision term $s (T_A , T_B) \propto T_A T_B$. This linear entropy scaling is quite similar to the trend of CGC based simulations producing linear energy scaling \cite{Nagle:2018ybc, Lappi:2006hq, Chen:2015wia, Romatschke:2017ejr}. To further develop this comparison, we treat the nuclear profiles from \code{trento} for $T_A T_B$ as proportional to energy and convert to entropy using a conformal equation of state\footnote{A conformal equation of state is an approximation where $e=3p$.}. While a simplification of what should be done, this will be sufficient to make direct comparisons between the phenomenological entropy scaling and CGC energy scaling. The precise definition of this comparison is that we take the linear energy scaling relationship from CGC to be $s (T_A , T_B) \propto (T_A T_B)^{3/4}$ in \code{trento}. The fact that this binary collision term can agree with data may indicate that the flaw in the two-component Glauber model could have less to do with multiplicity scaling but rather be due to the spatial profile with which the binary collision density was assumed to be deposited.

\code{trento} makes assumptions in its formulation that exclude potentially viable models such as a CGC description of the initial state. The choice of generalized mean Eq.~\ref{e:genmean}, does not include linear scaling in its parameter space despite that also being able to fit experimental data. Another strict assumption is that of the one-parameter $\Gamma$ distribution for the multiplicity fluctuations. The IP-JAZMA model \cite{Nagle:2018ybc} combined a CGC-like linear scaling of the initial energy density $\epsilon \propto T_A T_B$ with a log-normal distribution for the functional form of the multiplicity fluctuations and found that, with appropriate choices of parameters and multiplicity fluctuations, these models can describe the data reasonably well.

Both \code{trento} and CGC-like models are used extensively in the field of heavy-ion collisions to compare to experimental data and make predictions and generally work quite well. However, it is important to find the regimes where both of these models break down because that is where we will see new physics. To this goal, the analysis done in this chapter only utilizes \code{trento}, which has been modified to include the CGC-like behavior, in order to keep the comparison clean since CGC-like models have a lot more going on under the hood that could muddy the comparison.

It will also be important to study these choices of multiplicity fluctuations and functional forms in large and small systems. Many analyses use different assumptions when looking at different scales of heavy-ion collisions and so it is useful to try and further constrain our models by fitting across many different system sizes. 

In this chapter, I will systematically study the impact of these choices for the functional form of the entropy scaling and for multiplicity fluctuations, in the phenomenological \code{trento}, on the initial state eccentricities. I will look at two collision systems, $dAu$ and $AuAu$, which were chosen based on the previous work done in exploring their multiplicity distributions. Specifically, the system focused on in Ref.~\cite{Nagle:2018ybc} was $dAu$ and so that is of special interest while $AuAu$ is a very popular system of interest. It will be shown that there are identifiable differences between the two approaches that could be measured in experimental data through the fluctuations of $v_n$. This chapter reproduces and refines the work from Ref.~\cite{Carzon:2021tif}.

%^^^^^^^^^^^^^^^^^^^^^^^^^^^^^^^^^^^^^^^^^^^^^^^^^^^^^^^^^^^^^^^^^^^^^^^^^^^
\section{Multiplicity Fluctuations} \label{sec:MultiplicityFluctuations}
%^^^^^^^^^^^^^^^^^^^^^^^^^^^^^^^^^^^^^^^^^^^^^^^^^^^^^^^^^^^^^^^^^^^^^^^^^^^

To generate an initial condition, \code{trento} \cite{Moreland:2018gsh} creates nuclear thickness functions for each colliding nuclei using Eq.~\ref{eq:FluctuatingThicknessFunction}, where $\omega$ is a multiplicity weight factor, free to fluctuate event-by-event and nucleon-by-nucleon. 

The $\omega$ parameter should come from a distribution with mean 1 and allow large single nucleon fluctuations. This behavior is observed when fitting to data but the physical interpretation is unresolved. I will describe a way of interpreting these features but it is not necessarily the only way. The location of the mean at 1 could be because on average any "fluctuations" should disappear, otherwise they aren't fluctuations, and this reflects the fact that nucleons are a useful degree of freedom (there is also use for further refined degrees of freedom as in substructure, but that is likely less important in larger systems). The room for large fluctuations greater than the mean could be connected to the fact that we see a dramatic increase in gluon contributions when looking at high energy protons from DIS experiments (See Chap.~\ref{chap:ICCINGAlgorithm}). Another feature we want to look for is that the distribution is well behaved when $\omega$ goes to zero. Several distributions have historical and physical reasons for being used to sample multiplicity fluctuations, but the only requirement is the distribution has to contain these general features. It is, however, possible to differentiate different distributions in respect to experiment, as will be shown below, when looking at the extreme collisions.

The exact distribution that the multiplicity weights come from is an assumption made by \code{trento} to be a $\Gamma$ distribution, Eq.~\ref{eq:Gamma}, where the $k$ parameter controls the shape of the distribution \cite{Moreland:2018gsh}.
%__________________________________________________________________________
%
\begin{figure}[ht]
    \begin{center}
    \includegraphics[width=0.48\textwidth]{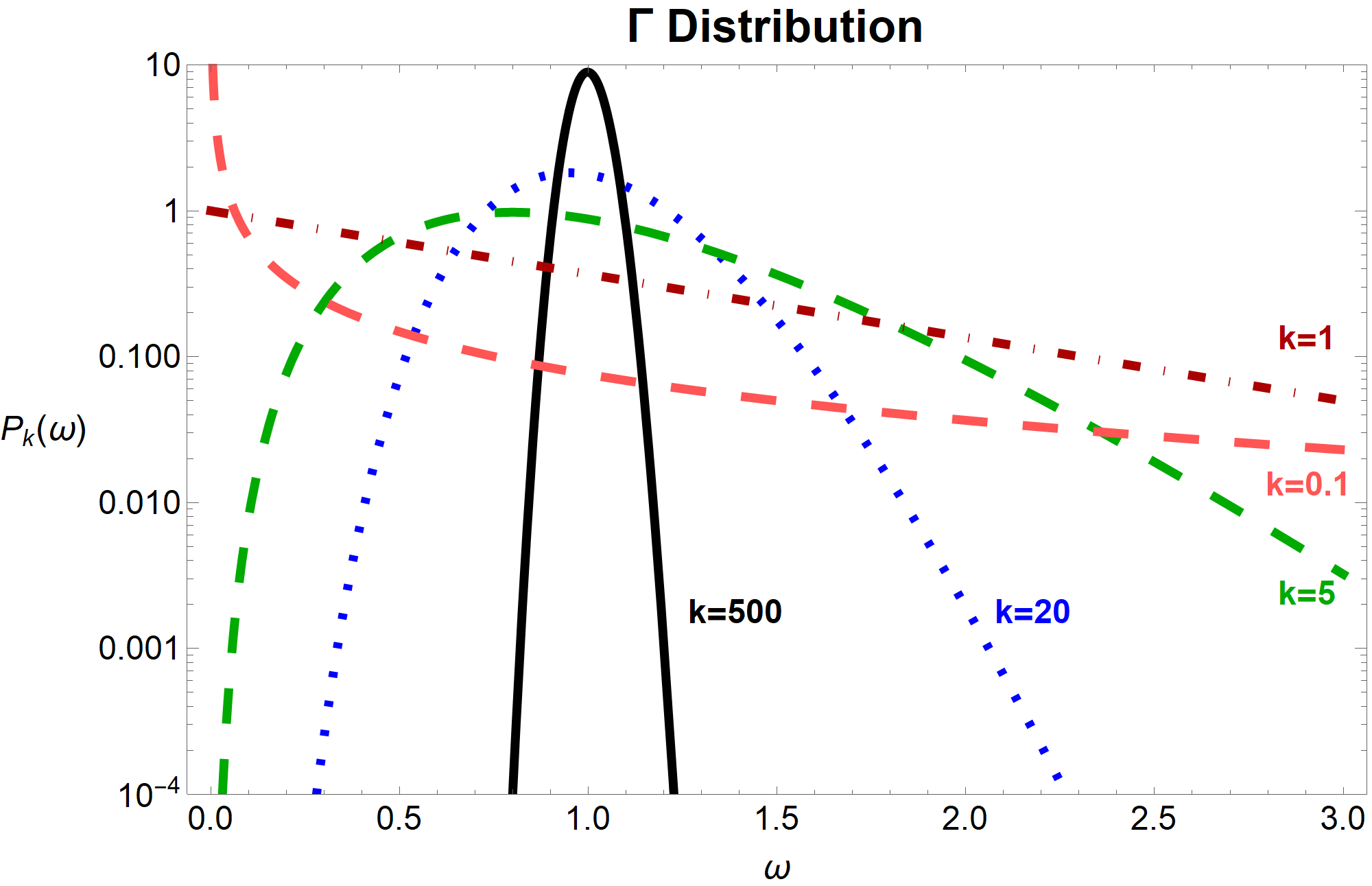}
    \includegraphics[width=0.48\textwidth]{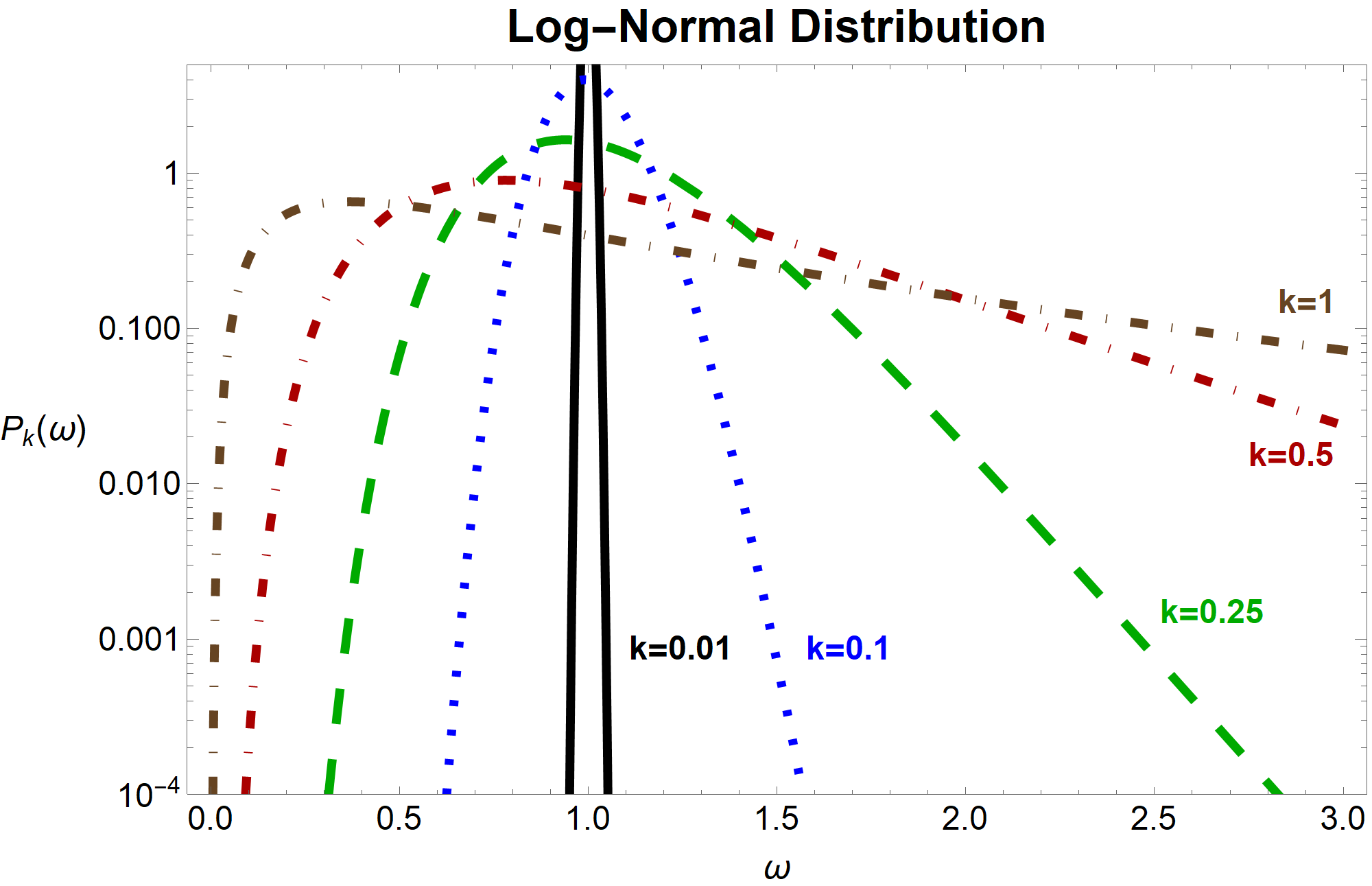}
    \caption{Illustration of the probability distributions of $\Gamma$ fluctuations Eq.\ \ref{eq:Gamma} (left) and log-normal Eq.\ \ref{eq:lognormflucs} (right) across several values of $k$ in order to highlight the difference in behavior of the two distributions. Figure from Ref.~\cite{Carzon:2021tif}.}
    \label{fig:flucdists}
    \end{center}
\end{figure}
%__________________________________________________________________________
%
It is important to have an idea of how the $k$ parameter will effect the $\Gamma$ distribution and this is illustrated in Fig.~\ref{fig:flucdists} across several values. As $k \rightarrow \infty$, the $\Gamma$ distribution, Eq.~\ref{eq:Gamma}, behaves like a delta function $\delta(\omega-1)$. In the case that $k\rightarrow1$, the width of the distribution grows with a preference for larger weights. The $\Gamma$ distribution behaves badly for values of $k < 1$ where we see a large divergence develop at $\omega = 0$ that acts as a counterbalance to the extremely long large-$\omega$ tail. 

This choice of distribution by \code{trento} does not have unique behavior. The assumption of $\Gamma$ distribution controlled multiplicity fluctuations could bias the Bayesian analysis to select on other parameters, such as the entropy scaling relationship of $p=0$, which translates to $s \propto \sqrt{T_{A}T_{B}}$. This is not just a conjecture since other models that lie outside these assumptions and the functional forms have been shown to fit experimental data. The one I focus on here is linear energy scaling from CGC, $\epsilon \propto T_A T_B$, which was coupled with multiplicity fluctuations using a log-normal distribution Eq.~\ref{eq:lognormflucs}. Now let's look at the behaviour of the log-normal distribution across values for $k$, in Fig.~\ref{fig:flucdists}, in order to compare against the $\Gamma$ distribution. As $k \rightarrow 0$, we see the log-normal distribution approach a delta function $\delta(\omega-1)$, while the width increases for $k \sim \mathcal{O}(1)$. This is an opposite trend in terms of the value of $k$ as compared to the $\Gamma$ distribution, which does not necessarily mean anything important but is key for intuition. A nice feature of the log-normal distribution is that it does not have the divergent behavior around $\omega = 0$ which we saw in the $\Gamma$ distribution. It is important to note that appropriate choices of $k$, for example: $k_\Gamma=5$ and $k_{log-normal}=0.5$, both distributions can produce similar behavior with slight differences, specifically in the width and large fluctuation tail of the distribution. These differences won't have a large effect for most centrality classes but, as we will see, become important when looking at the most central collisions which we know are the most sensitive to these small differences (See Sec.~\ref{sec:HistoryOfInitialState} and Chap.~\ref{chap:V2toV3Puzzle}).

%^^^^^^^^^^^^^^^^^^^^^^^^^^^^^^^^^^^^^^^^^^^^^^^^^^^^^^^^^^^^^^^^^^^^^^^^^^^
\section{Multiplicity Distributions} \label{sec:MultiplicityDistributions}
%^^^^^^^^^^^^^^^^^^^^^^^^^^^^^^^^^^^^^^^^^^^^^^^^^^^^^^^^^^^^^^^^^^^^^^^^^^^

\begin{figure}[ht]
    \begin{center}
    \includegraphics[width=0.48\textwidth]{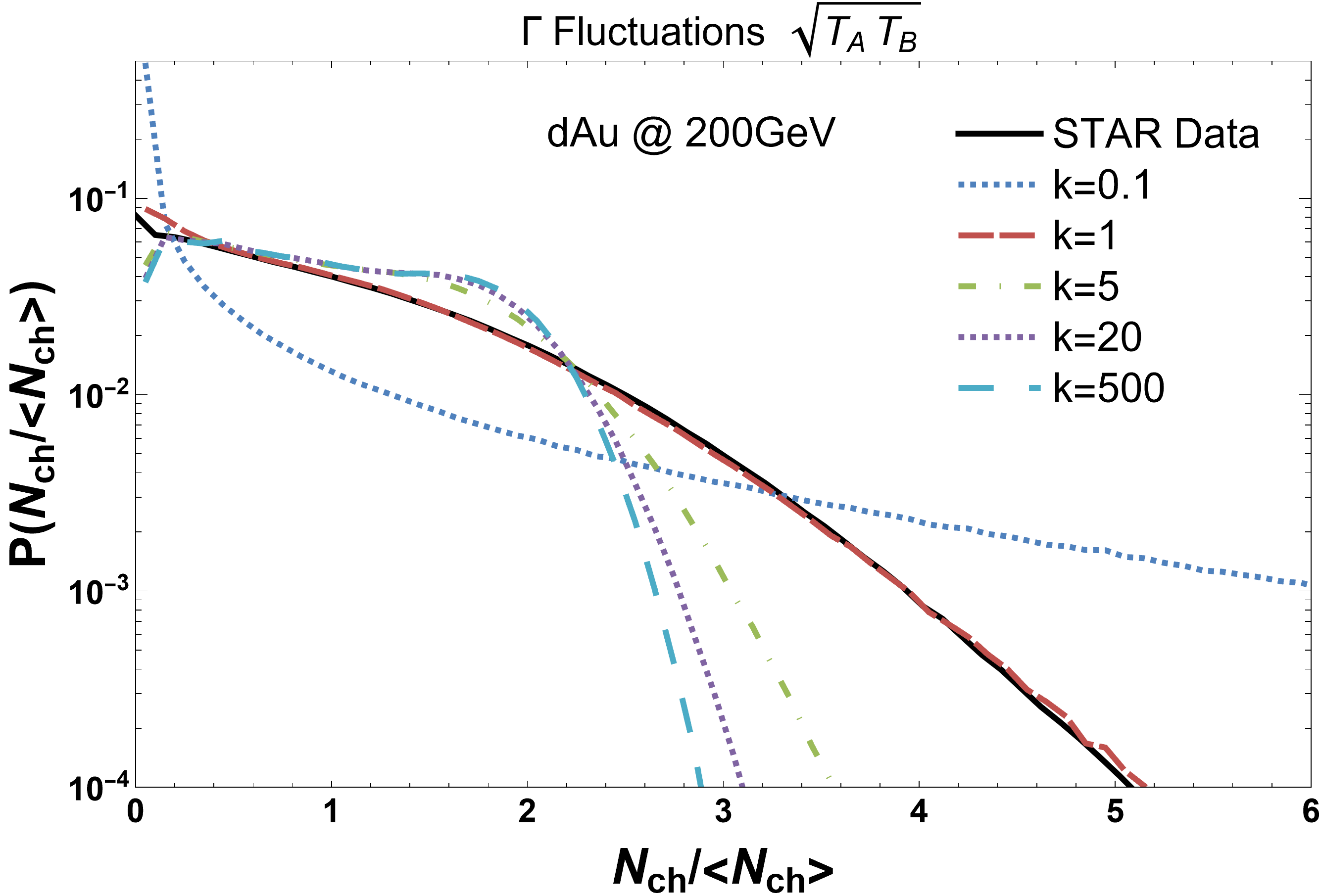}
    \includegraphics[width=0.48\textwidth]{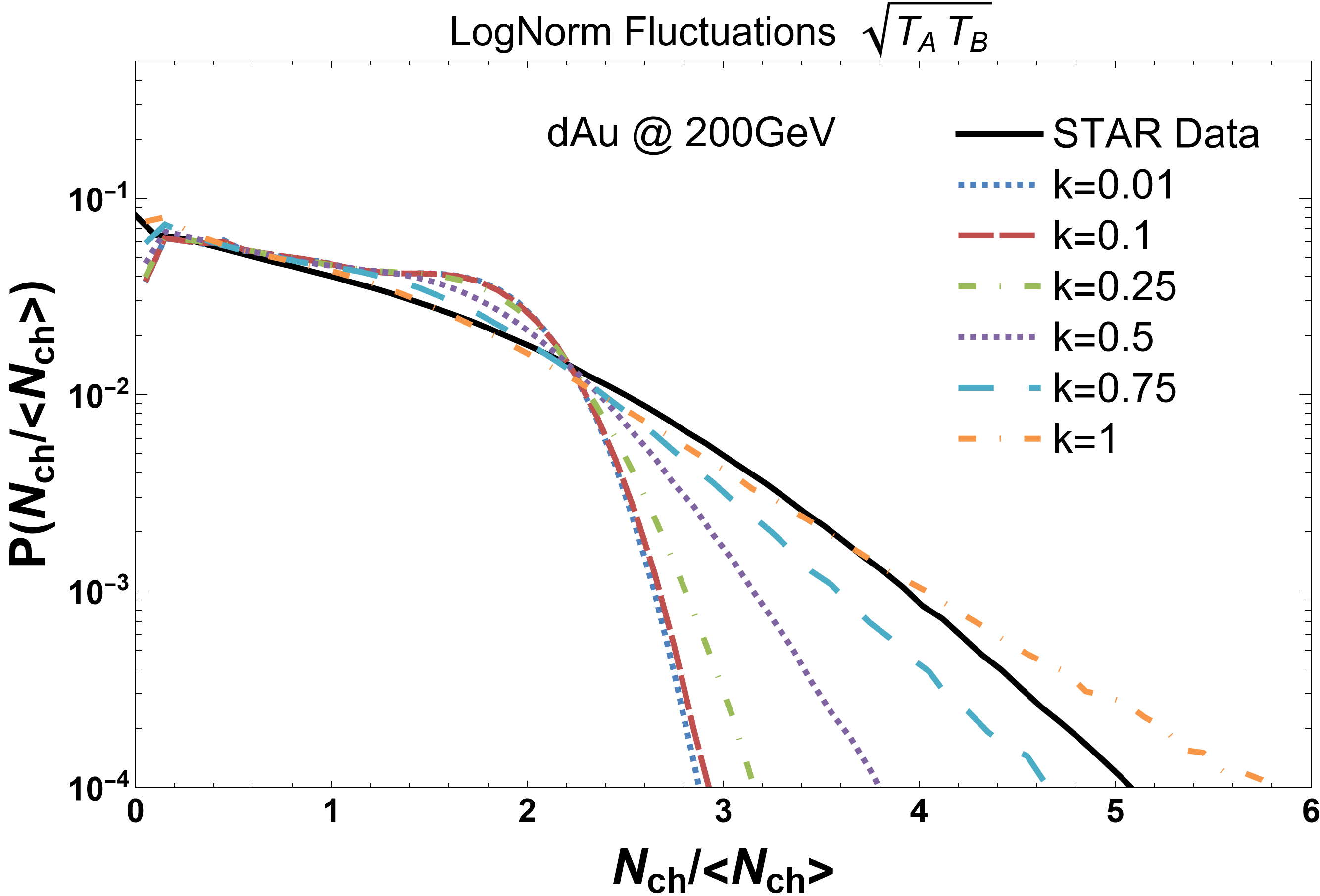}
    \caption{Plots of the $\sqrt{T_{A}T_{B}}$ entropy scaling using $\Gamma$ fluctuations Eq.\ \ref{eq:Gamma} (left) and log-normal Eq.\ \ref{eq:lognormflucs} (right) across all considered values of $k$. STAR data from Ref.~\cite{Abelev:2008ab}.}
    \label{fig:dAuQuarticSweep}
    \end{center}
\end{figure}

\begin{figure}[ht]
    \begin{center}
    \includegraphics[width=0.48\textwidth]{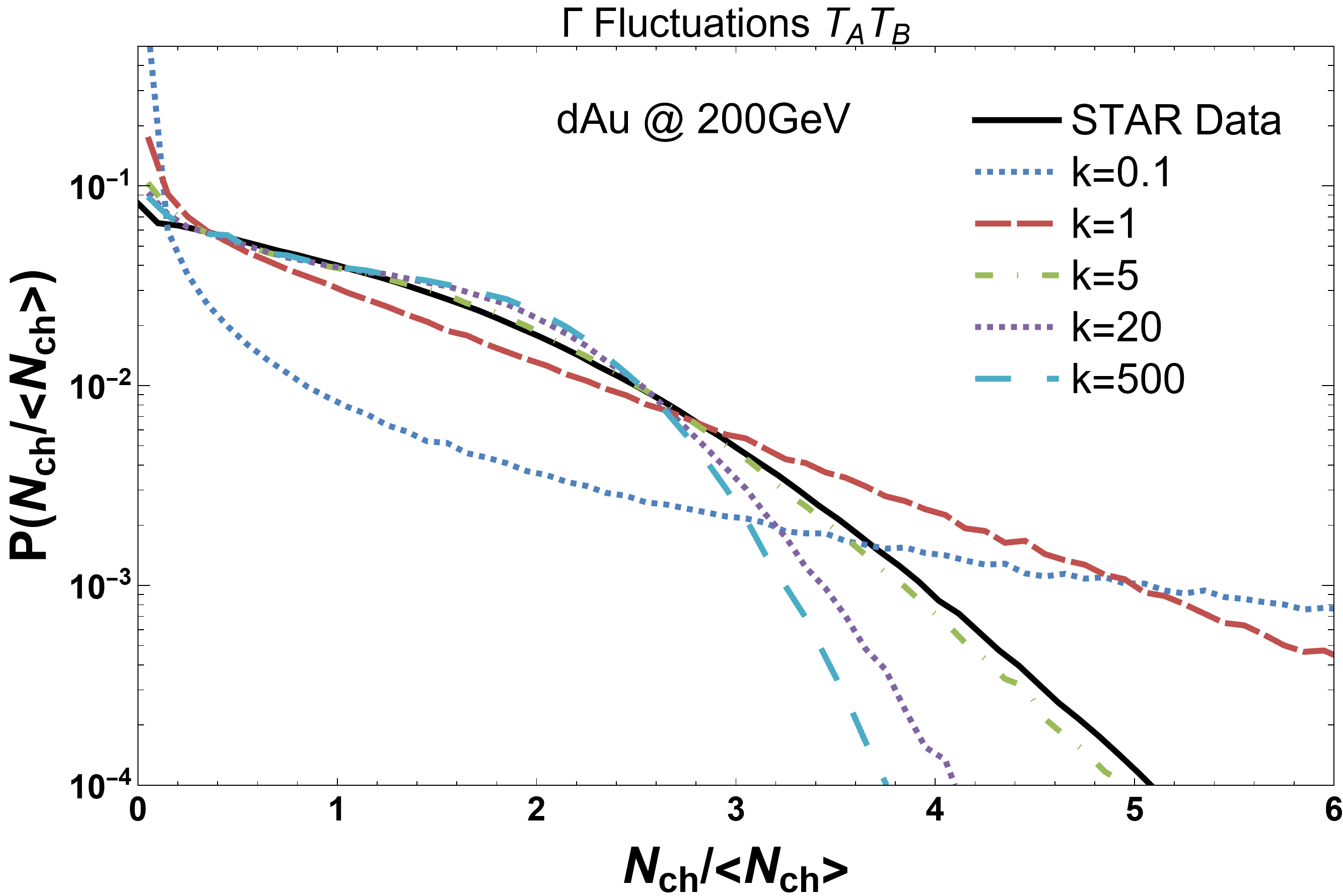}
    \includegraphics[width=0.48\textwidth]{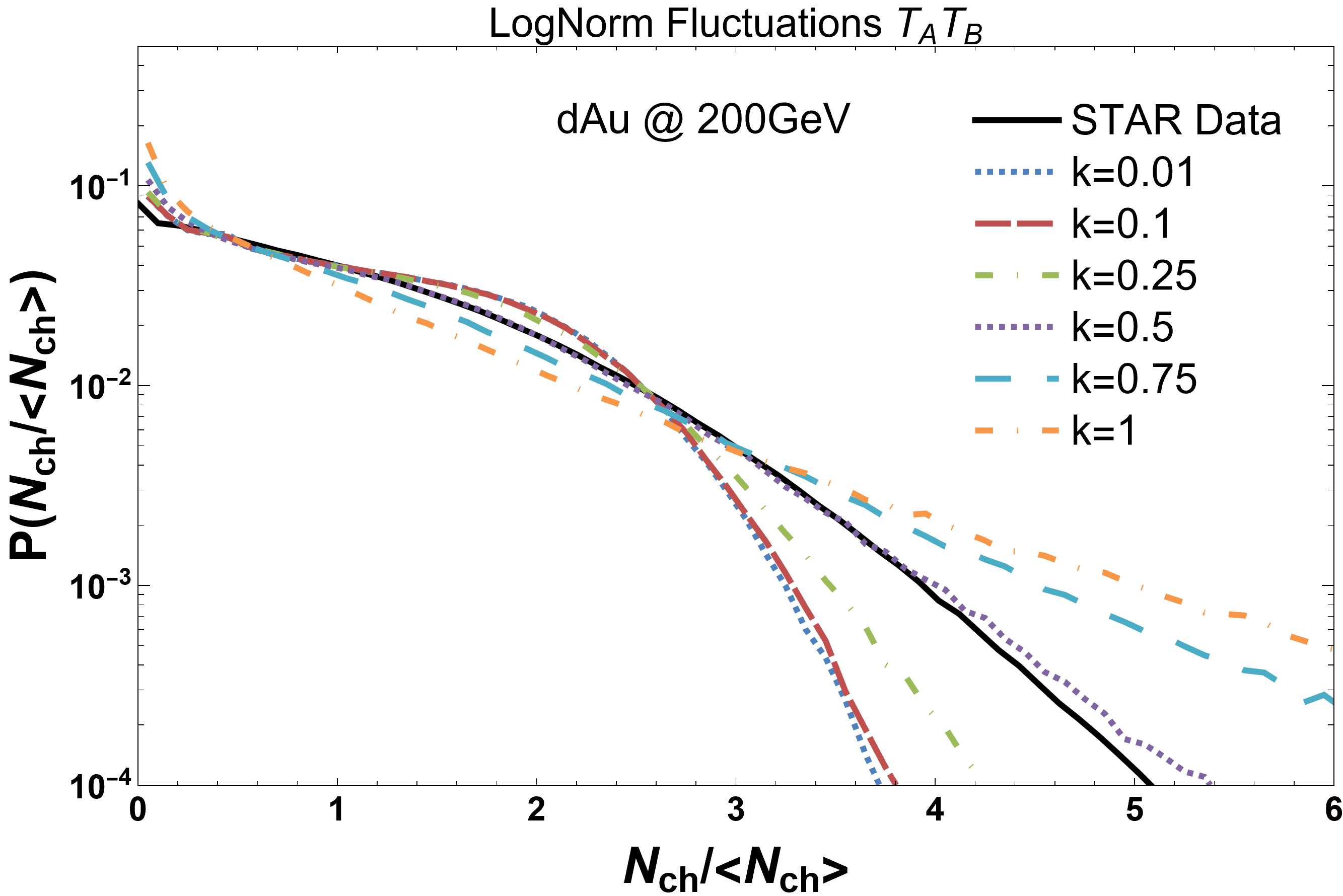}
    \caption{Plots of the $T_{A}T_{B}$ energy scaling using $\Gamma$ fluctuations Eq.\ \ref{eq:Gamma} (left) and log-normal Eq.\ \ref{eq:lognormflucs} (right) across all considered values of $k$. STAR data from Ref.~\cite{Abelev:2008ab}.}
    \label{fig:dAuLinearSweep}
    \end{center}
\end{figure}

For the sake of consistency and to remove any other contributing factors, we added the linear $T_A T_B$ scaling and log-normal fluctuations to \code{trento}. An assumption that $dN/dy \propto S_0$, with $S_0$ being the initial total entropy of the event, must be made when comparing estimates from the initial state to experimental multiplicity distributions. This is a reasonable assumption if there is entropy conservation throughout the evolution of the system (See Sec.~\ref{Sec:Centrality} for more details). We look at a sweep across $k$ for each of the entropy scalings and multiplicity fluctuation distributions combinations. This is a lot of different data sets, so we first look at the multiplicity distributions, which to some extent define the centrality classes, and select on the values of $k$ that best fit the experimental data \cite{Abelev:2008ab}. This selection is illustrated in Figs.~\ref{fig:dAuQuarticSweep} and \ref{fig:dAuLinearSweep}, where all values of $k$ considered are shown for the dAu collision system. Looking at the behavior of the $\Gamma$ fluctuations, we see that for both the quartic scaling, Fig.~\ref{fig:dAuQuarticSweep} (left), and linear scaling, Fig.~\ref{fig:dAuLinearSweep} (left), there is a change in character for $k < 1$, which is likely due to the bad behavior at $\omega=0$ and the flattening of the high fluctuation tail. We also see that the linear scaling needs a similar amount of fluctuations as quartic to agree with experiment. On the other hand, the log-normal fluctuations have a consistent behavior for different values of $k$ across both the quartic scaling, Fig.~\ref{fig:dAuQuarticSweep} (right), and linear scaling, Fig.~\ref{fig:dAuLinearSweep} (right). 

%__________________________________________________________________________
%
%__________________________________________________________________________
%

%__________________________________________________________________________
%
%__________________________________________________________________________
%

\begin{figure}[ht]
    \centering
    \includegraphics[width=0.6\textwidth]{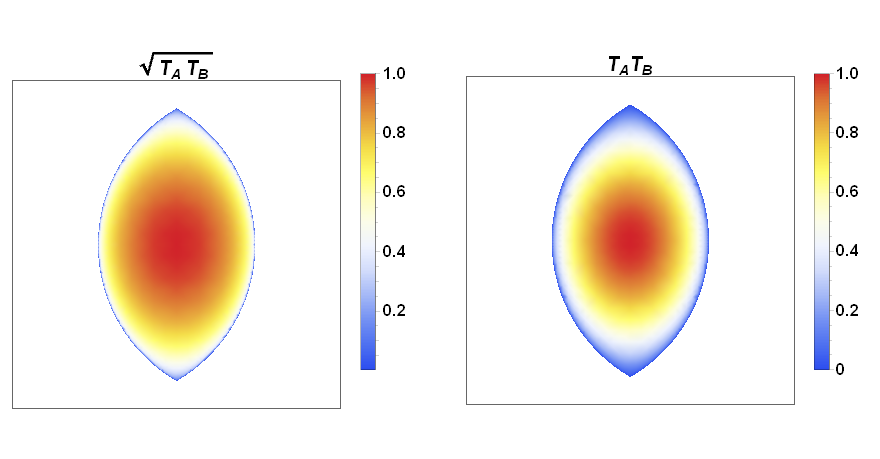}
    \caption{Cartoon of the difference between quartic and linear entropy scaling using smooth profiles for the two colliding nuclei. Figure from Ref.~\cite{Carzon:2021tif}.}
    \label{fig:SmoothingVisualization}
\end{figure}

The values of $k$ that best fit the experimental data generate distributions that satisfy the previously stated qualities that are needed; mean of 1 and large multiplicity fluctuations. When looking at the comparison between the quartic and linear scalings, we see that regardless of the fluctuation distribution, the quartic scaling consistently underestimates the high multiplicity tail while the linear scaling is more flexible. This difference can be understood by looking at the quartic and linear scalings, respectively, for optical Glauber collisions as illustrated in Fig.~\ref{fig:SmoothingVisualization}. We see that the quartic scaling relation creates a smoother distribution while the linear is sharper. This would explain why the quartic scaling underestimates the multiplicity distribution.

%__________________________________________________________________________
%
%__________________________________________________________________________
%

\begin{figure}[ht]
    \centering
    \includegraphics[width=0.48\textwidth]{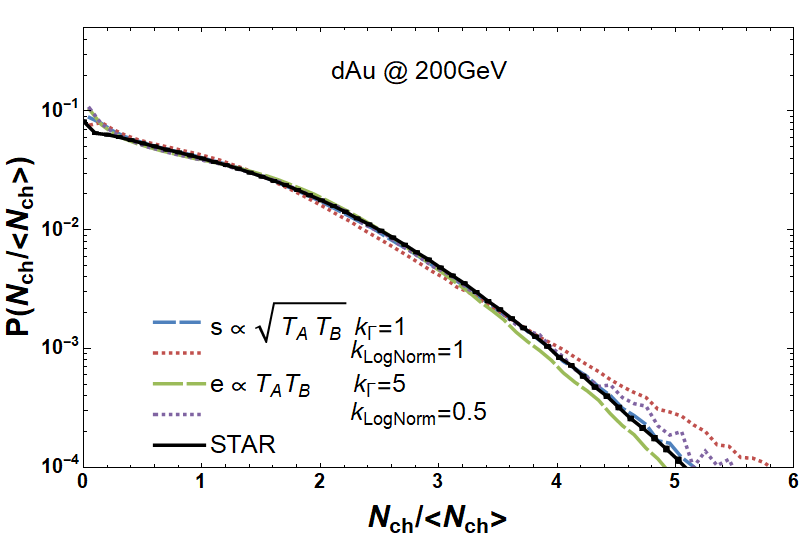}
    \caption{Multiplicity Distribution of dAu for functional forms $\sqrt{T_{A}T_{B}}$ and $T_{A}T_{B}$ using best fits for $\Gamma$ and lognormal multiplicity fluctuation distributions. Figure from Ref.~\cite{Carzon:2021tif}.}
    \label{fig:dAuMultiplicity}
\end{figure}

\begin{figure}[ht]
    \centering
    \includegraphics[width=0.48\textwidth]{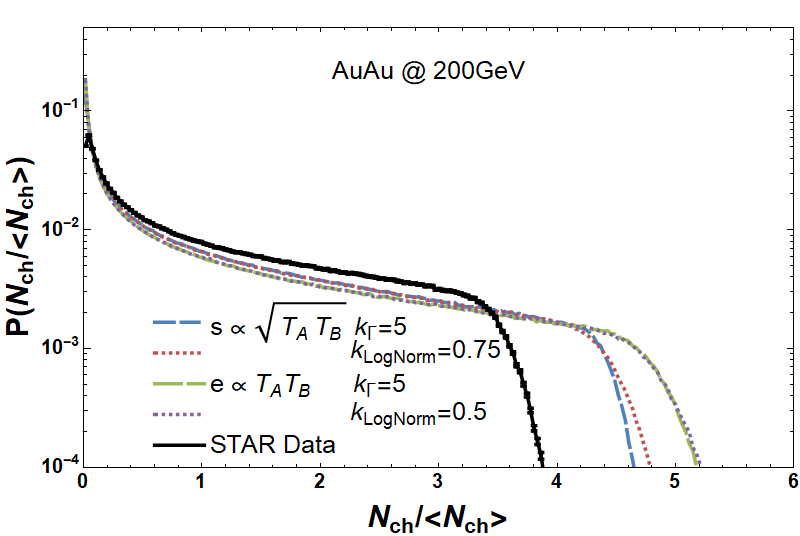}
    \caption{Multiplicity Distribution of AuAu for functional forms $\sqrt{T_{A}T_{B}}$ and $T_{A}T_{B}$ using best fits for $\Gamma$ and lognormal multiplicity fluctuation distributions. Figure from Ref.~\cite{Carzon:2021tif}.}
    \label{fig:AuAuMultiplicity}
\end{figure}

For the rest of the analysis, we select the values of $k$ for each combination of functional form and multiplicity fluctuation distribution to best match the multiplicity distributions of STAR data for dAu and AuAu at 200 GeV \cite{Abelev:2008ab} in Figs. \ref{fig:dAuMultiplicity} and \ref{fig:AuAuMultiplicity}, respectively. 

%__________________________________________________________________________
%
%__________________________________________________________________________
%

Revisiting the multiplicity distribution of dAu and comparing to STAR data from Ref.~\cite{Abelev:2008ab}, we see that all combinations in Fig.~\ref{fig:dAuMultiplicity}, which are the best fits, are nearly identical below $N_{ch}/\langle N_{ch}\rangle < 4$, which indicates that the different choices for entropy scaling and multiplicity fluctuations only have an effect on the multiplicity in ultra-central collisions. Linear entropy scaling can describe the ultra-central multiplicity tail using either $\Gamma$ or log-normal distributions for multiplicity fluctuations. The shapes of fluctuation distributions are similar for both the linear, $T_{A}T_{B}$, and phenomenological, $\sqrt{T_{A}T_{B}}$, scalings indicating that both functional forms need a similar amount of event-by-event fluctuations. This shows that both entropy scalings are capable of matching experimental data in dAu, in this observable, regardless of the choice of fluctuation distribution, indicating that they should be equally considered.

We can, also, look at the scaling and fluctuations choices for AuAu that best fit the experimental multiplicity distributions in Fig.\ \ref{fig:AuAuMultiplicity} to get another point of reference. Here there is a slight suppression, along the "back" of the distribution, as compared to the experimental data below $N_{ch}/\langle N_{ch}\rangle = 3.5$ that appears to be consistent for all combinations. When we look at the central collisions, both entropy deposition models do a poor job of fitting experimental data. The excess in the high multiplicity tail is larger for linear scaling which we also see reflected in the slightly larger suppression along the "back". This over prediction of high multiplicity events by both models could lead to mismatched centrality classifications between experiment and theory (See Sec.~\ref{Sec:Centrality} for more details on this process). For this observable, there is little difference between $\Gamma$ and log-normal fluctuations.

%__________________________________________________________________________
%
%__________________________________________________________________________
%

\begin{figure}[ht]
    \centering
    \includegraphics[width=0.48\textwidth]{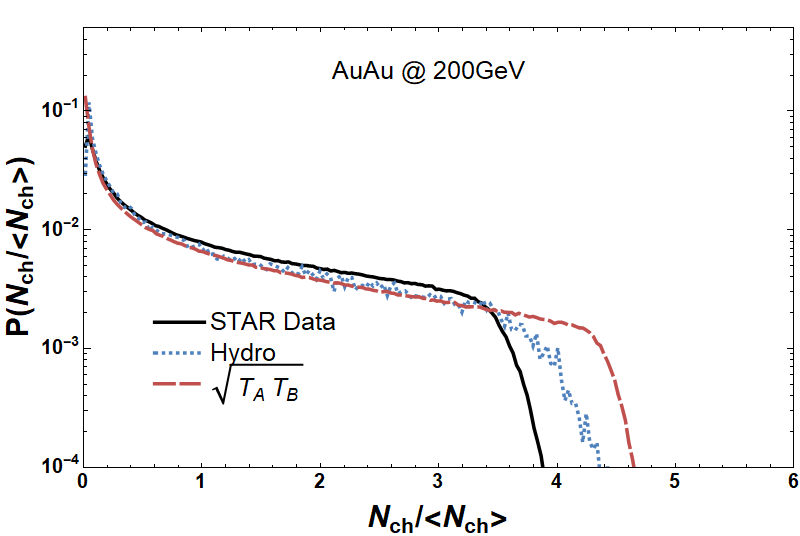}
    \caption{Illustration of the effect hydrodynamics has on the tail of the AuAu multiplicity distribution. Figure from Ref.~\cite{Carzon:2021tif}.}
    \label{fig:EffectOfHydro}
\end{figure}

The inability of the initial state models to match the high multiplicity tails of the experimental AuAu multiplicity distributions is troubling. One perspective might be that this disagreement arises because the effects of hydrodynamics on the multiplicity distribution are not taken into account. Investigating the argument, we use the hydrodynamic results from \cite{Alba:2017hhe,Rao:2019vgy}, which used \code{trento}+v-USPhydro \cite{Noronha-Hostler:2014dqa,Noronha-Hostler:2013gga} with a Lattice QCD based equation of state \cite{Alba:2017hhe} and the PDG16+ list \cite{Alba:2017mqu} for freeze out. In order to be as close a comparison as possible, only charged particles and the same kinematic cuts, as used by the experimental analysis, were implemented in the hydrodynamic results. The results from \cite{Alba:2017hhe,Rao:2019vgy} only used the $\sqrt{T_A T_B}$ scaling model and $\Gamma$ multiplicity fluctuations with $k=1.6$, which comes from the \code{trento} Bayesian analysis. We see that the hydrodynamic evolution suppresses the high multiplicity tail in Fig. \ref{fig:EffectOfHydro}. An important caveat is that the hydrodynamics curve in Fig. \ref{fig:EffectOfHydro} contains far fewer events due to the expensive nature of hydrodynamic simulations, but reflecting this lower number of events in the initial state distribution has no effect. This indicates that the dominant effect leading to this suppression of the high multiplicity tail is from hydrodynamics and not statistics. Though including the corrections from hydrodynamics brings the $\sqrt{T_A T_B}$ high multiplicity tail toward better agreement with experimental data it is not enough to fully resolve the issue. The case for the linear scaling model $T_A T_B$ is worse since it over predicts the high multiplicity tail more than $\sqrt{T_A T_B}$. This indicates an area in which these models are lacking and may point to the need for different physics contributions in ultra-central collisions, such as nuclear deformations (See Chap.~\ref{chap:V2toV3Puzzle} for further investigation of nuclear deformations). For this analysis, despite the linear scaling doing a worse job of fitting the high multiplicity tail than quartic scaling they both fail in the same way and so any solution to this disagreement may work for both models.

An important thing to consider when trying to reproduce these results, is that you have to bin in the same way the the experiment does. The number of bins can significantly shift the distributions up or down. The width of bins is chosen to be 0.1 for dAu and 0.03 for AuAu in order to be consistent with the experimental data.

%__________________________________________________________________________
%
%__________________________________________________________________________
%

%^^^^^^^^^^^^^^^^^^^^^^^^^^^^^^^^^^^^^^^^^^^^^^^^^^^^^^^^^^^^^^^^^^^^^^^^^^^
\section{Eccentricities}
%^^^^^^^^^^^^^^^^^^^^^^^^^^^^^^^^^^^^^^^^^^^^^^^^^^^^^^^^^^^^^^^^^^^^^^^^^^^

Utilizing the power of the linear response between the initial eccentricities and final flow harmonics (See Sec. \ref{sec:Eccentricities}), we can look at the effect these different choices of entropy scaling and multiplicity fluctuations have on the geometry of the initial state. Here, we will shift emphasis from dAu to AuAu, since in small systems linear response begins to break down \cite{Sievert:2019zjr,Schenke:2019pmk}. This makes it important to look at the full hydrodynamic simulations of dAu in order to compare with experimental data. Since there is an enormous cost to running hydrodynamic simulations and there are open questions \cite{Plumberg:2021bme,Cheng:2021tnq} about causality in small systems, we leave the full implementation to a future work. For the rest of this analysis, we choose to focus solely on the initial state eccentricities, which are important for any Bayesian analysis regardless if linear or non-linear scaling \cite{Noronha-Hostler:2015dbi} applies.

%^^^^^^^^^^^^^^^^^^^^^^^^^^^^^^^^^^^^^^^^^^^^^^^^^^^^^^^^^^^^^^^^^^^^^^^^^^^
\subsection{AuAu}
%^^^^^^^^^^^^^^^^^^^^^^^^^^^^^^^^^^^^^^^^^^^^^^^^^^^^^^^^^^^^^^^^^^^^^^^^^^^

\begin{figure}[ht]
    \begin{center}
    \includegraphics[width=0.48\textwidth]{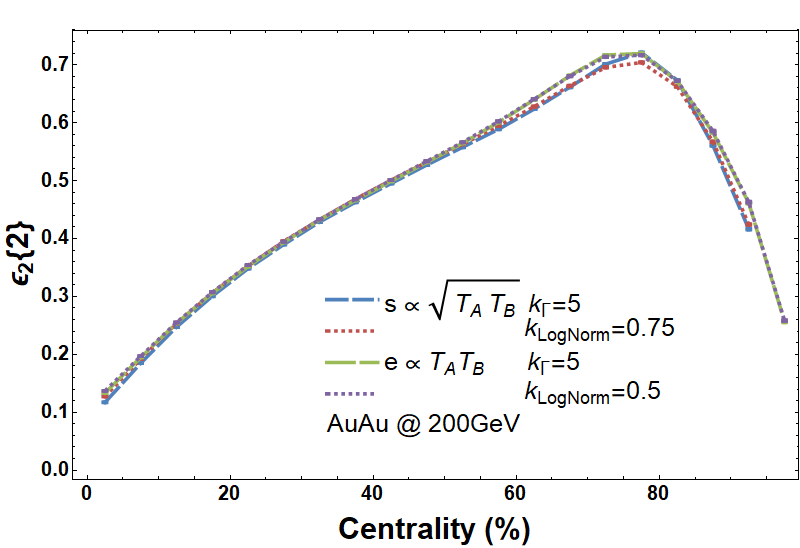}
    \includegraphics[width=0.48\textwidth]{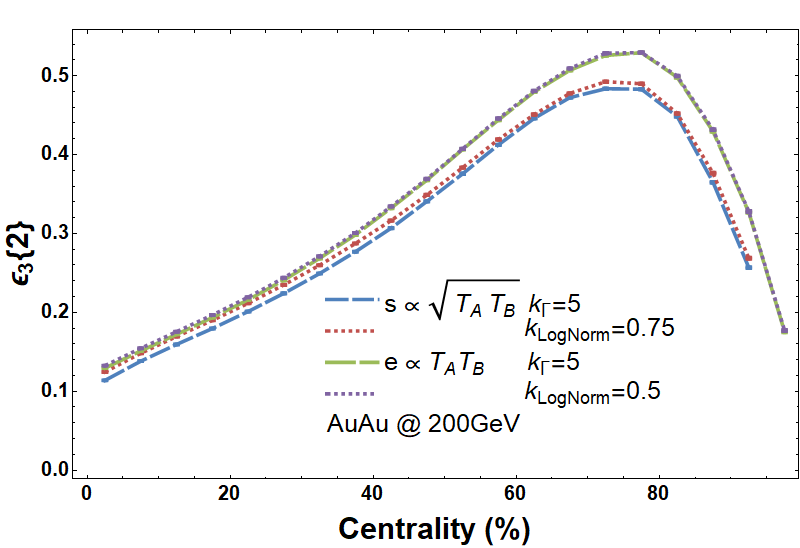}
    \caption{Two particle eccentricities of AuAu for functional forms $\sqrt{T_{A}T_{B}}$ and $T_{A}T_{B}$ using best fits for $\Gamma$ and lognormal multiplicity fluctuation distributions. Figure from Ref.~\cite{Carzon:2021tif}.}
    \label{fig:AuAueccs}
    \end{center}
\end{figure}

In Fig.~\ref{fig:AuAueccs}, we show $\varepsilon_{2}\{2\}$ (left) and $\varepsilon_{3}\{2\}$ (right) for each combination of scaling model and multiplicity fluctuation distribution with the value of $k$ that best matches the experimental multiplicity distribution. We see that, for $\varepsilon_{2}\{2\}$, there is good agreement between the two entropy scalings across all centralities, with only slight differences from 60-100\% centrality.  Triangularity, on the other hand, has the most variation between models with some slight separation between model combinations from 0-30\% centrality and significant differences in the more peripheral collisions. We see that for $\varepsilon_{3}\{2\}$, from 30-100\% centrality, linear scaling has a larger magnitude than quartic. There seems to be no significant sensitivity to the multiplicity fluctuation distribution for the geometry, except the triangularity of central collisions where log-normal fluctuations enhance $\varepsilon_{3}\{2\}$ for $\sqrt{T_A T_B}$ scaling. Central events are exactly where linear response is most applicable so this small sensitivity could be important.

%__________________________________________________________________________
%
%__________________________________________________________________________
%

\begin{figure}[ht]
    \begin{center}
    \includegraphics[width=0.48\textwidth]{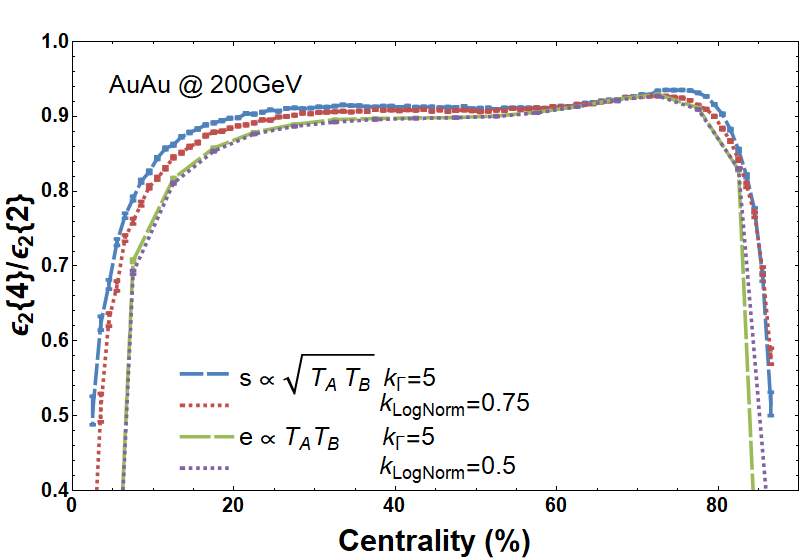}
    \includegraphics[width=0.48\textwidth]{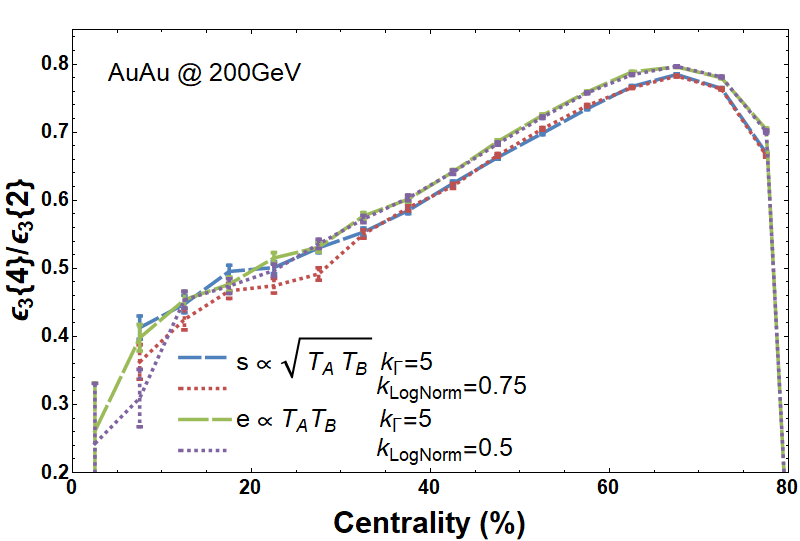}
    \caption{Four particle to two particle eccentricity ratios of AuAu for functional forms $\sqrt{T_{A}T_{B}}$ and $T_{A}T_{B}$ using best fits for $\Gamma$ and lognormal multiplicity fluctuation distributions. Figure from Ref.~\cite{Carzon:2021tif}.}
    \label{fig:AuAueccratios}
    \end{center}
\end{figure}

While the two-particle cumulants of the initial state eccentricity tells us about the average geometry, the ratio of four-particle to two-particle cumulants can quantify the fluctuations in that geometry (See Sec.~\ref{sec:Cumulants}). The elliptical, $\varepsilon_{2}\{4\}/\varepsilon_{2}\{2\}$ (left), and triangular, $\varepsilon_{3}\{4\}/\varepsilon_{3}\{2\}$ (right), geometry fluctuations are shown in Fig.~\ref{fig:AuAueccratios} for AuAu. This measure of the geometry fluctuations shows that the ellipticity, Fig.~\ref{fig:AuAueccratios} (left), of linear $T_A T_B$ has more fluctuations, as compared to quartic $\sqrt{T_A T_B}$, and is largely indifferent to the choice of multiplicity fluctuation distribution. There does appear to be sensitivity to the type of multiplicity fluctuations for $\sqrt{T_A T_B}$, which may lead to a distinction when compared to experimental data, though that data would require a lot of precision. The largest differences in elliptical fluctuations occur in central to mid-central collisions which correlates with the area of most variability in the multiplicity distributions. This makes sense, since the multiplicity distribution is used to define centrality and so the largest discrepancies, lacking any differentiation in the average ellipticity, would occur in the most central events.

We can also look at the fluctuations in the average triangularity, $\varepsilon_{3}\{4\}/\varepsilon_{3}\{2\}$, presented in Fig. \ref{fig:AuAueccratios} (right). Again focusing on the regime where linear response works best (See Sec.~\ref{sec:Cumulants}), central collisions, we see the most sensitivity comes from the multiplicity fluctuations with $\Gamma$ multiplicity fluctuations having a larger $\varepsilon_{3}\{4\}/\varepsilon_{3}\{2\}$ than log-normal. The error bars are large for these most central points, so that sensitivity may not be significant. For more mid-central to peripheral collisions (30-100\% centrality), the multiplicity fluctuations no longer have a discernible effect but there is a difference in the entropy scaling with linear estimating fewer fluctuations than quartic.

%__________________________________________________________________________
%
%__________________________________________________________________________
%

\begin{figure}[ht]
    \begin{center}
    \includegraphics[width=0.48\textwidth]{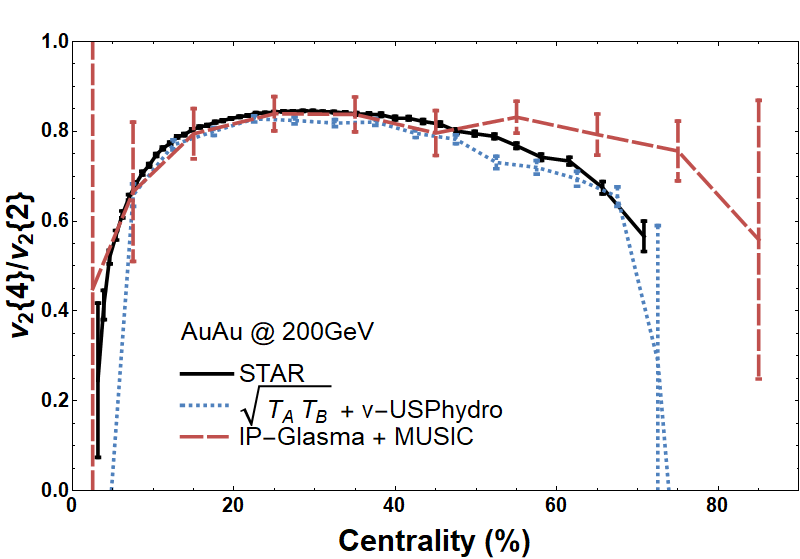}
    \caption{Comparison of the experimental measurement of $v_{2}\{4\}/v_{2}\{2\}$ from STAR \cite{STAR:2015mki} to two different hydrodynamic simulations \cite{Rao:2019vgy,Schenke:2019ruo} that reflect the different scaling models investigated in this paper. Figure from Ref.~\cite{Carzon:2021tif}.}
    \label{fig:EffectOfHydroOnV24V22}
    \end{center}
\end{figure}

We can estimate how the observable $\varepsilon_{2}\{4\}/\varepsilon_{2}\{2\}$ compares to experimental data by looking to hydrodynamic simulations that use quartic, \code{trento}+vUSPhydro, and linear, IP-Glasma+MUSIC, initial state entropy scalings. In Fig. \ref{fig:EffectOfHydroOnV24V22}, the four-particle to two-particle cumulant ratio of the second-order final state flow harmonics, $v_{2}\{4\}/v_{2}\{2\}$, is plotted against data from the STAR experiment \cite{STAR:2015mki}. The theoretical hydrodynamic curves are taken from Ref.~\cite{Rao:2019vgy}, which used $\sqrt{T_A T_B}$ + v-USPhydro, and Ref.~\cite{Schenke:2019ruo}, which used IP-Glasma + MUSIC and contains the CGC-like linear entropy scaling. From 0-50\% centrality, both models match the STAR data well, while above that centrality range, the linear scaling diverges from the data and under estimates the fluctuations. The peripheral disagreement in Fig. \ref{fig:EffectOfHydroOnV24V22} is not a surprise since this regime can be considered a small system and our understanding of the dynamics starts to break down. This observable further supports that both entropy deposition scalings are able to match experimental data despite their differences in origin. 

%__________________________________________________________________________
%
%__________________________________________________________________________
%

%^^^^^^^^^^^^^^^^^^^^^^^^^^^^^^^^^^^^^^^^^^^^^^^^^^^^^^^^^^^^^^^^^^^^^^^^^^^
\subsection{dAu}
%^^^^^^^^^^^^^^^^^^^^^^^^^^^^^^^^^^^^^^^^^^^^^^^^^^^^^^^^^^^^^^^^^^^^^^^^^^^

\begin{figure}[ht]
    \begin{center}
    \includegraphics[width=0.48\textwidth]{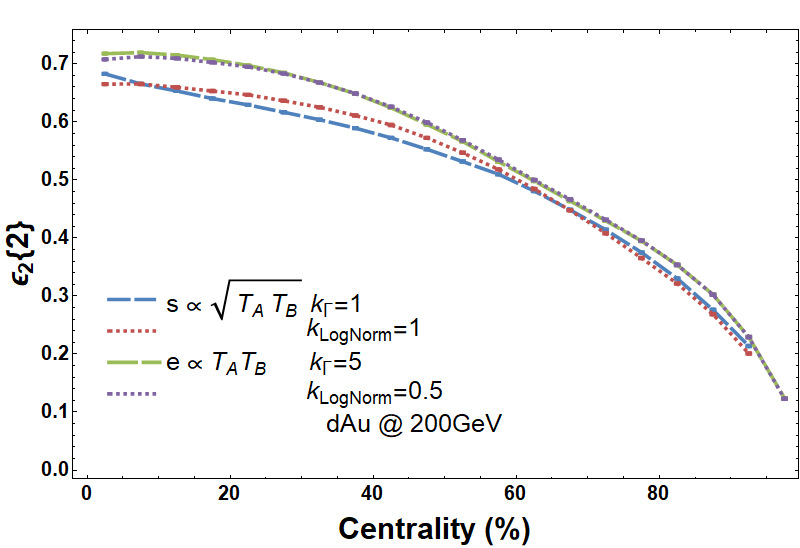}
    \includegraphics[width=0.48\textwidth]{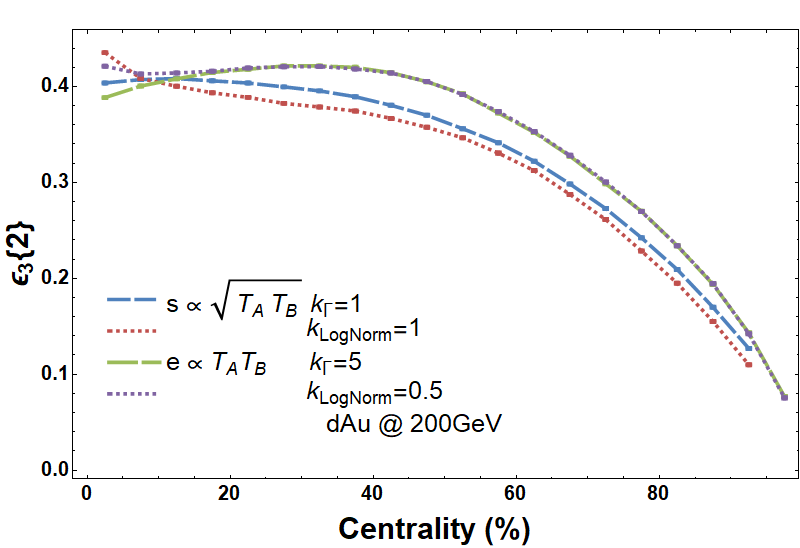}
    \caption{Two particle eccentricities of dAu for functional forms $\sqrt{T_{A}T_{B}}$ and $T_{A}T_{B}$ using best fits for $\Gamma$ and lognormal multiplicity fluctuation distributions. Figure from Ref.~\cite{Carzon:2021tif}.}
    \label{fig:dAueccs}
    \end{center}
\end{figure}

The smaller system of dAu is more sensitive to the choice of entropy scaling and multiplicity fluctuation distribution. This is seen clearly in the average geometry, $\varepsilon_{2}\{2\}$ (left) and $\varepsilon_{3}\{2\}$ (right), plotted in Fig.~\ref{fig:dAueccs} where significant differences are seen between the different choices.

%__________________________________________________________________________
%
%__________________________________________________________________________
%

The entropy scaling is the most dominant effect on the initial state eccentricities of dAu, with linear scaling having a consistently larger magnitude than quartic, except in ultra-central triangularity. The choice of multiplicity fluctuations is most significant in ultra-central collisions where the slope of the average geometry can change behavior even for the same entropy scaling. The relationship of linear scaling having a greater magnitude than quartic is broken by the multiplicity fluctuations in ultra-central $\varepsilon_{3}\{2\}$. Looking at the ellipticity, Fig.~\ref{fig:dAueccs} (left), we see that all of the curves have a flat slope for the ultra-central points except $\sqrt{T_A T_B}$ with $\Gamma$ fluctuations that curves up. The triangularity, Fig.~\ref{fig:dAueccs} (right), for ultra-central collisions curves up for log-normal fluctuations while $\Gamma$ fluctuations have slight downward curves. If we look back at the multiplicity distributions for dAu, Fig.~\ref{fig:dAuMultiplicity}, we see this ultra-central upward curve in $\varepsilon_{3}\{2\}$ corresponds to the distributions that overshot the high multiplicity tail. While the ultra-central collisions are really interesting, the general difference in magnitude between the two scalings is also important since it would lead to a difference in the extracted shear viscosity between the two entropy scalings in order to satisfy the same experimental constraints.

\begin{figure}[ht]
    \begin{center}
    \includegraphics[width=0.48\textwidth]{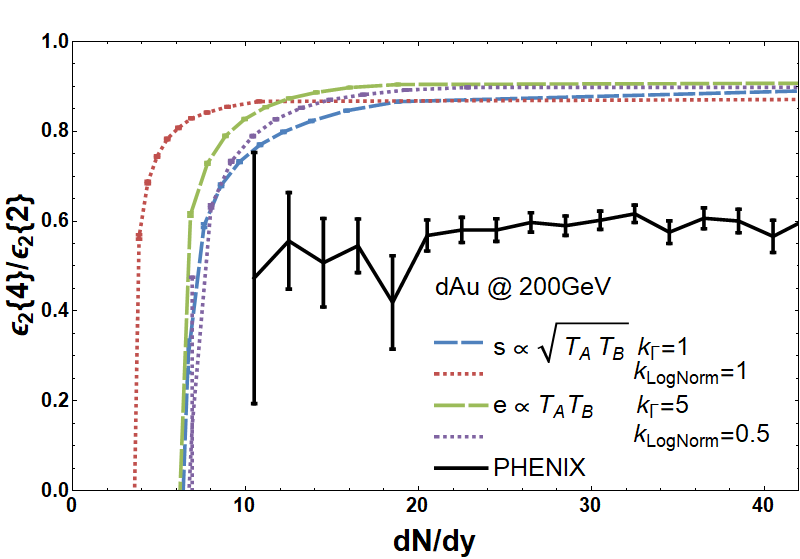}
    \includegraphics[width=0.48\textwidth]{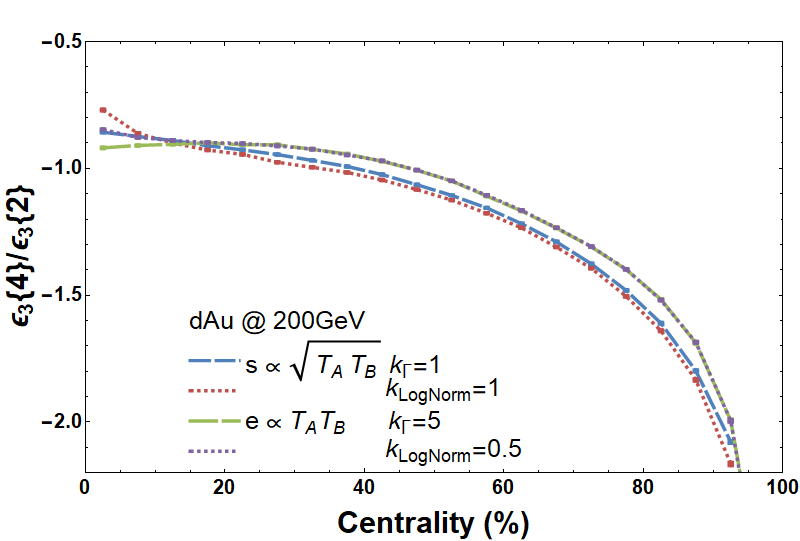}
    \caption{Four particle to two particle eccentricity ratios of dAu for functional forms $\sqrt{T_{A}T_{B}}$ and $T_{A}T_{B}$ using best fits for $\Gamma$ and lognormal multiplicity fluctuation distributions. PHENIX data taken from Ref.~\cite{Aidala:2017ajz}. Figure from Ref.~\cite{Carzon:2021tif}.}
    \label{fig:dAueccratios}
    \end{center}
\end{figure}

The STAR experiment does not have a measurement of $\varepsilon_{2}\{4\}/\varepsilon_{2}\{2\}$ for dAu at $\sqrt{200} GeV$, so we will compare our theoretical models against data from the PHENIX detector \cite{Aidala:2017ajz} in Fig. \ref{fig:dAueccratios} (left). This comparison comes with a decent amount of ambiguity since PHENIX reports this observable in relation to $N_{track}$ which does not easily map onto centrality. Thus, we plot the initial state ellipticity fluctuations with respect to $dN/dy$ (see e.g for more discussion on the topic \cite{ATLAS:2021kty}), which while not perfect we will assume here to be the same as $N_{track}$.

%__________________________________________________________________________
%
%__________________________________________________________________________
%

For $\varepsilon_{2}\{4\}/\varepsilon_{2}\{2\}$, we see a significant difference between the linear and quartic entropy scalings with a significant contribution from the choice of  multiplicity fluctuations in $\sqrt{T_{A}T_{B}}$ that was not seen in the average geometry. An important feature of this observable is where it changes sign (See Sec.~\ref{sec:Cumulants}), which we observe to be the same for all combinations except $\sqrt{T_A T_B}$ with log-normal fluctuations. Along those lines, there is a significant separation in the quartic scaling curves caused by the multiplicity fluctuation distribution. This separation is also seen in the linear scaling but to a much lesser extent. Because this is a small system, the effect from hydrodynamics would be large and non-linear making it difficult to draw specific conclusions from this comparison to experimental data. It would be helpful, however, to have a calculation of this observable in respect to $dN/dy$ or percent centrality which could give a more precise idea of where the sign change occurs and consequently could help in ruling out combinations of $\sqrt{T_A T_B}$ scaling with the different multiplicity fluctuation distributions.

On the left side of Fig. \ref{fig:dAueccratios}, $\varepsilon_{3}\{4\}/\varepsilon_{3}\{2\}$ is plotted for dAu and is imaginary for all centrality classes (See Sec.~\ref{sec:Cumulants}). We chose to plot the fluctuations of triangularity for dAu with respect to centrality instead of $dN/dy$, since it is easier to interpret and there is no experimental data yet for this observable. The same trends observed in $\varepsilon_{3}\{2\}$, Fig.~\ref{fig:dAueccs} (right), are seen in the 4-particle to 2-particle ratio, with linear scaling having a higher magnitude than quartic except for ultra-central events and log-normal fluctuations pushing the $\sqrt{T_A T_B}$ scaling upward for the ultra-central events while the $\Gamma$ distribution case remains flat.

%^^^^^^^^^^^^^^^^^^^^^^^^^^^^^^^^^^^^^^^^^^^^^^^^^^^^^^^^^^^^^^^^^^^^^^^^^^^
\section{Summary} \label{sec:FluctuationSummary}
%^^^^^^^^^^^^^^^^^^^^^^^^^^^^^^^^^^^^^^^^^^^^^^^^^^^^^^^^^^^^^^^^^^^^^^^^^^^

In Bayesian analyses of \code{trento}, the form from which multiplicity fluctuations are sampled is assumed to be a $\Gamma$ distribution. This restrictive choice may contribute to the preference of such analyses toward a reduced thickness functional form of $T_R = \sqrt{T_A T_B}$. There are functional forms outside of \code{trento}'s considered parameter space, specifically linear scaling $T_R \propto T_A T_B$, that have been shown to be consistent with both experimental multiplicity distributions in dAu and AuAu systems and the phenomenologically-preferred $\sqrt{T_A T_B}$. Works using linear scaling, for the reduced thickness function, paired this with multiplicity fluctuations derived from a log-normal distribution, which, while similar to the $\Gamma$ distribution, does contain features that are outside of \code{trento}'s parameter space.

The foundation of this work is the inclusion of linear scaling, $T_A T_B$, and log-normal multiplicity fluctuations in \code{trento} and allowing for specification of any combination of reduced thickness functional form and fluctuation distribution. The linear scaling relationship is seen in models that use energy density to define the initial state, while \code{trento} uses a characterization with respect to entropy, thus requiring a translation of $\epsilon \propto T_A T_B$ to one for entropy. This is accomplished through a conformal equation of state, although a more precise implementation is warranted, since the translation is dependent on the details of the particular equation of state used, and left for future work. Since the analysis here is done using boost-invariant simulations, multiplicity fluctuations in the longitudinal direction are not explored, though they also are interesting \cite{Jia:2015jga,Jia:2020tvb}, and left for future work.

We confirm that the reduced thickness scaling of $\sqrt{T_A T_B}$ does indeed prefer multiplicity fluctuations derived from a $\Gamma$ distribution, while the linear functional form paired with log-normal fluctuations is also able to match experimental data for the multiplicity distributions. Significant differences are observed in the elliptic and triangular geometry of AuAu and dAu, with linear $T_A T_B$ enhancing these geometries. This difference would lead to the extraction of different shear viscosities for each reduced thickness functional form, especially in small systems. The response of initial geometry to these choices is greatly enhanced in small systems making a correct description all the more important. A further analysis could show that \code{trento}'s extraction of the viscosities of the QGP from small systems may contain a systematic uncertainty controllable by increasing the allowed functional space. 

It is costly to calculate cumulant ratios in hydrodynamic models, when running Bayesian analyses, but due to response theory these observables can be directly obtained from initial state eccentricities, with $\varepsilon_{3}\{4\}/\varepsilon_{3}\{2\}$ in ultra-central collisions having less than $1\%$ error \cite{Carzon:2020xwp}. Analysis of cumulant ratios of initial state eccentricities, $\varepsilon_{2}\{4\}/\varepsilon_{2}\{2\}$ and $\varepsilon_{3}\{4\}/\varepsilon_{3}\{2\}$, are important since they are able to constrain parameters in the initial state through experimental data. We see that $\varepsilon_{2}\{4\}/\varepsilon_{2}\{2\}$ changes sign at different centralities with respect to models in dAu, indicating experimental data may be able to distinguish between parameter choices.

%%%%%%%%%%%%%%%%%%%%%%%%%%%%%%%%%%%%%%%%%%%%%%%%%%%%%%%%%%%%%%%%%%%%%%%%%%%
%
\chapter{Exploration of Nuclear Deformability and the $v_2 -to- v_3$ Puzzle} \label{chap:V2toV3Puzzle}
%
%%%%%%%%%%%%%%%%%%%%%%%%%%%%%%%%%%%%%%%%%%%%%%%%%%%%%%%%%%%%%%%%%%%%%%%%%%%

Our understanding of the initial state has undergone rapid development as more structure has been added (See Sec.~\ref{sec:HistoryOfInitialState}). The foundational assumption of the initial state is that nuclei are well approximated by a Woods-Saxon distribution in the transverse plane. While this formulation is able to fit experimental data well, there are areas in which it fails, specifically in ultra-central collisions. The failure of our models in describing heavy-ion collisions in ultra-central collisions is concerning, since this is the regime in which hydrodynamics should work the best \cite{Luzum:2012wu} and where linear response is the most applicable, especially for multi-particle cumulants. These ultra-central systems are the most sensitive to fluctuations since the dominant structure from the impact parameter is minimized. This provides a fertile ground for refinement of physics in the initial state and a unique opportunity for the exclusion of models.

Two systems that have seen significant development from disagreement in central collisions with experiment are $^{129}$Xe$^{129}$Xe and $^{238}$U$^{238}$U. When assuming a spherical nucleus for $^{129}$Xe in theoretical modeling, experiments measure an excess of $v_2\{2\}$ for central collisions. Xenon was theorized to have a small quadrapole deformation, at the time, and including this in initial state models was enough to describe the discrepancy later measured by experiments \cite{Giacalone:2017dud, Acharya:2018ihu, Acharya:2018eaq, CMS:2018jmx, ATLAS:2018iom}. A similar story featured ultra-central $^{238}$U collisions, where including a substantial quadrapole and a small hexadecapole deformation was able to enhance the elliptic flow and bring models closer to the experimental measurements \cite{Adamczyk:2015obl, Goldschmidt:2015kpa, Dasgupta:2016qkq, Giacalone:2018apa}, although there is still some disagreement.

In this chapter, I will explore nuclear deformations in ultra-central collisions of $^{208}$Pb in an attempt to solve the $v_2 -to- v_3$ puzzle. As such, the discussion in this chapter won't be an exhaustive review of all aspects of nuclear deformation in the initial state. For information on other interesting developments of the initial state using deformed nuclei see Sec.~\ref{sec:Intro:FurtherDevelopmentOfInitialState}. This chapter reproduces and refines the work from Ref.~\cite{Carzon:2020xwp}.

\section{Description of the $v_2 -to- v_3$ Puzzle}

\begin{figure}[ht]
    \centering
    \includegraphics[width=0.75\textwidth]{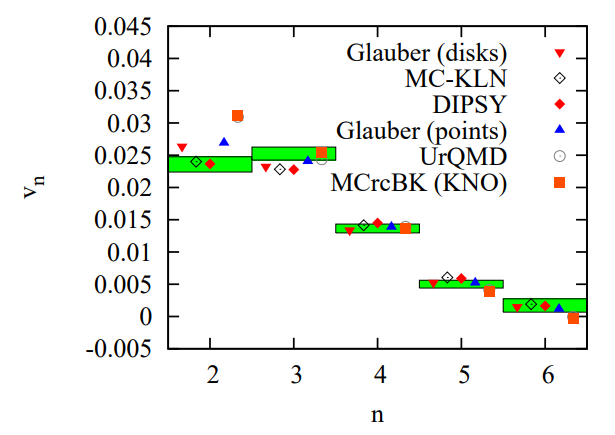} \\
    \caption{Comparison between experimental calculation of $^{208}$Pb$^{208}$Pb integrated $v_n$ for $n=2-6$ at $0-1\%$ centrality and the corresponding predictions from several initial state models. Green bars are from the ATLAS collaboration \cite{ATLAS:2012at}. Figure from Ref.~\cite{Luzum:2012wu}.}
    \label{fig:V2V3Puzzle}
\end{figure}

Experimental measurements \cite{ATLAS:2012at} of $^{208}$Pb$^{208}$Pb show that the elliptic, $v_2\{2\}$, and triangular, $v_3\{2\}$, flow in ultra-central collisions should be of the same magnitude. Theoretical models, that assume $^{208}$Pb is spherical, do not reflect this ordering of the flow coefficients \cite{Luzum:2012wu,CMS:2012xxa,Shen:2015qta,Rose:2014fba}, and predict that $v_2\{2\} > v_3\{2\}$. This means that all models either over predict elliptic flow or under predict triangular flow. An illustration of this puzzle is shown in Fig.~\ref{fig:V2V3Puzzle}, where the green bars are experimental data from ATLAS for integrated $v_n$ in the $0-1\%$ centrality range and the points are several theoretical models that have different initial conditions. We see that it is difficult for any of the models to match both elliptic and triangular flow at the same time. 
%__________________________________________________________________________
%
%__________________________________________________________________________
%

There have been some previous attempts at solving this puzzle, ranging from: varying the initial conditions \cite{Luzum:2012wu,Shen:2015qta, Gelis:2019vzt}, varying the transport coefficients \cite{Luzum:2012wu,Rose:2014fba}, and varying the equation of state \cite{Alba:2017hhe}. All attempted solutions have failed to fully resolve the discrepency between theory and experiment. A more recent study \cite{Giannini:2022lbj} used Bayesian parameter estimation from the state-of-the-art models and showed that we are still unable to describe the difference from experiment. When considering the $0-5\%$ centrality class, the models agree with experiment and an argument could be made that we have all but solved the issue. However, experimental data is available for much more central centrality classes, such as $0-0.02\%$, and the discrepancy remains at those centralities even when accounting for error. Another important detail is that $v_2\{2\}$ and $ v_3\{2\}$ have a dependence on the range of transverse momentum, which means that experiment does not always have $v_2\{2\}\approx  v_3\{2\}$. However, this effect is present in both theory and experiment, thus, not affecting the puzzle.

In order to better understand the $v_2 -to- v_3$ puzzle and see what is needed in the initial state, I will walk though the logic step-by-step.

\begin{itemize}
    
    \item We start with the powerful linear response of the final state flow harmonics to the initial state geometry (See Sec.~\ref{sec:Eccentricities}), $v_n = \kappa_n \varepsilon_n$, where $\kappa_n$ are the response coefficients. These response coefficients encode the information about the hydrodynamic evolution. There are non-linear contributions to the initial to final state mapping but they are negligible in ultra-central collisions \cite{Noronha-Hostler:2015dbi, Sievert:2019zjr, Rao:2019vgy}.

    \item For an ultra-central collision of two spherical nuclei, the initial state geometry should be close to circular and thus elliptic and triangular spatial eccentricities would vanish, or $\varepsilon_2 = \varepsilon_3 = 0$ on average.

    \item When you include the event-by-event fluctuations, non-zero ellipticity and triangularity are generated through they should still be approximately the same $\varepsilon_2 \{2\} \approx \varepsilon_3 \{2\}$.

    \item Now we look at what the effect off hydrodynamics will be on this initial state relationship. Viscosity affects different orders of structure differently, killing of higher-order structure more quickly. This means triangular geometry is dampened more than elliptic geometry. This translates to an ordering of the response coefficients such that $\kappa_3 < \kappa_2$.

    \item Thus, initial conditions for round nuclei have a geometry relationship of $\varepsilon_2 \{2\} \approx \varepsilon_3 \{2\}$ and this will lead to an ordering of the final state flow harmonics such that $v_2 \{2\} > v_3 \{2\}$.

    \item This means that if experiment measures $v_2 \{2\} \approx v_3 \{2\}$, we want the initial state ordering of geometry to be $\varepsilon_2 \{2\} < \varepsilon_3 \{2\}$ to offset the viscosity effects.
    
\end{itemize}

There are other discrepancies in ultra-central $^{208}$Pb collisions that are related to the $v_2 -to- v_3$ puzzle. The first, concerns the four-particle correlation of triangular flow, $v_3 \{4\}$. Taking the ratio with the two-particle correlation, $v_3\{4\}/v_3\{2\}$, we can get an estimate of the event-by-event fluctuations in $v_3$ (See Sec.\ref{sec:Cumulants}). Hydrodynamic models underpredict this ratio and, thus, overestimate the width of triangular event-by-event fluctuations. Being able to fit $v_3\{4\}/v_3\{2\}$ and get the correct ordering of initial state geometry at the same time would help to constrain the puzzle. Another potentially related issue is the four-particle correlation of the fourth-order flow harmonic, $v_4\{4\}^4$, where theory is insufficient in describing the experimental measurement \cite{Giacalone:2016mdr}. This inconsistency occurs in central to mid-central events and is outside the range of this analysis, but complimentary solutions could be possible.

It is a natural extension of the recent advances in describing the initial state to ask if nuclear deformation might help us solve the $^{208}$Pb $v_2 -to- v_3$ puzzle. It is not obvious that this would work considering $^{208}$Pb has a doubly-magic atomic number, meaning it has a number of both protons and neutrons that make complete shells, and is a highly stable nucleus. Furthermore, nuclear structure tables list the octupole deformation, $\beta_3$, of $^{208}$Pb to be zero \cite{Moller:2015fba}. This may, however, not be the case with a recent paper predicting a finite $\beta_3$ using a Minimization After Projection  model \cite{Robledo:2011nf}. The existence of a deformed $^{208}$Pb in heavy-ion collisions may be possible since the structure of nuclei can change with respect to energy and the regime in which nuclear structure experiments run is at much lower beam energies than relativistic heavy-ion collisions. Another factor to consider, is that heavy-ion collisions probe the color charge density and see the geometry of the whole system, while nuclear structure experiments only measure the electric charge density and its associated geometry. The electric charge density reflects the geometry of protons but not necessarily the full geometry of the nucleus, which could have features such as a neutron skin \cite{Tarbert:2013jze}. Since these two geometries do not have to the same, it plausible to think that the full geometry $^{208}$Pb might have a finite $\beta_3$, despite no deformation present in the electric charge density, which would be an important factor in heavy-ion collisions. 

This chapter will explore the possible octupole deformation of $^{208}$Pb and its capability of solving the $v_2$-to-$v_3$ puzzle. Since the puzzle was first discovered, new experimental data from the LHC has been released that was run at 5.02 TeV energy and had higher luminosity (more events per second), which has significantly reduced the error bars associated with this measurement. This reduction of error and advances in medium effects, such as improvements in the equation of state, has greatly reduced the disagreement between experiment and theory but has not fully resolved the issue. By using the correlated issue of theoretical misprediction of $v_3\{4\}/v_3\{2\}$ from ATLAS \cite{Aaboud:2019sma} we can constrain the range  of octupole deformation of $^{208}$Pb. 

Nested in this analysis are a few other comparisons that are important to introduce here. First, is an additional comparison between the effect of the octupole deformation on quartic entropy scaling, $\sqrt{T_A T_B}$, and linear entropy scaling, $T_A T_B$. The former comes from phenomenological analyses, while the latter is observed in Color Glass Condensate (CGC) theory. The motivation for including this extra parameter is to continue the discussions from Refs.~\cite{Lappi:2006hq, Nagle:2018ybc, Romatschke:2017ejr, Chen:2015wia}. The second addition, is looking at the octupole deformation of Pb at a lower collision energy of 2.76 TeV, which was used in the analysis of Ref.~\cite{Giacalone:2017uqx}. These inclusions connect this work to the larger discussions present in the field and provide useful comparisons for the validity of our model. Only Sec.~\ref{sec:OctupoleDeformedEccentricities} deals with these parameters in detail.

In this chapter, I will first review the relevant properties of the initial state with a focus on the deformation parameters $\beta_n$, in Sec.~\ref{sec:OctupoleDeformationInitialState}, and, in Sec.~\ref{sec:HydrodynamicModels}, the hydrodynamic models used in this analysis. In Sec.~\ref{sec:OctupoleDeformedEccentricities}, I look at the effect octupole deformation has on the initial state eccentricities. Finally, Sec.~\ref{sec:DeformedPbComparedToExperiment} will compare the full theoretical framework to experimental data.

\section{Theoretical Models} \label{sec:TheoreticalModels}

\subsection{Octupole Deformation in the Initial State} \label{sec:OctupoleDeformationInitialState}

Current phenomenological initial state models start with constructing nuclei by sampling each nucleon from a Woods-Saxon distribution on an event-by-event basis (See Sec.~\ref{sec:Intro:NucleonFluctuations}). This method provides a density profile that represents the geometrical structure of each heavy-ion that can then be used to get the collision's geometry through reduction of the projectile and target nuclei thickness functions. The Woods-Saxon distribution is a potential describing the distribution of nucleons in a nucleus and the two-parameter form is generically given in spherical coordinates by Eq.~\ref{eq:WoodsSaxon}, where $a$ is the skin thickness of the distribution and $R$ is the nuclear radius. This profile will be nearly constant for $r < R$ with an exponential tail for $r > R$. For a spherical nucleus, the nuclear radius is taken to be constant and deformations are introduced by making it a function of the angle $\theta$. This angular dependent nuclear radius can be decomposed into a complete sum of spherical harmonics, as described in Eq.~\ref{e:Rdef}, where $Y_{l m}$ are the spherical harmonics and the $\beta_l$ coefficients are the nuclear deformation parameters. For a spherical nucleus, all $\beta_l=0$ while deformations are introduced by setting the coefficients to be non-zero with the first three representing quadrupole ($\beta_2$), octupole ($\beta_3$), or hexadecupole ($\beta_4$) deformations. Though the summation in Eq.~\ref{e:Rdef} over $Y_{\ell m}$ is from $\ell \in [0,\infty)$, we only consider here the octupole deformation $\beta_3$, since, for the solution of the $v_2$-to-$v_3$ puzzle, we require an enhancement of triangular geometry which will be most affected by the octupole deformation. The $\beta_3$ deformation will not only have an effect on the triangular geometry but will also enhance the elliptic geometry, though the hope is that this secondary effect will be low enough that the $v_2$-to-$v_3$ discrepancy can be closed. 

For this analysis, we look at several different values of $\beta_3=0, 0.0375, 0.075,$ and $0.12$. The choice of these values is motivated by Ref.~\cite{Robledo:2011nf}, where they used a Minimization After Projection model to predict the octupole deformation $\beta_3$ of $^{208}$Pb, and found $\beta_3\approx 0.0375$ in the ground state and $\approx 0.075$ in the first excited state. This same work also used a Hartree-Fock-Bogoliubov calculation, where the deformation was found to be $\beta_3=0$. An expected error of $\pm 25\%$ for the comparison of these deformation parameters between theory and experiment was also provided. This leads to a constraint on the octupole deformation of $^{208}$Pb to be $\beta_3 \lesssim 0.1$ from nuclear structure calculations at low energies. We include the value of $\beta_3= 0.12$, which is outside the nuclear structure range, as an upper limit and to better quantify its effect when comparing to experimental data.

The initial state model we use here is the \code{trento} initial condition \cite{Moreland:2014oya} that has been modified to include a non-zero $\beta_3$ as well as linear entropy scaling. For the majority of the analysis, specifically the hydrodynamic simulations, we focus on the entropy scaling relationship, $p=0$ or $s\propto\sqrt{T_A T_B}$, that was obtained from the \code{trento} Bayesian analysis \cite{Bernhard:2016tnd}. Further parameter choices include: multiplicity fluctuations using a $\Gamma$ distribution with $k=1.6$, and a nucleon width  of $\sigma=0.51 \, \mathrm{fm}$. While looking at the effect of deformation on the initial state in Sec.~\ref{sec:OctupoleDeformedEccentricities}, we include a comparison to linear entropy scaling, $s\propto{T_A T_B}$, as an extension of the discussions from Refs.~\cite{Lappi:2006hq, Nagle:2018ybc, Romatschke:2017ejr, Chen:2015wia}. For this chapter, we explicitly use $s\propto{T_A T_B}$ which is not exactly the scaling found from CGC which is linear in proportion to energy, $\epsilon\propto{T_A T_B}$. This failure in definition was discovered in the analysis done in Chap.~\ref{chap:ExplorationOfMultiplicityFluctuations}, which occurred after this work. This change in definition is very small for large systems and is unlikely to change the results of this work. Additionally, we use multiplicity fluctuations from a $\Gamma$ distribution with $k=20$ for the linear scaling.

\subsection{Hydrodynamic Models} \label{sec:HydrodynamicModels}

To compare directly to experimental data, we run full hydrodynamic simulations using two different frameworks to ensure the results are not unique to a particular model.

The first framework (Parameter 1) uses v-USPhydro \cite{Noronha-Hostler:2013gga,Noronha-Hostler:2014dqa}, which solves the equations of motion using Smoothed Particle Hydrodynamics (SPH), for the hydrodynamic stage. For this framework, the shear viscosity to entropy density ratio is set to a constant, $\eta/s=0.047$, while the bulk viscosity to entropy density ratio is set to zero, $\zeta/s=0$, and the freeze-out temperature is $T_{FO}=150$ MeV. We use the equation of state WB21/PDG16+ \cite{Alba:2017hhe}, since it has been verified against experimental data across system sizes and beam energies \cite{Alba:2017hhe,Giacalone:2017dud} though other predictions in ArAr and OO systems have yet to be confirmed \cite{Sievert:2019zjr}. The hydrodynamic evolution is followed by direct decays \cite{Sollfrank:1991xm,Wiedemann:1996ig} using all known resonances from the PDG16+ \cite{Alba:2017mqu}. This is a particularly important feature of this model with the intention being to use the most state-of-the-art list of particles which would most closely match Lattice QCD results \cite{Alba:2017mqu}. Since this analysis, further investigation \cite{SalinasSanMartin:2022inb} has found that PDG16+ is still missing some resonances, specifically ones with strange baryons. An improved analysis could be done using this new PDG21+ in future studies. Unfortunately, a side effect of choosing a detailed treatment of feed-down decays is that hadronic transport is not included in this framework since many of the resonances decay into 3 or 4 body decays and those cases are not covered by the state-of-the-art hadron transport codes SMASH \cite{Ono:2019ndq} and URQMD. This, too, could be rectified in a future analysis thanks to Ref.~\cite{SalinasSanMartin:2022inb} where the authors converted the 3 and 4 body decays into systems of 2 body decays that can be used in SMASH.

\begin{figure}[ht]
    \centering
	\includegraphics[width=0.6\linewidth]{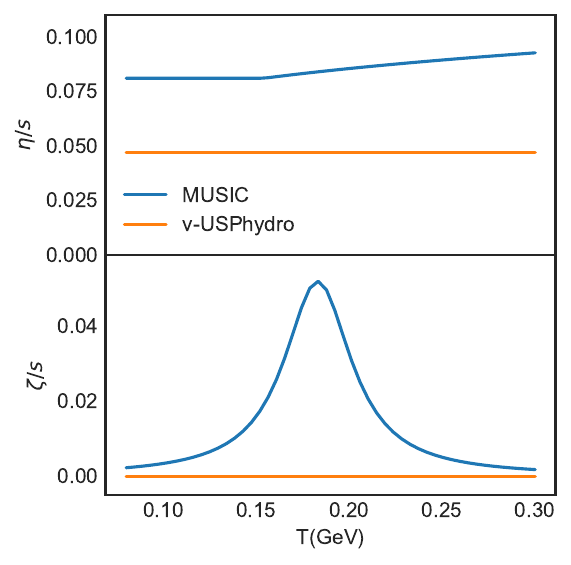} \\
	\caption{Comparison of the bulk and shear viscosities entering the two hydrodynamic models using MUSIC and v-USPhydro. Figure from Ref.~\cite{Carzon:2020xwp}.}
	\label{f:viscosities}
\end{figure}

The second framework (Parameter 2) uses  MUSIC \cite{Schenke:2010nt,Schenke:2010rr}, a grid-based code, for the hydrodynamic stage. All parameters for this framework are the Maximum a Posteriori (MAP) values which come from the Bayesian analysis of momentum integrated observables reported in Table 5.9 of  Ref.~\cite{Bernhard:2016tnd}. Here, both the shear and bulk viscosities depend on temperature and are shown in Fig.~\ref{f:viscosities}. The equation of state used here is \textit{s95p-v1.2 }\cite{Huovinen:2009yb}. At $T=151 MeV$ the fluid freezes out into hadrons and resonances which are evolved by the hadron cascade code UrQMD \cite{Bass:1998ca, Bleicher:1999xi}. This choice, in contrast to the first framework, gives up the full particle list for the inclusion of hadron transport. The parameters for this framework were tuned for a \code{trento} initial condition of energy density initialized at $\tau=0$ and given a pre-equilibrium evolution until $\tau_{\rm fs}=1.16$ fm/$c$ using free streaming. Here we use \code{trento} initial conditions of entropy density at finite time $\tau_0 = 0.6$ fm/$c$ with no initial transverse flow. Despite this difference in the initialization, the fit to data is expected to be reasonable at most centralities \cite{NunesdaSilva:2018viu, NunesdaSilva:2020bfs}.

%__________________________________________________________________________
%
%__________________________________________________________________________
%

In both frameworks, the hydrodynamic simulations, v-USPhydro and MUSIC, pass analytical solutions (the Gubser test) \cite{Marrochio:2013wla} so there is confidence in the fact that any differences arise from the physical ingredients and not from numerics. To summarize, the physics differences between the two frameworks are: structure of the code (SPH vs grid), equation of state (WB21/PDG16+ vs \textit{s95p-v1.2 }), shear and bulk viscosities (constant vs temperature dependent), list of hadrons and resonances (inclusion of 3 and 4 body vs only 2 body decays), and hadronic transport (none vs UrQMD), respectively. Furthermore, the equations of motion are different between the two models; v-USPhydro uses phenomenological Israel-Stewart while MUSIC uses DNMR with the second order transport coefficients from \cite{Denicol:2014vaa}.

\section{Results}

\subsection{Eccentricities} \label{sec:OctupoleDeformedEccentricities}

A measurement of the impact parameter of a collision is impossible experimentally and so the number of produced particles is used to determine the centrality of collisions, since it is anti-correlated with impact parameter (See Sec.~\ref{Sec:Centrality}). The most central events will be those with high multiplicity and low impact parameter, with the most central collisions ($\approx0-1\%$) having a vanishing impact parameter. Finer binning of events into ultra-central collisions means you are sensitive to the nuclear deformation and fluctuations in particle production \cite{Noronha-Hostler:2019ytn}.  

In the initial state we can calculate the impact parameter exactly, but to better compare to data we choose to use initial entropy for centrality binning. This is possible due to the number of particles produced in the final state being controlled by the entropy of the system at freeze-out. Since the QGP has a low viscosity, the total entropy of the system is nearly conserved throughout the evolution and means that using the initial entropy to determine centrality is a good approximation of the final state classification.

For an individual event, we use the definition of eccentricity, as described in Sec.~\ref{sec:Eccentricities}, to quantify the geometry of the initial state. The important details of the characterization of initial state eccentricities, to consider for this analysis, are detailed as follows. The ellipticity of an event is quantified using the $\varepsilon_2$ coefficient, which is at a maximum in mid-central to peripheral collisions of Pb and a minimum in central collisions. This geometry is highly tied to the overlap region, or centrality of a collision. The other important geometry, we will consider, is the triangularity which is represented by $\varepsilon_3$. This higher-order geometry has no dependence on the overlap region and is driven entirely by event-by-event fluctuations. Given that geometry induced by fluctuations is small compared to centrality dependant geometry, $\varepsilon_3$ is expected to have a lower magnitude as compared to $\varepsilon_2$ at most centralities of large collision systems.

\begin{figure}[ht]
    \centering
    \includegraphics[width=0.75\textwidth]{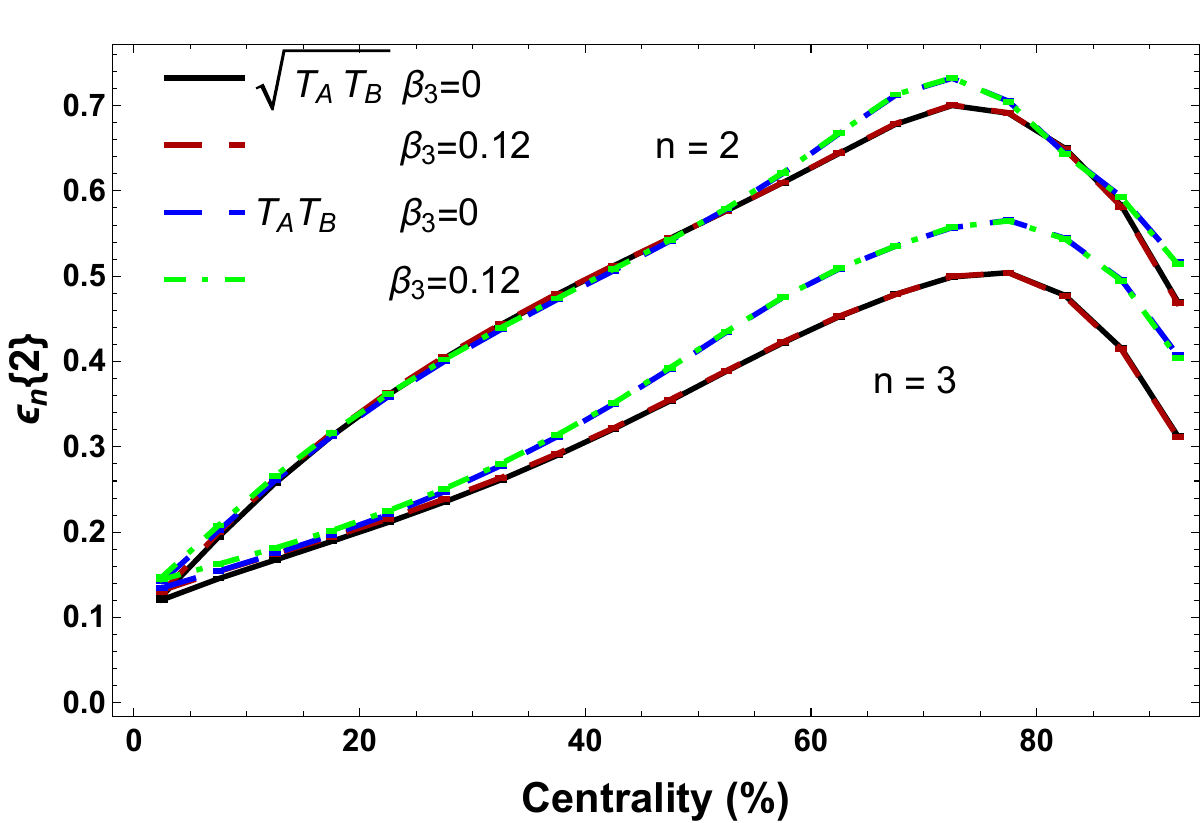} \\
    \caption{The RMS eccentricities $\varepsilon_n\{2\} = \sqrt{\langle \varepsilon_n^2 \rangle}$ for the elliptic $n=2$ and triangular $n=3$ harmonics as a function of collision centrality for $\sqrt{s_{NN}}=5.02$ TeV Pb Pb collisions.  Different curves reflect the choices of initial entropy deposition ($\sqrt{T_A T_B}$ vs $T_A T_B$) and octupole deformation $\beta_3$. Figure from Ref.~\cite{Carzon:2020xwp}.}
    \label{fig:ecc2}
\end{figure}

There are three important sources of fluctuations to consider in the initial state of deformed Pb collisions; nucleon, multiplicity, and deformation induced fluctuations. We can get an estimate of the contribution from these sources by looking at the two-particle cumulants (See Sec.~\ref{sec:Cumulants}) of initial state geometry, which is shown in Fig.~\ref{fig:ecc2}. The comparison in Fig.~\ref{fig:ecc2} is between the two entropy scalings with and without octupole deformation for each, where we take use the maximum considered value of $\beta_3$. For the ellipticity, there is no noticeable difference between the entropy scalings, except in peripheral collisions, and no sensitivity to the nuclear deformation with the exception of the most central bins, although it is hard to see in Fig.~\ref{fig:ecc2} and requires further investigation. The triangular geometry is more interesting since we can see the contribution of each type of fluctuation. The influence of nucleon fluctuations is seen in the fact that $\varepsilon_3\{2\}$ is non-zero and is, thus, the largest contribution to triangularity. Next, we see there is a difference between the entropy scaling relationships, with linear scaling having a larger triangularity than $\sqrt{T_A T_B}$, which is somewhat related to multiplicity fluctuations (See Chap.~\ref{chap:ExplorationOfMultiplicityFluctuations}) and is the second largest source of fluctuations. Finally, we see that turning on octupole deformation has no effect on triangularity except in the most central events where we start to see sensitivity. Nuclear deformation is, thus, the smallest source of geometry fluctuations and entirely unnecessary when considering the geometry of mid-central to peripheral centralities in large systems. This comparison further supports the observation that ultra-central collisions are the most sensitive to small fluctuations, particularly deformations of the Woods-Saxon distribution. 

%__________________________________________________________________________
%
%__________________________________________________________________________
%

\begin{figure*}[ht]
    \begin{tabular}{c c}
        \includegraphics[width=0.5\textwidth]{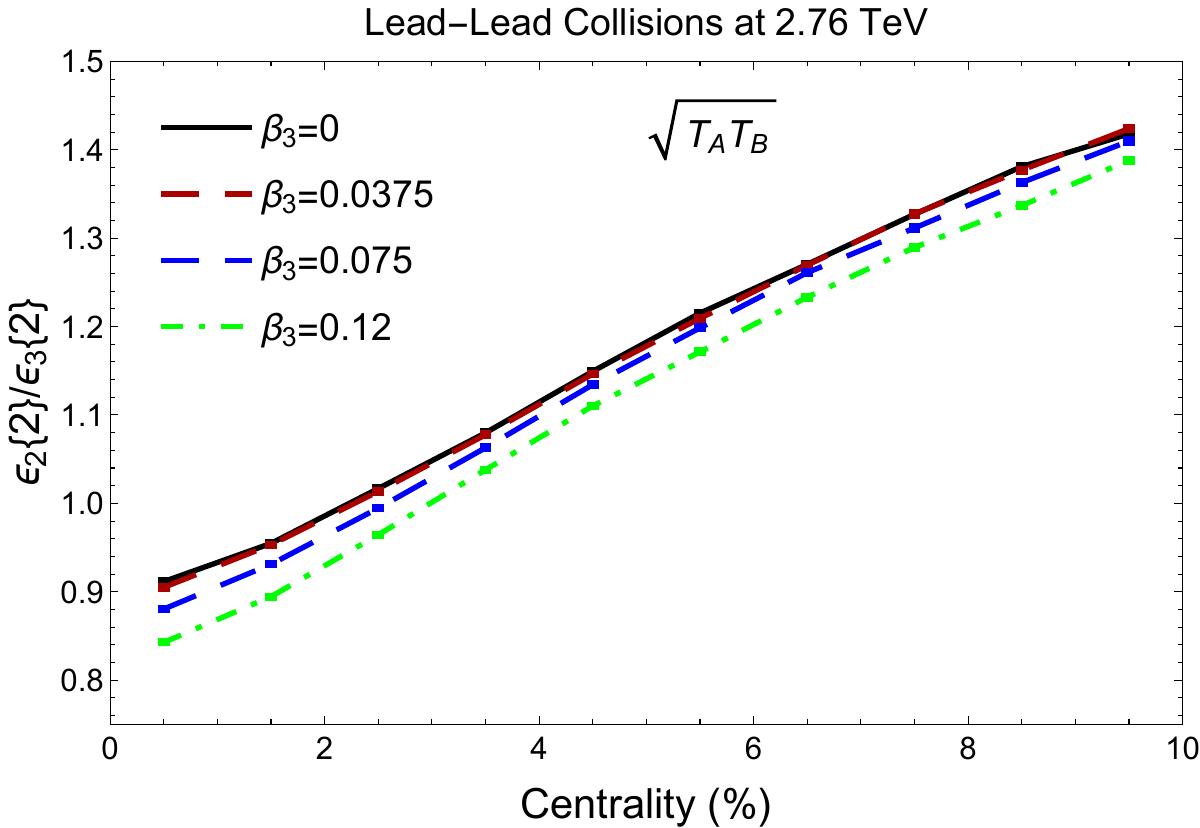} &
        \includegraphics[width=0.5\textwidth]{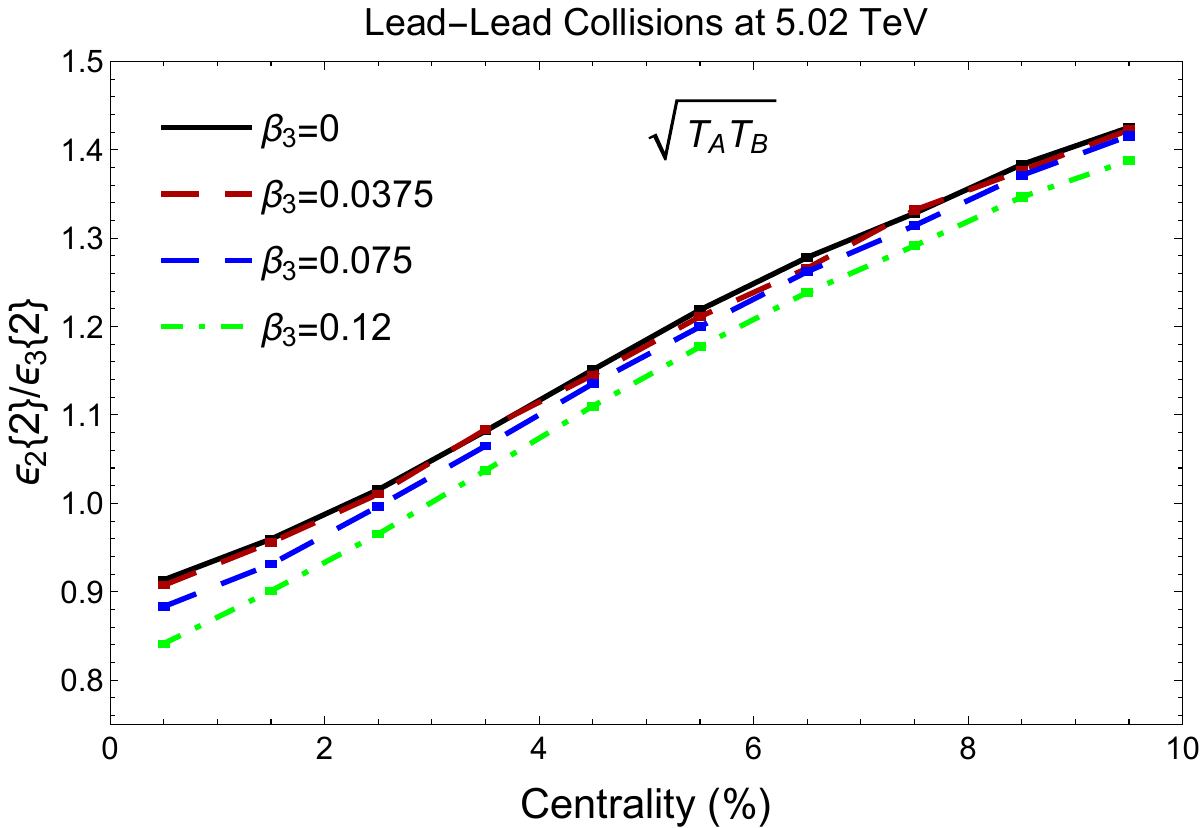} \\
        \includegraphics[width=0.5\textwidth]{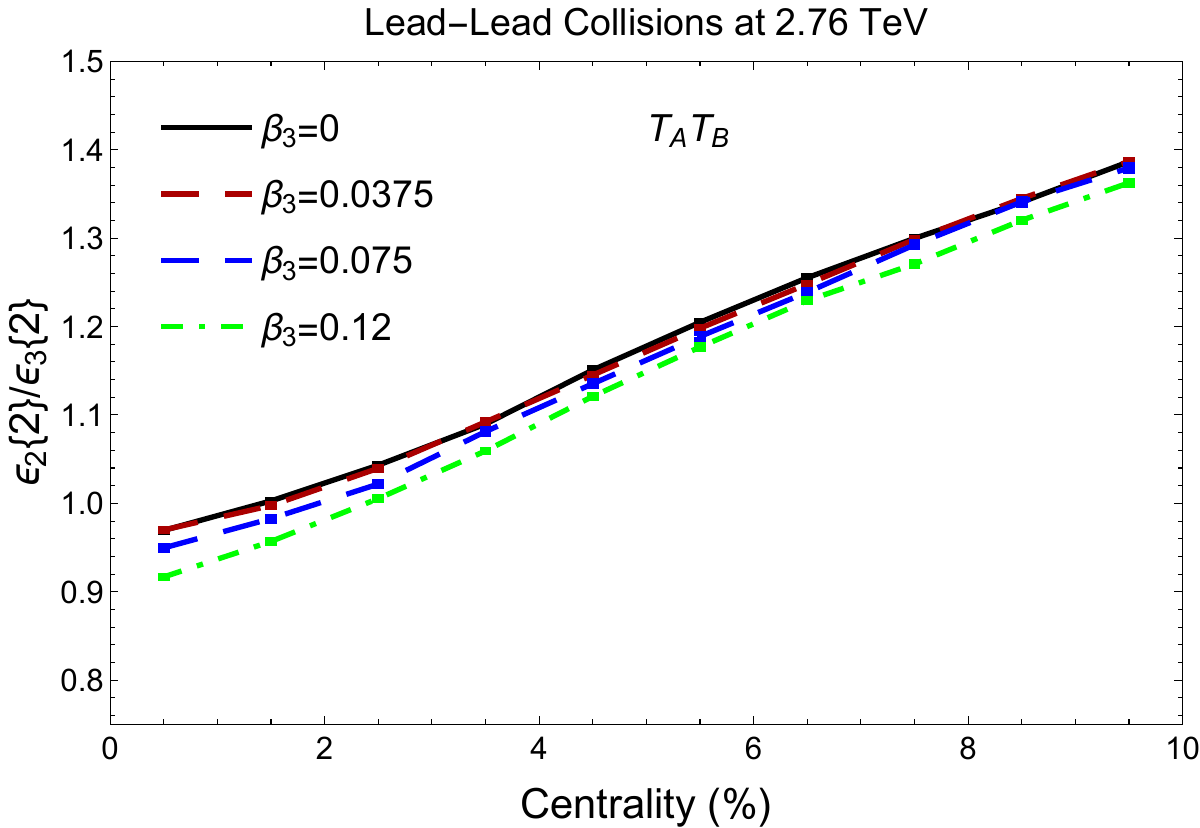} & 
        \includegraphics[width=0.5\textwidth]{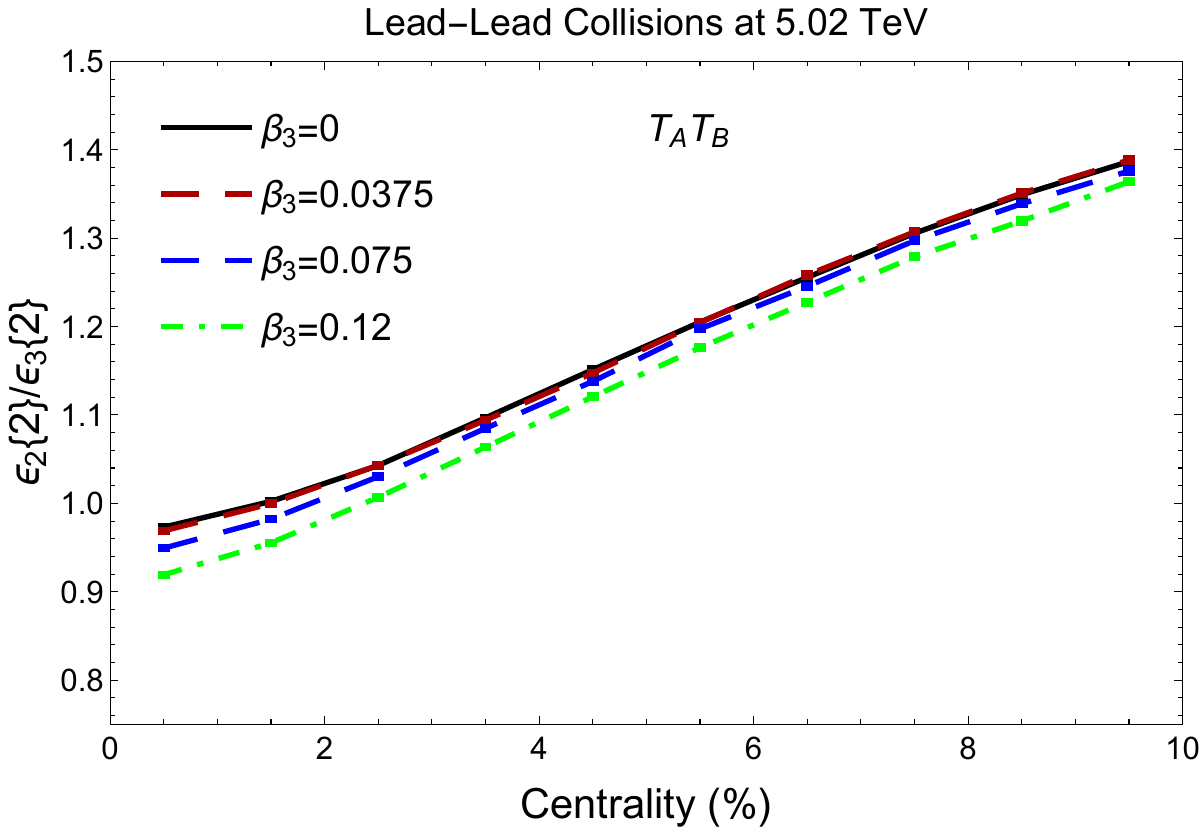} 
    \end{tabular}
    \caption{Ratio of $\varepsilon_2\{2\}/\varepsilon_3\{2\}$ for PbPb collisions at $2.76$ TeV (left) and $5.02$ TeV (right) for varying ocutpole deformations $\beta_3$. We also compare two models for the initial-state entropy deposition: $s \approx \sqrt{T_AT_B}$ (top row) and $s \approx T_A T_B$ (bottom row). Figure from Ref.~\cite{Carzon:2020xwp}.}
    \label{fig:rat}
\end{figure*}

Since we are looking for an ordering of the initial state geometry such that $\varepsilon_2 < \varepsilon_3$, we can look at the ratio of the two-particle cumulants of the ellipticity to triangularity to better quantify the relationship. Due to the difference in order of the linear response coefficients, the ratio $\frac{\varepsilon_2 \{2\}}{\varepsilon_3 \{2\}}$ is not comparable to the same ratio of final state flow harmonics since the $\kappa_n$ in Eq.~\ref{eq:LinearResponse} will not cancel. This observable is still useful for quantifying the initial state, though, since we know what the ordering of initial geometry should be. Now focusing on central events in bins of 1\% from 0-10\%, we show the $\frac{\varepsilon_2 \{2\}}{\varepsilon_3 \{2\}}$ observable in Fig.~\ref{fig:rat} for two collision energies, 2.76 TeV (left) and 5.02 TeV (right), and both entropy scalings, $\sqrt{T_A T_B}$ (top) and $T_A T_B$ (bottom). Intuitive understanding of this ratio observable is tied to its relation to unity, which would indicate that the ellipticity and triangularity are equal in magnitude, and how it deviates from that. An enhancement of triangularity, which is needed to solve the $v_2$-to-$v_3$ puzzle, correlates to the ratio $\frac{\varepsilon_2 \{2\}}{\varepsilon_3 \{2\}}$ being less than 1. Looking at how introducing the octupole deformation effects this ratio, we see that across all parameter spaces there is a relative enhancement of $\varepsilon_3$ which increases with large deformation. We further observe, that quartic scaling is naturally more triangluar than linear for ultra-central collisions and is more sensitive to the nuclear deformation. This leads us to the conclusion that $\sqrt{T_A T_B}$ entropy scaling is better positioned at solving this puzzle with a $10-20\%$ effect coming from the octupole deformation as opposed to $T_A T_B$ only seeing an enhancement of triangularity of a few percent. A final observation, is that there is no strong dependence of this cumulant ratio on the beam energy, which is consistent with other studies \cite{Rao:2019vgy}.

%__________________________________________________________________________
%
%__________________________________________________________________________
%

The ratio of $v_2\{2\}$ and $v_3\{2\}$ is most emphasized in the search for a solution to the $v_2$-to-$v_3$ puzzle since it encapsulates the crux of the issue. However, there are other discrepancies, between theory and experiment, in ultra-central $^{208}$Pb collisions that could be used to constrain the puzzle, namely the 4-particle to 2-particle ratio of $v_3$, $v_3\{4\}/v_3\{2\}$. The authors of Ref.~\cite{Giacalone:2017uqx} pointed out that, while $\frac{\varepsilon_2 \{2\}}{\varepsilon_3 \{2\}}$ is not comparable to the final state due to the hydrodynamic response not cancelling, the 4-particle to 2-particle ratio of $\varepsilon_3$ would have nearly-linear response in central collisions:  
\begin{equation}
\frac{v_3\{4\}}{v_3\{2\}}\approx\frac{\varepsilon_3\{4\}}{\varepsilon_3\{2\}}.
\end{equation}
In order to test this relationship, we compared the initial state ratio against the final state ratio using 11,000 events from 0-1\% centrality. The difference between $\frac{v_3\{4\}}{v_3\{2\}}$ and $\frac{\varepsilon_3\{4\}}{\varepsilon_3\{2\}}$ was less than $1\%$, confirming the claim of Ref.~\cite{Giacalone:2017uqx} and allowing us to compare this ratio of eccentricities to experimental data directly with confidence. This observable happens to be statistically 'hungry', requiring on the order of 15 million events to reduce error bars enough to resolve differences. Running full hydrodynamic simulations on so many events is quite costly, given the current infrastructure, and so using the initial state quantity allows us to reach the required statistics for this observable.

\begin{figure*}[ht]
    \begin{tabular}{c c}
        \includegraphics[width=0.5\textwidth]{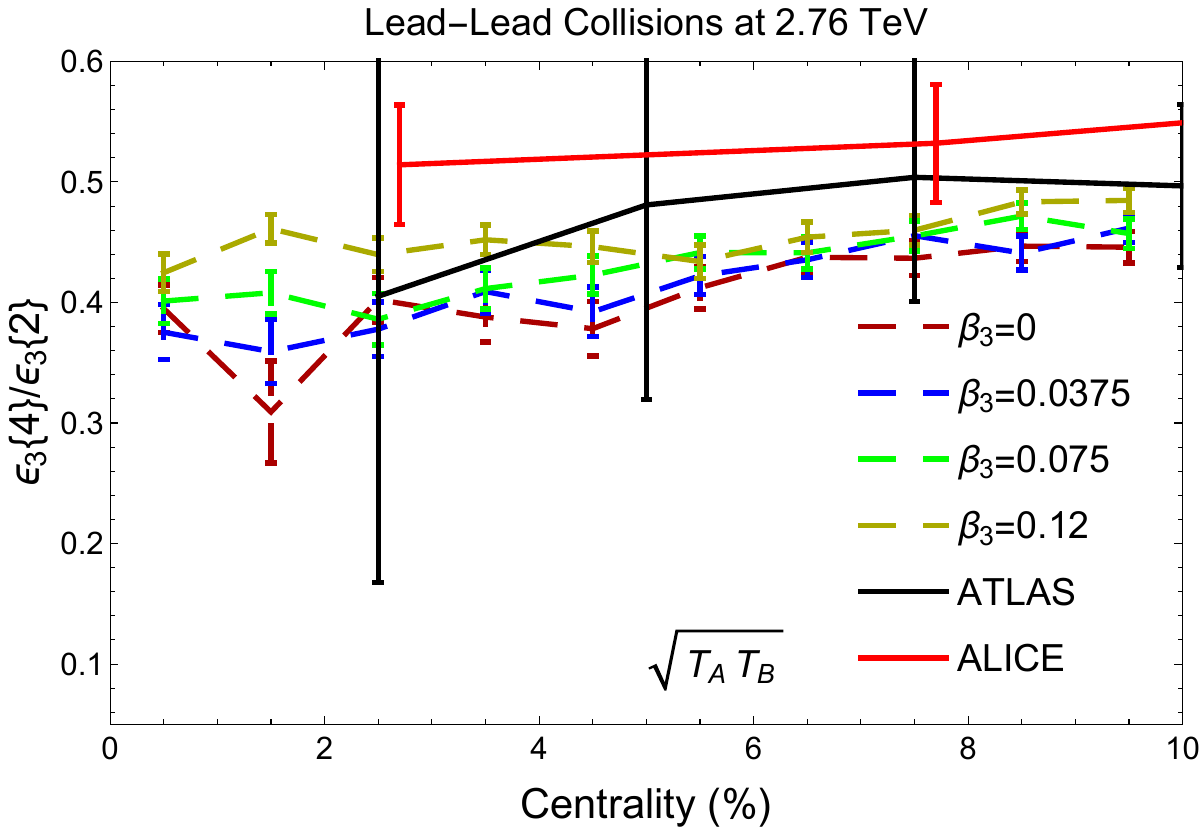} &
        \includegraphics[width=0.5\textwidth]{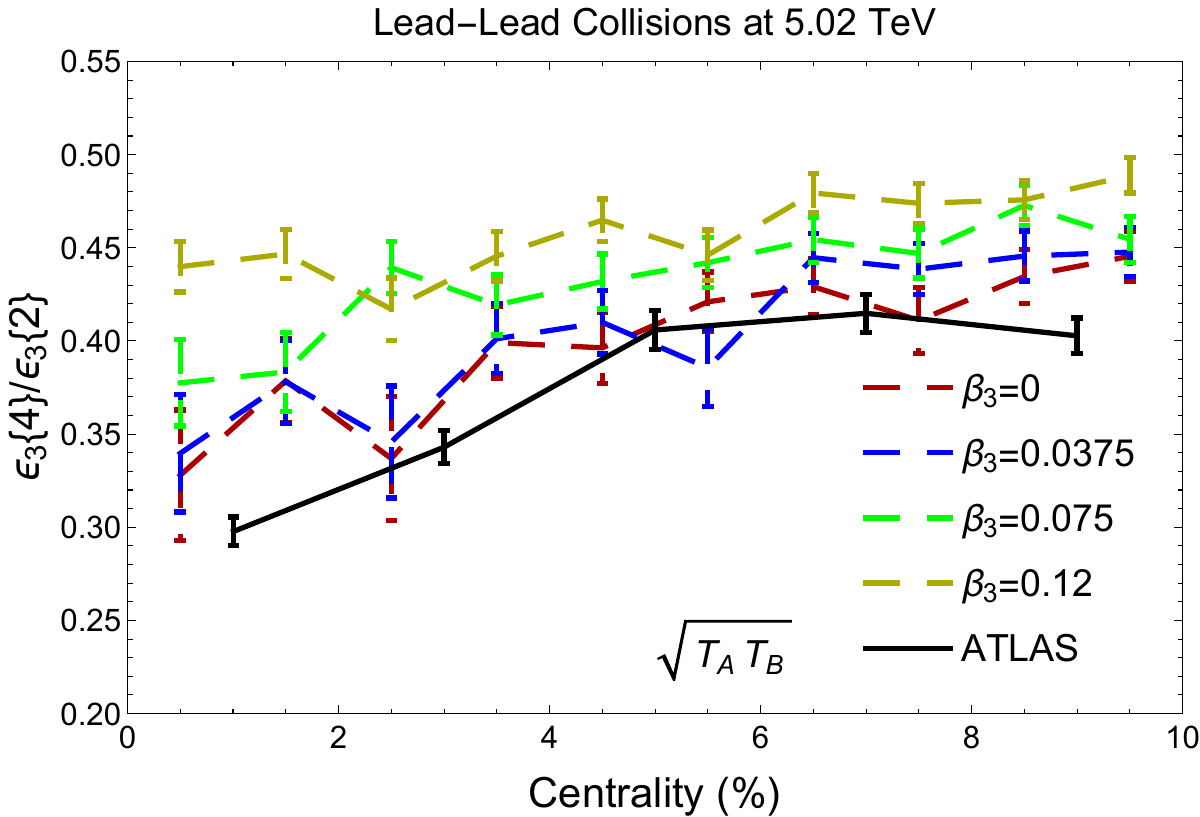} \\
        \includegraphics[width=0.5\textwidth]{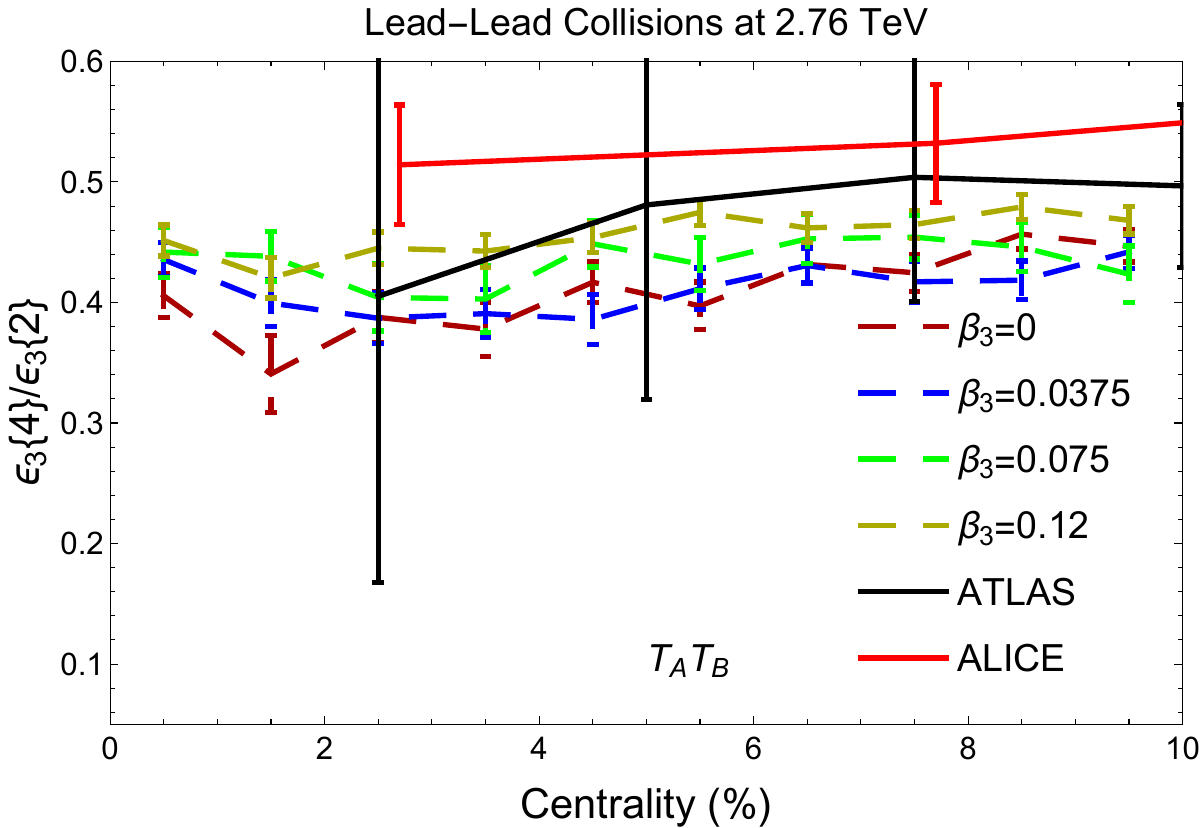} &
        \includegraphics[width=0.5\textwidth]{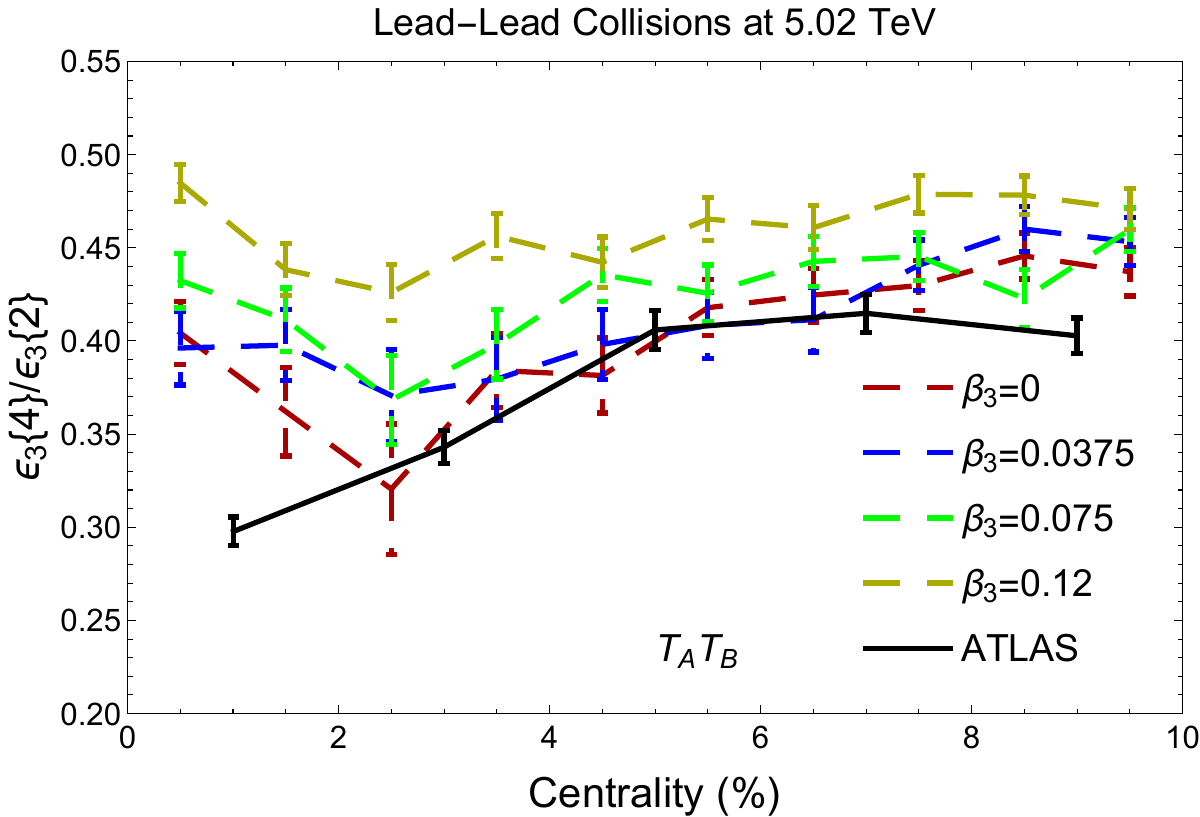} 
    \end{tabular}
    \caption{Ratio of $\varepsilon_3\{4\}/\varepsilon_3\{2\}$ for Pb Pb collisions at $2.76$ TeV (left) and $5.02$ TeV (right) for varying octupole deformations $\beta_3$. We also compare two models for the initial-state entropy deposition: $s \approx \sqrt{T_AT_B}$ (top row) and $s \approx T_A T_B$ (bottom row).  We compare against data at 2.76 TeV from ALICE \cite{ALICE:2011ab} and ATLAS \cite{Aad:2014vba} and at 5.02 TeV from ALICE \cite{Acharya:2018lmh, Acharya:2019vdf}. Figure from Ref.~\cite{Carzon:2020xwp}.}
    \label{fig:v34v32}
\end{figure*}

The cumulant ratio $\frac{\varepsilon_3\{4\}}{\varepsilon_3\{2\}}$ is plotted in Fig.~\ref{fig:v34v32} for two collision energies, 2.76 TeV (left) and 5.02 TeV (right), and both entropy scalings, $\sqrt{T_A T_B}$ (top) and $T_A T_B$ (bottom). The enhancement of octupole deformation translates to an increase of this ratio in all cases, with the greatest response coming from the higher energy collisions. We have looked at the lower collision energy of 2.76 TeV for the sake of consistency, since the previous comparisons for this observable \cite{Giacalone:2017uqx} were for Pb Pb collisions at this energy, where the estimated experimental errors were quite large. One contribution to these large error bars, is because the covariance was not reported by the experimental collaborations and leads to an over counting of the correlated errors for this observable. When looking at how $\beta_3$ effects this ratio, we see there is not enough variation to differentiate between different deformations when compared to experimental measurements. If anything, we see a slight preference for a large $\beta_3$ of 0.12 when compared to ALICE results \cite{Acharya:2019vdf}. 

%__________________________________________________________________________
%
%__________________________________________________________________________
%

Thankfully, the ATLAS experiment did a run of $^{208}$Pb collisions at 5.02 TeV at a higher luminosity and reported the complete uncertainty on the ratio of $\frac{v_3\{4\}}{v_3\{2\}}$, which will allow us an opportunity to constrain the octupole deformation, $\beta_3$. These reduced error bars on the experimental data for ultra-central collisions rule out linear scaling entirely along with $\sqrt{T_A T_B}$ for $\beta_3 \geq 0.075$. The linear scaling also appears to be less sensitive to the octupole deformation, which is consistent with the behavior seen in Fig.~\ref{fig:rat}. 

\begin{figure}[ht]
    \centering
    \includegraphics[width=0.48\textwidth]{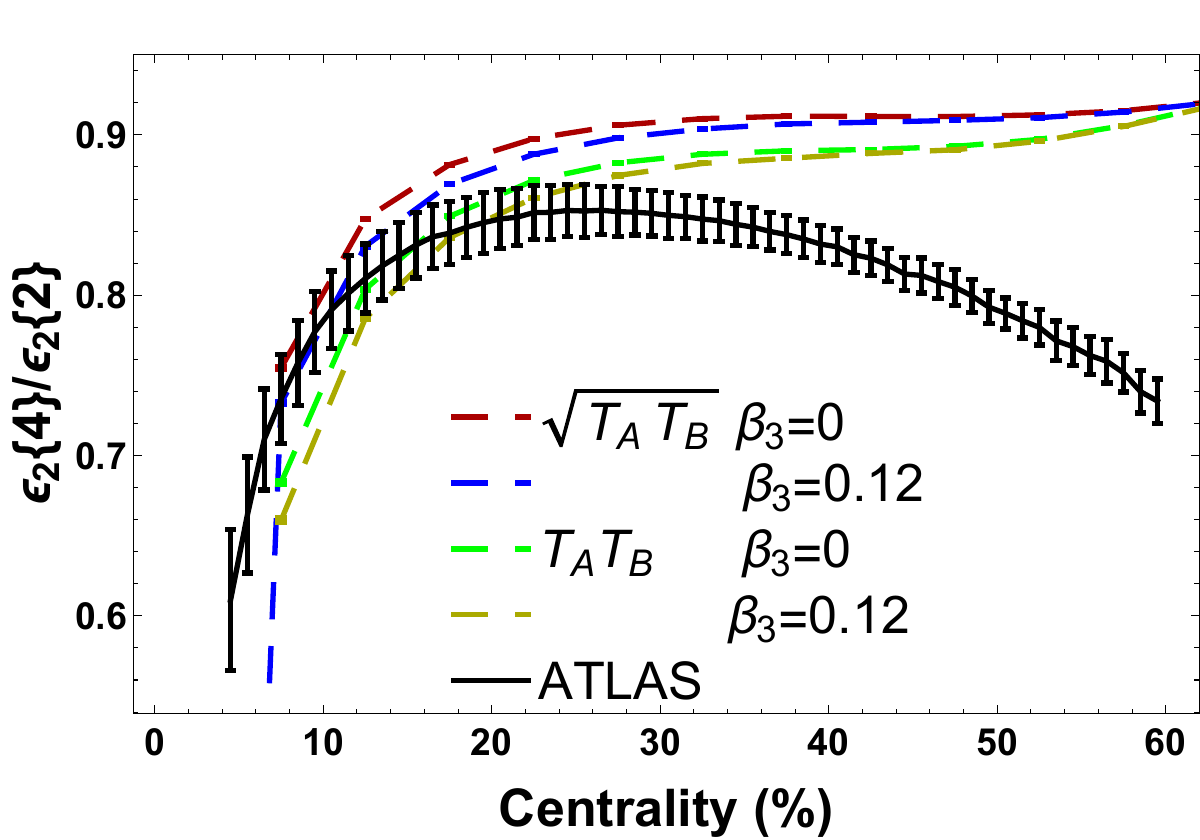} \\
    \caption{Ratio of $\varepsilon_2\{4\}/\varepsilon_2\{2\}$ for PbPb collisions at $5.02$ TeV for varying octupole deformations, $\beta_3$, and $\sqrt{T_AT_B}$ and $T_AT_B$ scaling.  We compare against data at 2.76 TeV from ATLAS \cite{Aad:2014vba}. Figure from Ref.~\cite{Carzon:2020xwp}.}
    \label{fig:e24e22}
\end{figure}

Though not important for this analysis, we can also look at the cumulant ratio for elliptic geometry, $\frac{\varepsilon_2\{4\}}{\varepsilon_2\{2\}}$, in Fig.~\ref{fig:e24e22}, where we focus on $\beta_3$ of 0 and 0.12 for both scalings. For central collisions, 0-20\% centrality, the initial state with no octupole deformation agrees with the data from ALICE \cite{Acharya:2019vdf}, while the large deformation of $\beta_3=0.12$ is in tension with the experimental measurement. For centralities greater than 30\%, all of the models deviate significantly from experimental data though this is not unexpected due to the influence of non-linear response increasing in mid-central collisions.

%__________________________________________________________________________
%
%__________________________________________________________________________
%

Summarizing the results of the eccentricity analysis, we see that while the $\frac{\varepsilon_2 \{2\}}{\varepsilon_3 \{2\}}$ observable in Fig.~\ref{fig:rat} is served better by a large deformation, the cumulant ratio $\varepsilon_3\{4\}/\varepsilon_3\{2\}$ is constrained to $\beta_3 \leq 0.0375$. The tension between these two observables is extremely useful for solving the $v_2$-to-$v_3$ puzzle, since it highly constrains the value of $\beta_3$ in this analysis and can be used for verification of future analyses.

\subsection{Deformed Pb Compared to Experiment} 
\label{sec:DeformedPbComparedToExperiment}

Now we shift the analysis to the final state comparison with experimental data using 5,000-10,000 PbPb events from each hydrodynamic framework detailed in Sec.~\ref{sec:HydrodynamicModels}. For the full hydrodynamic analysis, we simplify the range of parameters, due to the significant run time required, and focus on the quartic entropy scaling, $T_R = \sqrt{T_A T_B}$ ($p=0$), since linear scaling was unable to match with data in Sec.~\ref{sec:OctupoleDeformedEccentricities}. Furthermore, we only look at the higher energy collision of 5.02 TeV due to the increased sensitivity to the octupole deformation and much smaller error bars. Comparison with experiment, here, will primarily focus on the ALICE Run 2 \cite{Acharya:2019vdf} measurement since there is significantly more data than the previous CMS measurement from LHC Run 1. Specifically, we compare to the LHC Run 2 using $0.2 \, \mathrm{GeV} < p_T < 3 \, \mathrm{GeV}$ \cite{Acharya:2018lmh}. 

\begin{figure}[ht]
    \includegraphics[width=0.5\textwidth]{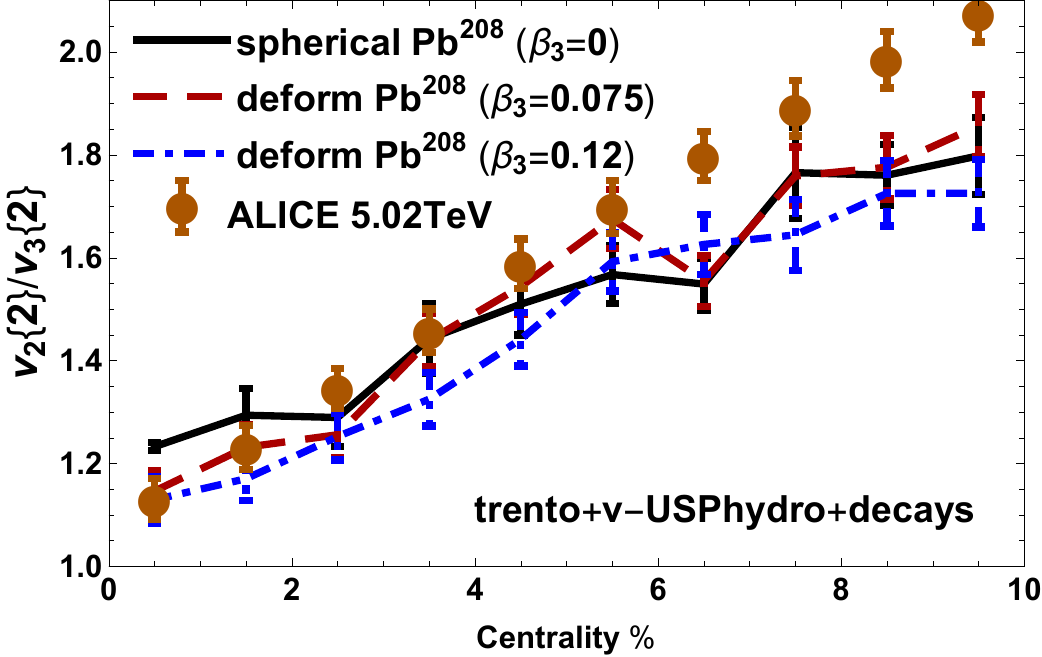} 
    \includegraphics[width=0.5\textwidth]{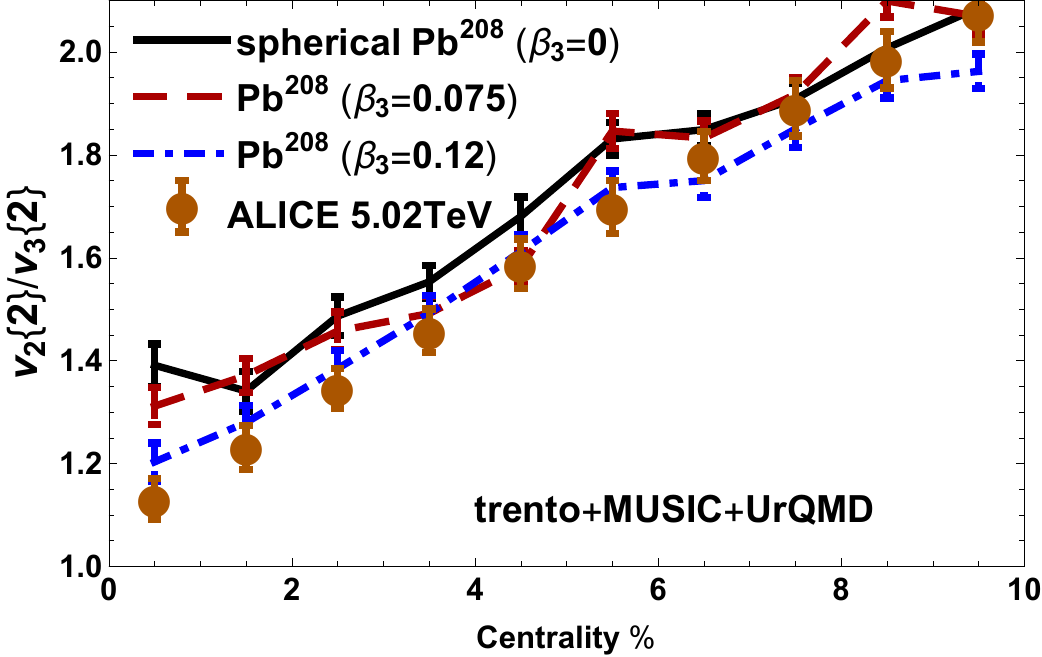} 
    \caption{Ratio of $v_2\{2\}/v_3\{2\}$ in PbPb 5.02TeV collisions compared to experimental data from ALICE \cite{Acharya:2018lmh, Acharya:2019vdf}. Figure from Ref.~\cite{Carzon:2020xwp}.}
    \label{fig:v2/v3}
\end{figure}

First, the ratio of $v_2 \{2\} \, / \, v_3 \{2\}$ is plotted in Fig.~\ref{fig:v2/v3} for \code{trento}+v-USPhydro+decays (left) and \code{trento}+MUSIC+UrQMD (right) against the ALICE data. In both frameworks, we see that a finite $\beta_3$ significantly improves the fit to experimental data. The values of $\beta_3$ that work best are 0.075, for \code{trento}+v-USPhydro+decays, and 0.12, for \code{trento}+MUSIC+UrQMD. An interesting difference between the two frameworks is that the absolute magnitudes of $\frac{v_2\{2\}}{v_3\{2\}}$ are smaller in v-USPhydro than in MUSIC. Though the $\frac{v_3 \{4\}}{v_3 \{2\}}$ ratio should cancel out most medium effects, the ratio $\frac{v_2\{2\}}{v_3\{2\}}$ contains a significant dependence on viscosity. This indicates that the lower magnitude of the v-USPHydro ratio is due to a smaller shear viscosity, which dampens $v_3\{2\}$ less than the MUSIC triangular flow.

%__________________________________________________________________________
%
%__________________________________________________________________________
%

\begin{figure*}[ht]
     \includegraphics[width=\linewidth]{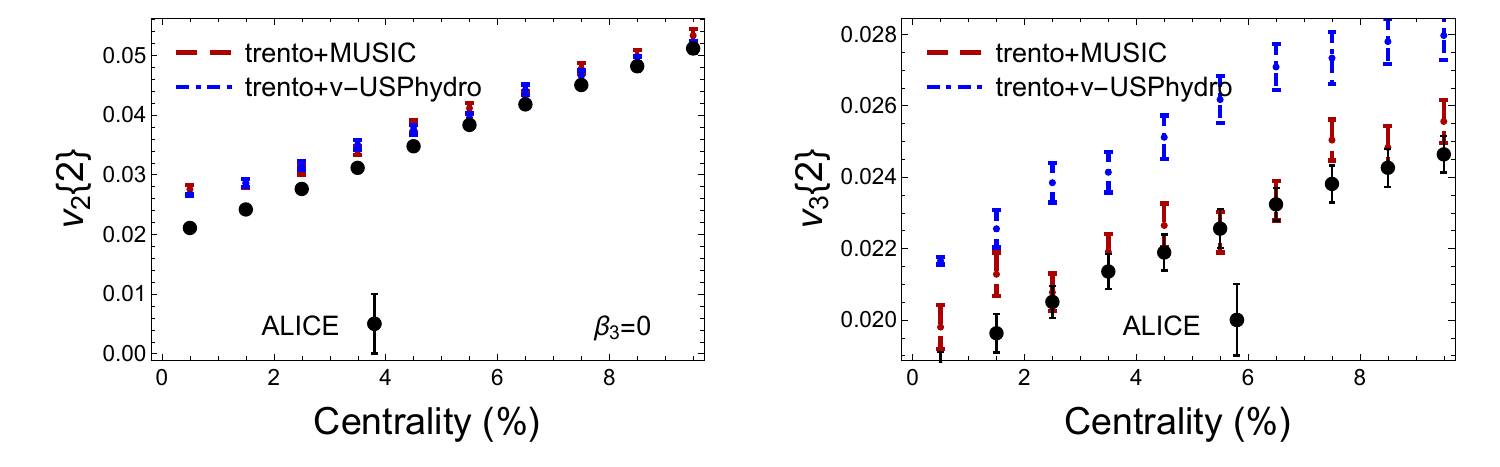} 
    \caption{Comparison of $v_2\{2\}$ and $v_3\{2\}$ in PbPb 5.02TeV collisions from v-USPhydro and MUSIC compared to experimental data from ALICE \cite{Acharya:2018lmh, Acharya:2019vdf}. Figure from Ref.~\cite{Carzon:2020xwp}.}
    \label{fig:vnabs}
\end{figure*}

The ratio $\frac{v_2\{2\}}{v_3\{2\}}$ is only one piece of the story, with the next piece being the absolute magnitudes of the elliptic and triangular flows individually. A comparison of $v_2\{2\}$ (left) and $v_3\{2\}$ (right) against the ALICE measurement is shown in Fig.~\ref{fig:vnabs} using spherical Pb for the initial state. Both frameworks fit $v_2\{2\}$ well for centralities greater than 10\%, though in central collisions they both overshoot the experimental data. Looking at the triangular flow, \code{trento}+v-USPhydro, again, overpredicts the data, while \code{trento}+MUSIC does a good job of fitting the data. This implies that, for these parameters, the weight of the $v_2$-to-$v_3$ puzzle falls on over prediction of elliptic flow $v_2\{2\}$ and not under prediction of $v_3\{2\}$, and is generally characterized by the tension between $v_2$ and $v_3$. This bodes ill for the attempt of using an octupole deformation to boost $\varepsilon_3$ in order to better agree with data.

%__________________________________________________________________________
%
%__________________________________________________________________________
%

\begin{figure}[ht]
    \centering
     \includegraphics[width=0.45\linewidth]{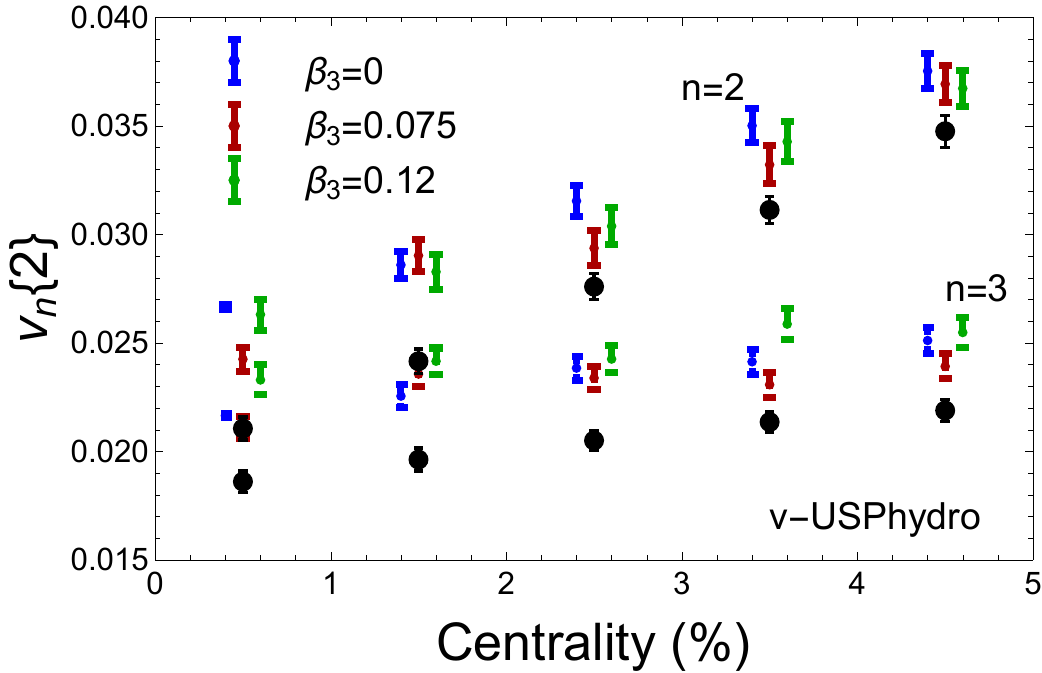}    \includegraphics[width=0.45\linewidth]{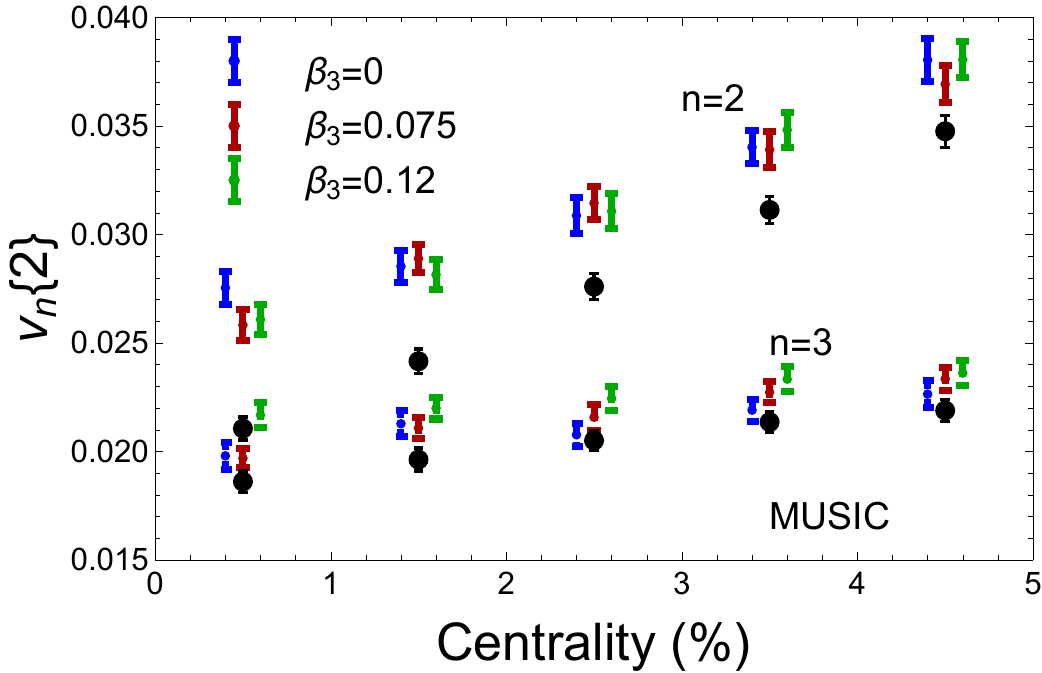} 
    \caption{Absolute values of $v_2\{2\}$ and $v_3\{2\}$ in PbPb 5.02TeV collisions from v-USPhydro and MUSIC compared to experimental data from ALICE \cite{Acharya:2018lmh, Acharya:2019vdf} varying $\beta_3$. Figure from Ref.~\cite{Carzon:2020xwp}.}
    \label{fig:both_b3}
\end{figure}

Looking at the results including a non-zero $\beta_3$ deformation, this omen is born true in Fig.~\ref{fig:both_b3}, where on the left is the framework based on v-USPHydro and the right is MUSIC. We do see an enhancement of $v_3$ tied to an increased $\beta_3$ in most cases. However, the effect of $\beta_3$ has a non-trivial effect on both $v_2$ and $v_3$ where we see non-monotonic behavior with respect to octupole deformation. 

%__________________________________________________________________________
%
%__________________________________________________________________________
%

\begin{figure}[ht]
    \centering
    \includegraphics[width=0.7\textwidth]{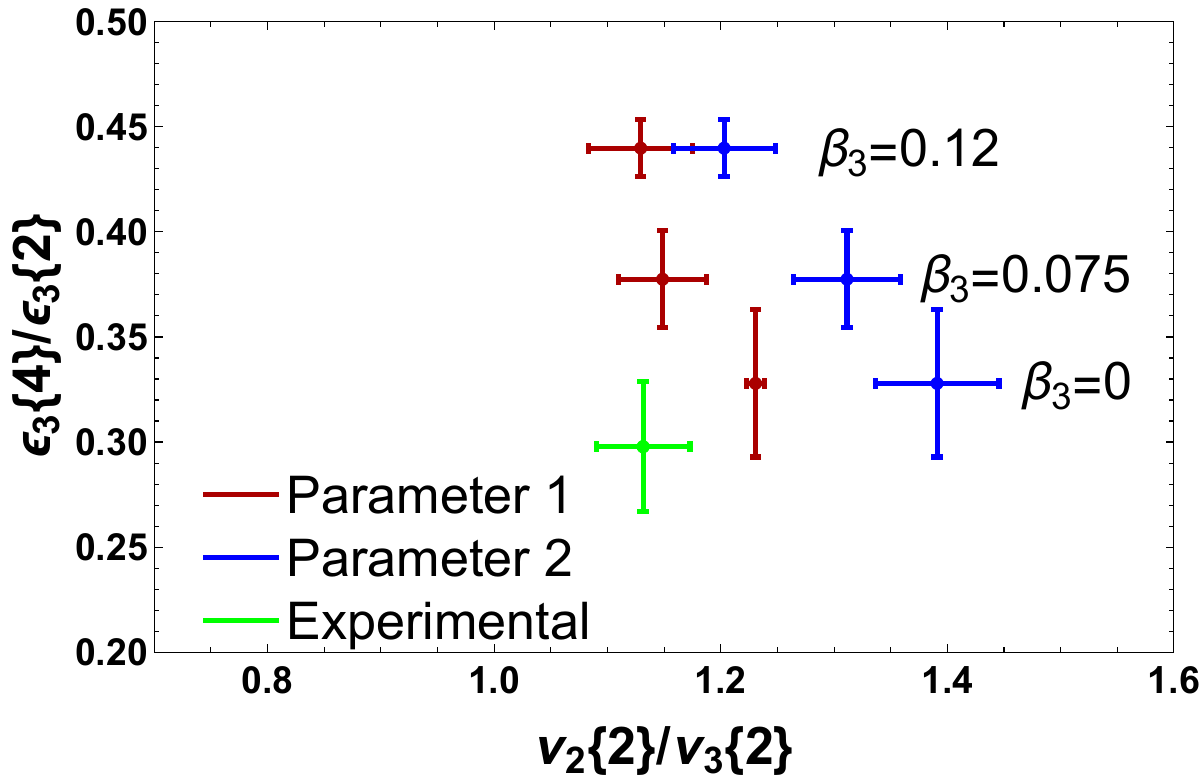}\\
    \caption{Influence of the octupole deformation $\beta_3$ on multiple observables. Parameter 1 is the ratio of $\varepsilon_3\{4\} \, / \, \varepsilon_3\{2\}$ from \code{trento} initial conditions  with $\varepsilon_2\{2\} \, / \, \varepsilon_3\{2\}$ the parameter tune from v-USPhydro. Parameter 2 is the ratio of $\varepsilon_3\{4\} \, / \, \varepsilon_3\{2\}$ from \code{trento} initial conditions with   $\varepsilon_2\{2\} \, / \, \varepsilon_3\{2\}$ with the parameter tune from MUSIC. Clearly adding a nonzero $\beta_3$ does not improve the overall agreement with data.  Details of this plot are discussed in the text. Figure from Ref.~\cite{Carzon:2020xwp}.}
    \label{f:2obs}
\end{figure}

The ability of an octupole deformation to solve the $v_2$-to-$v_3$ puzzle is still not fully resolved, though there are signs that a finite $\beta_3$ may improve agreement with data. A factor that is in tension with increasing the deformation is the $\varepsilon_3\{4\} \, / \, \varepsilon_3\{2\}$ ratio, where, in Sec.~\ref{sec:OctupoleDeformedEccentricities} and plotted in Fig.~\ref{fig:v34v32}, we saw an upper bound of $\beta_3 \leq 0.0375$ enforced by incompatibility with data for larger values. The benefit of increasing $\beta_3$ for $v_2 \{2\} \, / \, v_3 \{2\}$ in Fig.~\ref{fig:v2/v3} must be balanced against the negative impact on agreement of $v_3 \{4\} \, / \, v_3\{2\}$ with experimental data. Though imperfect, this tension can be illustrated by plotting these two ratios independently against each other as done in Fig.~\ref{f:2obs}. Several caveats concerning the comparison of Fig.~\ref{f:2obs} are essential to address immediately. What is plotted as the ``experimental'' point represents a combination of the ATLAS data for $v_3 \{4\} \, / \, v_3 \{2\}$ \cite{Aad:2014vba} and the ALICE data for $v_2 \{2\} \, / \, v_3 \{2\}$ \cite{Acharya:2018lmh, Acharya:2019vdf}, since both observables are not simultaneously measured in either experiment. On the theoretical side, the vertical axis represents the initial state $\varepsilon_3 \{4\} \, / \, \varepsilon_3 \{2\}$ (without running hydrodynamics) while the horizontal axis is $v_2 \{2\} \, / \, v_3 \{2\}$ in the full hydrodynamic frameworks. Using the initial state observable $\varepsilon_3 \{4\} \, / \, \varepsilon_3 \{2\}$ as an approximation of $v_3 \{4\} \, / \, v_3 \{2\}$ should be sufficient for the reasons described in Sec.~\ref{sec:OctupoleDeformedEccentricities}. Furthermore, in Fig.~\ref{f:2obs}, Parameter 1 refers to the framework with v-USPHydro and Parameter 2 the framework with MUSIC. Despite these caveats, this results of this analysis are well illustrated by Fig.~\ref{f:2obs}, which shows that as we improve agreement of $v_2 \{2\} \, / \, v_3 \{2\}$ with experimental data, $v_3 \{4\} \, / \, v_3 \{2\}$ is detrimentally effected.

%__________________________________________________________________________
%
%__________________________________________________________________________
%

\section{Summary}
\label{sec:DeformedSummary}

Since the formulation of the $v_2$-to-$v_3$ puzzle, many avenues toward resolution have been explored and exhausted. Attempts include variations in initial conditions of spherical Pb nuclei \cite{Shen:2015qta}, transport coefficients \cite{Rose:2014fba}, and equation of state \cite{Alba:2017hhe}. All models of the initial state, that assume spherical Pb ions, fail to produce the required triangular flow fluctuations in ultra-central PbPb despite accuratly reproducing elliptic flow fluctuations \cite{Giacalone:2017uqx}. A recent work, Ref.~\cite{Giannini:2022bkn}, provided a current assessment of the $v_2$-to-$v_3$ puzzle by extending state-of-the-art Bayesian models, tuned to reproduce observables at typical centralities, to ultra-central collisions. Their main conclusion is that despite the increased accuracy of these models in describing central collisions, as one goes to ultra-central collisions tension remains between theory and experiment. Moreover, current resolutions of this tension only lead to a worsening of agreement in other centrality classes.

The $v_2$-to-$v_3$ discrepancy is not the only problem in heavy-ion collisions, with issues in reproducing triangular fluctuations, $v_3 \{4\} \, / \, v_3 \{2\}$, and the four-particle correlation of the fourth-order flow harmonic, $v_4\{4\}^4$, in mid-central collisions. Though any one of these issues may be solved in isolation, seeking a simultaneous fit of these observables provides an extremely useful constraint on the introduction of new physics in the initial state. An important aspect of this analysis was including the quantification of the triangular flow fluctuations in assessing the success of a non-zero octupole deformation of ${}^{208}$Pb in describing the $v_2$-to-$v_3$ puzzle. The result is best summarized by Fig.~\ref{f:2obs}, where we see that a finite $\beta_3$ improves the fit of $v_2 \{2\} \, / \, v_3 \{2\}$ to data, but deteriorates agreement with $v_3 \{4\} \, / \, v_3 \{2\}$. This tension between observables helps to better constrain the puzzle, which mirrors the conclusions of Ref.~\cite{Giannini:2022bkn}.

Another layer of comparison is done through the individual absolute magnitudes of $v_2 \{2\}$ and $v_3 \{2\}$ in Fig.~\ref{fig:both_b3}. This dimension of the analysis, elucidates that the largest discrepancy, in both hydrodynamic models, comes from an over-prediction of $v_2 \{2\}$ rather than an under-prediction of $v_3 \{2\}$.

As such, an octupole deformation of ${}^{208}$Pb is not a viable resolution of the $v_2$-to-$v_3$ puzzle in ultra-central collisions, at least in these models. Though both v-USPhydro and MUSIC are hydrodynamic models that do a good job in describing bulk flow observables in heavy-ion collisions, the details of the implementation are different and emphasize different physics. A particular difference is in viscosity, with the larger value from MUSIC leading to a larger suppression of higher harmonics, namely $v_3$, which brings it into better quantitative agreement with $v_3 \{2\}$ than v-USPhydro, which overpredicts the data, though v-USPhydro matches $v_2 \{2\} \, / \, v_3 \{2\}$ better than MUSIC. 

An important correlation, between hydrodynamic and initial state parameters, is illustrated by the interplay of viscosity, $\eta / s$, and $\beta_3$ as seen in this analysis. Adjustments of $\beta_3$ can be used to dial in the ratio $v_2 \{2\} \, / \, v_3 \{2\}$, with no control over the absolute magnitude of either individual quantity. While the magnitudes of $v_2 \{2\}$ and $v_3 \{2\}$ are directly affected by the viscosity in correlation to the prior choice of $\beta_3$. It is reasonable to conclude that a simultaneous fit of initial state properties, such as quadrapole and octupole deformations, and hydrodynamic transport parameters would be useful in resolving the $v_2$-to-$v_3$ puzzle. With the significant effect that viscosity seems to have on the $v_2$-to-$v_3$ puzzle, a more sophisticated treatment of out-of-equilibrium effects may be important. Another improvement could be made in implementing a temperature dependent $\eta / s$ in the v-USPHydro model.

%%%%%%%%%%%%%%%%%%%%%%%%%%%%%%%%%%%%%%%%%%%%%%%%%%%%%%%%%%%%%%%%%%%%%%%%%%%
%
\chapter{ICCING Algorithm} \label{chap:ICCINGAlgorithm}
%
%%%%%%%%%%%%%%%%%%%%%%%%%%%%%%%%%%%%%%%%%%%%%%%%%%%%%%%%%%%%%%%%%%%%%%%%%%%

Until recently, the simulations of heavy-ion collisions at high energies have been entirely focused on the evolution of the energy density, which operates under the assumption that this is the dominant component of the system. As such, the initial condition has seen development almost exclusively in its construction of the initial energy density, although in truth it should consist in a complete specification of the initial energy-momentum tensor and currents of conserved charges. Much work has been done in initializing the initial momentum components of the energy-momentum tensor \cite{Gardim:2011qn, Gardim:2012yp, Gale:2012rq, Schenke:2019pmk, Liu:2015nwa, Kurkela:2018wud,Plumberg:2021bme,Chiu:2021muk} mainly through the use of pre-equilibrium simulations. 

The conserved charge currents are a much more recent development, and this has been focused on finite net baryon density \cite{Werner:1993uh, Itakura:2003jp, Shen:2017bsr, Akamatsu:2018olk, Mohs:2019iee} at low beam energies. Including baryon density has primarily been through some type of 'baryon stopping' which allows for participating nucleons or their valence quarks, depending on the model, to be decelerated and remain within the QGP. At high energies, the assumption was that the nuclei have high enough momentum that they do not contribute to the QGP beyond the initial collision and, thus, there is zero net baryon density. This is supported by the baryon stopping models, which see small contribution from this source at high energies. With the search for the critical point in the QCD phase diagram \cite{Karpenko:2015xea, Borsanyi:2018grb, Noronha-Hostler:2019ayj, Monnai:2019hkn, Monnai:2016kud, Bazavov:2018mes, Critelli:2017oub, Parotto:2018pwx, Demir:2008tr, Denicol:2013nua, Denicol:2018wdp, Kadam:2014cua, Stephanov:1999zu, Stephanov:2017ghc, Jiang:2015hri, Mukherjee:2016kyu, Nahrgang:2018afz, An:2019osr, Du:2019obx, Batyuk:2017sku, Shen:2017bsr}, description of systems at lower collision energies became more important, which is where baryon stopping models see a significant contribution. 

Describing the QGP, at high energies, with the mean field assumption that the total charge vanishes is consistent with a leading-order picture in pQCD, which describes the initial state as composed entirely of gluons. However, through the historical development of the initial state in Sec.~\ref{sec:HistoryOfInitialState}, we have seen the importance of event-by-event fluctuations and that they are essential for an accurate description of the initial conditions of heavy-ion collisions. In this way, assuming the initial total charge density vanishes is a parallel to the optical Glauber initial state and ignores local fluctuations of charge. In addition to baryon density, other conserved charges may be present in the initial state such as strangeness, electric charge, spin, and chirality. Initializing several of these charges with the same method would provide some additional constraints on the modeling. 

\begin{figure}[h!]
    \centering
	\includegraphics[width=0.55\textwidth]{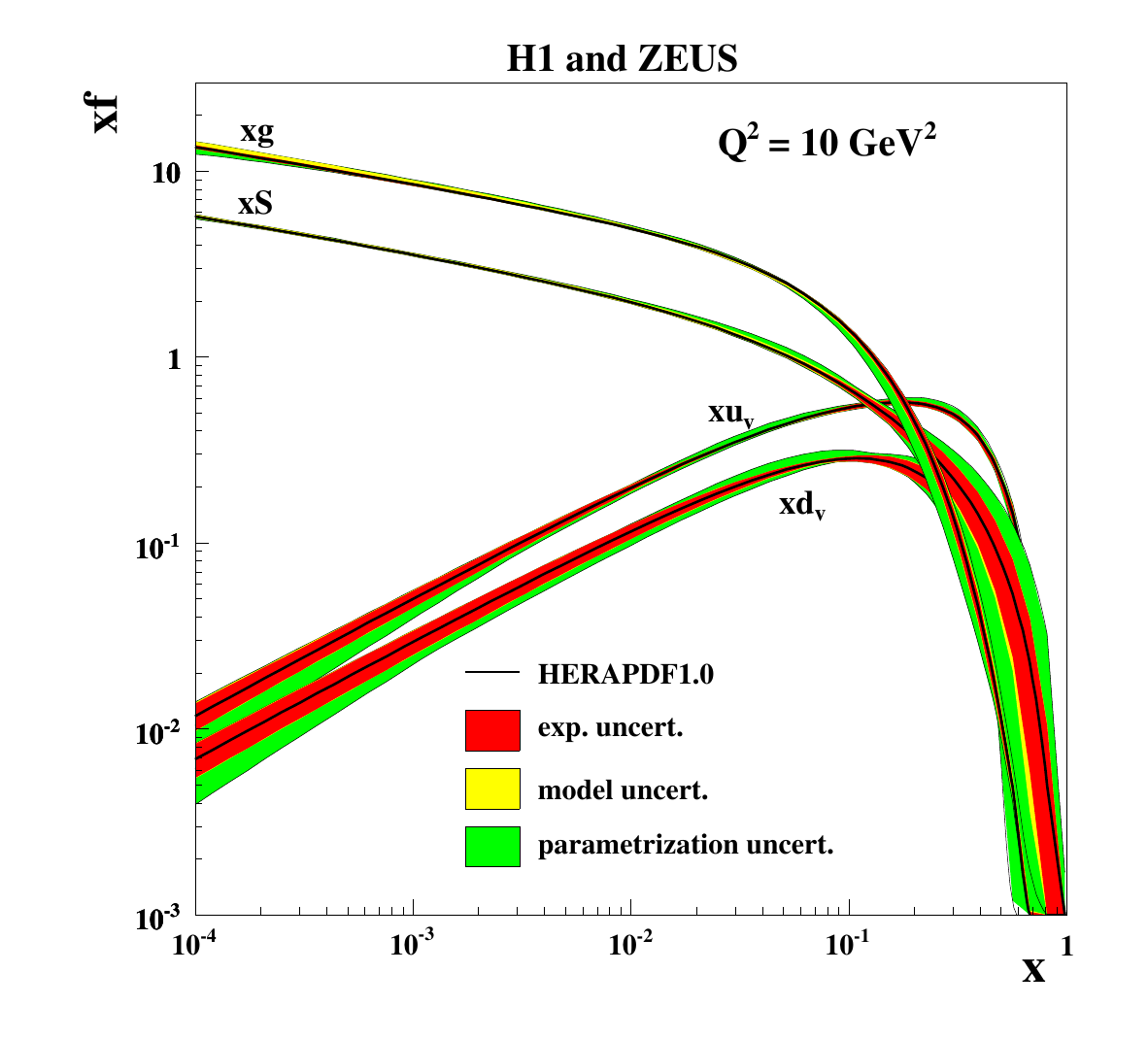}
	\caption{
	    PDFs extracted by the H1 and ZEUS Collaborations measuring sea quarks at small $x$, including cold nuclear matter. An important observation, is that sea quark abundances $(xS)$ at small $x$ are a sizeable contribution as compared to gluons $(xg)$.  This figure reproduced following Creative Commons Attribution License guidelines from Fig.~19(b) of Ref.~\cite{Aaron:2009aa}.
	}
	\label{f:HERAPDF}
\end{figure}

Intuition of the source for local conserved charge fluctuations at high collision energies can be obtained by looking at the structure of the proton as probed in DIS experiments. The PDF, as extracted by H1 and ZEUS collaborations, is shown in Fig.~\ref{f:HERAPDF}, which is measured as a function of Bjorken-x, where $x\propto p_T\/\sqrt{s}$. The parameter $x$ is proportional to the inverse of the collision energy so that small-x corresponds to high energies and large-x is comparable to rest energy. As one goes to small-x, there is an exponential enhancement of the gluon contribution, $xg$, and a similar increase in sea quark abundances, $xs$, which are quark/anti-quark pairs produced by gluon splitting. While the saturation of gluons is consistent with leading order pQCD, the sea quark enhancement can be thought of as next-to-leading order and come from gluons splitting into quark/anti-quark pairs. This source of conserved charges will preserve the net zero charge assumption while allowing for local fluctuations in the charge density. The relationship between the gluon and sea quark abundances at small-x, from Fig.~\ref{f:HERAPDF} is consistent with the perturbative expectation of $x S \approx \alpha_s \, x g$, where $\alpha_s$ is the strong coupling constant. There are also non-perturbative mechanisms of sea quark fluctuations \cite{Shuryak:2002qz}.

%__________________________________________________________________________
%
%
%__________________________________________________________________________

The inclusion of $q \bar{q}$ pairs in an initial condition generator would be a rich source of physics since it provides information about baryon, strange, and electric (BSQ) conserved charges and any other properties of the quarks, such as spin, isospin, and chirality. These splitting probabilities for $g \rightarrow q \bar{q}$ were recently derived by the authors of Ref.~\cite{Martinez:2018ygo} in several CGC models, and this work is concerned with the application of them in an initial state generator. This conserved charge initial state would allow study of charge transport at top collider energies. In practice, any perturbative splitting could occur, although at the energies in the initial state heavy quark production would be highly suppressed.

In this chapter, I will discuss the implementation of these splittings in a Monte-Carlo initial state supplemental framework titled \code{iccing}, Initial Conserved Charges in Nuclear Geometry. The \code{iccing} model was constructed to be modular and use any initial energy density as input for the sampling of $g \rightarrow q \bar{q}$, though, in this work I focused on \code{trento} initial conditions as input to illustrate the physical consequences of this model. For simplicity, only the BSQ conserved charges are considered, although this framework makes it trivial to include new degrees of freedom. The results of the \code{iccing} model are a modified energy and new charge densities that can be read directly into hydrodynamic simulations, such as the BSQ hydrodynamic model \code{ccake} (Conserved ChArges in hydrodynamiK Evolution) \cite{Plumberg:inPrep}. This chapter reproduces and refines the work from Ref.~\cite{Carzon:2019qja} along with elements from Ref.~\cite{Carzon:2023zfp} and Ref.~\cite{Plumberg:inPrep}.

%---------------------------------------------------------------------------
%
\section{Code structure}
%
%---------------------------------------------------------------------------

%__________________________________________________________________________
%
\begin{figure}[h!]
    \centering
	\includegraphics[width=0.65\textwidth]{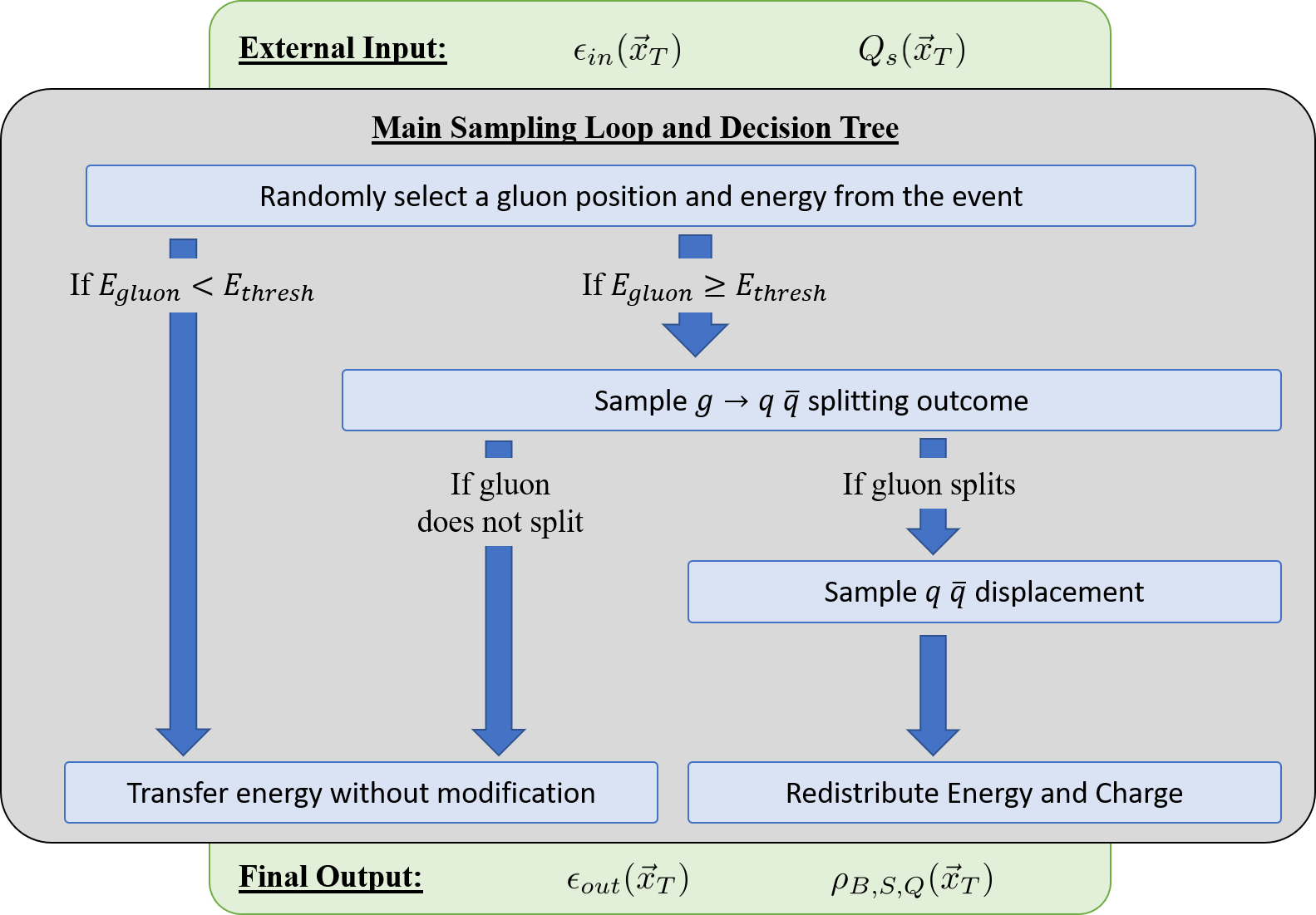}
	\caption{Decision tree of the \code{iccing} algorithm. Figure from Ref.~\cite{Carzon:2019qja}.}
	\label{f:Algorithm}
\end{figure}
%
%__________________________________________________________________________

\begin{figure}[h!]
    \centering
	\includegraphics[width=0.85\textwidth]{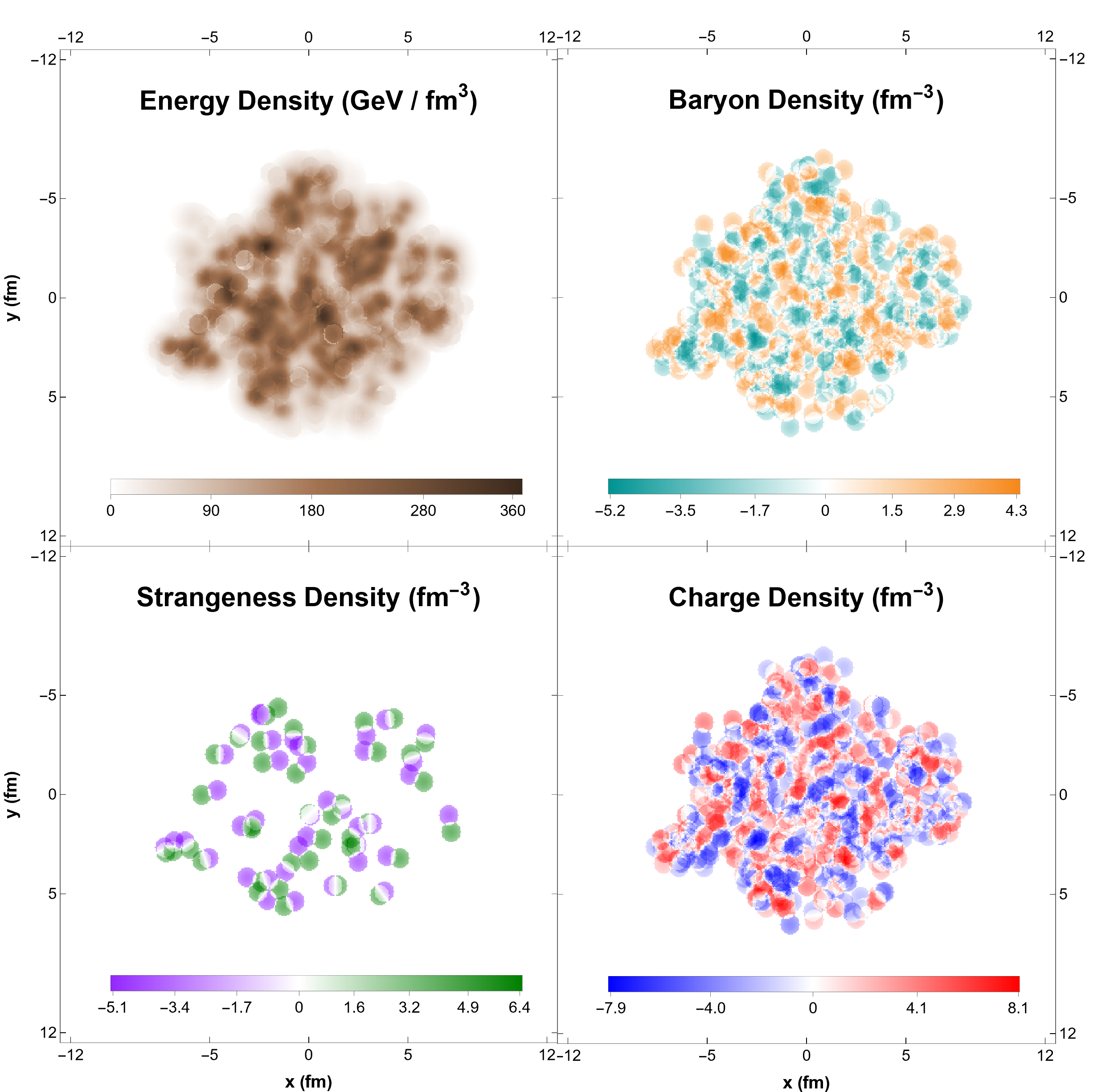}
    \caption{An event after being fully resampled by the \code{iccing} algorithm, resulting in a reconstructed energy density as well as new distributions of the three conserved charges.  Note that some artifacts of the energy redistribution can be seen in the modified energy density. Figure from Ref.~\cite{Carzon:2019qja}.
	}
	\label{f:Events}
\end{figure}

Here I present a high level overview of the \code{iccing} algorithm which will provide context for the more detailed discussion below. In Fig.~\ref{f:Algorithm}, a visualization of the algorithmic structure is presented highlighting the broad strokes of the process. On an event basis, the first step is reading in an externally provided energy density profile $\epsilon (\vec{x}_\bot)$ and some profile that may be used as the saturation scale (this is required for the splitting probabilities and in this case is taken to be the target nucleus entropy profile, $T_B$, see Sec.~\ref{subsec:SplittingProbabilities}). From this initial profile, an energy site is chosen randomly to act as the center of a possible gluon. With respect to this gluon center, the total energy available $E_{R}$ inside a circle of a predefined radius $r_{gluon}$, is determined. A fraction of this energy $E_G$ is chosen to be associated with the gluon by sampling an energy dependent distribution $dP\/dE_G$. Then a Monte Carlo sampling of the quark flavor ratios, provided by the integrated quark production probabilities in Fig.~\ref{f:multratio}, is used to determine whether the gluon will split into a quark/anti-quark pair and, if the split was successful, their flavor. Another sampling is done to determine both the distance between the two quarks $r_\bot$ and the fraction of energy $\alpha$ that is shared by each, using the differential quark production probabilities in Sec.~\ref{subsec:SplittingProbabilities}. Finally, the energy, baryon, strange, and charge densities of the quark and anti-quark are distributed in the output density grids with radius $r_q$. If the gluon did not undergo a splitting, its energy is copied over without modification. This process is repeated until all of the energy from the input density has been transferred to the output grids. A plot of the densities from a fully processed event is shown in Fig.~\ref{f:Events}.

%__________________________________________________________________________
%
%
%__________________________________________________________________________

The \code{iccing} algorithm is written in C++ and utilizes a modular design that separates the event variable handling, IO functions, and gluon splitting functions into different classes. A myriad of tests are included, with a simple input parameter as a switch, in the code that can be used to isolate different key functions that may need to be monitored. These tests specifically target areas of the code that are left open to change such as the splitting probability functions. Smaller parts of the algorithm, for example the spatial correlation functions used to determine the location of the quark/antiquark pair and required interpolation functions, are further separated out into their own classes and structured in such a way that allows them to be easily updated and changed. The location and processing of input parameters for \code{iccing} are highlighted by notes throughout the code that can be used for the easy introduction of new variables as needed. External dependencies are kept to a minimum to ensure that new users have an easier install and usage experience. Where multiple choices could be anticipated by the user, options are provided to accommodate different use cases, for example the input and output of densities can be done in the form of a full grid with filler zeros for empty grid points or in the form of only valued points.

In the rest of this chapter, I will lay out in detail what each part of the algorithm does in turn. Reference is made, though epigraph, to where in the code each operation is located. All tests are explained and included where relevant.

%..........................................................................
%
\section{Input and Initialization}
\label{sec:InputInitialization}
%
%..........................................................................

Before detailing the algorithmic steps of \code{iccing}, it is important to explain the input to the model and how the initialization process is carried out. The parameters of the model are sorted into three categories: parameters that are dependent on external input (Table~\ref{table:ExternalParameters}), general options (Table~\ref{table:ICCINGGeneralParameters}), and physical options (Table~\ref{table:ICCINGPhysicalParameters}). These can be specified in a configuration file and used to run \code{iccing}. The only parameters that require specification, and do not have default values, are those that identify the directories where the initial conditions, theory input, and \code{iccing} output are located. The full parameter list is contained in Tables~\ref{table:ExternalParameters}, \ref{table:ICCINGGeneralParameters}, and \ref{table:ICCINGPhysicalParameters}, where relevant options are provided with any default values and units as well as short descriptions of each parameter. Further description of the model parameters and their effect on the algorithm is presented throughout this chapter and Chaps.~\ref{chap:ICCINGResults} and \ref{chap:PreEquilibriumEvolution}.

%__________________________________________________________________________
%
\begin{table}[h!]
    \centering
    \includegraphics[width=0.69\textwidth]{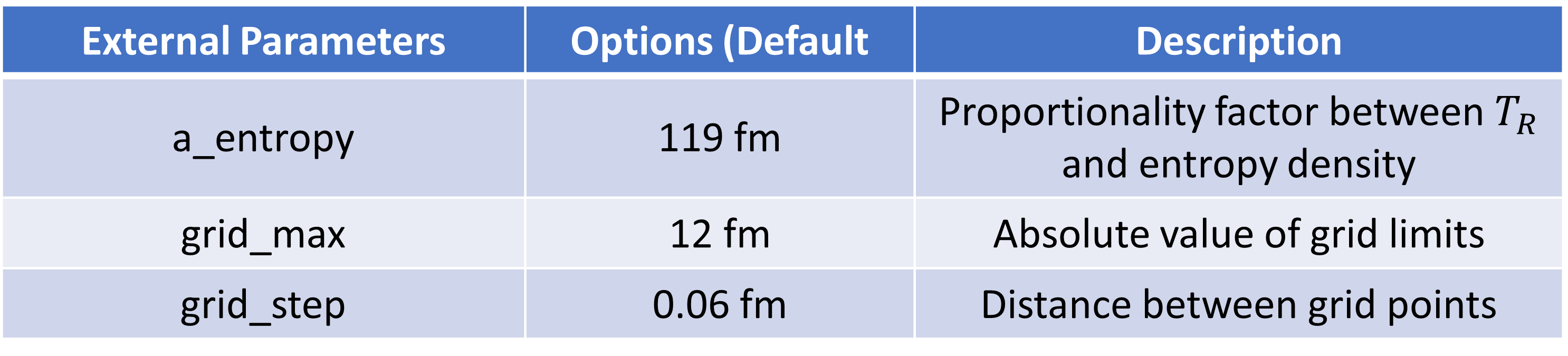}

    \caption{Table of parameters that \code{iccing} requires as input from initial state code. 'a{\_}trento' is a conversion factor between the reduced thickness function and the entropy density.}
    \label{table:ExternalParameters}
\end{table}
%
%__________________________________________________________________________

%__________________________________________________________________________
%
\begin{table}[h!]
    \centering
    \includegraphics[width=0.69\textwidth]{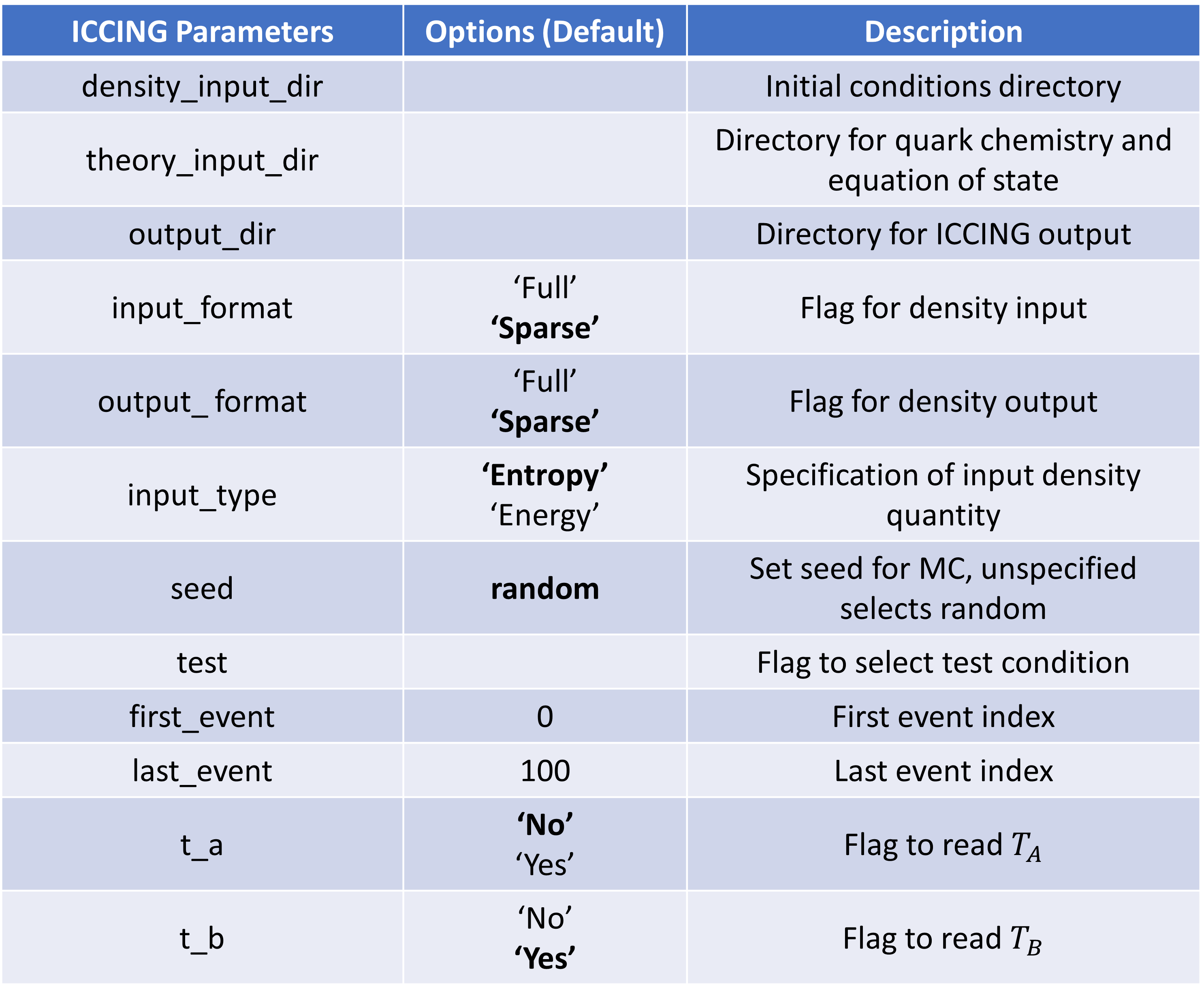}
    
    \caption{General parameters for the \code{iccing} code.}
    \label{table:ICCINGGeneralParameters}
\end{table}
%
%__________________________________________________________________________

%__________________________________________________________________________
%
\begin{table}[h!]
    \centering
    \includegraphics[width=0.69\textwidth]{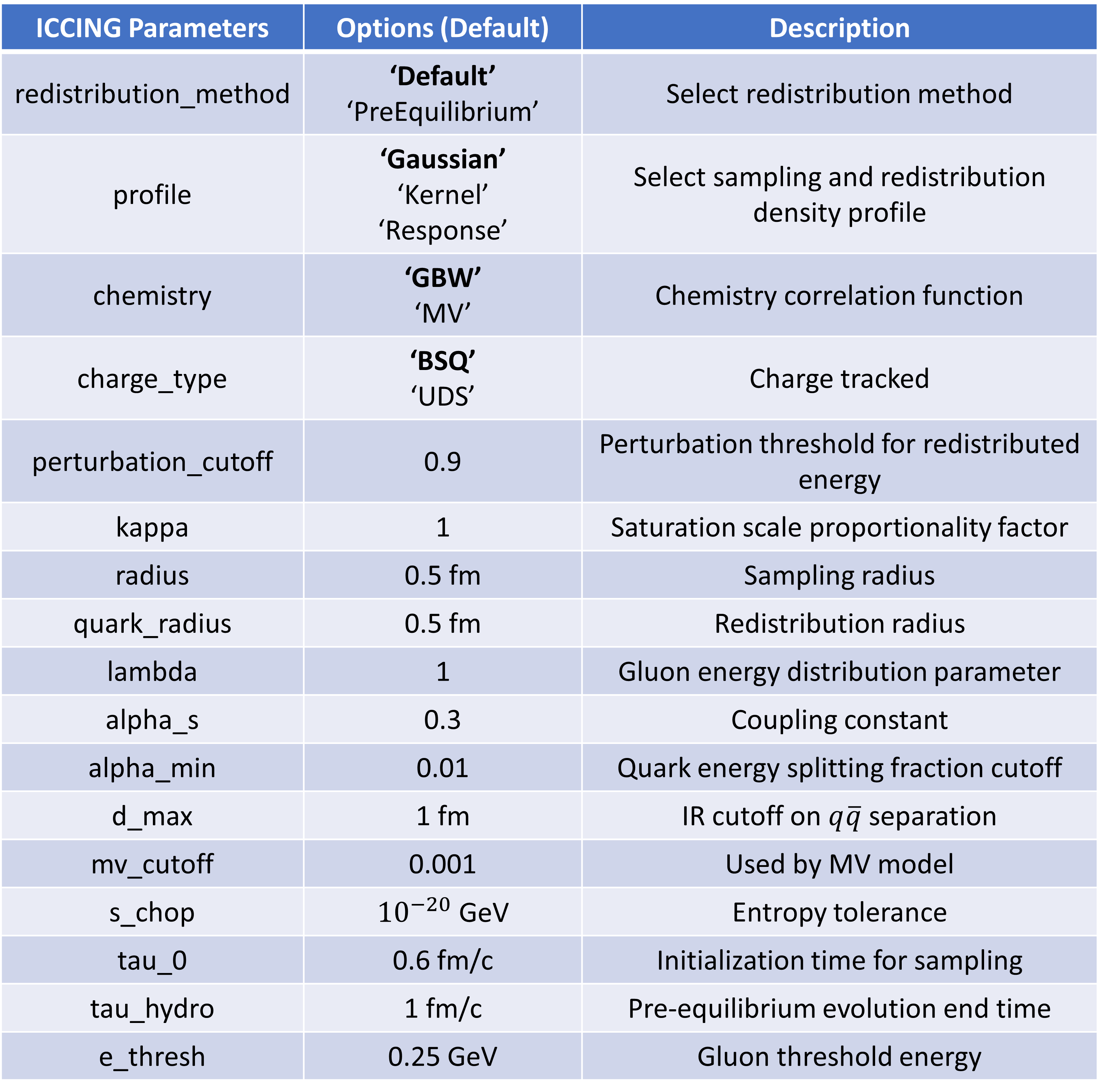}
    
    \caption{Physical parameters for the \code{iccing} algorithm.}
    \label{table:ICCINGPhysicalParameters}
\end{table}
%
%__________________________________________________________________________

%..........................................................................
%
\subsection{Initial Condition}
\label{subsec:InitialCondition}
%
%..........................................................................

\epigraph{IO::ReadEvent()}{}

As it is currently formulated, the \code{iccing} model requires a 2-D initial geometry profile from with to sample gluons and a profile representing the saturation scale. Along with these profiles, the density type and exact dimensions must be specified. If a profile proportional to entropy is provided, then the proportionality constant must be included. The model is derived for sampling over an initial energy density, so if the initial entropy density is provided, then this is converted to energy using the specified equation of state (see Sec.~\ref{subsec:ConvertInitialCondition}). 

For the work done here, the \code{iccing} algorithm uses and is optimized for initial conditions from the \code{trento} model, due to its model agnostic formulation and parameterized generalization of MC-Glauber (see Sec.~\ref{sec:Trento}). However, any 2-D initial condition can be used as input albeit with careful consideration for any differences. 

For the analysis done throughout Chaps.~\ref{chap:ICCINGAlgorithm}, \ref{chap:ICCINGResults}, and \ref{chap:PreEquilibriumEvolution}, \code{trento} initial conditions for PbPb collisions at $\sqrt{s_{NN}} = 5.02 \, \mathrm{TeV}$ are used, with parameters motivated by Refs.~\cite{Moreland:2014oya, Bernhard:2016tnd, Alba:2017hhe, Giacalone:2017dud,Bernhard:2019bmu} (specifically: nucleon width, $w = 0.51 \, \mathrm{fm}$, and the multiplicity fluctuation parameter, $k=1.6$). For the saturation scale $Q_s (\vec{x}_\bot)$, we use the conventional formulation with respect to the target nucleus, $T_B (\vec{x}_\bot)$, that is dependent on a proportionality factor: 
\begin{align}   \label{e:kappadef}
    Q_s (\vec{x}_\bot) = \kappa \sqrt{T_B (\vec{x}_\bot)} .
\end{align}

The value of $\kappa$ is dependent on the model and the normalization used to define the saturation scale. As a default value, we choose $\kappa = 1 \, \mathrm{GeV} \, \mathrm{fm}$. As a benchmark, we can compare this to the value of $\kappa$ as computed in pQCD using the normalization of 
Ref.~\cite{Kovchegov:2012mbw}. Two saturation scales can be obtained, by treating the nucleons as single quarks or gluons and are given by:
\begin{align}
    Q_s^2 (\vec{x}_\bot) =
        \begin{cases}
            \frac{4\pi \alpha_s^2 C_F}{N_c} \, T_B (\vec{x}_\bot) \qquad &\mathrm{if \, nucleon \rightarrow quark} 
            \\
            4\pi \alpha_s^2 \, T_B (\vec{x}_\bot) \qquad &\mathrm{if \, nucleon \rightarrow gluon}
        \end{cases} .
\end{align}
Using $\alpha_s \approx 0.3$, the proportionality factor for each case is: 
\begin{align} \label{e:kappapQCD}
    \kappa =
        \begin{cases}
            0.140 \, \mathrm{GeV \, fm} \qquad &\mathrm{if \, nucleon \rightarrow quark} 
            \\
            0.210 \, \mathrm{GeV \, fm} \qquad &\mathrm{if \, nucleon \rightarrow gluon}
        \end{cases} .
\end{align}
From these estimates, we see the chosen default of $\kappa = 1 \, \mathrm{GeV \, fm}$ is roughly 5 times larger than $\kappa$ for a nucleon assumed to be composed of a single parton, and seems to be a reasonable ballpark. An measure of the quantitative dependence of the geometry on $\kappa$ is explored in Chap.~\ref{chap:ICCINGResults}.

The quark chemistry and splitting correlation functions, as described in Sec.~\ref{subsec:SplittingProbabilities}, are strongly dependent on the saturation scale. The analytic expressions, in Sec.~\ref{subsec:SplittingProbabilities}, are obtained for the (semi)dilute/dense regime of CGC effective theory, which asymmetrically treats the two colliding nuclei. This assumption treats one nuclei as 'dense' and the other as '(semi)dilute', where the second leads to a linear dependence that ultimately cancels out its contribution. The inherent asymmetry of the formulation allows us to describe the saturation scale using only one of the colliding nuclei, which is here taken to be $T_B$. As \code{trento} does not provide this profile out-of-box, we made modifications to access $T_A$ and $T_B$, since the saturation scale is generally dependent on both. The correlations and probabilities obtained in Sec.~\ref{subsec:SplittingProbabilities} are valid for asymmetric systems, such as CuAu, although in this analysis we extend them beyond the range of validity, specifically the symmetric PbPb, since these systems are better understood. As such, future work should be dedicated to including alternative probability distributions tailored for dense/dense and dilute/dilute collisions, both from CGC theory and other sources. The framework of \code{iccing} was created with these extensions in mind. 

%..........................................................................
%
\subsection{Convert Initial Condition}
\label{subsec:ConvertInitialCondition}
%
%..........................................................................

\epigraph{IO::InitializeEOS()}{}

If the provided initial state is specified as proportional to entropy, we multiply the density by the $a_{entropy}$ parameter to get the initial entropy density. Then, the equation of state, provided by Ref.~\cite{Alba:2017hhe} at $\mu_B = 0$, is used to map the entropy density $s(\vec{x}_\bot)$ into energy density $\epsilon(\vec{x}_\bot)$ in units of $\mathrm{GeV} / \mathrm{fm}^3$. The conversion to energy density is required since the \code{iccing} framework depends on probability distributions dependent on energy rather than entropy. Efficiency of the resampling algorithm is increased by discarding grid points that contain numerically infinitesimal entropy densities. This process is controlled by a grooming variable $S_\mathrm{chop}$ with the default value of $10^{-20} \, \mathrm{fm}^{-3}$. 

This process can be generalized by including a new equation of state class that could switch between tabular and conformal equations of state for this conversion process.

%..........................................................................
%
\subsection{Splitting Probabilities}
\label{subsec:SplittingProbabilities}
%
%..........................................................................
\epigraph{IO::InitializeSplitter() and Functions.h}{}

The purpose of the \code{iccing} algorithm is to generate conserved charge densities in the initial state through $g \rightarrow q \bar{q}$ processes, which necessitates some theoretical input for the gluon splitting process. Practically, all that is required, for the algorithm, are chemistry ratios of quark flavor production and a function from which to sample kinematic properties of the quark/anti-quark pair. The code was designed to allow for great flexibility in these theoretical inputs, though here we use specific calculations of these processes under certain assumptions. As currently implemented, the chemistry ratios are read in from a data file and the differential probability distribution for sampling kinematics is defined in the "Correlation" class. For the remainder of this section, I will review the theoretical calculation, used for these inputs, from Refs.~\cite{Martinez:2018ygo,Carzon:2019qja}.

The calculation, presented in Ref.~\cite{Martinez:2018ygo} and with application in Ref.~\cite{Carzon:2019qja}, is obtained from CGC theory and produces the differential probability distribution, $\frac{dP}{dr_\bot \, d\alpha}$, for a gluon to split into a $q \barq$ pair with given kinematics. The relevant degrees of freedom come from $\tvec{r}$, which is the separation vector between quark and anti-quark pairs, and $\alpha$, which is the fraction of light-front momentum (i.e., energy) carried by the quark (the anti-quark carries $1-\alpha$). While this provides a source for sampling kinematics, an integral over $r_\bot$ and $\alpha$ is used to obtain the total probability of a gluon splitting into quark/anti-quark pair of a given flavor. The calculation from Ref.~\cite{Martinez:2018ygo} expresses this probability distribution in terms of \textit{dipole scattering amplitudes}, the natural degrees of freedom in CGC effective theory. To quantify the effect this choice in theory has on the algorithm, we evaluated the probability distribution under two different models of the dipole amplitude: the Golec-Biernat-Wusthoff (GBW) \cite{Golec-Biernat:1998zce} and McLerran-Venugopalan (MV) \cite{McLerran:1993ni, McLerran:1993ka, McLerran:1994vd} models. Their dipole amplitudes are defined as:
\begin{align} \label{e:dipoles}
D_2 (\tvec{r}) =
    \begin{cases}
        \exp\left(-\tfrac{1}{4} |\tvec{r}|^2 Q_s^2\right) & \mathrm{GBW} 
        \\
        \exp\left(-\tfrac{1}{4} |\tvec{r}|^2 Q_s^2 \ln\tfrac{1}{|\tvec{r}| \Lambda} \right) & \mathrm{MV}
    \end{cases} ,
\end{align}
where $\Lambda$ is an infrared cutoff. We see that the GBW model describes the dipole amplitude as a Gaussian. This is model is able to describe the nonlinear effects in the deep saturation regime at large dipoles, but is insufficient in capturing the transition to power-law behavior in the dilute regime of small dipoles \cite{Martinez:2018ygo}. This deficiency is more properly captured in the MV model by the inclusion of a logarithmic infrared cutoff.

The differential probability distributions for these two methods are given by:
\begin{align} \label{e:prob1_first}
    \left.\frac{dP}{dr_\bot d\alpha}\right|_{GBW} &= 
    \frac{\alpha_s}{4 \pi}  m^2 r_\bot
    \Big[ 1 - \exp\Big( - \tfrac{1}{4} \Big\{ \alpha^2 + (1-\alpha)^2 \Big\} \, r_\bot^2 Q_s^2 \Big) \Big]
    \notag \\ &\hspace{1cm} \times
    \Big[ \big(\alpha^2 + (1-\alpha)^2 \big) K_1^2 ( m r_\bot) + K_0^2 (m r_\bot) \Big]
\end{align}
and
\begin{align} \label{e:prob2_first}
    \left.\frac{dP}{dr_\bot d\alpha}\right|_{MV} &=
    \frac{\alpha_s}{4 \pi}  m^2 r_\bot
    \Big[ 1 - \exp\Big(
    - \tfrac{1}{4} \Big\{ \alpha^2 \, \textcolor{red}{\ln\tfrac{1}{\alpha r_\bot \Lambda}}
    - (1-\alpha)^2 \, \textcolor{red}{\ln\tfrac{1}{(1-\alpha) r_\bot \Lambda}} \Big\} r_\bot^2 Q_s^2
    \Big) \Big]
    \notag \\ &\hspace{1cm} \times
    \Big[ \big(\alpha^2 + (1-\alpha)^2 \big) K_1^2 ( m r_\bot) + K_0^2 (m r_\bot) \Big] ,
\end{align}
where $K_{0,1}$ are Bessel functions. Highlighted in red are the terms that the infrared cutoff of the MV model introduce to the differential probability distribution. 

The integrated probabilities for each model are given by:
\begin{align} \label{e:multratio1_first}
    P_{tot}\Big|_{GBW} &=
    \frac{\alpha_s}{4 \pi}
    \int\limits_0^1 d\alpha \, \int\limits_0^\infty d\zeta \, \zeta \,
    \Big[ 1 - \exp\Big( - \tfrac{1}{4} \Big\{\alpha^2 + (1-\alpha)^2\Big\} \, \tfrac{Q_s^2}{m^2} \, \zeta^2 \Big) \Big]\,
    \notag \\ &\hspace{1cm} \times
    \Big[ \big(\alpha^2 + (1-\alpha)^2 \big) K_1^2 (\zeta) + K_0^2 (\zeta) \Big]
\end{align}
and
\begin{align} \label{e:multratio2_first}
    P_{tot}\Big|_{MV} &=
    \frac{\alpha_s}{4 \pi}
    \int\limits_0^1 d\alpha \, \int\limits_0^\infty d\zeta \, \zeta \,
    \Big[ 1 - \exp\Big(
    - \tfrac{1}{4} \Big\{ \alpha^2 \, \textcolor{red}{\ln\tfrac{1}{\alpha \zeta \, \Lambda/m}}
    - (1-\alpha)^2 \, \textcolor{red}{\ln\tfrac{1}{(1-\alpha) \zeta \, \Lambda/m} }\Big\} \, \tfrac{Q_s^2}{m^2} \, \zeta^2
    \Big) \Big]
    \notag \\ &\hspace{1cm} \times
    \Big[ \big(\alpha^2 + (1-\alpha)^2 \big) K_1^2 (\zeta) + K_0^2 (\zeta) \Big] .
\end{align}
It is important to note that the saturation scale, $Q_s$, depends on the position in the transverse plane, $\tvec{x}$, and introduces that dependence into the differential and integrated probabilities. We see that these probabilities in the MV model (Eq.~\ref{e:prob2_first} and Eq.~\ref{e:multratio2_first}) only differ from GBW (Eq.~\ref{e:prob1_first} and Eq.~\ref{e:multratio1_first}) by the inclusion of logarithmic terms in the exponential (denoted by red highlights), but are otherwise unmodified. Here $\Lambda \rightarrow 0$ is an IR regulator scale which must be chosen to be smaller than the lightest physical scale in the problem -- in this case, the up quark mass $\Lambda \lesssim 2 \: \mathrm{MeV}$.

\begin{figure}
\begin{center}
	\includegraphics[width=0.7\textwidth]{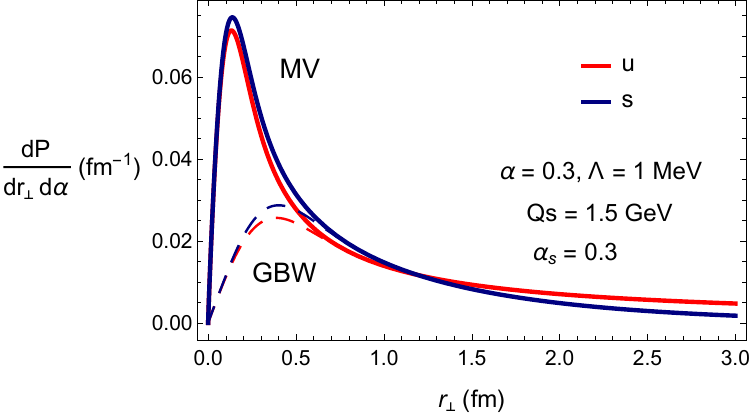}
	\caption{
	    Differential $q \barq$ splitting probability as a function of the distance $r_\bot$ between them for both the GBW~(Eq.~\ref{e:prob1_first}, dashed curves) and MV~(Eq.~\ref{e:prob2_first}, solid curves) models.  We plot both the up quark and strange quark probabilities for representative parameter values. The down quark distribution is indistinguishable from the up quark distribution and is not shown explicitly. Figure from Ref.~\cite{Carzon:2019qja}.
	}
	\label{f:probabilities}
	\end{center}
\end{figure}

The differential probability distributions, with respect to the separation distance $r_\bot$, is plotted in Fig.~\ref{f:probabilities} for the up and strange quarks in each model. We see that there is agreement between GBW and MV models in the long distance behavior, which corresponds to the deep saturation regime. However, at short distances the MV model reflects the correct transition to a power-law behavior in the dilute regime, while the simple Gaussian behavior, at large distances, is continued by the GBW model. This results in a more pronounced peak, in the MV model as compared to GBW, at short distances and, thus, a greater probability to produce quarks overall. For both models, the only dependence on flavor arises from the quark mass and leads to small differences in the probability distribution between quarks of different flavor.

%__________________________________________________________________________
%
%
%__________________________________________________________________________
%

\begin{figure}
    \centering
    \includegraphics[width=0.49\textwidth]{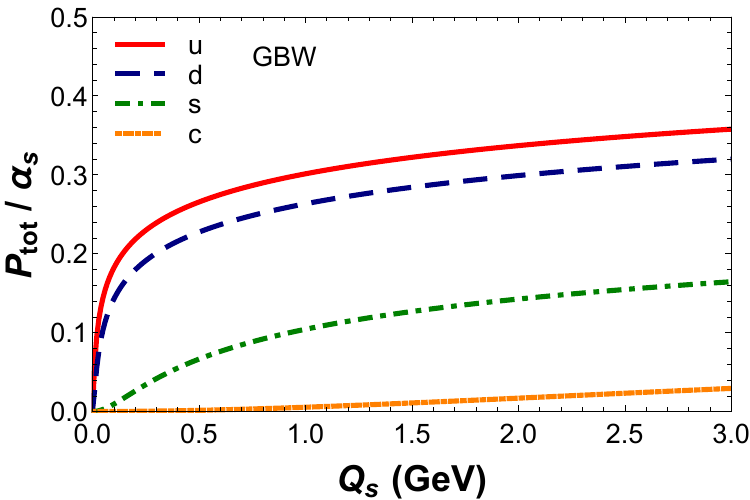}
    \includegraphics[width=0.49\textwidth]{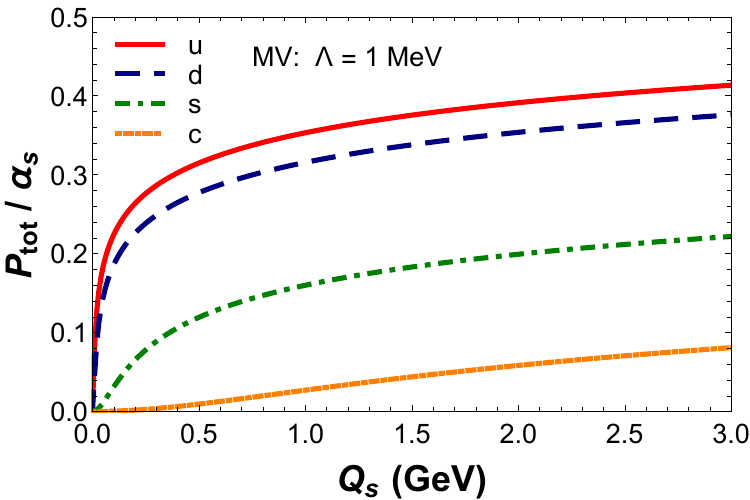}
	\caption{Total quark production probabilities (scaled by the strong coupling $\alpha_s$) as a function of the target saturation scale $Q_s$ for various flavors in the GBW (left) and MV (right) models. Figure from Ref.~\cite{Carzon:2019qja}.}
	\label{f:multratio}
\end{figure}

The integrated probabilities for each model are presented in Fig.~\ref{f:multratio}, with respect to the saturation scale, from which we observe an increase in probability as a function of $Q_s$. These probabilities are ratios of the quarks produced to the gluon density and set the overall chemistry of the initial state. We see a clear reflection of the mass ordering $P_\mathrm{tot} (u) > P_\mathrm{tot} (d) > P_\mathrm{tot} (s) > P_\mathrm{tot} (c)$ arising from the splitting functions. The total probabilities are consistent with the observation, from Fig.~\ref{f:probabilities}, that the MV model will produce more quarks than the GBW. An interesting difference between the two models can be seen in the strange quark probabilities, where the MV model has a steeper increase with respect to $Q_s$ than GBW. This may have an influence on the location of strange production, pushing more production to low energy regions in the MV model. 

%__________________________________________________________________________
%
%
%__________________________________________________________________________

It is important to note that if one extends the model to $Q_s$ or $\alpha_s$ that is too large, then it becomes possible that the quark probabilities may sum to a number larger than 1. This would be an indication of a violation of the weak coupling assumption $\alpha_s \ll 1$ and is beyond the regime of validity for our model.

%..........................................................................
%
\subsection{Density Mask}
\label{subsec:DensityMask}
%
%..........................................................................

\epigraph{IO::InitializeEvent() and Mask.h}{}

The final step in a $g \rightarrow q \bar{q}$ process is the redistribution of the energy density and associated charge of each particle. Quarks, as far as we know, are more or less point-like objects though we cannot treat them this way since that description is incompatible with computational methods for hydrodynamics. Instead, we distribute the energy and charge of the quarks as a density with some radius. The specification of this density profile introduces a new degree of freedom to the algorithm. By default, we chose to describe the quark density profiles as Gaussian blobs,
\begin{align}
    \rho_{Gaussian} (\vec{r}) =
    \frac{\rho_{tot}}{(\Delta x)^2 \tau_0} \frac{e^{-\frac{\vec{r}^2}{2\|r\|^2}}}{\int\limits_R e^{-\frac{\vec{r}^2}{2\|r\|^2}} d\vec{r}}
\end{align}
where $\rho_{tot}$ is the total energy or charge, $\Delta x$ is the grid spacing, $\tau_0$ is the initialization time, and the radius of the distributed density is the width of the Gaussian, cut off at the width of the distribution. For this work, we chose the radius of this density profile to be $r_q=0.5 fm$, so that the quarks were on the order of 1 fm in size. A more sophisticated parameterization can be done using generalized parton distriubtions (GPDs) \cite{Ji:1996ek, Radyushkin:1996nd, Ji:1996nm, Radyushkin:1996ru, Collins:1996fb, Diehl:2003ny}, although $r_q$ is expected to be some non-perturbative scale, which is reflected in our choice. 

\begin{figure}
    \centering
    \includegraphics[width=0.5\linewidth]{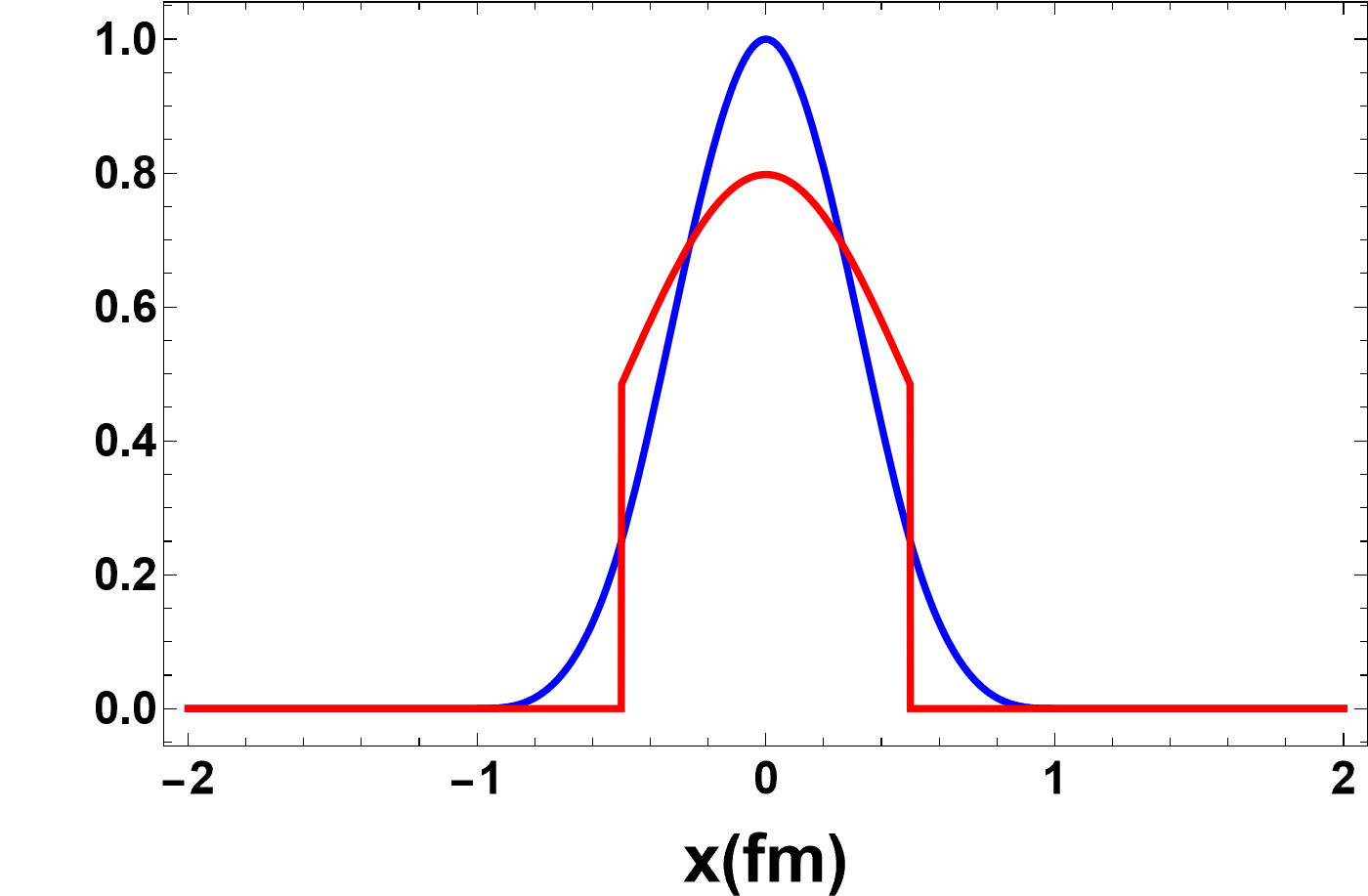}
    
    \caption{Comparison of kernel to Gaussian quark/anti-quark density profile. The cutoff for the Gaussian profile is $0.5 fm$ while the kernel function goes to zero at $1 fm$.}
    \label{fig:KernelVsGaussian}
\end{figure}

In the process of connecting \code{iccing} to BSQ hydrodynamic simulations, namely \code{ccake} \cite{Plumberg:inPrep}, a more subtle understanding of the importance of the density profile developed. In specific, it was discovered that the edge of the quark charge densities, using the Gaussian profile, created a large gradient that posed a problem for connection to hydrodynamics, which requires smooth gradients. To resolve this issue and provide a better generalization of the model, I added the "Mask" class to \code{iccing}, which contains the definition of the density profile used for distribution and implements both Gaussian and Kernel function profile options. The Kernel function is defined piece-wise as:
\begin{align}
    \rho_{Kernel} (\vec{r}) =
    \begin{cases}
        0 \: , & \: , \frac{\sqrt{\vec{r}^2}}{0.5 \, r_q} \geq 2
        \\
        \mathcal{K} \frac{1}{4} \left[2 - \frac{\sqrt{\vec{r}^2}}{0.5 \, r_q} \right]^3 \: , & \frac{\sqrt{\vec{r}^2}}{0.5 \, r_q} \geq 1
        \\
        \mathcal{K} \left[1 - 1.5 \left(\frac{\sqrt{\vec{r}^2}}{0.5 \, r_q}\right)^2 + 0.75 \left(\frac{\sqrt{\vec{r}^2}}{0.5 \, r_q}\right)^3 \right] \: , & \frac{\sqrt{\vec{r}^2}}{0.5 \, r_q} < 1
    \end{cases} \: ,
\end{align}
where $\mathcal{K}$ is the normalization constant defined as
\begin{align}
    \mathcal{K} = \frac{10}{7\pi \Delta x^2 \tau_0} \frac{1}{(0.5 r_q)^2} \:,
\end{align}
and the function goes to 0 at $r_q$. This was chosen as an alternative since it is used by \code{ccake} \cite{Plumberg:inPrep} and is known to work with hydrodynamic requirements. The profile is defined so that the density goes smoothly to zero at the edge, while the Gaussian profile has infinite tails and always requires a cutoff. A comparison between the two profiles is illustrated in Fig.~\ref{fig:KernelVsGaussian}. The Gaussian profile is defined with $r_q=0.5 fm$, while the radius of the Kernel function profile is chosen as $1 fm$, to ensure consistent size behavior with the Gaussian profile. 

Further analysis, of the effect of the profile definition on the event geometry, is presented in Sec.~\ref{SubSec:LocalGlobalImpactDensityProfile}. The implementation of pre-equilibrium evolution for the quark densities, required another extension of this formulation to include a time dependent radius, which is explored in Chap.~\ref{chap:PreEquilibriumEvolution}.

%---------------------------------------------------------------------------
%
\section{Monte Carlo} \label{sec:MonteCarlo}
%
%---------------------------------------------------------------------------

Having established the required input and initialization of the \code{iccing} model, I will describe the sampling process in detail.

%---------------------------------------------------------------------------
%
\subsection{Select Location of Gluon} \label{subsec:SelectGluon}
%
%---------------------------------------------------------------------------

\epigraph{Event::SampleEnergy() and Event::GetGlue()}{}

A founding assumption of this method is that the initial state, $\epsilon (\vec{x}_\bot)$, is composed by a density of gluons, and individual gluons have the opportunity to split into $q \barq$ pairs. Practically, this is accomplished by selecting gluons from the provided initial energy density. We start by randomly choosing a valued point from the initial condition to serve as the "seed" for a gluon. In this formulation, any point that contains non-zero energy density has an equal probability of being chosen, although this could be further parameterized by introducing energy dependence to the probability. Next, the total energy in some circular region, defined by the "seed" and radius $r_{gluon}$, is calculated. For much of this work, we set the sampling radius to be equal to that of the redistributed density profile, although these could vary independently.

From this circular region, centered on the gluon seed, the total energy available, $E_R$, is obtained by integrating over the energy density. This integration is done over the hypersurface defined by the initialization time, $\tau_0$, since the initial condition is a three-dimensional energy density in units of $\mathrm{GeV} / \mathrm{fm}^3$. The default value of $\tau_0=0.6 fm/c$ is taken from the \code{TRENTO} Bayesian analysis. For an initial condition, the scale of the density must be set such that it is compatible with the total multiplicity as measured by experiment. This scaling is controlled by the entropy proportionality factor, $a_{Entropy}$ (See Sec.~\ref{Sec:Centrality}), and the initialization time. Explicitly, the total energy available to a gluon sampling is calculated by integrating over some region $R$:
\begin{align}
E_R &= \int\limits_{R} d^3 x \, \epsilon (\vec{x}) = \tau_0 \int\limits_{R} d\eta \, d^2 x_\bot \, \epsilon(\vec{x}_\bot) \\
\frac{dE_R}{d\eta} &= \tau_0 \int\limits_R d^2 x_\bot \, \epsilon(\vec{x}_\bot) = \tau_0 \Delta x \Delta y \sum\limits_{i \in R} \epsilon_i ,
\end{align}
where $\Delta x$ and $\Delta y$ are the grid spacings in units of fermi. 

The current structure of \code{ICCING} is 2D and boost-invariant and as such assumes the provided initial condition is under the same constraints. This makes the quark production rates boost-invariant and, thus, they must be compared to the necessary thresholds to produce a $q \barq$ pair of a given flavor per unit rapidity. The structure of the model and theoretical calculations, make it possible to extend this process to 3D in future work.

%---------------------------------------------------------------------------
%
\subsection{Sample Energy of Gluon} \label{subsec:SampleGluonEnergy}
%
%---------------------------------------------------------------------------

\epigraph{Splitter::SplitSample() and Splitter::RollGlue()}{}

Having determined the total energy $E_R$ (per unit rapidity) available within the circular region $R$, some fraction of it is assigned to belong to a singular 'gluon'. The gluon distribution, from Ref.~\cite{Carzon:2019qja} and within the boost-invariant approximation, depends inversely on the light-front momentum $q^+$, which for an ultrarelativistic particle is equivalent to its energy ($q^+ \approx \sqrt{2} E_G$):
\begin{align}
    \frac{d\sigma^G}{d q^+} \propto \frac{1}{q^+}
    \qquad \rightarrow \qquad
    \frac{dP}{dE_G} \propto \frac{1}{E_G} ,
\end{align}
where $\sigma^{G}$ is the gluon cross section. The natural choice would be to sample the single gluon's energy from the boost-invariant distribution $\sim 1 / E_G$, though for the sake of flexibility we use the distribution: 
\begin{align}   \label{e:lambdaparam}
   \frac{dP}{dE_G} \propto  \left(\frac{1}{E_G} \right)^\lambda
\end{align}
and allow the specification of the exponent $\lambda$. One could use this flexibility to mimic the effects of linear small-$x$ quantum evolution by modifying the gluon distribution. Here we choose to contrast this parameterization against a uniform distribution of gluon energies, corresponding to $\lambda = 0$.

\begin{figure}
        \begin{centering}
    	\includegraphics[width=0.47\textwidth]{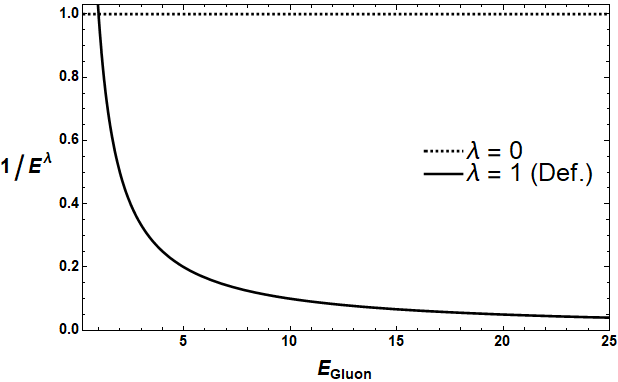} \,
    	\includegraphics[width=0.45\textwidth]{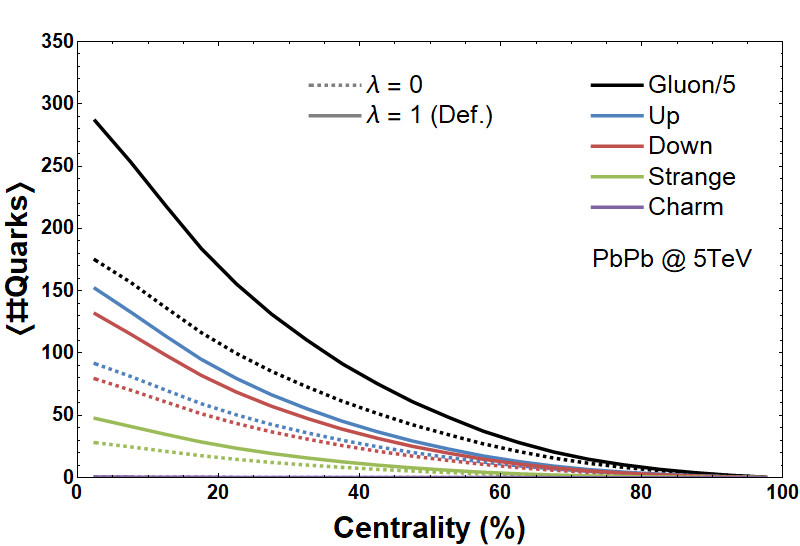}
    	\end{centering}
    	\caption{Illustration of the function Eq.~\ref{e:lambdaparam} used to sample the gluon energy (left) and its effect on the quark multiplicities (right).  Note that the gluon multiplicities have been scaled down by a factor of 5 for better comparison to the quarks. Figure from Ref.~\cite{Carzon:2019qja}.}
    	\label{f:lambdaEffect}
    \end{figure}

With this gluon energy distribution, defined by the $\lambda$ exponent, we select some portion of the total available energy, to be associated with a single gluon, and bound this sampling from below with some minimum energy threshold $E_G \in [E_\mathrm{thresh} , E_R]$. This threshold is used to restrict production in low energy areas, such as the periphery of the event, but has a non-trivial effect on relative flavor production if chosen to be larger than the minimum required to produce two light quarks. The $\lambda$ parameter weights the distribution to favor the production of fewer gluons in total, with more hard gluons, or more gluons by increasing the soft sector. This is reflected in the quark production since the quark multiplicity is proportional to the number of opportunities available for splitting. We can see this illustrated in Fig. \ref{f:lambdaEffect} by comparing the uniform gluon energy distribution, $\lambda = 0$, with the boost-invariant case, $\lambda = 1$. The bias of gluon energies, to the soft sector, by the steeply-falling energy distribution, in the boost-invariant case, significantly enhances the average multiplicity for all $q \barq$ pairs.

%__________________________________________________________________________
    %
   %
%__________________________________________________________________________

As energy is apportioned to gluons and quark/anti-quark pairs, it is transferred over to the output density and no longer considered in the sampling (See Sec.~\ref{sec:Redistribute}). For an individual gluon sampling, some fraction of the energy available, $\frac{E_G}{E_R} \leq 1$, is redistributed to the output density. This allows for multiple gluons to be sampled from the same region and naturally weights production toward areas with more energy. Furthermore, this redistribution is done in a proportional way, as to reduce artifacts in the energy distribution from the sampling process.

%---------------------------------------------------------------------------
%
\subsection{Sample Flavor} \label{subsec:SampleFlavor}
%
%---------------------------------------------------------------------------

\epigraph{Splitter::SplitSample() and Splitter::RollFlavor()}{}

Having sampled a gluon appropriate for splitting, we determine the flavor of the quark/anti-quark pair under consideration. This is done through a Monte-Carlo sampling of the integrated splitting probabilities from Eqs.~\ref{e:multratio1_first} and \ref{e:multratio2_first}. The probability for a splitting is set by the $Q_s$ of the sampled region, with the possibility of remaining a gluon determined by $1-P_{up}-P_{down}-P_{strange}-P_{charm}$. Top and bottom quarks are not considered since their occurrence would be rare, due to their large masses, and the main production channel is hard scatterings. Charm production is also ignored since it occurs infrequently only in the most central events and does not provide sufficient statistics for analysis. A comparison of quark production for hard and soft processes would be an interesting area of investigation since it could provide some insight into the significance of the quark production implemented in this model, but this is left for future work.

For a given flavor outcome, a check is made to determine if the gluon has enough energy to actually produce two quarks of that flavor, $E_G < 2 m$, where $m$ is taken to be the rest mass of the chosen quark. If this test is not passed, then the splitting is rejected and the procedure starts from the beginning with a new gluon seed.

%---------------------------------------------------------------------------
%
\subsection{Sample Location} \label{subsec:SampleLocation}
%
%---------------------------------------------------------------------------

\epigraph{Splitter::SplitSample() and Splitter::RollLocation()}{}

Once a successful $g \rightarrow q \barq$ splitting is identified, the kinematic properties of the quark/anti-quark pair must be determined. This procedure is accomplished through a Monte-Carlo sampling of the differential splitting probability distributions from Eqs.~\ref{e:prob1_first} and \ref{e:prob2_first} to select appropriate values for the separation distance, $r_\bot$, and relative light-front momentum fraction, $\alpha$, for the pair. The distributions are dependent on the saturation scale and quark mass. Sampling is done simultaneously for both $r_\bot$ and $\alpha$. 

An interval, $[\alpha_\mathrm{min}, 1 - \alpha_\mathrm{min}]$, is specified for the random sampling of the energy fraction $\alpha$. The values of $0$ and $1$ are excluded as options, by using $\alpha_\mathrm{min}$ as a small offset, since these would indicate that one of the products of the splitting carries so little energy that the underlying assumptions of the calculation would be violated. Again, we choose to parameterize this option by allowing for the specification of $\alpha_\mathrm{min}$, which we choose here to be $0.01$ by default. 

Similarly, the separation distance of the quark/anti-quark pair is sampled from the interval $[0, d_\mathrm{max}]$, where $d_\mathrm{max}$ puts a limit on the distance since the underlying perturbative calculation breaks down at long distances. The maximum separation is another accessible parameter in the model, although it should be chosen on the order of $\ord{1/\Lambda_{QCD}}$ and our default choice is $d_\mathrm{max} = 1 \, \mathrm{fm}$, which is consistent.

\begin{figure}
    \centering
	\includegraphics[width=0.3\textwidth]{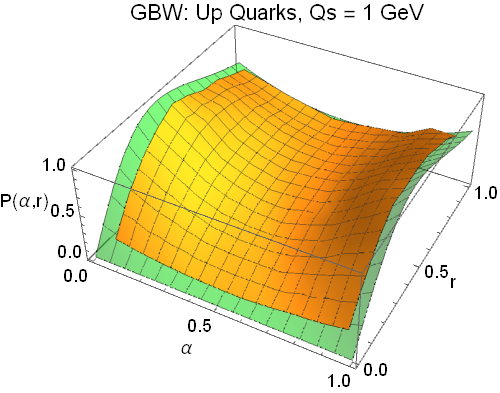}
    \includegraphics[width=0.3\textwidth]{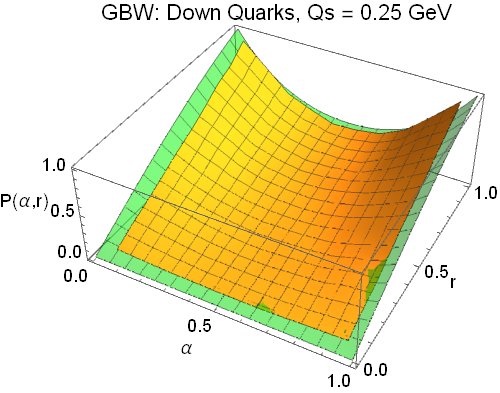}
    \includegraphics[width=0.3\textwidth]{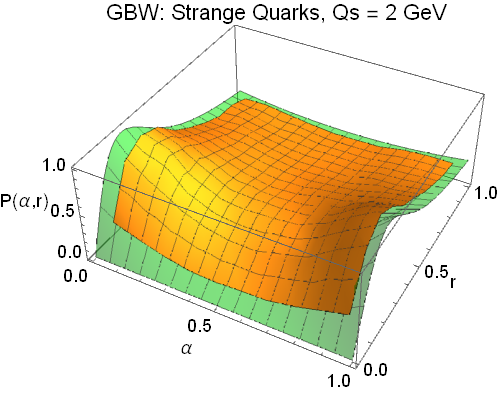}
    \\
    \includegraphics[width=0.3\textwidth]{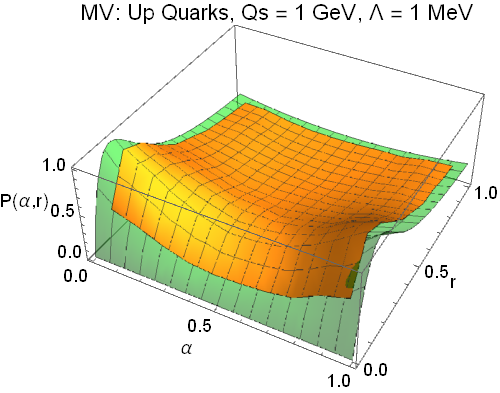}
    \includegraphics[width=0.3\textwidth]{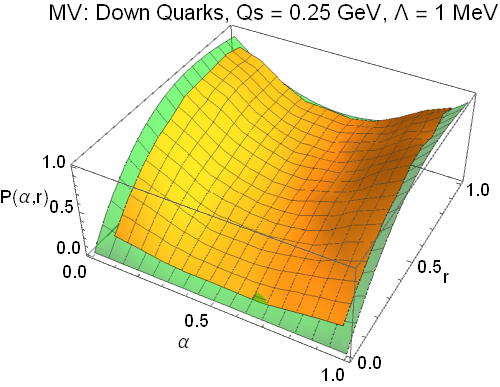}
    \includegraphics[width=0.3\textwidth]{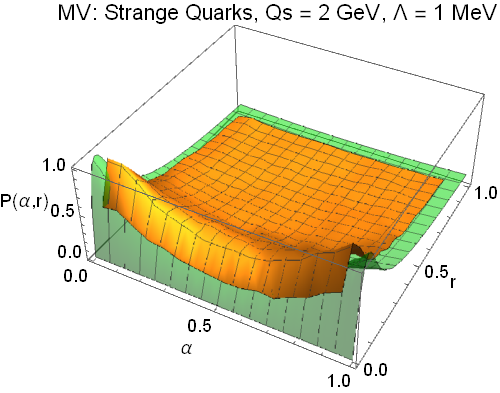}
    
	\caption{
	    Plot of the input and output of the Monte Carlo sampling, validating that it successfully reproduces the theoretical probability distribution for the GBW and MV models. Figure from Ref.~\cite{Carzon:2019qja}.
	}
	\label{f:MCvalidate}
	
\end{figure}

Using these sampling intervals, the pair values of $\alpha$ and $r_\bot$ are determined through standard rejection sampling techniques. An illustration is provided in Fig.~\ref{f:MCvalidate}, where the probability distribution is plotted for the analytic solution of Eqs.~\ref{e:prob1_first} and \ref{e:prob2_first} (represented by the green surface) and compared to a reconstruction by way of Monte-Carlo sampling (orange surface). This validation is done for several combinations of quark flavor and $Q_s$ in both the GBW and MV models.

%__________________________________________________________________________
%
%
%__________________________________________________________________________
 
The final kinematic ingredient is a random angle $\phi$, chosen from $[0, 2\pi]$, that is used to orient the displacement vector $\vec{r}_\bot$ in the transverse plane with respect to the coordinate axes. Now the properties of the quark/anti-quark pair are fully determined, with the displacement vectors specified as $\Delta \vec{x}_\bot^{q} = (1-\alpha) \vec{r}_\bot$ and $\Delta \vec{x}_\bot^{\barq} = -\alpha \vec{r}_\bot$. These definitions preserve the center of momentum of the $q \barq$ pair and is an explicit feature of the $g \rightarrow q \barq$ light front wave function \cite{Kovchegov:2006qn}.

%---------------------------------------------------------------------------
%
\section{Energy and Charge Redistribution}
\label{sec:Redistribute}
%
%---------------------------------------------------------------------------

There are two positive resolutions of the Monte-Carlo sampling detailed in Sec.~\ref{sec:MonteCarlo}: a gluon is selected and remains a gluon or the chosen gluon successfully splits into a quark/anti-quark pair. In any case, the next step of the algorithm is redistributing the energy density and any associated charge according to the properties determined in Sec.~\ref{sec:MonteCarlo} using the initialized mask of Sec.~\ref{subsec:DensityMask}. 

\begin{figure}
\begin{center}
	\includegraphics[width=0.7\textwidth]{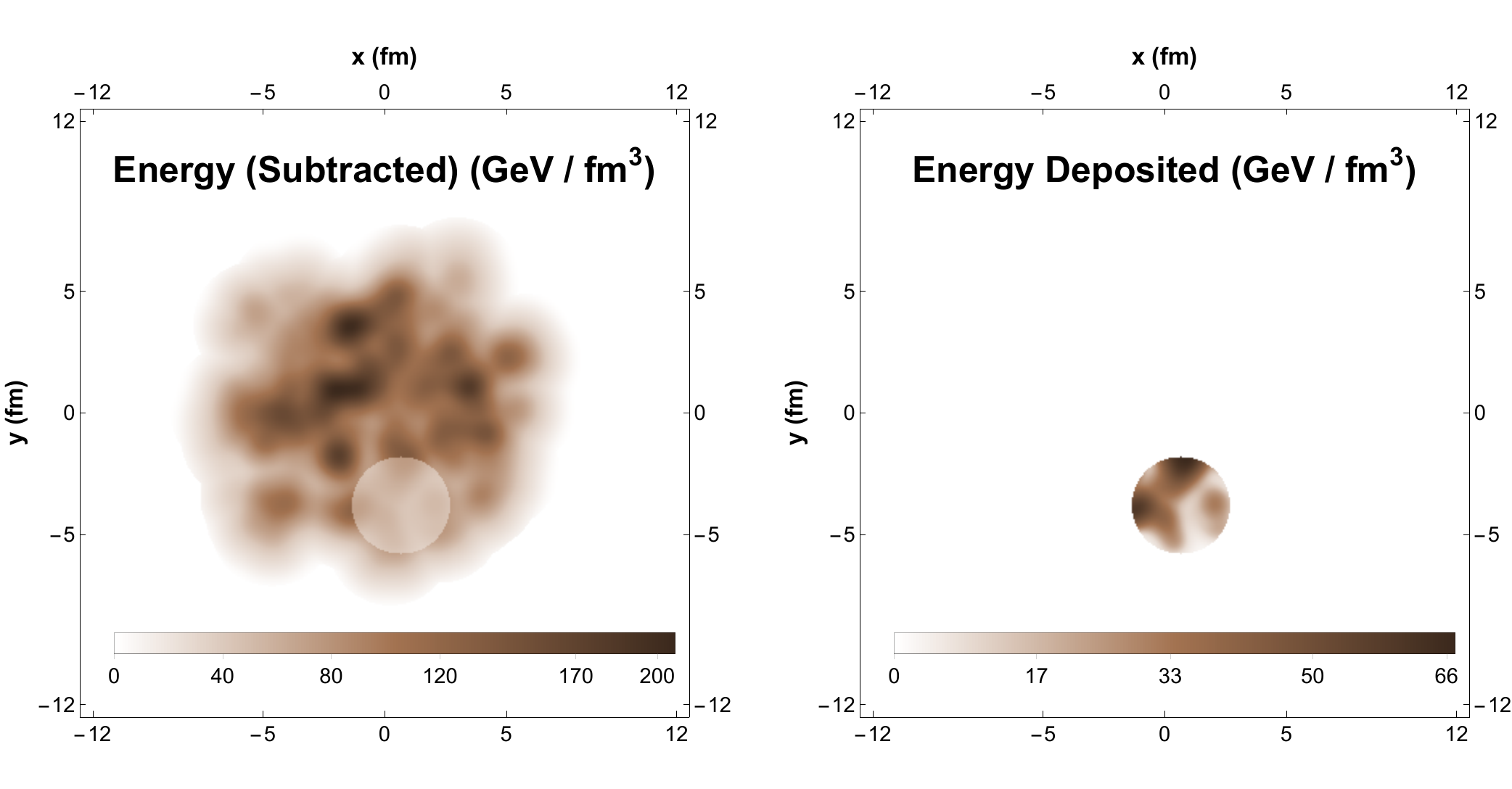}
	\caption{Illustration of how the \code{iccing} algorithm transfers energy when the gluon {\it{does not}} split into a $q \barq$ pair.  The energy is deducted from the input grid (left) and deposited in the output grid (right) as shown.  	The energy transfer is done point by point and proportionately to the total enclosed energy.  As a result, the transferred energy retains the underlying geometric structure of the original energy density, as seen in both the input energy grid after subtraction and the output energy grid after deposition.  The gluon radius here has been greatly increased to clearly show these details. Figure from Ref.~\cite{Carzon:2019qja}.
	}
	\label{f:cutpaste}
	\end{center}
\end{figure}

In the case that a sampled gluon remains a gluon, the energy density associated with it must be removed from the input density and transferred to the output density. While not a requirement, we choose to do this redistribution in such a way as to preserve the underlying geometry of the initial state. This is accomplished by subtracting, from the input density, and adding, to the output density, the energy of the gluon in a proportional manner by transferring a fraction of energy, $\frac{E_G}{E_R}$, at each point in the gluon. This ensures that no gluon will fully deplete an energy site in the input grid. An area can, however, be removed from the sampling process when a region is selected that has total energy below $E_{thresh}$. An illustration, where the gluon radius has been greatly increased for clarity, of the gluon redistribution process is presented in Fig.~\ref{f:cutpaste}.

%__________________________________________________________________________
%
%
%__________________________________________________________________________

The process of redistributing densities associated with quark/anti-quark pairs is more involved. The subtraction of the generating gluon from the initial density is executed in the same manner as above, proportionally to ensure the sampling geometry is preserved. The distribution of the energy and charge of the quarks requires more attention. The quark and anti-quark will receive a portion of the energy from the generating gluon dictated by the relative momentum fraction $\alpha$: $\alpha E_G$ and $(1-\alpha) E_G$, respectively. These energies are then distributed at the appropriate locations defined by the separation distance $r_\bot$, light-front momentum fraction $\alpha$, and transverse orientation $\theta$ with respect to the center of the generating gluon. The profile, with which the energy is distributed, is chosen by the user with the 'profile' parameter, whose implementation is described in Sec.~\ref{subsec:DensityMask} and effect explored in Sec.~\ref{SubSec:LocalGlobalImpactDensityProfile} and Chap.~\ref{chap:PreEquilibriumEvolution}. The default behavior is using a Gaussian profile to distribute the quark/anti-quark densities.

One final constriction may be put on the \code{iccing} model through the specification of a pertubative cutoff. Up to this point, measures were taken to avoid any splittings that may be nonphysical, but these were done implicitly. These measures are controlled by $E_{thresh}$ and the mass threshold $(E_G \geq 2 m)$. Though these should be enough for most cases, a more explicit enforcement is required for others. The perturbative cutoff allows the specification of some limit on the magnitude of a quark's energy density as compared to the background. If a splitting violates this cutoff, it is thrown out and the Monte-Carlo process repeated. A detailed description of this pertubative cutoff is provided in Chap.~\ref{chap:PreEquilibriumEvolution}, where it is an integral part of that analysis, with further exploration in Sec.~\ref{SubSec:LocalGlobalImpactDensityProfile}.

%__________________________________________________________________________
%
\begin{figure}
\begin{center}
	\includegraphics[width=\textwidth]{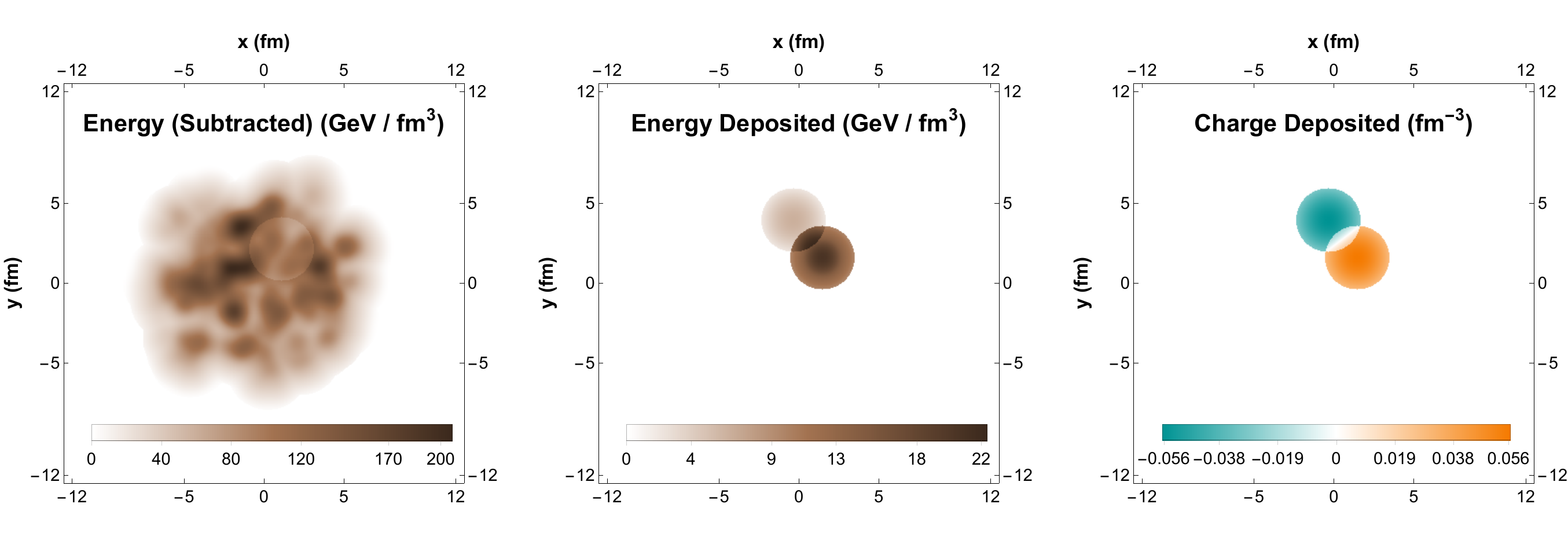}
	\caption{
	Illustration of how the \code{iccing} algorithm transfers energy when the gluon {\it{does}} split into a $q \barq$ pair.  The energy is deducted from the input grid (left plot) proportionately, preserving the underlying geometry in the input grid.  But it is deposited in Gaussian blobs for the $q \barq$ pair, which are displaced relative to the original gluon position. This modifies the energy density distribution (center plot) and also leads to a net displacement of positive and negative charge (here baryon density, right plot).  Note that the energy is in general shared unequally between the quark and antiquark; in this case, the quark carried about 75\% of the original gluon energy, as visible in the center plot.  Here both the radii and overall $q \barq$ displacement have been greatly increased to clearly show these details. Figure from Ref.~\cite{Carzon:2019qja}.
	}
	\label{fig:QuarkRedistribution}
	\end{center}
\end{figure}
%
%__________________________________________________________________________

Finally, the new initial state charge densities are distributed. For both quark and anti-quark, charge density is deposited in the final output using the same kinematic variables and density profile as the energy density. This mirroring of behavior for the energy and charge densities of the $q \barq$ pair ensures that every site with charge is associated with some energy density. An illustration of the distribution of quark/anti-quark densities is shown in Fig.~\ref{fig:QuarkRedistribution}, where the quark is associated with positive charge and the anti-quark with negative.

%---------------------------------------------------------------------------
%
\section{End and Output}
\label{subsec:EndAndOutput}
%
%---------------------------------------------------------------------------

The Monte-Carlo sampling of gluons and redistribution of energy and charge from the input density to the output densities continues until the total energy of the input density is less than $E_{thresh}$, at which point all remaining energy is transferred to the output. For each splitting, preservation of energy conservation is monitored and there are checks to ensure redistribution does not exceed the dimensions of the system. 

The main output from \code{iccing} consists of grids of the redistributed energy density $\epsilon^\mathrm{(output)}$ along with the charge densities $\rho_B$, $\rho_S$, and $\rho_Q$ of baryon number, strangeness, and electric charge, respectively. The full density grids can be printed in several formats or completely suppressed. Additionally, the Fourier coefficients of the energy and charge densities are printed to a selection of files in the specified output directory (See Sec.~\ref{sec:Eccentricities} for a full description and explicit calculation). The multiplicities of quark pairs of each flavor are also printed.

%%%%%%%%%%%%%%%%%%%%%%%%%%%%%%%%%%%%%%%%%%%%%%%%%%%%%%%%%%%%%%%%%%%%%%%%%%%
%
\chapter{ICCING Results} \label{chap:ICCINGResults}
%
%%%%%%%%%%%%%%%%%%%%%%%%%%%%%%%%%%%%%%%%%%%%%%%%%%%%%%%%%%%%%%%%%%%%%%%%%%%

This chapter reproduces and refines the work from Ref.~\cite{Carzon:2019qja}.

\begin{figure}[h!]
        \centering
    	\includegraphics[width=\textwidth]{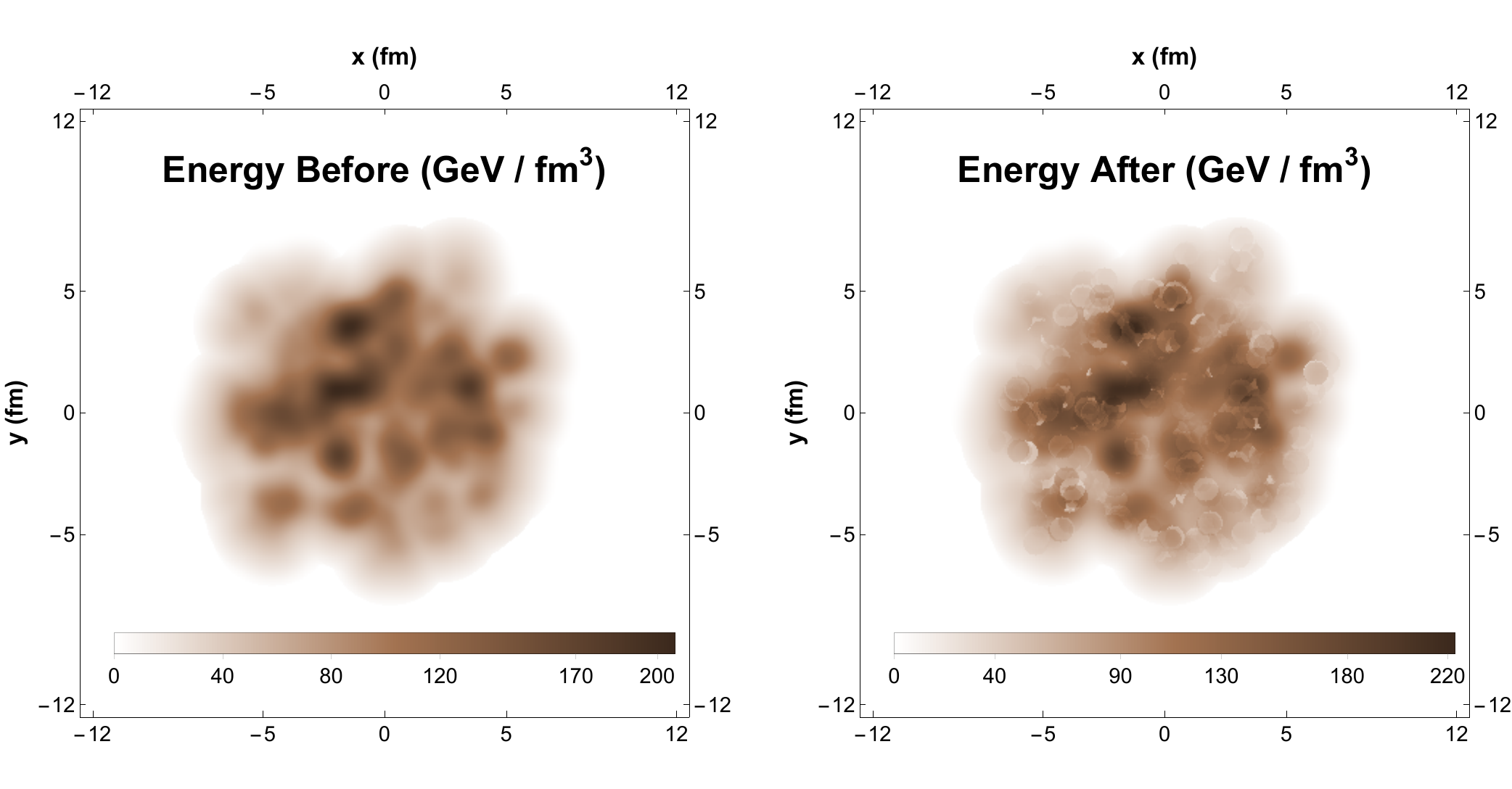}
    	\caption{
    	Comparison of the energy density before (left plot) and after (right plot) running the \code{iccing} algorithm.  As a result of redistributing the energy density from the $g \rightarrow q \barq$ splitting, the energy density profile is somewhat modified, including visible artifacts associated with the model implementation. Figure from Ref.~\cite{Carzon:2019qja}.}
    	\label{f:BeforeAfterPlot}
    \end{figure}

It is important to quantify the extent to which the \code{iccing} algorithm modifies the input energy density through the redistribution of energy due to the quantum mechanical $g\to q\bar{q}$ splitting process. There is a nontrivial modification of the energy density by this process as seen in Fig.~\ref{f:BeforeAfterPlot}, where it is possible to see circular artifacts in the energy density where gluons were sampled and split into quarks using the Gaussian profile. Though the artifacts are highly dependent on artificial choices made by the selection of parameter values, i.e. the gluon radius $r$, they are a real effect of the partonic splitting physics. The sampling artifacts are important to quantify since \code{trento}, by itself, has already been shown to accurately describe final state flow harmonics and fluctuations of bulk particle production at the LHC \cite{Moreland:2014oya, Bernhard:2016tnd, Alba:2017hhe, Giacalone:2017dud}. Consistency with experimental data could be disrupted if \code{iccing} introduces new charge distributions at the cost of modifying the energy profile too much.

%__________________________________________________________________________
    %
    %
%__________________________________________________________________________

\begin{figure}[h!]
        \centering
    	\includegraphics[width=0.45 \textwidth]{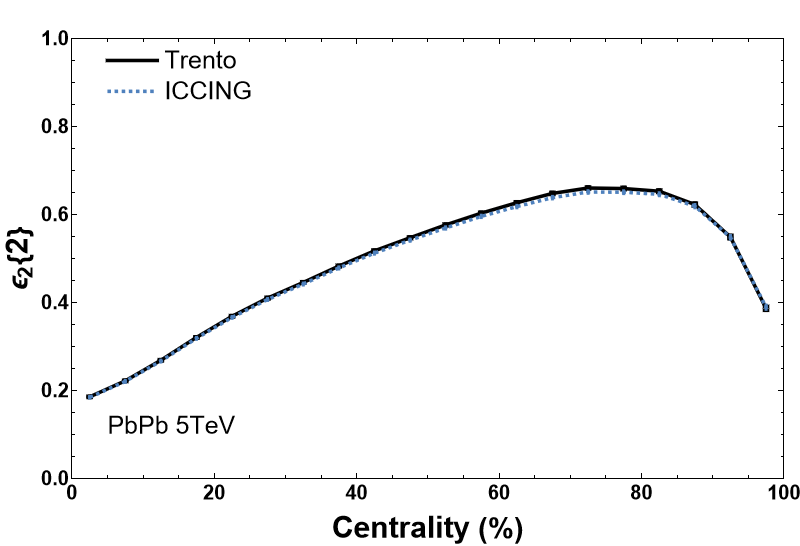} \,
    	\includegraphics[width=0.45\textwidth]{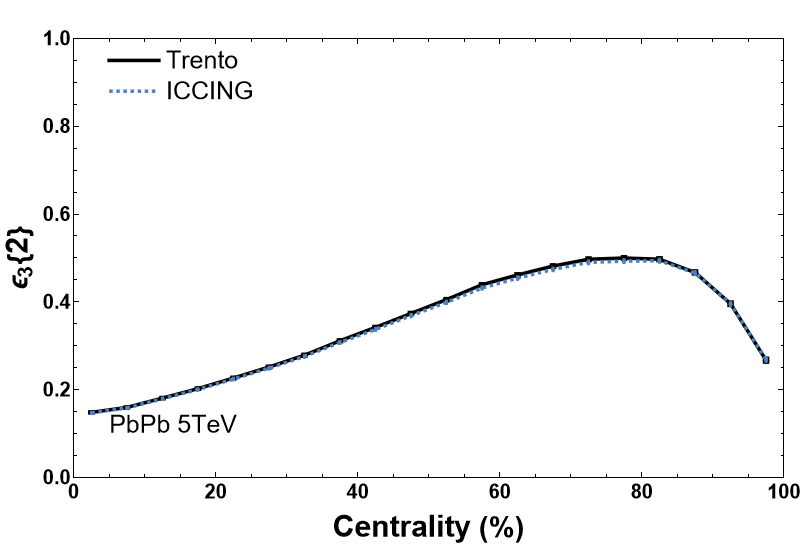}
    	
    	\caption{Eccentricities of the energy density as a function of centrality before (solid curve) and after running \code{iccing} (dotted curve). Figure updated from Ref.~\cite{Carzon:2019qja}.}
    	\label{f:EccInOut}
    \end{figure}

The 2-particle cumulants of ellipticity and triangularity of the energy density as a function of centrality, shown in Fig.~\ref{f:EccInOut}, are a good way to quantify the effect of \code{iccing} on the original initial condition. These observables see little to no modification by the \code{iccing} process. This is quite interesting, since, from the comparison in Fig.~\ref{f:BeforeAfterPlot}, there are clear artifacts from the \code{iccing} process in the energy density but these are only seen at the 'microscopic' level and have no effect on event averaged 'macroscopic' quantities. These 'microscopic' fluctuations are further investigated in Sec.~\ref{SubSec:LocalGlobalImpactDensityProfile} and play an important part in Chap.~\ref{chap:PreEquilibriumEvolution}. While the insensitivity of Fig.~\ref{f:EccInOut} means agreement with previous experimental comparisons is preserved, the introduction of these new small scale fluctuations may be important in the connection to hydrodynamic simulations.

%__________________________________________________________________________
    %
   %
%__________________________________________________________________________

Many of the results, presented here, are taken from Ref.~\cite{Carzon:2019qja}, but have received several important updates. In the original analysis, a bug occurred, in the calculation of the cumulants, that failed to account for events with vanishing triangularity and treated them as having significant $\varepsilon_3$. This error is only significant in peripheral centralities for charge densities. The correct behavior can be verified by looking at the most peripheral collisions and seeing that the geometry goes to zero since there are many events that do not produce quark/anti-quark pairs. 

Another issue was discovered in the handling of the energy density left over after the completion of the \code{iccing} sampling process. When the total energy, left to sample from, is less than the threshold to make a quark/anti-quark pair, then the remaining energy should be transferred over to the output energy density. In the original analysis, this end behavior was not carried through. The only significant contribution from this was that the most peripheral energy geometry saw a spike upward due to the only contribution coming from quark/anti-quark splittings. This error was also corrected for all results presented here.

In this chapter, I begin with a description of the estimators of charge geometry and the subtleties involved with using the standard eccentricities for the conserved charge density in Sec.~\ref{Sec:ChargeGeometryEstimators}. This is followed by: an analysis of the \code{iccing} model to investigate the use of strangeness as a probe of the initial state, Sec.~\ref{Sec:StrangenessAsProbe}, an analysis of the model's sensitivity to its parameters, Sec.~\ref{Sec:SensitivityAnalysis}, and finally a description of the coupling of \code{iccing} to the hydrodynamics code \code{ccake}.

%..........................................................................
%
\section{Charge Geometry Estimators} \label{Sec:ChargeGeometryEstimators}
%
%..........................................................................
 
The characterization of the initial state using the standard eccentricities from Sec.~\ref{sec:Eccentricities} is well established when the density being described is positive definite, as is the case with energy and entropy density. However, if $f(\bm{r}) = \rho(\bm{r})$ is a charge density, in particular, one where the total net charge is zero and local fluctuations exist with positive and negative contributions, then the situation becomes more complicated. To understand this issue, we first define the net charge and net dipole moment (as a complex vector),
\begin{align}
	q_{tot} &= \int d^2 \bm{r} \, \rho(\bm{r})
\end{align}
and
\begin{align}
     \bm{d} &= \int d^2 \bm{r} \, \bm{r} \, \rho(\bm{r}) \:,
\end{align}
respectively.
At top collider energies, $q_{tot} = 0$, since net baryon stopping is suppressed, and the center of charge
\begin{align}
\bm{r}_{COC} \equiv
\frac{\int d^2 r \, \bm{r} \, \rho(\bm{r})}
{\int d^2 r \, \rho(\bm{r})}
= \frac{1}{q_{tot}} \bm{d}
\end{align}
becomes undefined. A consequence of vanishing $q_{tot}$ is that the dipole moment is the same with respect to any origin of coordinates. If we shift the distribution to an arbitrary origin at $\bm{R}$,
\begin{align}
\int d^2 r (\bm{r} - \bm{R}) \rho(\bm{r}) =
\bm{d} - \underbrace{q_{tot}}_{= \: 0} \bm{R} = \bm{d}
\end{align}
then the dipole moment is unchanged. This makes it impossible to define a corresponding frame such that $\bm{\mathcal{E}_1} = 0$ when the total charge vanishes, meaning there is always a finite directed eccentricity proportional to the dipole moment. This means the standard eccentricities, Eq.~\ref{e:ecc1} and Eq.\ref{e:ecc2}, must be modified, since it is not possible to construct a center-of-charge frame for the total charge density and ensure that $\bm{\mathcal{E}_1}$ vanishes for a conserved charge with $q_{tot} = 0$.
 
This can be resolved, though, by treating the positive and negative charge density separately: 
\begin{align}
\rho_{\mathcal{X}} \equiv \rho^{(\mathcal{X}^+)} \: \theta(\rho_{\mathcal{X}}) +
\rho^{(\mathcal{X}^-)} \: \theta(-\rho_{\mathcal{X}}) ,
\end{align}
where $\mathcal{X}$ represents the specific BSQ density and we have suppressed the position argument $\bm{r}$ for brevity.  Then the corresponding eccentricities of the positive and negative charge densities are
\begin{align} \label{eq:ChargeEccentricities}
\varepsilon_n^{(\mathcal{X}^{\pm})} &\equiv
\left|
\frac{
    \int d^2 \bm{r} \, \left( \bm{r} - \bm{r}_{COC}^{(\mathcal{X}^{\pm})} \right)^n \,
    \rho^{(\mathcal{X}^{\pm})}(\bm{r})
    }{
    \int d^2 \bm{r} \, \left| \bm{r} - \bm{r}_{COC}^{(\mathcal{X}^{\pm})} \right|^n \,
    \rho^{(\mathcal{X}^{\pm})}(\bm{r})
    }
\right| ,
\end{align}
where
\begin{align}
\bm{r}_{COC}^{(\mathcal{X}^{\pm})} \equiv \frac{\int d^2 \bm{r} \, \bm{r} \, \rho^{(\mathcal{X}^{\pm})}(\bm{r})}{\int d^2 \bm{r} \, \rho^{(\mathcal{X}^{\pm})}(\bm{r})}
\end{align}
is the center of charge for either the positive or negative densities.  

\begin{figure}[h!]
    \centering
    \includegraphics[width=0.48\textwidth]{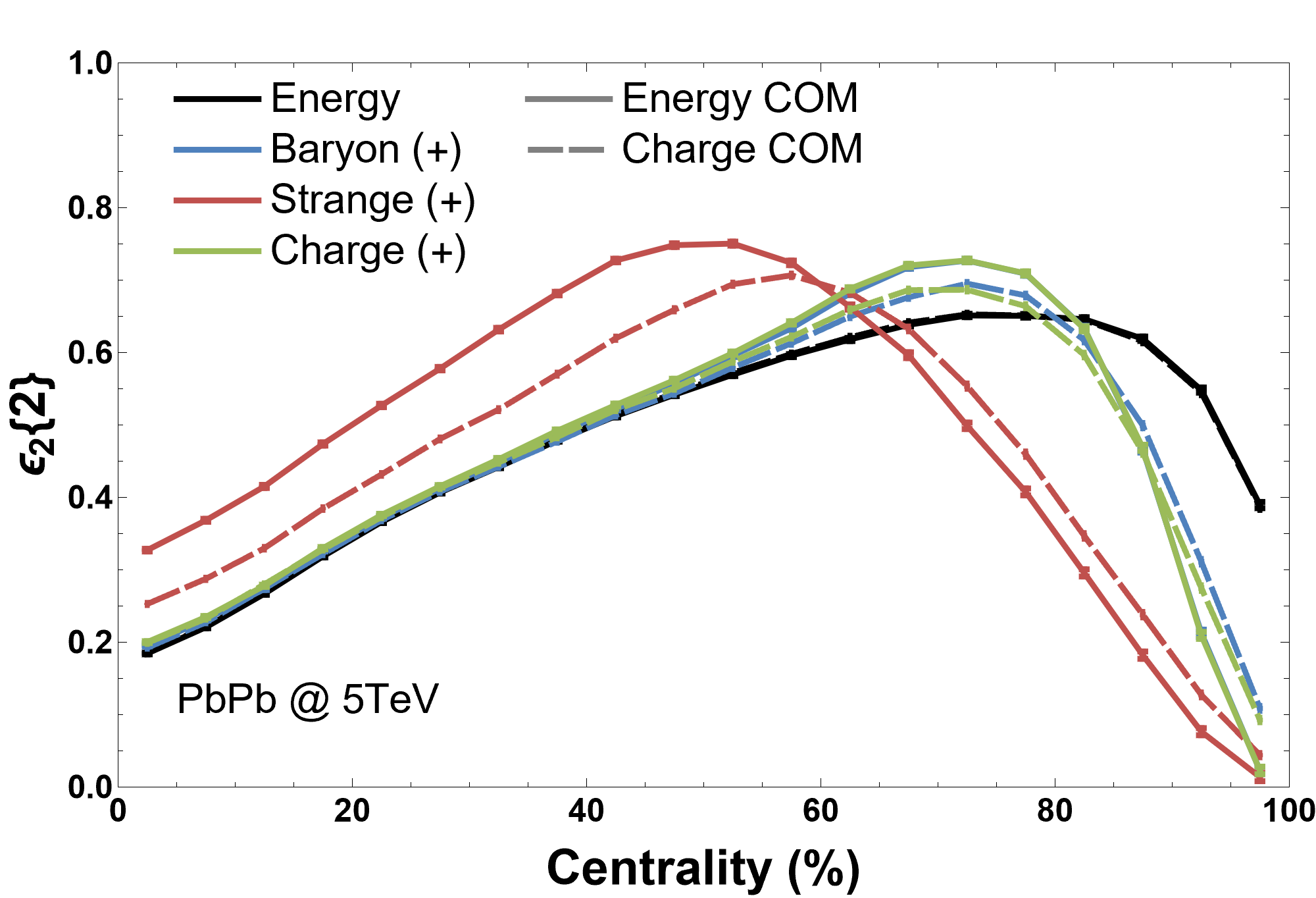}
    \includegraphics[width=0.48\textwidth]{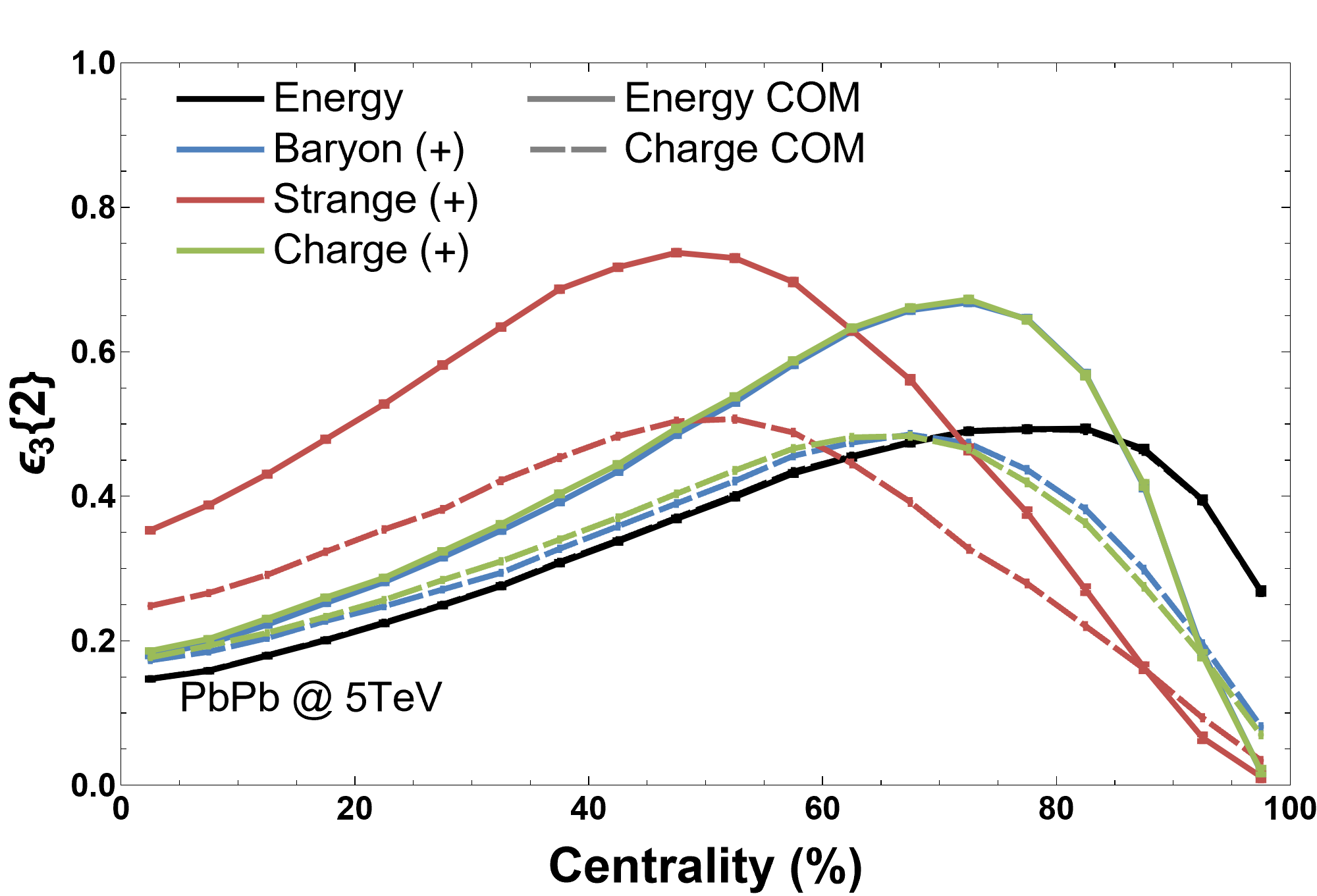}
    \caption{Comparison of the charge eccentricities when calculated with reference to the energy center of mass (COM) or charge center of mass (COC).}
    \label{fig:COMvsCOC}
\end{figure}

Alternatively, one can choose to define the charge eccentricities with respect to the center of mass of the energy density, providing a common reference point for the positive and negative charge geometries. For cumulants (see Sec.~\ref{sec:Cumulants}) of the charge eccentricities, using both COC and COM, there is no difference between the positive and negative event averaged geometries, thus, for all results presented herein only the positive charge density geometry is reported. The difference between charge eccentricities calculated with respect to COC or COM is illustrated in Fig.~\ref{fig:COMvsCOC}. For both reference points, the centrality dependence of the geometry is similar although the magnitude is suppressed for COC eccentricities, most notably for triangularity. In the rest of this work, all calculations of the initial state charge geometry are performed with respect to the center of mass of the energy density. 

%__________________________________________________________________________
%
%__________________________________________________________________________
%

While using the modified eccentricities, from Eq.~\ref{eq:ChargeEccentricities}, as estimators of the initial state charge anisotropy is not shown here to be the optimal solution, it is the best option available, at least until a full hydrodynamic framework with conserved charges is completed and one can systematically check the best charge estimator for the final flow harmonics. This is left to future work but a good direction would be a derivation of charge estimators from the cumulant generating functions.

%..........................................................................
%
\section{Strangeness As a Distinct Probe of the Initial State} \label{Sec:StrangenessAsProbe}
%
%..........................................................................

\begin{figure}[h!]
        \begin{centering}
    	\includegraphics[width=0.45 \textwidth]{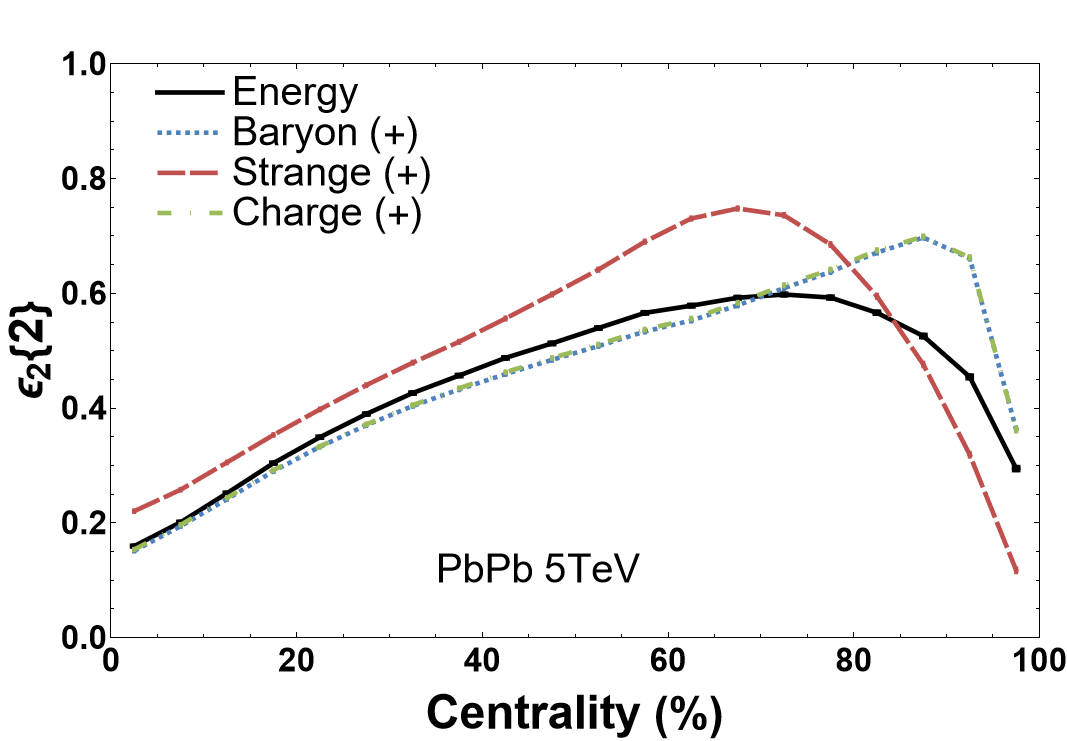} 
    	\\
    	\vspace{.5cm}
    	\includegraphics[width=0.45 \textwidth]{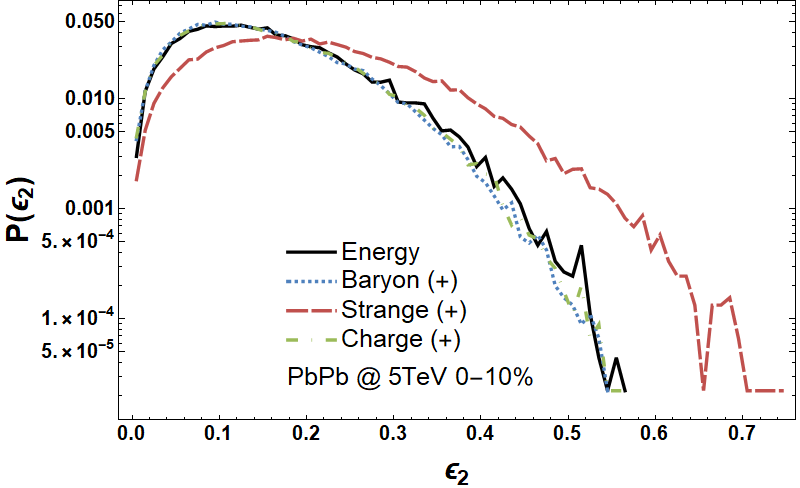} \,
    	\includegraphics[width=0.45 \textwidth]{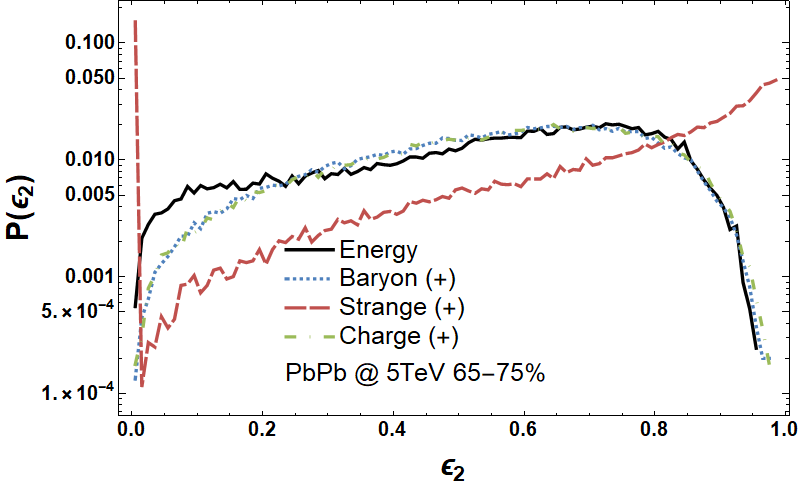}
    	\end{centering}
    	\caption{
    	\textbf{Top:}  RMS ellipticity $\varepsilon_2 \{2\}$ versus centrality.
    	\textbf{Bottom Left:} Distribution of ellipticity in the $0-10\%$ bin.
    	\textbf{Bottom Right:} Distribution of ellipticity in the $65-75\%$ bin. Figure updated from Ref.~\cite{Carzon:2019qja}.
    	}
    	\label{f:IBSQCentrality}
    \end{figure}

Having established the effect of \code{iccing} on the initial condition input, we can look at the geometries of charge produced by the algorithm. The elliptic eccentricities of energy, baryon, strangeness, and electric charge are shown in Fig.~\ref{f:IBSQCentrality}. In the top plot of Fig.~\ref{f:IBSQCentrality} is $\varepsilon_2 \{2\}$, where, for $ < 65\%$ centrality, the baryon number and electric charge distributions track the energy distribution very closely, while the strangeness distribution is much more eccentric.  To gain a better intuition of this behaviour, we examine the $0-10\%$ centrality bin differentially by plotting the corresponding probability distribution in the second plot of Fig.~\ref{f:IBSQCentrality}. On an event-by-event basis, the $B, Q$ distributions produce the same geometry profile as the energy density; we can attribute this to the huge abundances of $u, d$ quarks produced in these central events. These nearly-massless quarks are produced in large quantities and are nearly homogeneous in central collisions, so that the resulting charge densities mirror the original energy density profile, quantitatively.  In contrast, the strangeness distribution is more eccentric than the bulk geometry, even in the most central collisions.  This indicates that the strange quarks do not saturate the collision, instead generating a lumpy and highly anisotropic geometry even when the energy is quite round.

%__________________________________________________________________________
    %
    %
%__________________________________________________________________________

Another area that would be important to analyze differentially are peripheral collisions at $65-75\%$ centrality, where we see the $B, Q$ ellipticities begin to deviate from the bulk energy density and the strangeness experiences its peak, as shown in the bottom right third panel of Fig.~\ref{f:IBSQCentrality}. Looking at the $B, Q$ distributions, we see that they qualitatively track the same geometry as the energy density, reflecting the still-dominant role of impact-parameter-driven geometry, but nontrivial differences start to appear. Events start to shift to larger $B, Q$ eccentricities when compared to the energy distributions that produced them, indicating the $u, d$ quark abundances have dropped low enough to no longer saturate the bulk geometry and are now characterizing a different, lumpier distribution. The strangeness distribution for this centrality window has a dramatic increase in events with strangeness ellipticity close to $1$ and a large spike about zero. These peaks in the distribution, at $0$ and $1$, most likely reflect events with few strange quarks, where exactly one or two $s \bar{s}$ pairs may be created, which leads to one or two blobs of positive strangeness that are innately round (zero eccentricity) or innately elliptical (maximal eccentricity). This double-peak behavior in $\epsilon_2^{(S+)}$ correlates with the location of the peak in the cumulant $\varepsilon_2^{(S+)} \{2\}$, again indicative that there is a transition to event geometries controlled by a small number of produced strange quarks.

%__________________________________________________________________________
    %
    \begin{figure}[h!]
        \centering
    	\includegraphics[width=0.45 \textwidth]{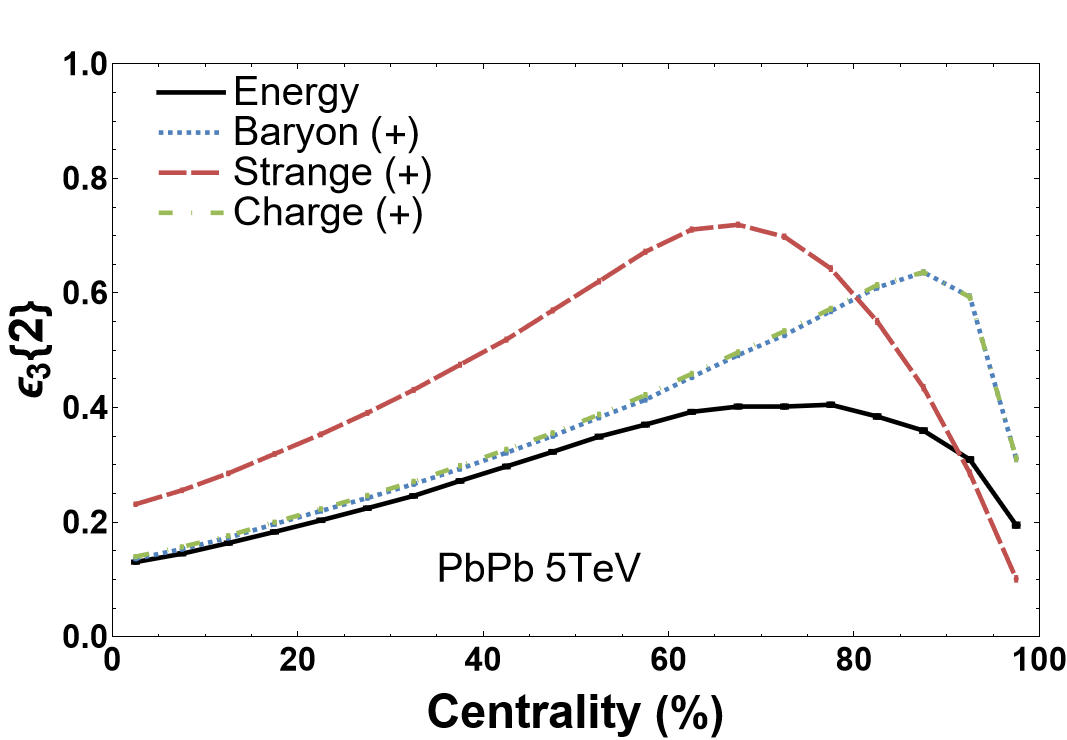}
    	
    	\caption{
         RMS triangularity $\varepsilon_3 \{2\}$ versus centrality. Figure updated from Ref.~\cite{Carzon:2019qja}.}
    	\label{f:E3_BSQCentrality}
    \end{figure}
    %
%__________________________________________________________________________

The triangularity, $\varepsilon_3 \{2\}$, of energy and $BSQ$ densities is shown in Fig.~\ref{f:E3_BSQCentrality}. Unlike $\varepsilon_2 \{2\}$ which is dominated by the mean-field elliptical geometry below $\lesssim 60\%$ centrality, $\varepsilon_3 \{2\}$ arises entirely from fluctuations, without this mean-field background.  Up to very peripheral collisions $> 80\%$ centrality, there is a clear hierarchy in the densities: $\varepsilon_3^{(S^+)} \{2\}  > \varepsilon_3^{(B^+ \, , \, Q^+)} \{2\} > \varepsilon_3^{(E)} \{2\}$. This is consistent with the introduction of new sources of sub-nucleonic fluctuations contributing to the $BSQ$ distributions, since strange quarks are produced the least and, therefore, their geometry fluctuates the most. Triangularity, $\varepsilon_3$, is the most sensitive probe into the differences between the various charge and energy geometries due to its direct sensitivity to the fluctuating charge distributions, independent of the mean-field background.

%__________________________________________________________________________
    %
    \begin{figure}[h!]
        \centering
    	\includegraphics[width=0.45 \textwidth]{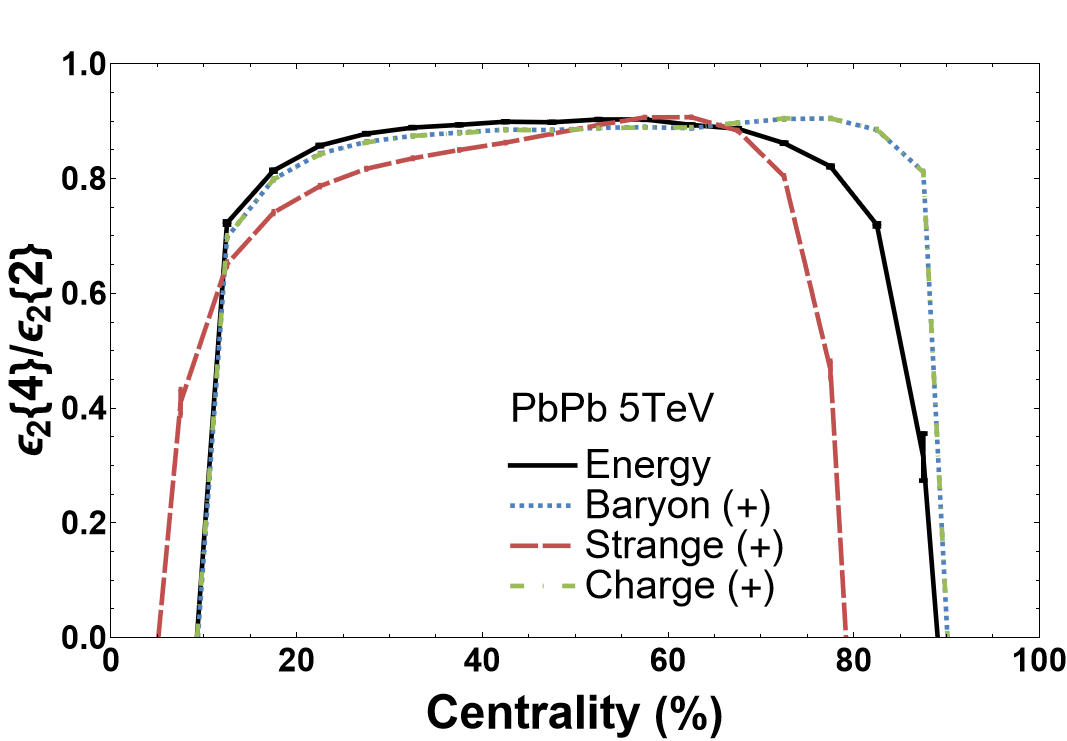}
    	\\
    	\vspace{.5cm}
    	\includegraphics[width=0.45 \textwidth]{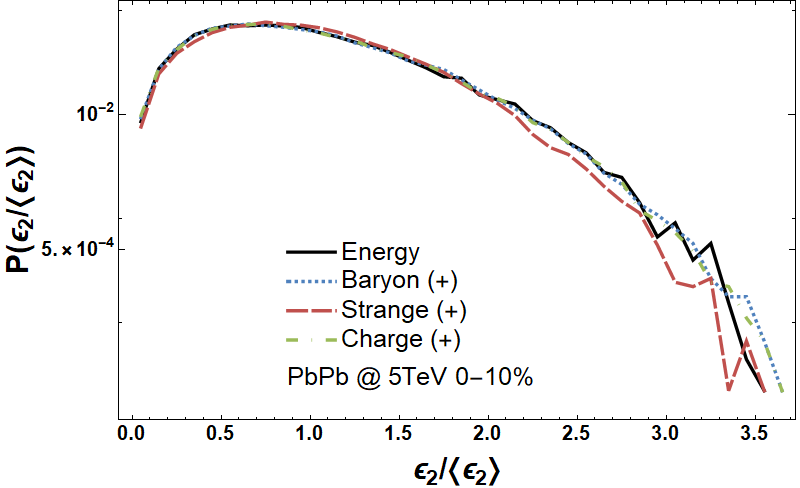}
    	\includegraphics[width=0.45 \textwidth]{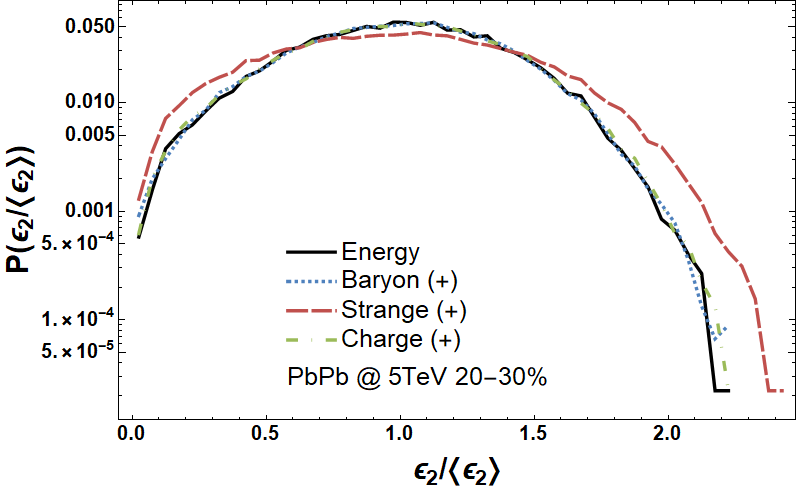}
    	\caption{
    	\textbf{Top:} Cumulant ratio $\varepsilon_2 \{4\} / \varepsilon_2 \{2\}$ versus centrality. 
    	\textbf{Bottom Left:} Probability distribution of the ellipticity $\varepsilon_2$ in the $0-10\%$ bin, scaled by the mean $\langle \varepsilon_2 \rangle$ to illustrate the width of the distribution.  Note that the strangeness distribution is \textit{narrower} than the bulk.
    	\textbf{Bottom Right:} Probability distribution of the ellipticity $\varepsilon_2$ in the $20-30\%$ bin, scaled by the mean to illustrate the width.  Note that the strangeness distribution is \textit{broader} than the bulk. Figure updated from Ref.~\cite{Carzon:2019qja}.
    	}
    	\label{f:IBSQ42Centrality100K}
    \end{figure}
    %
%__________________________________________________________________________

Looking at the cumulant ratio $\varepsilon_n \{4\} / \varepsilon_n \{2\}$ for the various charge and energy distributions, shown in Fig.~\ref{f:IBSQ42Centrality100K}, can tell us about the variance of the $\varepsilon_n$ distribution which is measured by the deviation of the ratio from unity. We can analyze this differentially by plotting the $\varepsilon_n$ probability distribution normalized by the mean to better reflect differences in width of the distributions corresponding to the deviation of the cumulant ratio from unity. For $< 70\%$ centrality in the first panel of Fig.~\ref{f:IBSQ42Centrality100K}, we see the baryon/electric charge distributions track the energy density profile quite closely, while strangeness is significantly different. There is an interesting crossing of curves at around $10\%$ centrality, where the strangeness distribution changes from having a ratio smaller than the bulk, in mid-central collisions, to a value larger than the bulk, in central collisions. We can investigate this more by plotting the distributions normalized by the mean for the $0-10\%$ centrality bin (bottom left of Fig.~\ref{f:IBSQ42Centrality100K}) and the $20-30\%$ range (bottom right of Fig.~\ref{f:IBSQ42Centrality100K}). For $20-30\%$, we see that the strangeness distribution is broader than the bulk, which reflects the increased number of event-by-event fluctuations due to a small number of strange quarks being produced. Event-by-event fluctuations of energy density are comparatively narrower and more peaked around their mean value, dictated by the strong mean-field background of the bulk elliptical geometry. In $0-10\%$ centrality, the strange quark distribution is narrower than the bulk, which reflects a qualitative change of the strangeness distribution relative to the bulk. Looking back at the $\varepsilon_2$ cumulant ratio, fluctuations of the energy density increase greatly when going to central collisions, which may be understood as the disappearance of mean-field elliptical geometry when the impact parameter goes toward zero. Event-by-event fluctuations, compared against the vanishing background, are greatly magnified which results in a broadening of the width of the $\varepsilon_2$ histogram. However, the strangeness distribution is broadened significantly less by this effect, resulting in a reduced slope for the strange $\varepsilon_2$ cumulant ratio and a crossing of the curves at $10\%$ centrality.  

The qualitative change in the strangeness distribution relative to the energy distribution as a function of centrality indicates, that in addition to a dependence on the number of quark pairs produced, the underlying geometries from which quarks are produced are different and act differently with respect to centrality. If the only dependence of the BSQ charge geometries was the number of particles produced, then we would expect to see the same hierarchy in $\varepsilon_2 \{4\} / \varepsilon_2 \{2\}$, that is, $E \approx B^+ \approx Q^+ > S^+$ maintained across mid-central and central collisions. The fact that the disappearance of the bulk elliptical geometry in central collisions affects the strangeness distribution differently from the others is a reflection of the geometry capable of pair producing strange quarks not being identical to the bulk geometry. It is reasonable to assume that since the strange quarks have a large mass threshold of $2 m_s \approx 200 \, MeV$, they may only be produced by hot spots in the collision that contain enough energy density to meet this mass threshold. This can explain why the cumulant ratio of strangeness responds in a less singular manner than the bulk in central collisions: while the geometry of the bulk is becoming very round in central collisions, the geometry of the hot spots capable of pair produced strange quarks is innately lumpier. Thus, the width of the strangeness distribution is less sensitive to the absence and the presents of a large elliptical bulk geometry.  

%__________________________________________________________________________
%
\begin{figure}[h!]
    \centering
    \includegraphics[width=0.45 \textwidth]{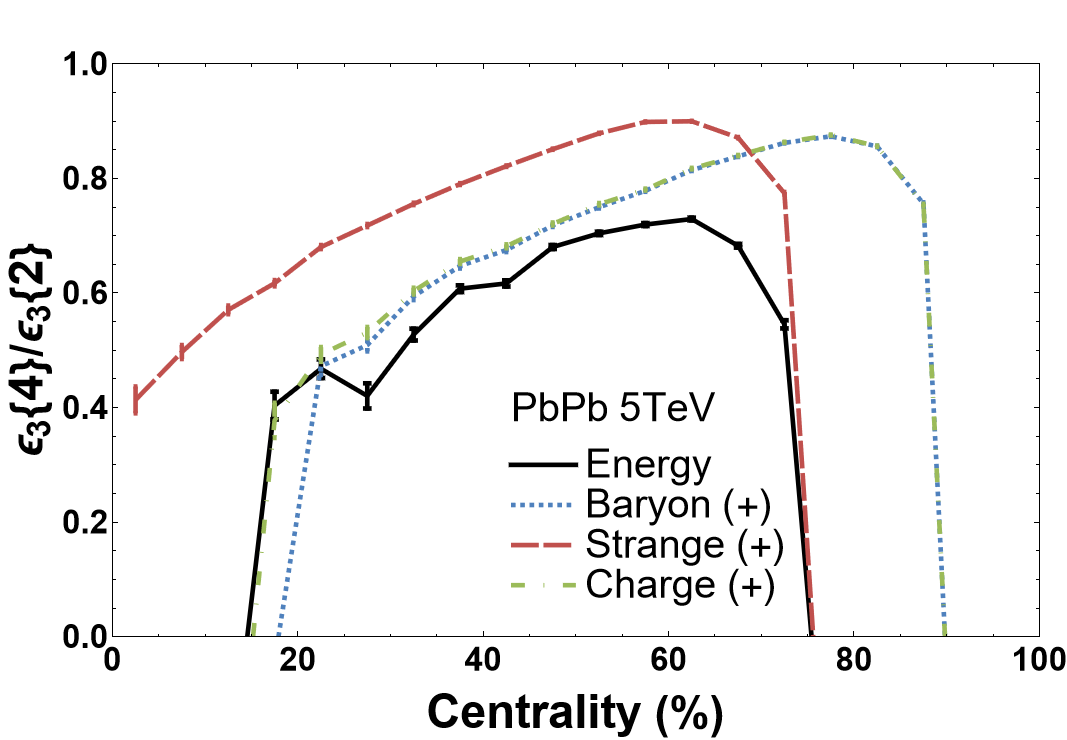}
    \includegraphics[width=0.45 \textwidth]{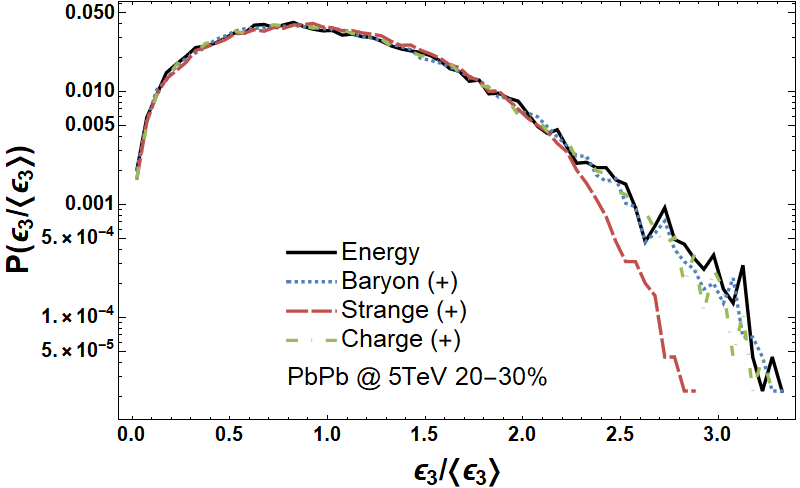}
   
    \caption{
        Cumulant ratio $\varepsilon_3 \{4\} / \varepsilon_3 \{2\}$ versus centrality (left) and the probability distribution for $\varepsilon_3$ for the centrality class $20-30\%$ (right). Figure updated from Ref.~\cite{Carzon:2019qja}.
    }
    \label{f:E3_IBSQ42Centrality100K}
\end{figure}
%
%__________________________________________________________________________

Looking at the cumulant ratio for triangularity, $\varepsilon_3 \{4\} / \varepsilon_3 \{2\}$ shown in the first panel of Fig.~\ref{f:E3_IBSQ42Centrality100K}, the absence of a mean-field background geometry produces an ordering hierarchy comparable to the one seen in the $0-10\%$ bin for $\varepsilon_2$: $S^+ > B^+ \approx Q^+ \approx E$. In central collisions, the strong elliptical shape of the event geometry, $\varepsilon_2$, disappears and the event-by-event fluctuations produce a large fractional change in the ellipticity, whose expectation value is close to zero. In a similar way, $\varepsilon_3$ has an expectation value close to zero, leading to large fractional changes compared to that small value. This is true for the gluons and light quarks which saturate the bulk geometry, but this is not the case for the strange quarks. Since fewer strange quarks are produced, they have an inherently lumpier hot spot distribution whose triangularity does not average automatically to zero. Thus both for $\varepsilon_2$ in the central bin of Fig.~\ref{f:IBSQ42Centrality100K} and for $\varepsilon_3$ across all centralities, the relative fluctuations of strange particles are smaller than the light quarks. We can further illustrate this through the probability distribution of $\varepsilon_3$ in the centrality window of $20-30\%$ in the second panel of Fig.~\ref{f:E3_IBSQ42Centrality100K} wherein strangeness, indeed, has a narrower distribution.  We argue that, in full context, these descriptors of the event-by-event fluctuations of the strangeness distribution and its centrality dependence provide systematic evidence for the coupling of strangeness to an independent underlying event geometry.

%---------------------------------------------------------------------------
%
\section{Sensitivity Analysis to Model Parameters}
\label{Sec:SensitivityAnalysis}
%
%---------------------------------------------------------------------------

An important aspect of introducing a new model is looking at the impact of all parameters on said model. The standard way to do this is through Bayesian analysis as is the case with \code{trento} \cite{Bernhard:2016tnd}. This method is not yet possible with \code{iccing} since an important aspect of applying Bayesian analysis would be the final state observables from experimental measurements. It is unclear which experimental observables would best quantify the charge geometry and be compatible with the information generated by \code{iccing}. In lieu of full Bayesian analysis, here I present a review of the sensitivity of the \code{iccing} model to it's various parameters. (If unspecified, all parameters are set to the defaults found in Sec.~\ref{sec:InputInitialization}). Most of the free parameters of \code{iccing} specifically influence the location and frequency of splittings with only the profile with which the charge perturbation is distributed and its radius having no direct effect on these aspects of the algorithm.

\begin{figure}[h!]
    \centering
    \includegraphics[width=0.8\textwidth]{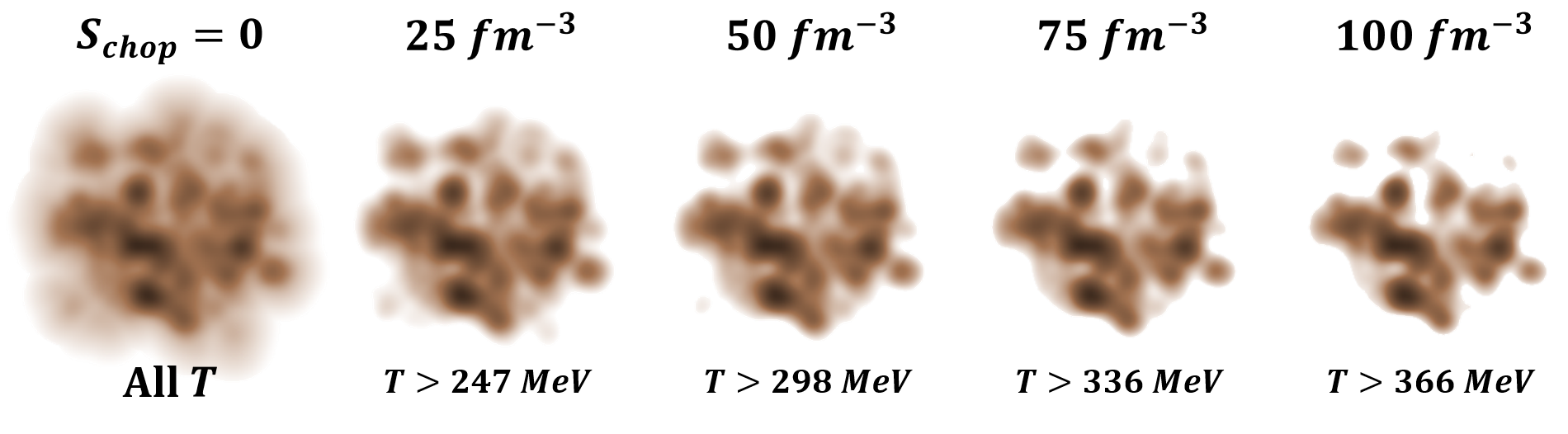}
    \\
    \includegraphics[width=0.45 \textwidth]{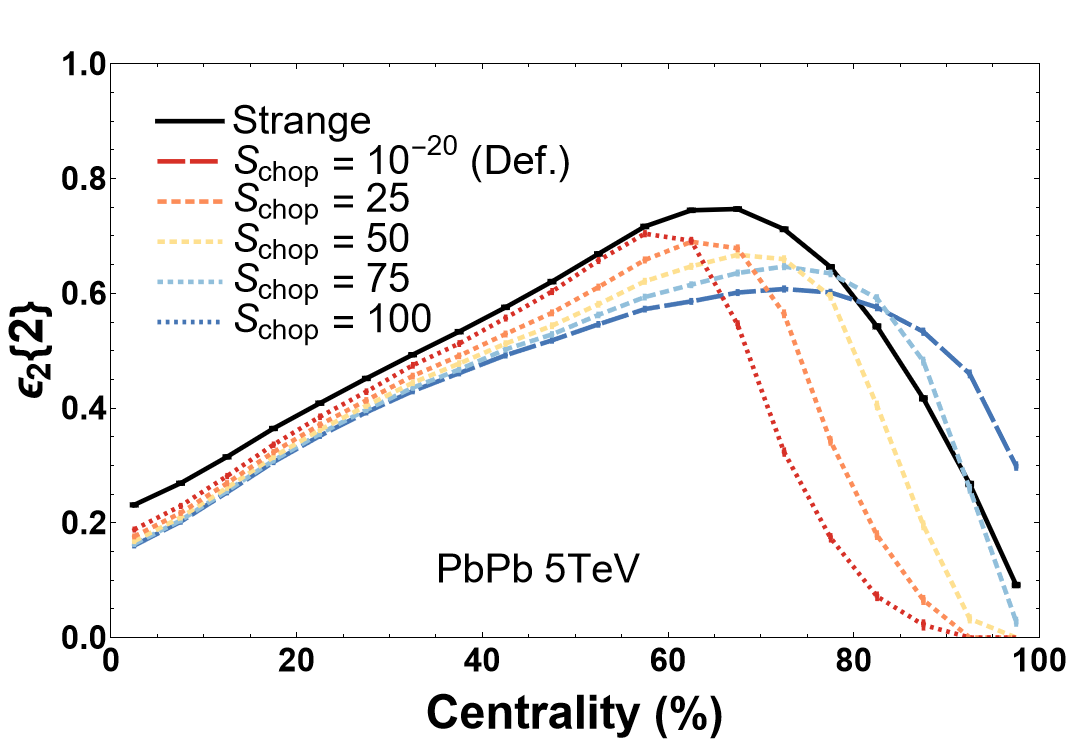} \,
    \includegraphics[width=0.45 \textwidth]{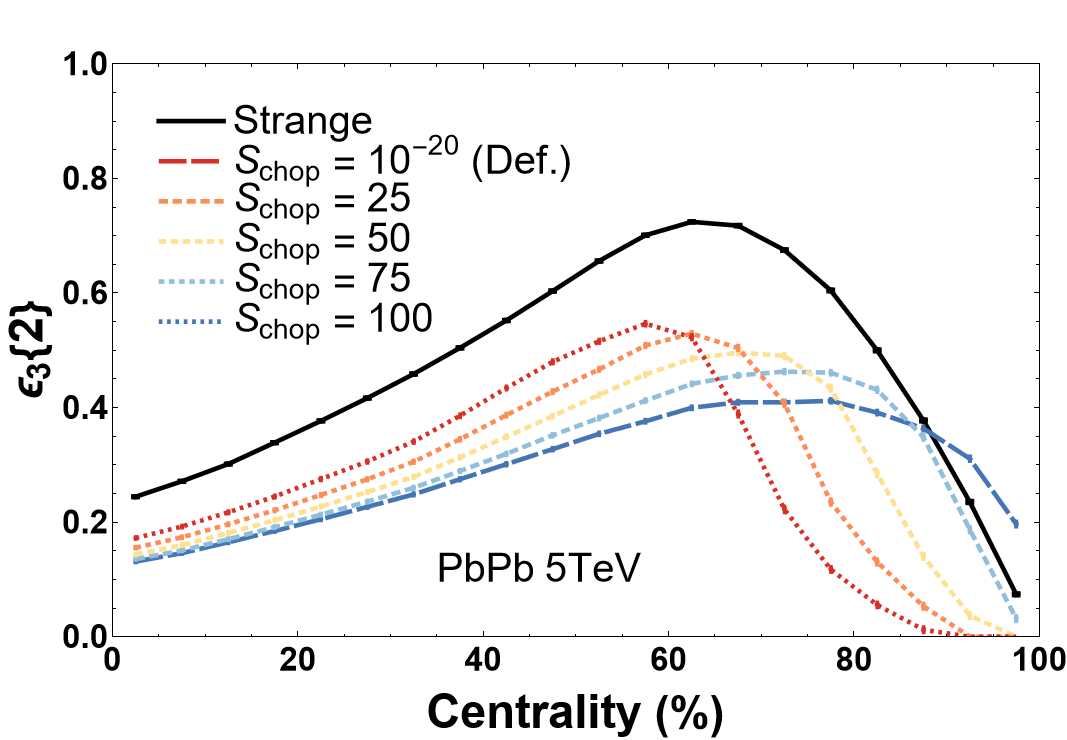}
    
    \caption{Comparison of the $S^+$ eccentricities (black) with the eccentricities of the energy distribution after applying a grooming cut $S_{\mathrm{chop}}$ between $0$ and $100 \, \mathrm{fm}^{-3}$ to the entropy density using the equation of state \cite{Alba:2017hhe}.  Top:  Increasing $S_\mathrm{chop}$ grooms the bulk geometry to select on the hot spots which dominate strangeness production.  Bottom: The geometry of the energy distribution after hot spot grooming converges toward the geometry responsible for the strangeness production. Figure updated from Ref.~\cite{Carzon:2019qja}.
    }
    \label{f:Echop}
\end{figure}

An important parameter that can further elucidate the idea of strangeness coupling to hot spot geometry is the grooming parameter $S_\mathrm{chop}$ described in Sec.~\ref{subsec:ConvertInitialCondition}. I compare the ellipticity and triangularity of the strangeness (black curve) against the initial energy geometry at different values of this grooming parameter in Fig.~\ref{f:Echop}. The effect of $S_\mathrm{chop}$ is to throw out any points with initial entropy density below the specified cutoff, which allows me to impose tighter and tighter cuts on hot spot geometry present in the initial condition which \code{iccing} samples. As the cutoff is increased, selecting on hot parts of the initial condition, the energy geometry approaches the strangeness distribution produced by \code{iccing}. Looking at the qualitative features, it appears that selecting on hot spots, in the energy density with $S_\mathrm{chop}$, increases the magnitude of eccentricities below the peak and shifts the location of the peak to the left, which is mirrored in the strangeness distribution when compared to the unmodified bulk energy density. This is by no means conclusive evidence for the connection between strangeness and hot spot geometry by itself, but in its consistency with the results presented in Sec.~\ref{Sec:StrangenessAsProbe} warrants further investigation to pin down the exact correlation.

%__________________________________________________________________________
%
%
%__________________________________________________________________________

%
%__________________________________________________________________________
%
\begin{figure}[h!]
    \centering
	\includegraphics[width=0.45 \textwidth]{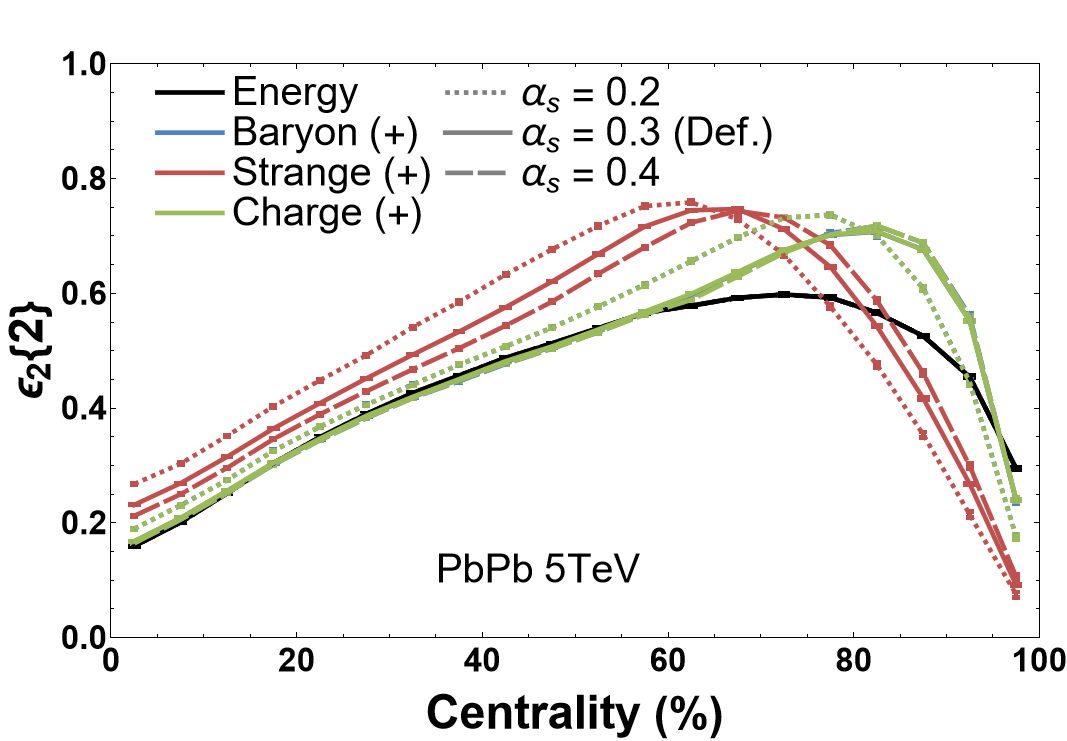} \,
	\includegraphics[width=0.45 \textwidth]{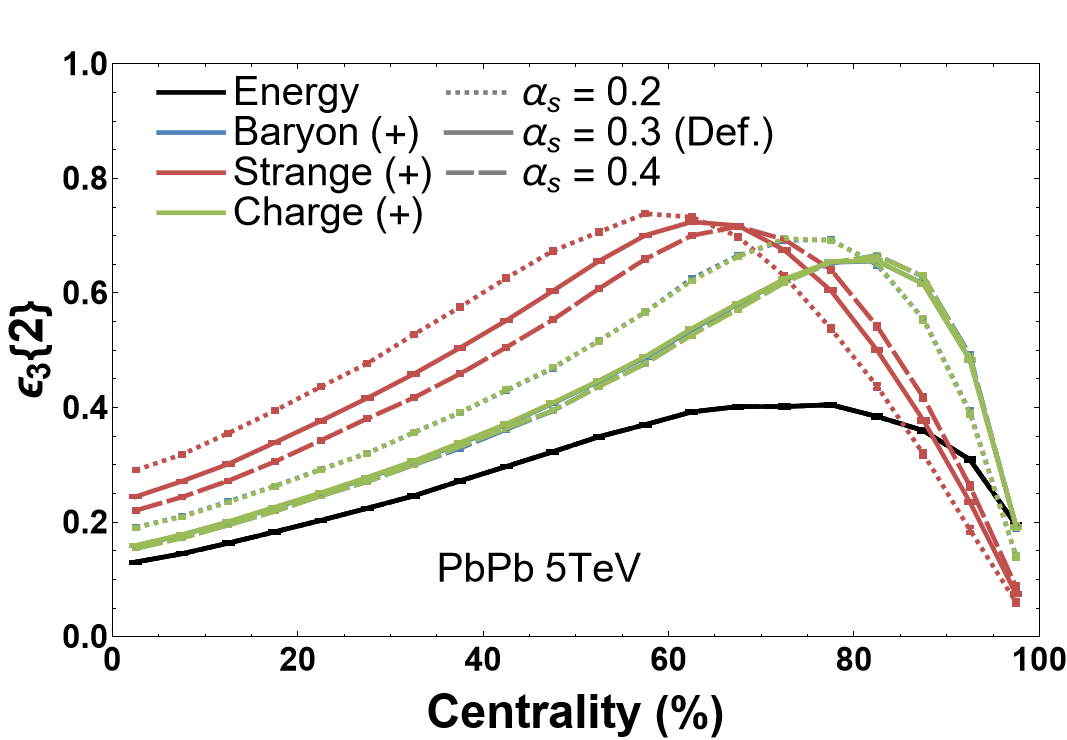}
	
	\caption{Comparison of the second cumulants $\varepsilon_n \{2\}$ for different values of the strong coupling $\alpha_s$. Figure updated from Ref.~\cite{Carzon:2019qja}.}
	\label{f:IBSQCentralityAlphaS}
\end{figure}
%
%__________________________________________________________________________
%

The flavor chemistry of the initial state is most effected by the QCD coupling constant $\alpha_s$, which controls the magnitude of $q \barq$ pair production rates while having no effect on the ratios. For different values of $\alpha_s$, the ellipticity $\varepsilon_2\{2\}$ and triangularity $\varepsilon_3\{2\}$ (left and right, respectively) of the BSQ charge distributions and the bulk energy density as a function of the centrality is shown in Fig.~\ref{f:IBSQCentralityAlphaS}.  In both observables, a significant dependence on the value of $\alpha_s$ is seen, which flattens the centrality dependence and shifts the peak left as $\alpha_s$ is increased.  These effects come from the increased probability to produce quarks regardless of flavor, which results in $BSQ$ distributions that are consistently less eccentric and closer to saturating the bulk geometry. The peak indicates a transition from a smooth geometry to a granular one and the rightward shift with larger $\alpha_s$ shows that the  BSQ distributions become granular at higher centralities. Since triangularity is driven by fluctuations, $\alpha_s$ has a larger effect on $\varepsilon_3 \{2\}$ than $\varepsilon_2 \{2\}$ since it directly effects how many BSQ fluctuations are created. Triangularity converges toward the bulk geometry faster, and it exhibits the same rightward shift of the peak as ellipticity.

%__________________________________________________________________________
%
\begin{figure}[h!]
    \centering
	\includegraphics[width=0.45 \textwidth]{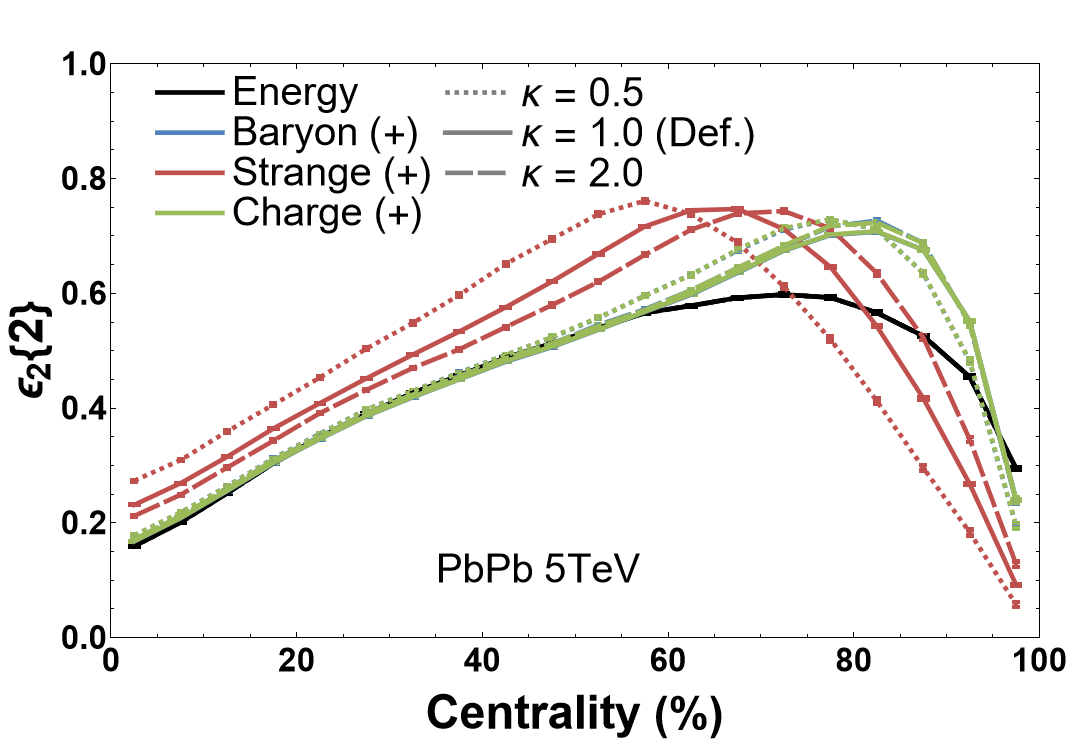} \,
	\includegraphics[width=0.45 \textwidth]{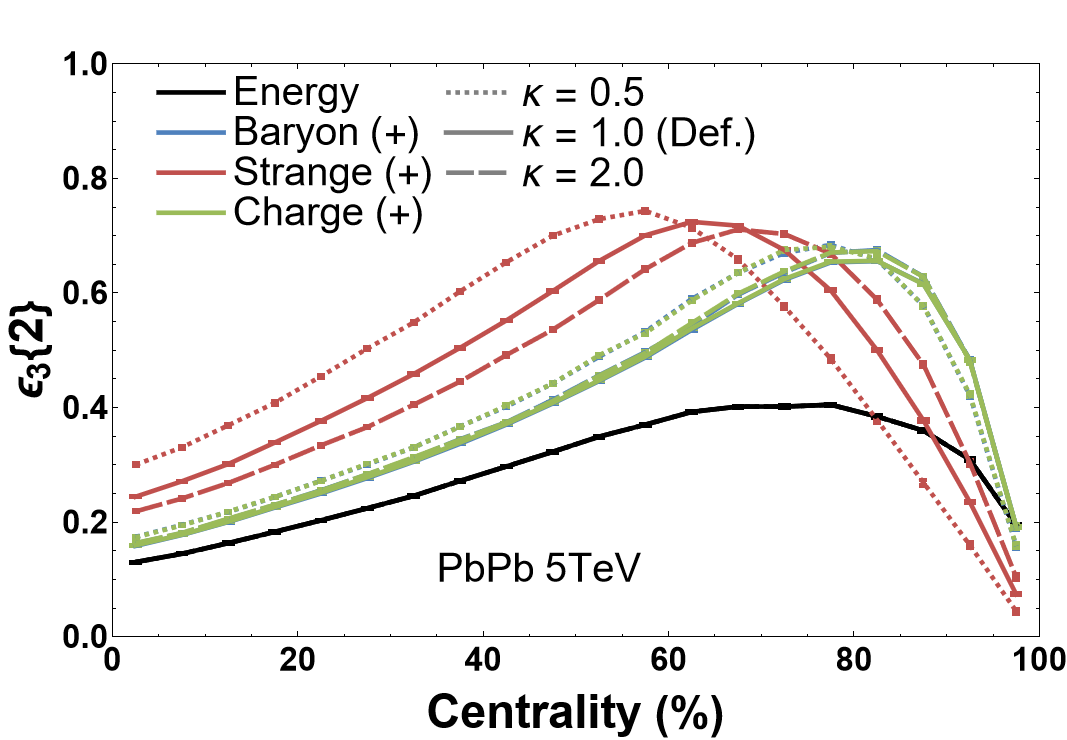}
	
	\caption{ Comparison of the second cumulants $\varepsilon_n \{2\}$ for different normalizations $\kappa$ of the saturation scale $Q_s$. Figure updated from Ref.~\cite{Carzon:2019qja}.}
	\label{f:IBSQCentralityQs}
\end{figure}
%
%__________________________________________________________________________

The saturation scale, $Q_s$, is controlled by the parameter $\kappa$ (see Eq.~\ref{e:kappadef}) and has little impact on the spatial correlations, which determine the momentum fraction and seperation of the quark/anti-quark pair, shown in Fig.~\ref{f:probabilities}, while having a significant affect on the quark chemistry, as seen in Fig.~\ref{f:multratio}. By varying $\kappa$ up and down by a factor of $2$ from its default value effectively shifts quark splitting probabilties from Fig.~\ref{f:multratio} toward the right or left, respectively. The response of $\varepsilon_2\{2\}$ and $\varepsilon_3\{2\}$ to this change, shown in Fig.~\ref{f:IBSQCentralityQs}, is similar to the dependence on $\alpha_s$ seen in Fig.~\ref{f:IBSQCentralityAlphaS}. This indicates that the effect of varying $\kappa$ comes primarily from its impact on the total quark multiplicities for the different flavors. While the overall behavior is the same as $\alpha_s$, it is interesting that $\kappa$ impacts the strangeness eccentricities more than the BQ geometries. This difference in effect is because the probability, with respect to $Q_s$, to create strange quark pairs has a different behavior as compared to up and down quark pairs. This is clear in Fig.~\ref{f:multratio}, where increasing the value of a gluon's $Q_s$ by a factor of 2 leads to a generally equal increase in the total probabilities to produce all quark flavors. However, the slope with which the probability increases corresponds to a greater \textit{percentage} increase in the abundance of strange quarks (~$+ 60\%$) versus up and down (~$+ 15\%$). This is due to the fact that the chemistry of the model we use here exhibits (exact or approximate) geometric scaling, depending only on the ratio $Q_s / m$ so that a change in $Q_s$ has a different effect on quarks of different mass \cite{Carzon:2019qja}. Thus, the strangeness geometry gets significantly closer to the bulk with an increase of $\kappa$, while the effect on light flavors is much smaller.

%__________________________________________________________________________
%
\begin{figure}[h!]
    \centering
	\includegraphics[width=0.45 \textwidth]{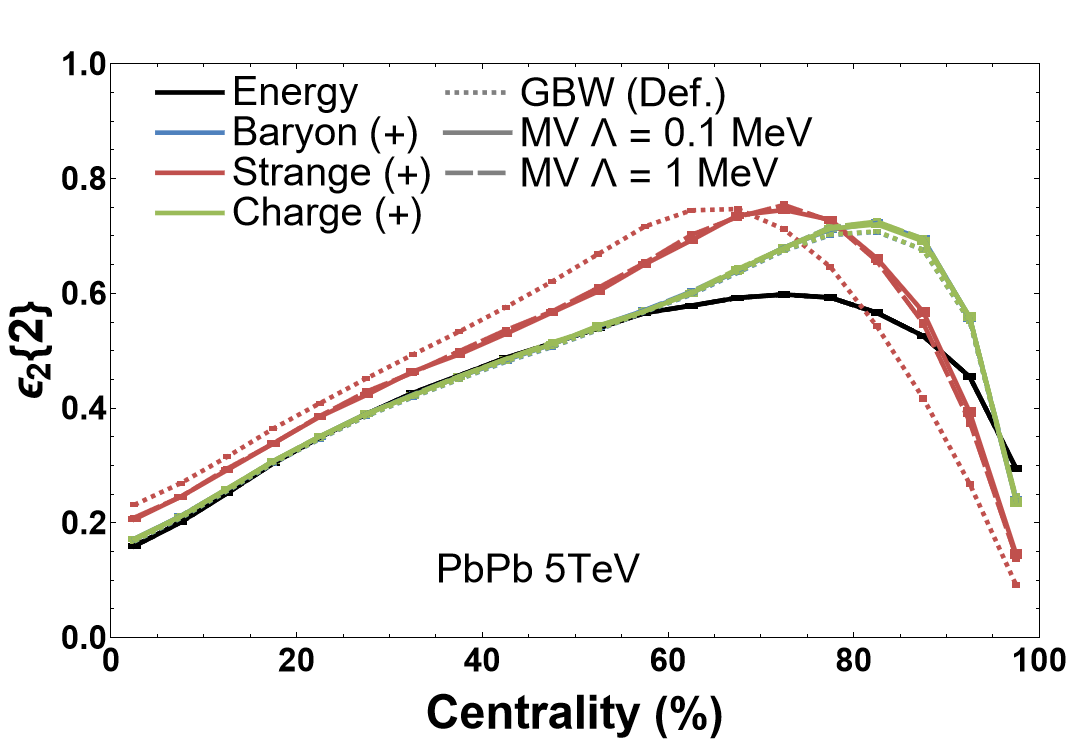} \,
	\includegraphics[width=0.45 \textwidth]{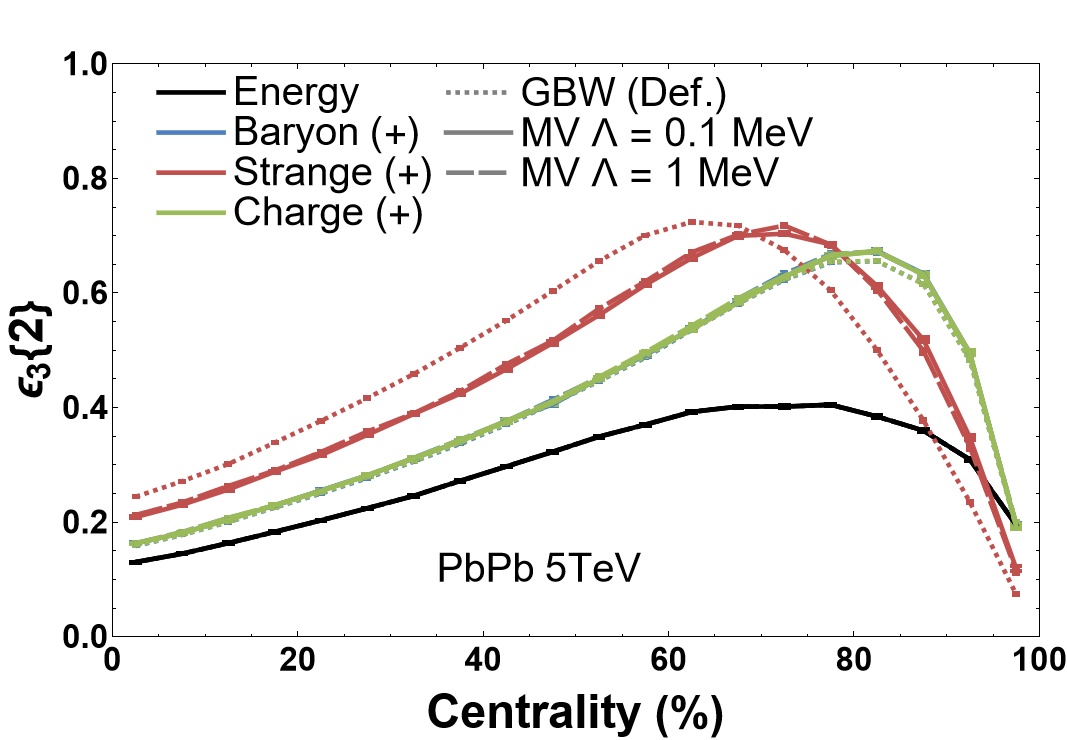}
	
	\caption{ Comparison of the second cumulants $\varepsilon_n \{2\}$ between the GBW and MV models, as well as for two different choices of the IR cutoff $\Lambda$ in the MV model. Figure updated from Ref.~\cite{Carzon:2019qja}.}
	\label{f:IBSQCentralityDipole}
\end{figure}
%
%__________________________________________________________________________
 
The model, with which the splitting kinematics and flavor chemistry are sampled from, is a free choice and here I compare the GBW to the MV model (Eqs.~\ref{e:multratio1_first} and Eq.~\ref{e:multratio2_first}, respectively). The difference between models is illustrated in Figs.~\ref{f:probabilities} and \ref{f:multratio}, where the MV model has a higher probability to produce quarks at short distances, which leads to a higher probability of producing quarks overall. The effect of these models on the eccentricities is seen in Fig.~\ref{f:IBSQCentralityDipole}: changing from the GBW to MV model leads to a flattening of the centrality dependence and a rightward shift of the eccentricity peaks. This is similar to the dependence on $\kappa$ from Fig.~\ref{f:IBSQCentralityQs}, where a roughly constant probability increase, in absolute terms, translates into a greater percentage increase for strange quark/anti-quark pairs, which leads to a larger shift in the strangeness distribution. For the MV model, the IR cutoff can be varied and we see in Fig.~\ref{f:IBSQCentralityDipole} that this choice has a negligible effect on the eccentricities. 

%__________________________________________________________________________
%
\begin{figure}[h!]
    \centering
	\includegraphics[width=0.45 \textwidth]{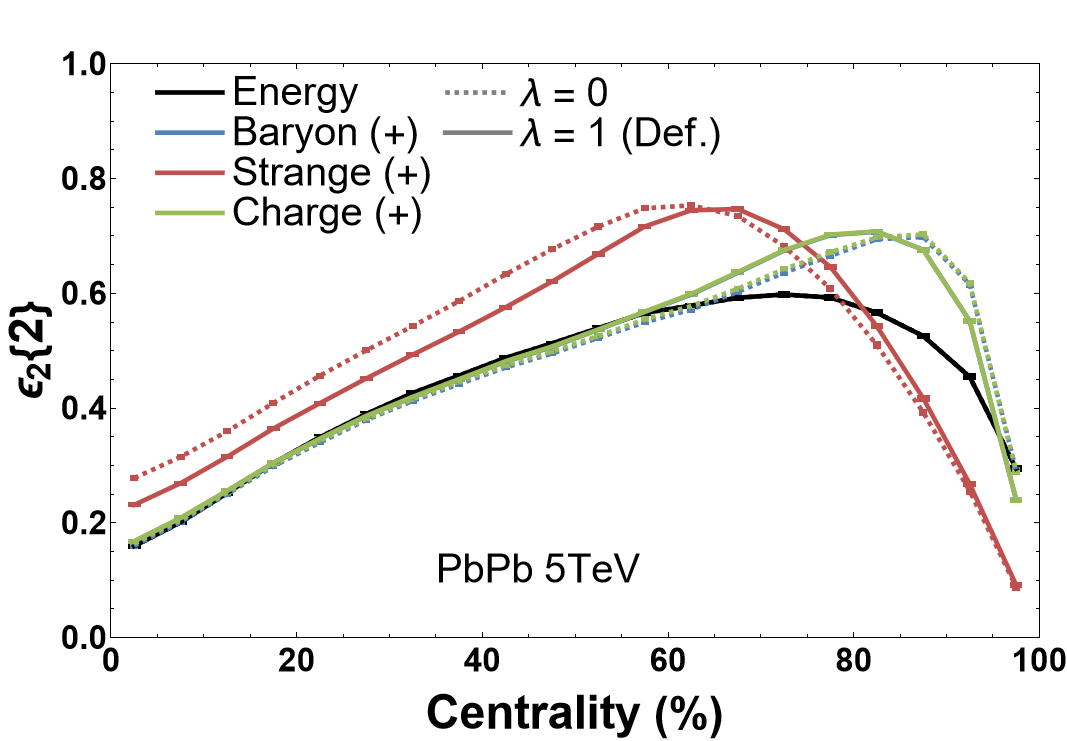} \,
	\includegraphics[width=0.45 \textwidth]{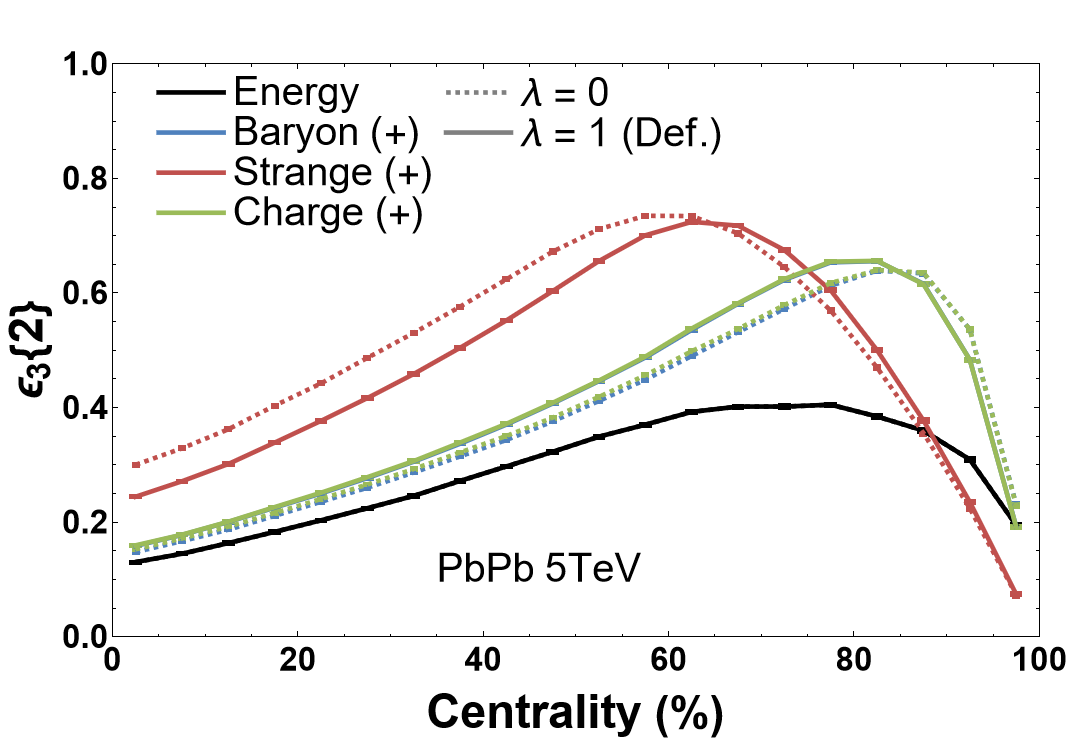}
	
	\caption{Comparison of the second cumulants $\varepsilon_n \{2\}$ for different distributions of gluon energy controlled by the parameter $\lambda$. Figure updated from Ref.~\cite{Carzon:2019qja}.}
	\label{f:IBSQCentralityLambda}
\end{figure}
%
%__________________________________________________________________________

The exponent that controls the energy dependence of the gluon spectrum (Eq.~\ref{e:lambdaparam}), $\lambda$, is expected to have a significant impact on the initial-state chemistry. Some exploration of the effect $\lambda$ was shown in Fig.~\ref{f:lambdaEffect}, where the boost-invariant spectrum, $\lambda = 1$, is compared with the constant spectrum, $\lambda = 0$, and leads to a significant \textit{decrease} in overall quark production since more hard gluons are sampled and thus fewer opportunities are available to create quark/anti-quark pairs. This change in quark multiplicities is reflected by a corresponding change in the BSQ eccentricities as evidenced in Fig.~\ref{f:IBSQCentralityLambda}. Moving from $\lambda = 1$ to $\lambda = 0$ decreases quark multiplicities and results in an increase in the strangeness eccentricity with its peak shifting to the left, as expected. On the other hand, a shift in the opposite direction is observed in the baryon number and electric charge eccentricities, despite all quark multiplicities decreasing when going to $\lambda = 0$, as seen previously in Fig.~\ref{f:lambdaEffect}.

%__________________________________________________________________________
%    
\begin{figure}[h!]
    \centering
	\includegraphics[width=0.45 \textwidth]{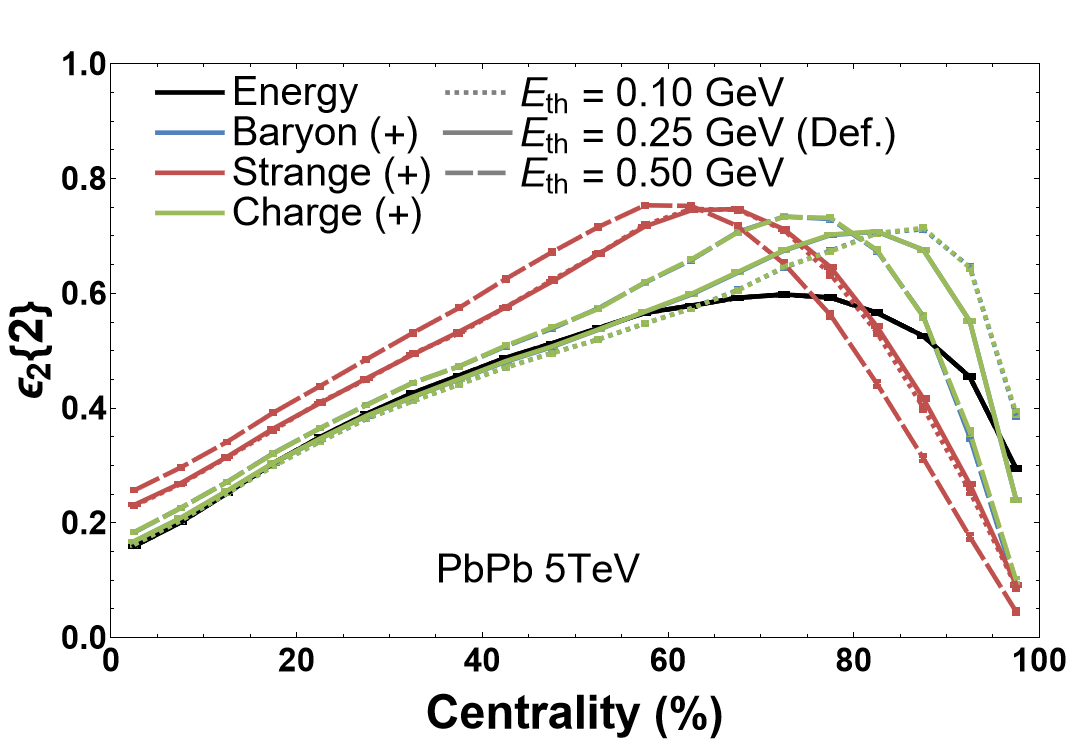} \,
	\includegraphics[width=0.45 \textwidth]{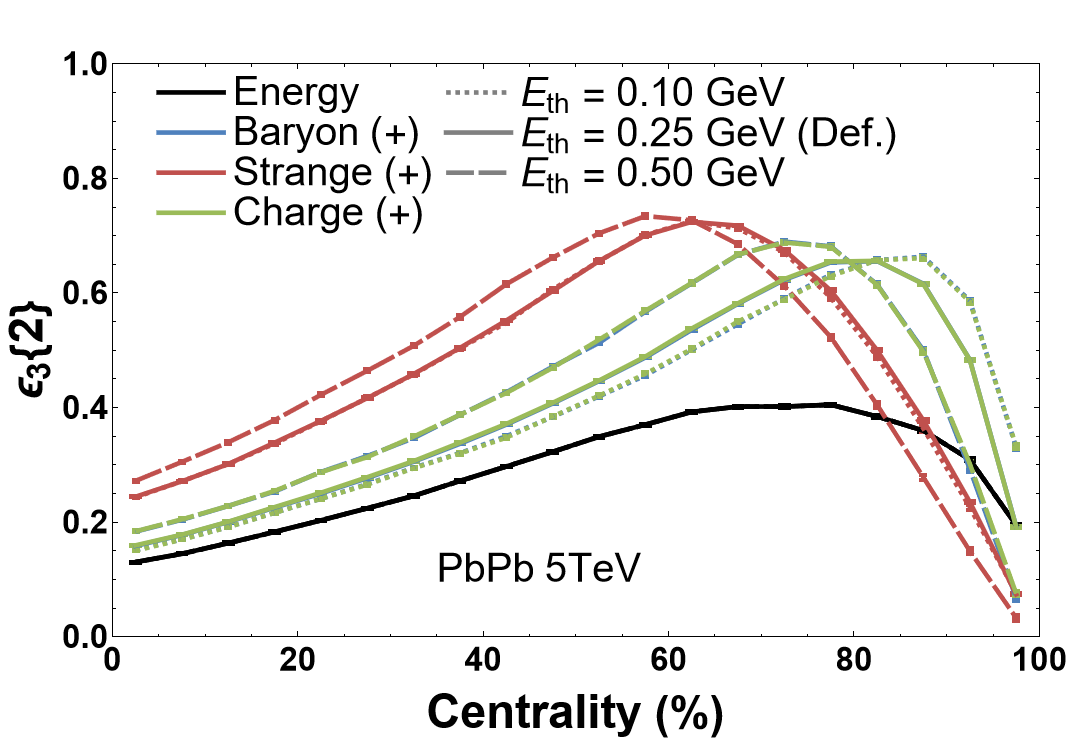}
	
	\caption{  Comparison of the second cumulants $\varepsilon_n \{2\}$ for different values of the threshold energy $E_{\mathrm{thresh}}$. Figure updated from Ref.~\cite{Carzon:2019qja}.}
	\label{f:IBSQCentralityEth}
\end{figure}
%
%__________________________________________________________________________

The final parameter that has a noticeable impact on the charge geometry is the threshold energy $E_\mathrm{thresh}$, which determines when a gluon is eligible to undergo $g \rightarrow q \bar{q}$ splitting. This threshold energy has a dual use in the algorithm, also acting as the termination condition: when the total energy available to be sampled is less than $E_\mathrm{thresh}$, then the algorithm exits and transfers any leftover energy to the output. The value of $E_\mathrm{thresh}$ can have a significant impact on the overall runtime of \code{iccing}, since it reduces extraneous sampling. Thus, $E_\mathrm{thresh}$ should be as large as possible without interfering with the quark/anti-quark production. In Fig.~\ref{f:IBSQCentralityEth}, we can observe that for too large of a choice for $E_\mathrm{thresh}$ splittings can be artificially suppressed, which has an impact on the initial geometry. We can look at the strange density geometry for a clear example of a problematic value of $E_\mathrm{thresh}$. For $E_\mathrm{thresh} < 2 m_s \approx 200 \, \mathrm{MeV}$, there is no effect on the geometry of the strangeness density. However, once $E_\mathrm{thresh} = 500 \, \mathrm{MeV}$, we see there is a significant effect on the strange geometry that is consistent with the suppression of strange quark/anti-quark pairs (the peak of the distribution shifts toward the left). Similar behavior is seen for the $B, Q$ eccentricities, with even a $E_\mathrm{thresh} = 100 \, \mathrm{MeV}$ showing suppression of light quarks. This suggests that the default value $E_\mathrm{thresh} = 250 \, \mathrm{MeV}$ is too large, suppressing light quark/anti-quark production, and needs to be set lower than the threshold corresponding to twice the mass of the up quark, $E_\mathrm{thresh} = 2 m_u \approx 5 \, \mathrm{MeV}$.

%__________________________________________________________________________
%
\begin{figure} [h!]
    \centering
    \includegraphics[width=0.45 \textwidth]{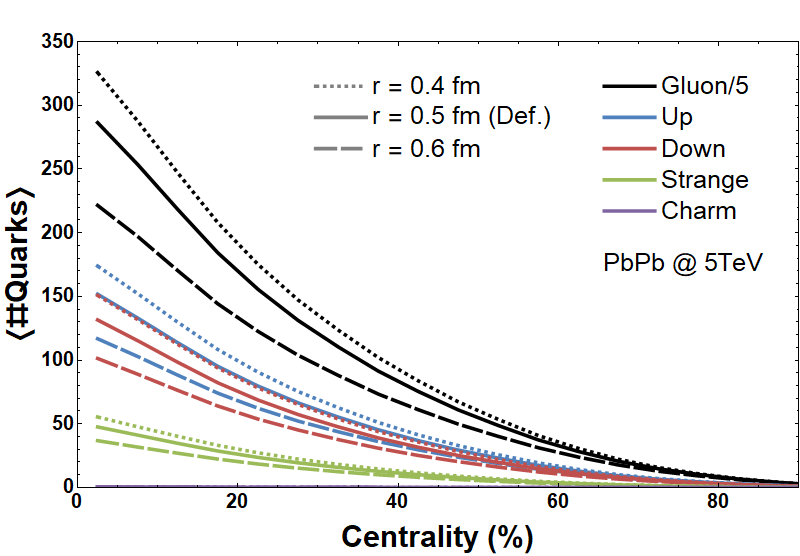} 
    \\
    \includegraphics[width=0.45 \textwidth]{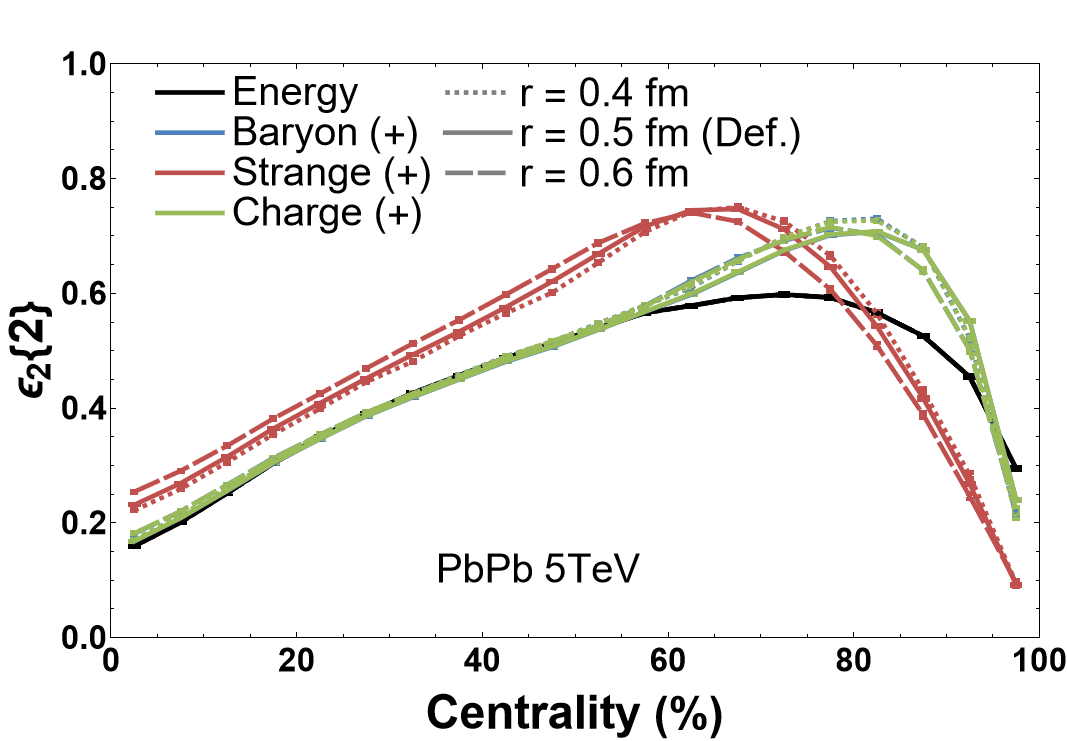} \,
	\includegraphics[width=0.45 \textwidth]{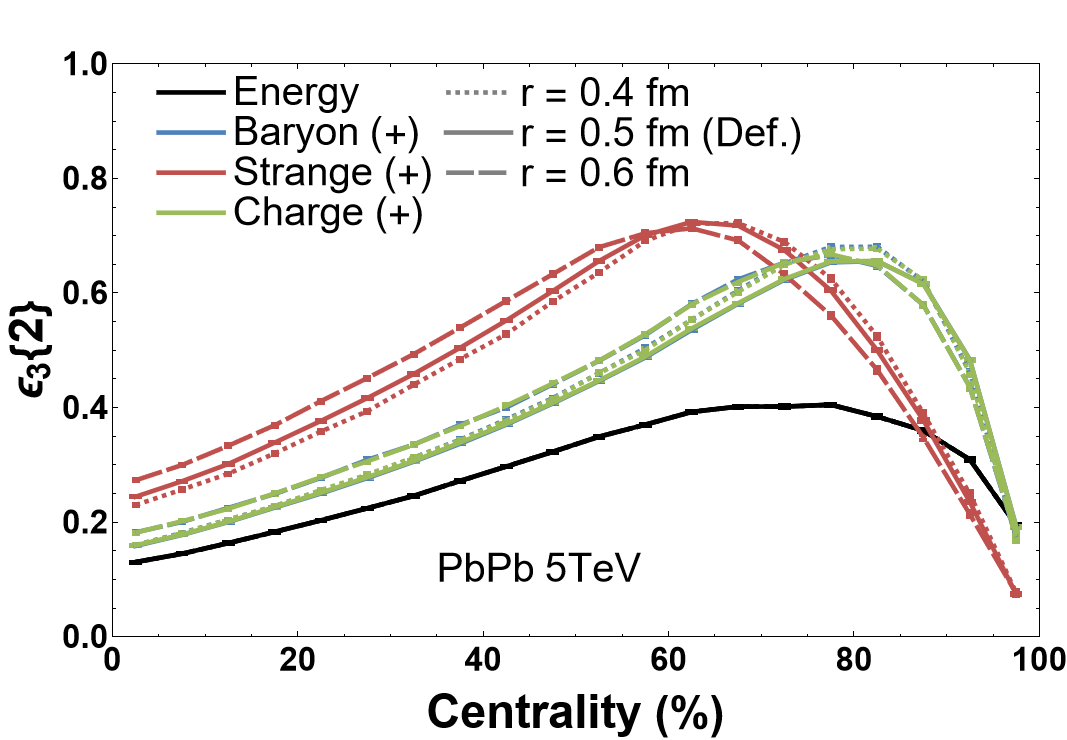}
	
	\caption{Comparison of the second cumulants $\varepsilon_n \{2\}$ for different radius parameters of the quarks and gluons. Figure updated from Ref.~\cite{Carzon:2019qja}.}
	\label{f:IBSQCentralityRad2}
\end{figure}
%
%__________________________________________________________________________

All of the previously discussed parameters have an effect on the location and probability of $g \rightarrow q \bar{q}$ splittings. There are two parameters that determine the profile of the charge perturbations created by \code{iccing}: the density profile used to distribute quark charge and energy and the radius of this distribution. The implementation of the density profile is discussed in Sec.~\ref{subsec:DensityMask} and further expounded upon and explored in Sec.~\ref{SubSec:LocalGlobalImpactDensityProfile} and Chap.~\ref{chap:PreEquilibriumEvolution}. In Chap.~\ref{chap:PreEquilibriumEvolution}, the gluon radius is fixed while the quark radius is allowed to change dynamically in the algorithm. 

Here I will focus on the impact of the radius for this profile on the eccentricities. There are actually two radius parameters that can be varied, one for the distribution of charge and energy from quark splittings and another for the gluon sampling for those splittings. In this section, I first set them equal to each other exploring their general impact on the eccentricities then look at them separately. Where the radius $r$ of the quark and gluon blobs comes into play in the algorithm is detailed in Sec.~\ref{sec:Redistribute}. In Fig.~\ref{f:IBSQCentralityRad2}, the effect of the charge perturbation radius on the eccentricities is relatively small, which is good since it is perhaps the most \textit{ad hoc} choice in the model. Crucial to understanding Fig.~\ref{f:IBSQCentralityRad2} is the fact that the gluon sampling radius and quark distribution radius are coupled here. You can see the effect of the gluon sampling radius in the quark multiplicity plot, where increasing the radius leads to fewer quarks, because there are more heavy gluons that transfer more energy to the output, while decreasing the radius inflates quark production because you have more gluons that provide more splitting opportunities. The eccentricity plots reflect both the effect from the change in quark production and the way the charge and energy is redistributed. These different effects seem to balance each other out so that the change to the eccentricities is very small. 

%---------------------------------------------------------------------------
%
\section{Coupling to Hydrodynamics}
\label{SubSec:LocalGlobalImpactDensityProfile}
%
%---------------------------------------------------------------------------

This chapter reproduces and refines the work from Ref.~\cite{Plumberg:inPrep}.

\begin{figure} [h!]
    \centering
    \includegraphics[keepaspectratio, width=\linewidth]{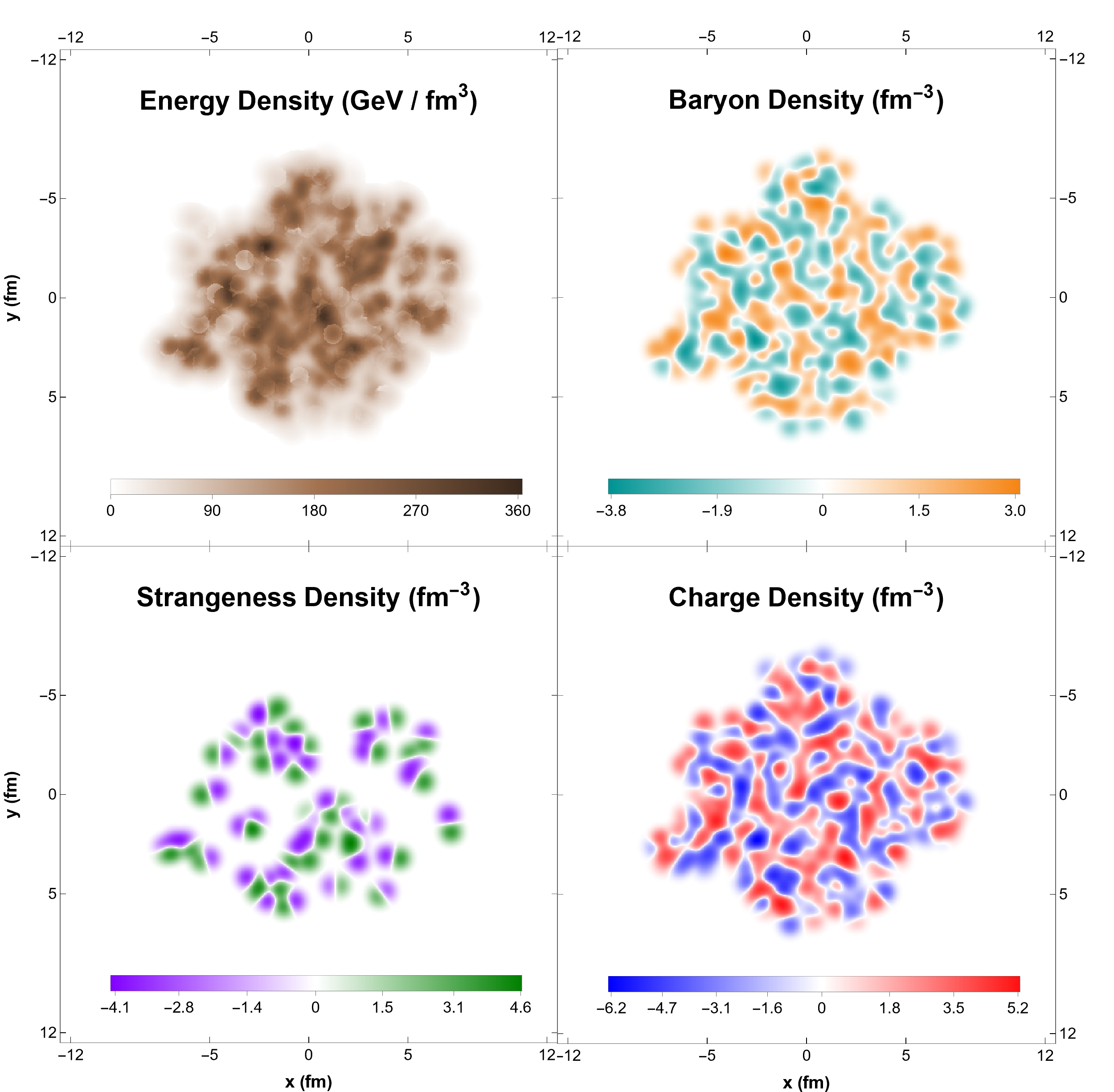}
    
    \caption{Energy and charge density profiles of an \code{iccing} event run with the Kernel function mask. Identical sampling to Fig.~\ref{f:Events} aside from the choice of mask.}
    \label{fig:FullEventMaskComparison}
\end{figure}

For the validity of the application of a hydrodynamic description of a medium, it must be near equilibrium, or alternatively, there must be a separation of scales between the interaction length and system size. Thus, initial conditions for hydrodynamic simulations of heavy-ion collisions must be near equilibrium and not contain large gradients. As constructed, \code{ICCING} uses a "mask" with which to redistribute energy and charge associated with the produced quark/anti-quark pairs. The default choice for this was a Gaussian distribution that gets cut off at the width and produces clearly defined charge regions but also produces large gradients near the edge of these charge perturbations. In an attempt to smooth out these large gradients, the \code{Mask.h} class was introduced, adding the option to specify the redistribution mask, and an alternative Kernel function mask was added to \code{ICCING}. This new mask forces the edges to go smoothly to zero at the radial distance from the center of the perturbation. To provide comparable profiles, with the Kernel function, for the quark/anti-quark pairs, the radius is doubled. An illustration of the difference between masks is shown in Fig.~\ref{fig:KernelVsGaussian} and further details of the implementation are contained in Sec.~\ref{subsec:DensityMask}. An example of a fully sampled event, using the Kernel function, is provided in Fig.~\ref{fig:FullEventMaskComparison} and can be compared to Fig.~\ref{f:Events}, which is the same event with the only difference being the use of the Gaussian profile. It is clear to see that the edges of the quarks and anti-quarks in Fig.~\ref{fig:FullEventMaskComparison} are less defined and the gradients are consequently smoother and more natural looking. 

\begin{figure} [h!]
    \centering
    \includegraphics[keepaspectratio, width=0.3\linewidth]{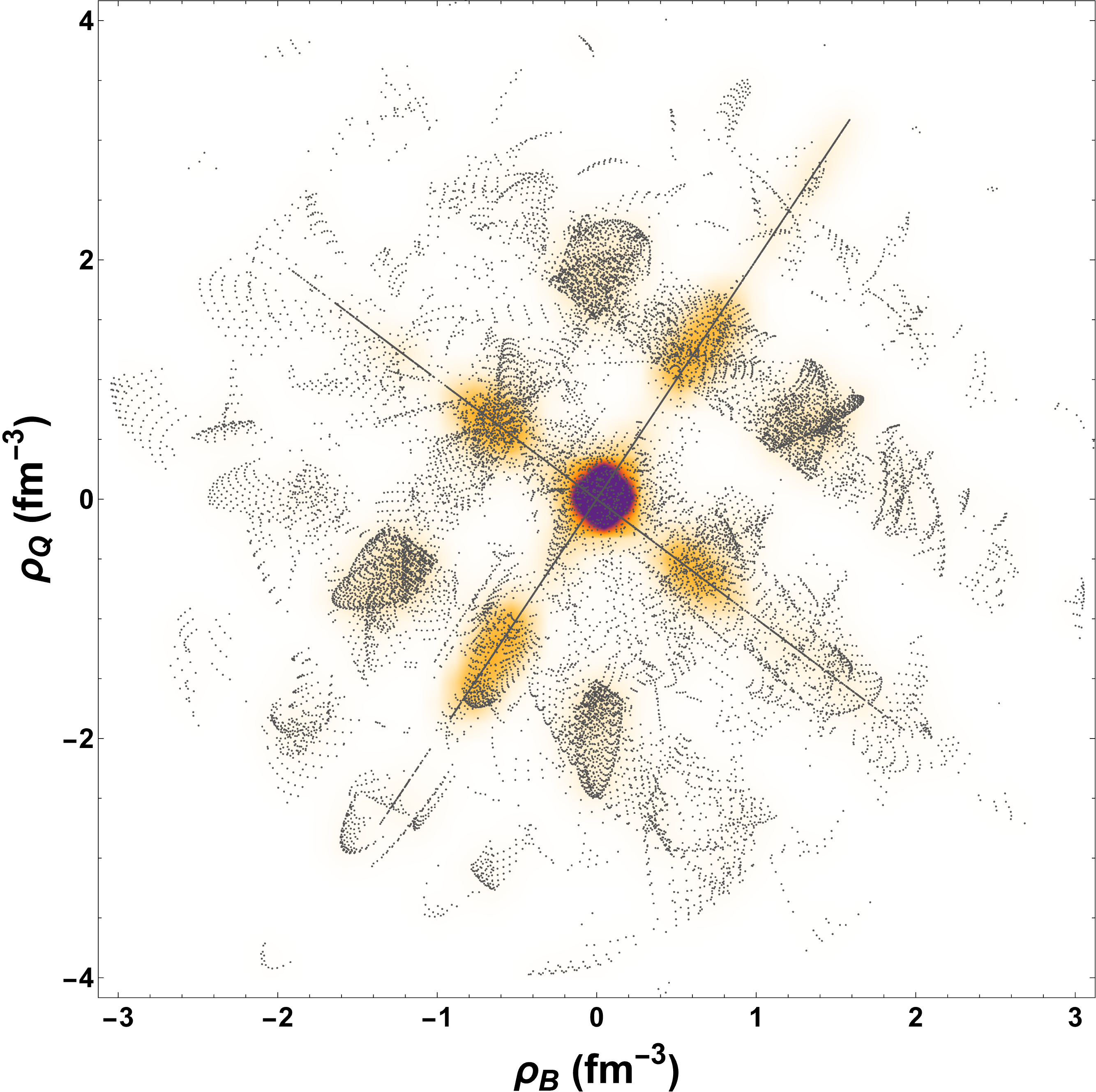}
    \includegraphics[keepaspectratio, width=0.3\linewidth]{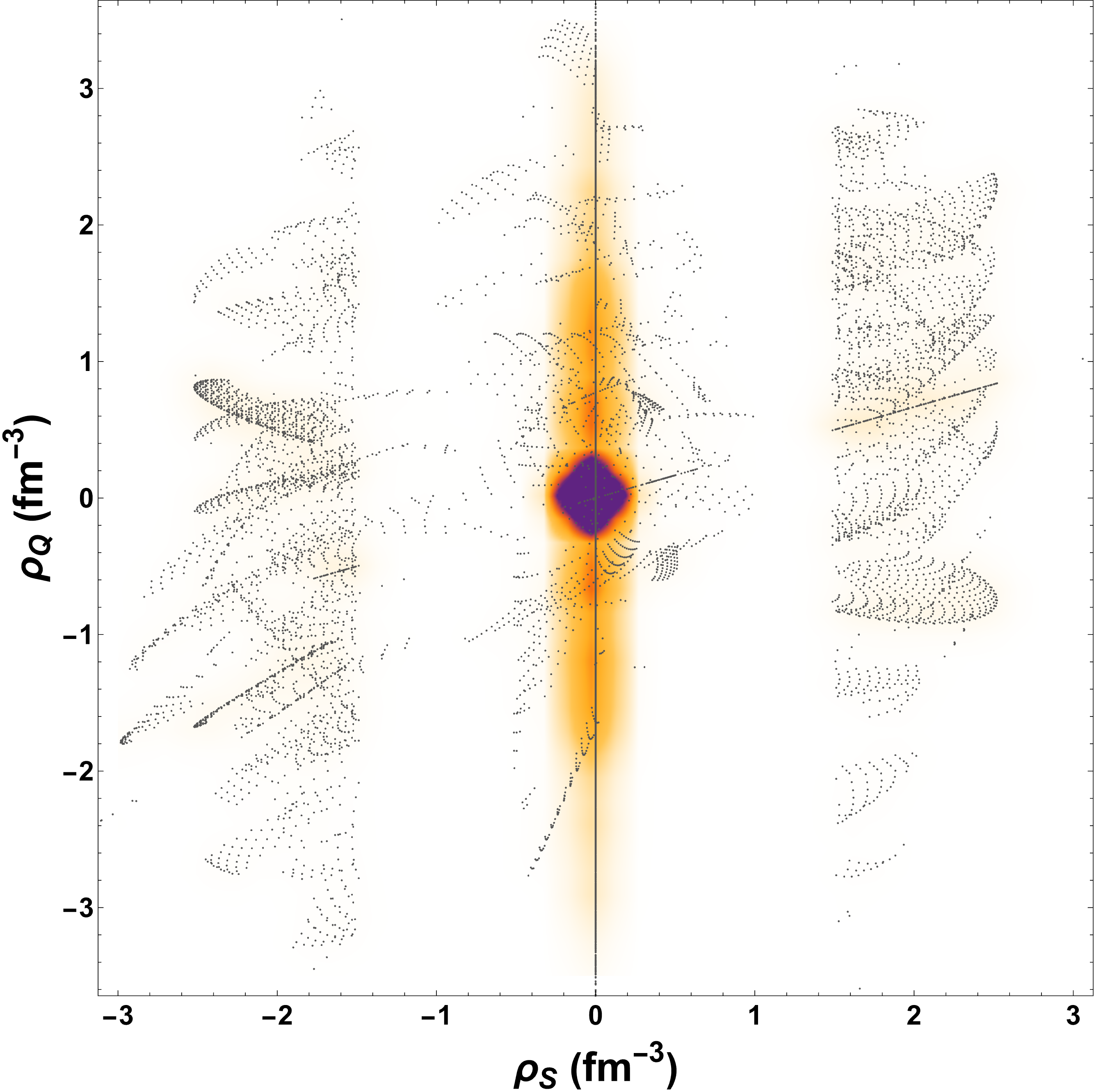}
    \includegraphics[keepaspectratio, width=0.3\linewidth]{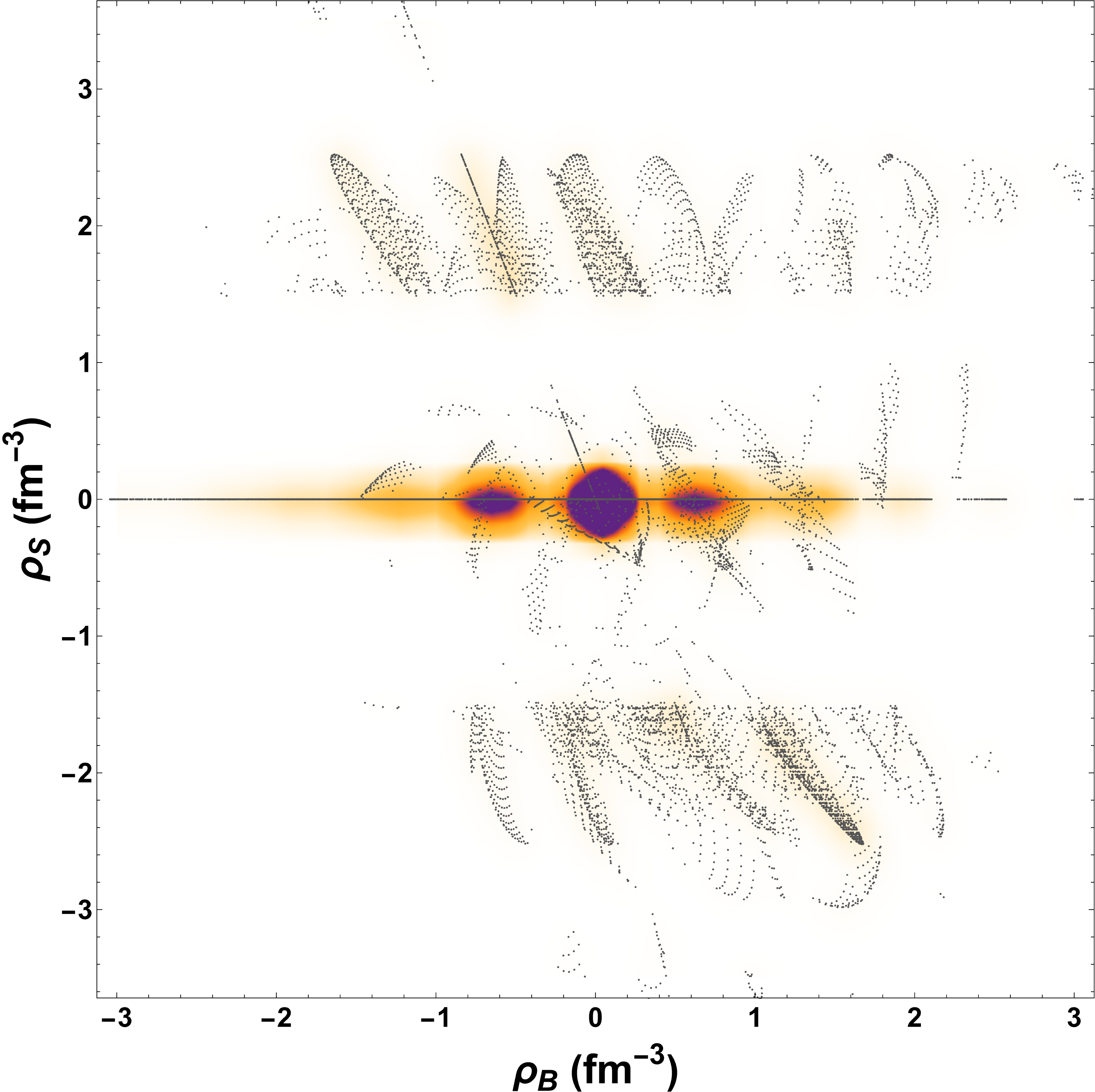}
    \\
     \includegraphics[keepaspectratio, width=0.3\linewidth]{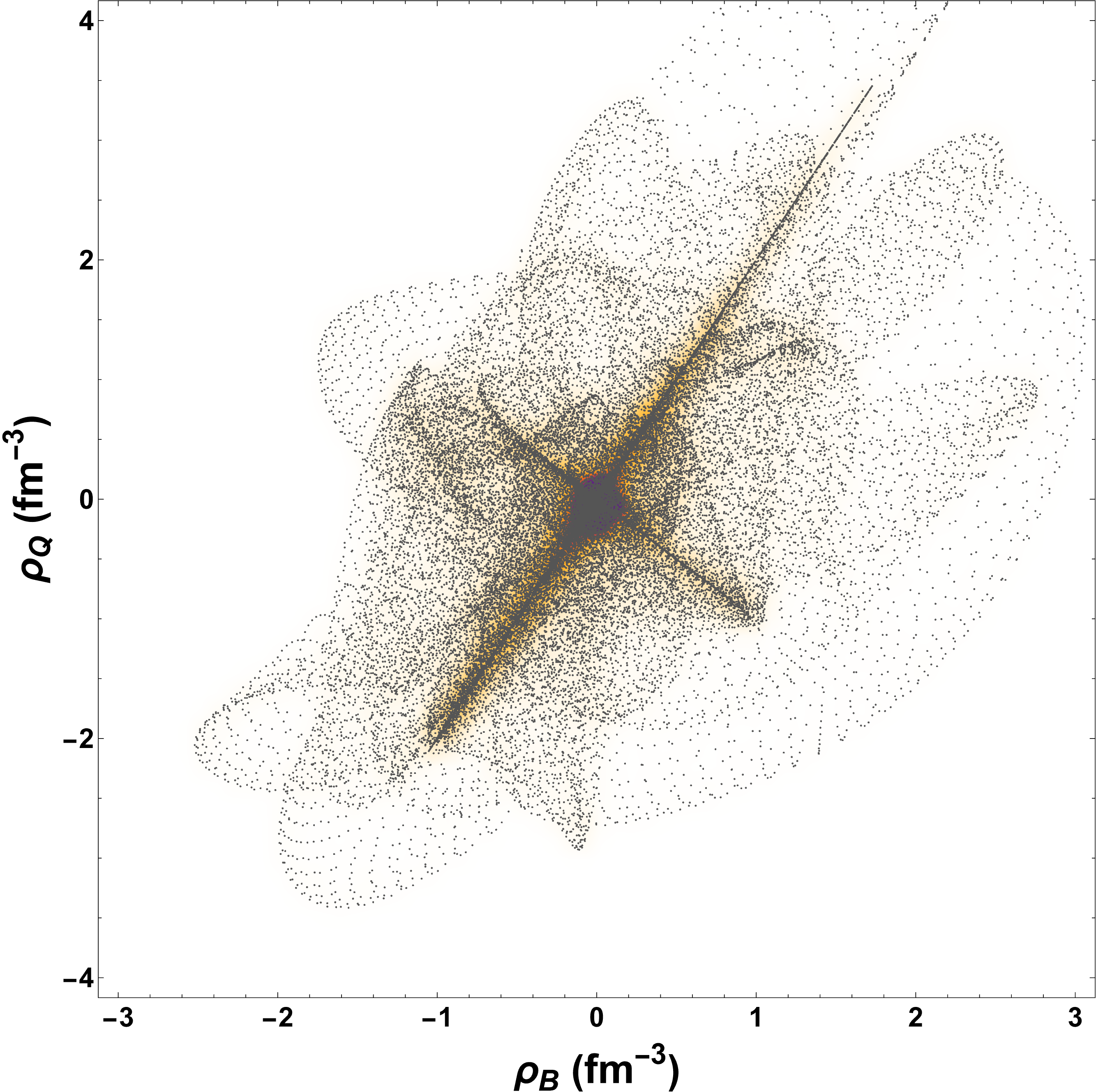}
    \includegraphics[keepaspectratio, width=0.3\linewidth]{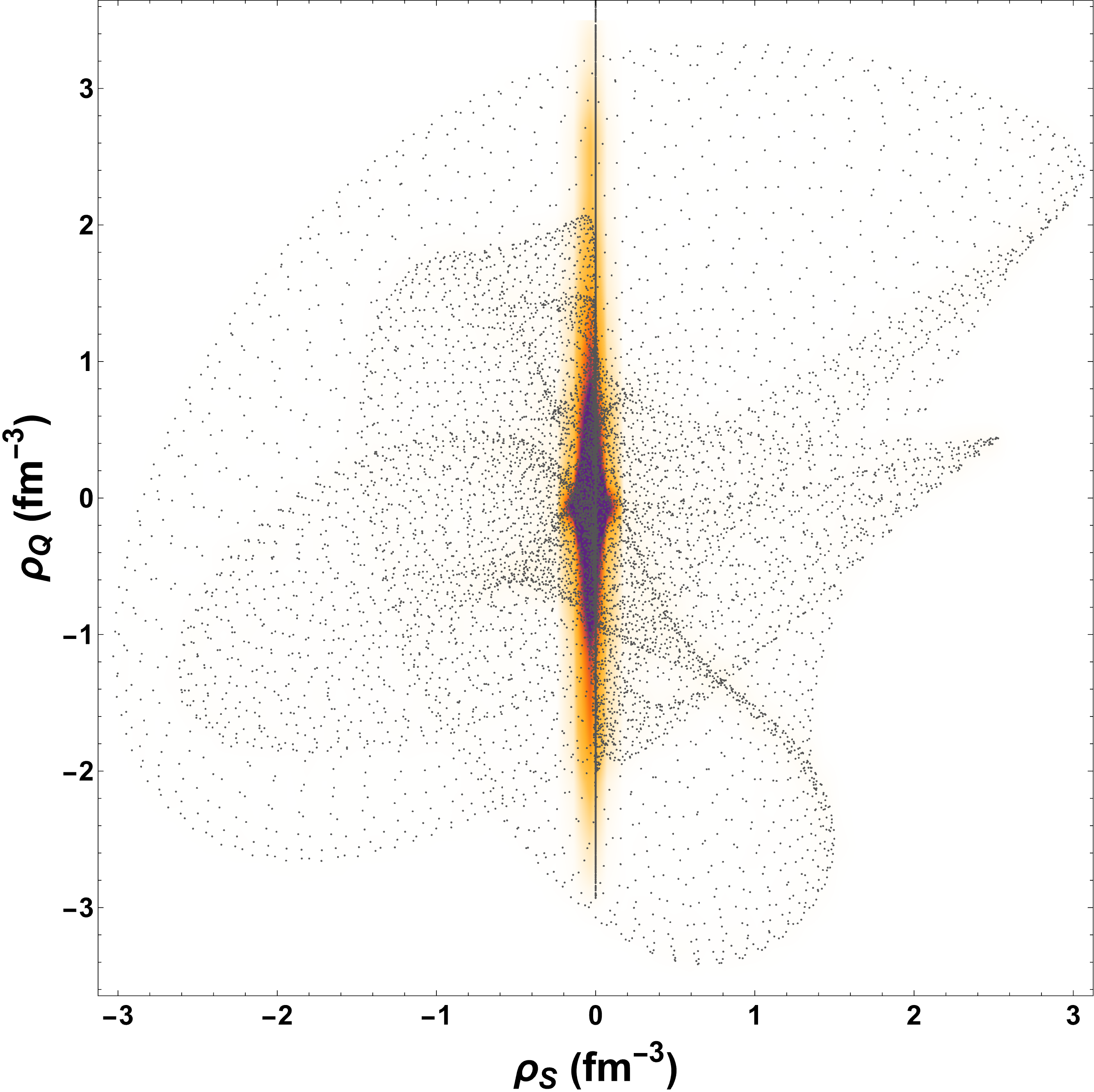}
    \includegraphics[keepaspectratio, width=0.3\linewidth]{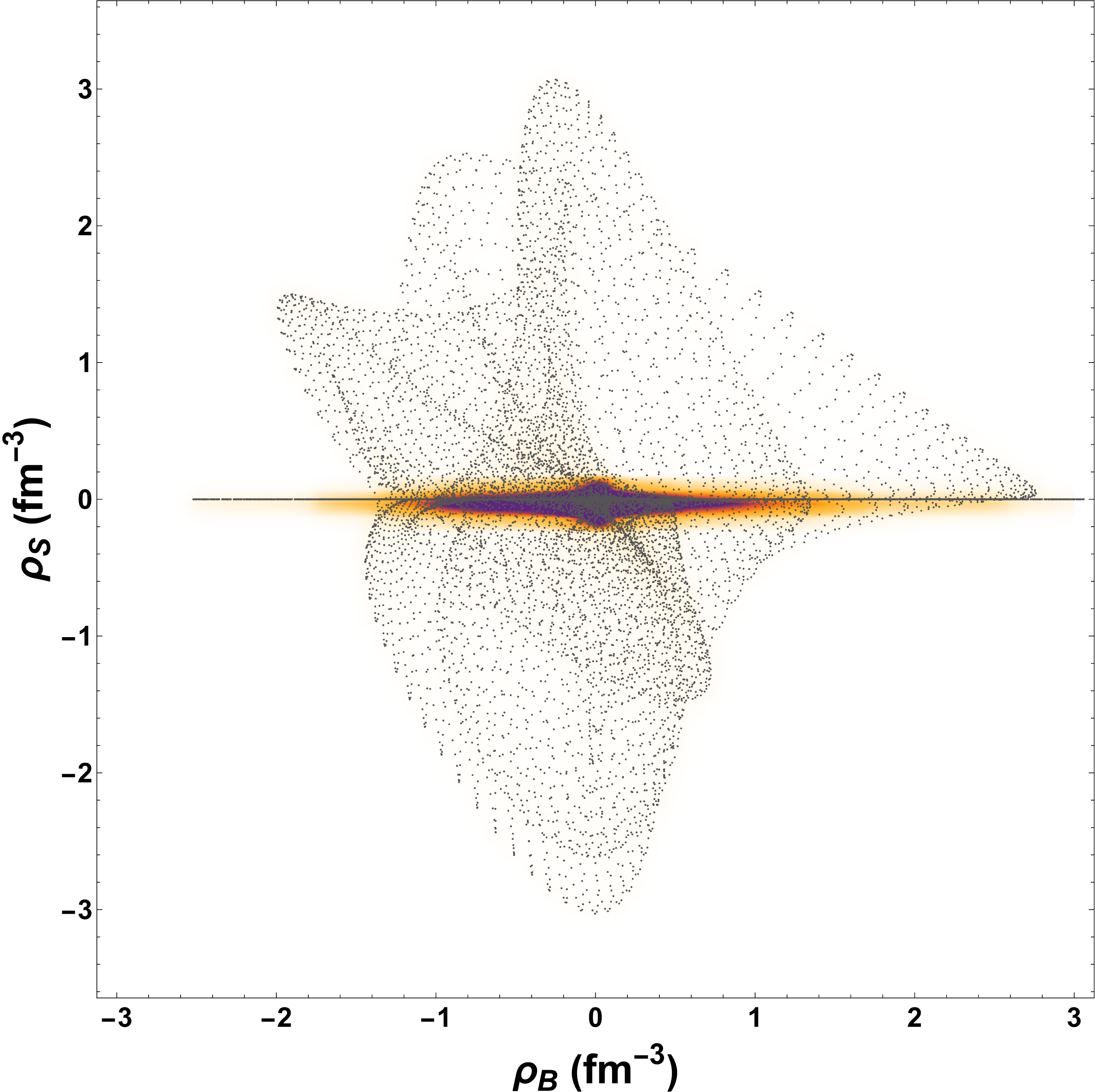}
    
    \caption{Distribution of points in the charge density sector of a single \code{iccing} event, where the Gaussian profile, along the top row, and the Kernel function, along the bottom row, were used to redistribute the quark/anti-quark densities.}
    \label{fig:MicroscopicAnalysis}
\end{figure}

Further investigation shows that the event averaged geometry of $\varepsilon_2\{2\}$ and $\varepsilon_3\{3\}$ is insensitive to the choice in mask profile. We can also look at the microscopic structure in the charge density sectors of a single event by plotting the distribution of baryon, strange, and electric charge densities with respect to each other in Fig.~\ref{fig:MicroscopicAnalysis}. Along the top row of Fig.~\ref{fig:MicroscopicAnalysis} are the density sectors for an event using the Gaussian profile and along the bottom, the same event, using the Kernel function profile. We see a significant sensitivity to the mask profile on the microscopic scale, with the Gaussian profile creating sharp structure in the charge distribution and the Kernel function having a smoother more natural look. This lack of sensitivity on the event averaged 'macroscopic' scale despite the significant modification of the microscopic structure reveals a useful degree of freedom in the \code{iccing} model. 

This is not the only issue encountered when coupling \code{iccing} to a hydrodynamic code. Additionally, the normalization of the initial condition required updating and a more explicit definition, Eq.~\ref{eq:PerturbativeCutoff}, of perturbative splittings was required.

\begin{figure} [h!]
    \centering
    \includegraphics[keepaspectratio, width=0.45\linewidth]{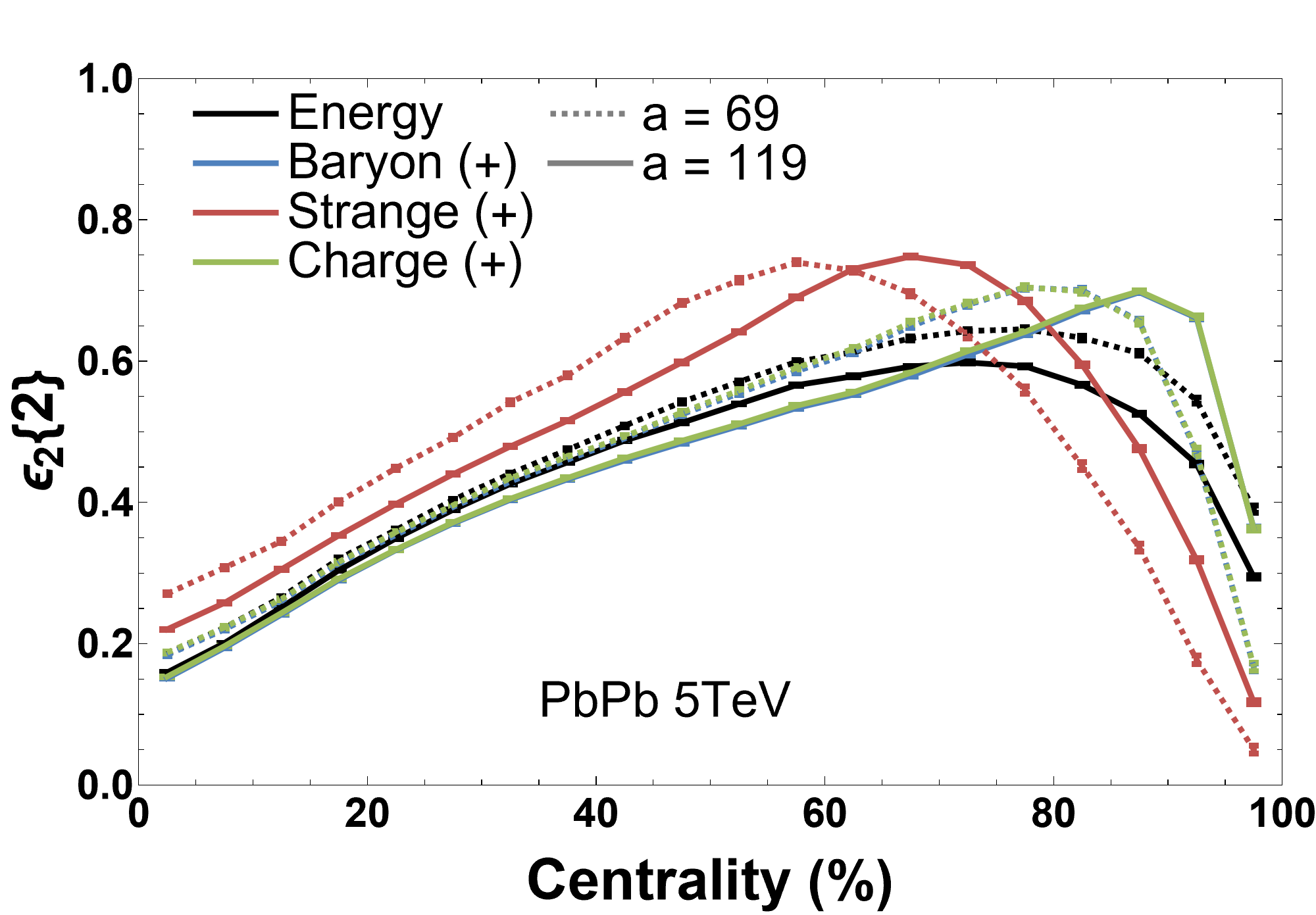}
    \includegraphics[keepaspectratio, width=0.45\linewidth]{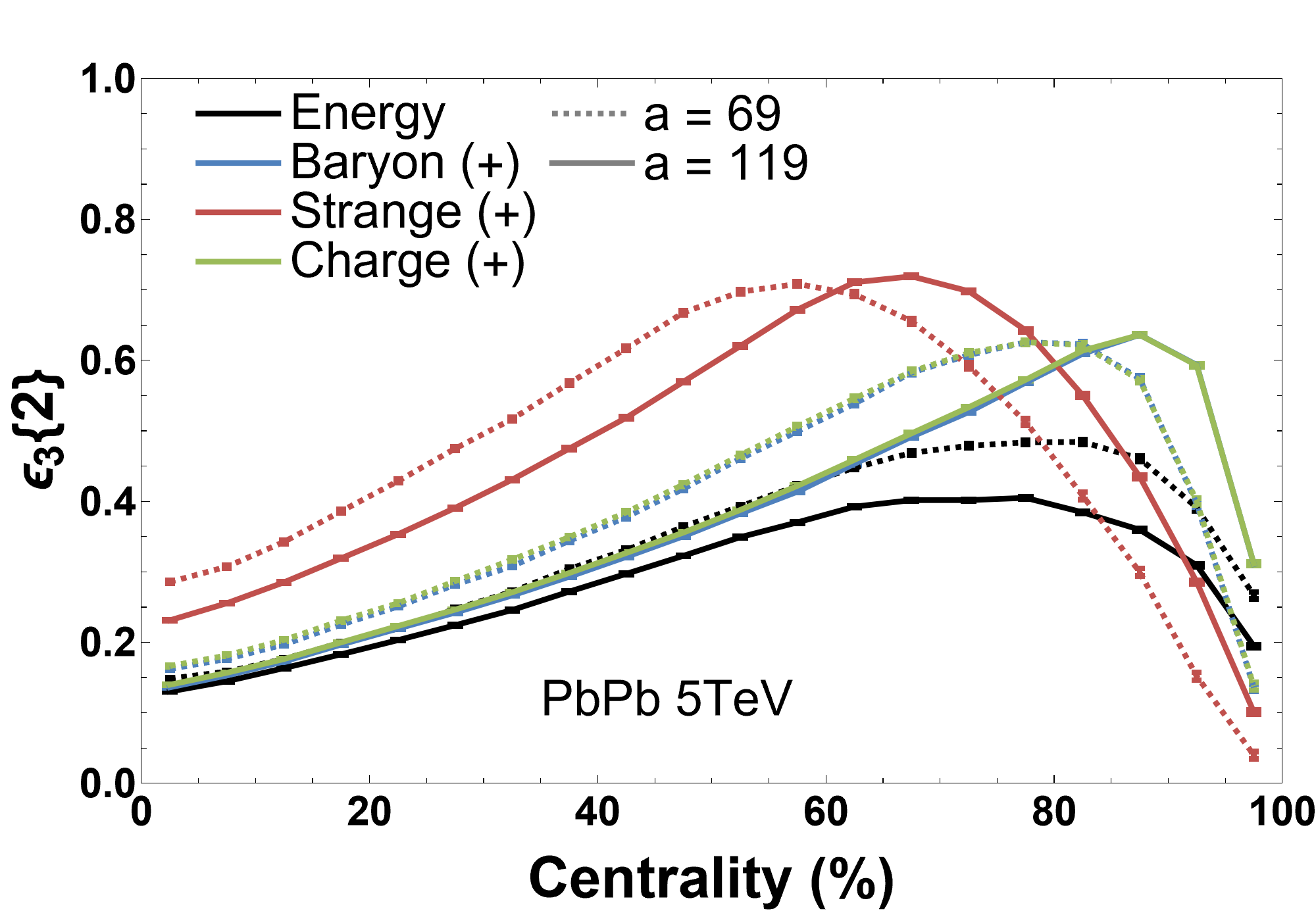}
    
    \caption{Elliptic and triangular geometries for two values of $a_{entropy}$: $119 fm$ and $69 fm$, corresponding to a fit of final state multiplicity using only energy density and energy+charge density, respectively.}
    \label{fig:NormEccentricities}
\end{figure}

The \code{iccing} algorithm initializes new charge densities and thus changes the thermodynamic properties of the initial state. Adding the new charge density information requires a thermodynamic source, which in this case, must by the entropy since \code{iccing} conserved energy. This is not a problem since the entropy of the system, while nearly conserved throughout the evolution, is set in the initial state by comparison to final state multiplicity (See Sec.~\ref{subsec:EstimationOfA}). However, this does mean that the entropy proportionality factor must be re-tuned, since the previous values were obtained assuming an initial state of only energy density and with the inclusion of charge densities now over-predicts the final state multiplicity. This re-tuning was performed with the resulting value of the proportionality factor being $a_{entropy}=69 fm$. This does have a significant effect on the energy and charge eccentricities as observed in Fig.~\ref{fig:NormEccentricities}.

The last issue, discovered in the process of coupling to hydro, is the failure of a significant number of grid points, from the \code{iccing} initial condition, to map onto the equation of state scheme used in the hydrodynamiccode. This equation of state scheme, first attempts to use a 4-D equation of state from lQCD to convert the energy+charge density of a given point into temperature and chemical potential, which is required for the evolution. If a given energy+charge combination does not exist in the lQCD equation of state, then several other equation of states are attempted that become less physical but widen the parameter space. Even in this scheme, that should catch almost all cases, there were issues with some combinations of energy and charge. 

One contributing factor, was the location of some quark/anti-quark splittings that violated a perturbative view of the model. Already present in \code{iccing}, there were some tests that determined the physical possibility of quark/anti-quark production in a given area, specifically the mass threshold test and the $E_{thresh}$ parameter. The mass threshold is a simple test that determines if a gluon selected for quark/anti-quark splitting contained enough energy to produce two quarks of the mass corresponding to their flavor. The $E_{thresh}$ parameter, on the other hand, is a lower bound on the available energy in a given region that is given the consideration of splitting. This threshold is used to speed up the code, and restrict production in low energy areas, by skipping any sampling that would not have been able to produce a quark/anti-quark pair to begin with. 

\begin{figure} [h!]
    \centering
    \includegraphics[keepaspectratio, width=0.45\linewidth]{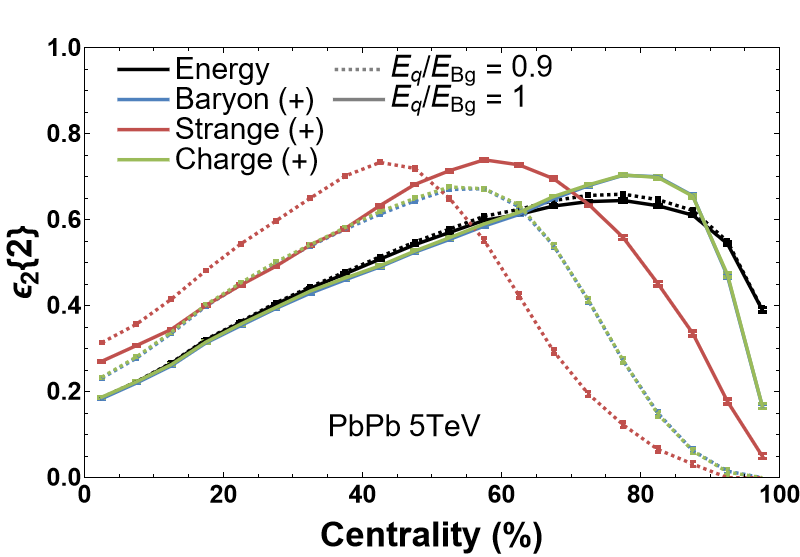}
    \includegraphics[keepaspectratio, width=0.45\linewidth]{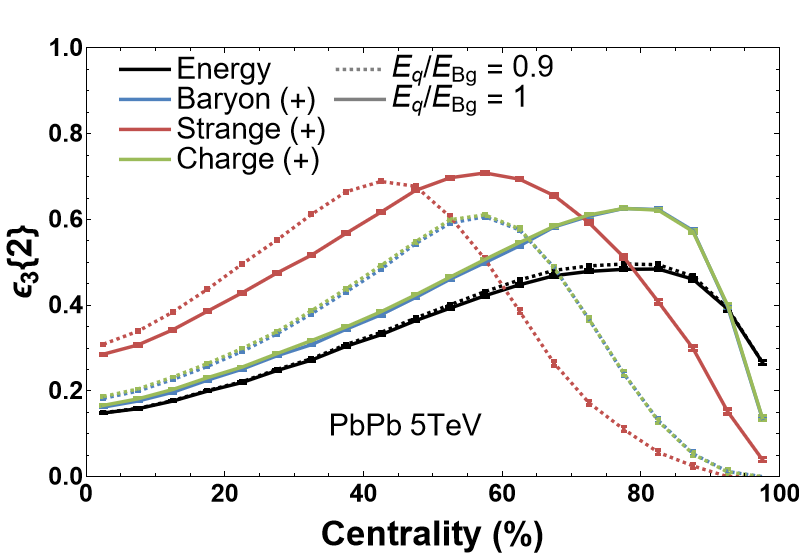}
    
    \caption{Effect of the strict perturbation cut off for quark/anti-quark production on the initial state geometry. Events, here, use $a_{entropy}=69 fm$ and the Kernel function for redistribution.}
    \label{fig:NormPertEccentricities}
\end{figure}

These physically motivated suppressions of production, however, are not sufficient in ensuring only perturbative splittings occur, an assumption made in the formulation of the \code{iccing} algorithm. A similar issue was observed in Chap.~\ref{chap:PreEquilibriumEvolution} and fixed by the enforcement of a perturbative cutoff to quark/anti-quark production. This cutoff, defined in Eq.~\ref{eq:PerturbativeCutoff}, compares each point in the quark/anti-quark energy densities and ensures that they are below some fraction of the energy density in that region from before the sampling process. The enforcement of this cutoff, with $P=0.9$, has a significant effect on the initial state geometry, as seen in Fig.~\ref{fig:NormPertEccentricities}, with a suppression of quark production that kills geometry in peripheral events and enhances geometry in central to mid-central collisions. The geometry represented by Fig.~\ref{fig:NormPertEccentricities} is that which is most compatible with hydrodynamic simulations.

%

%---------------------------------------------------------------------------
%
\chapter{Pre-equilibrium Evolution of Conserved Charges}
\label{chap:PreEquilibriumEvolution}
%
%---------------------------------------------------------------------------

The initial state of heavy-ion collisions is predominantly composed of saturated gluons emerging from the low-$x$ wave functions of the colliding nuclei (See Chap.~\ref{chap:ICCINGAlgorithm}). Methods used for the construction of the initial state focus on a characterization of the initial geometry in terms of the energy or entropy density. This initial condition can be far from equilibrium and outside the hydrodynamic description used for the QGP phase of the system. For this reason, the large gradients present in the initial state are often 'smoothed out' by a pre-equilibrium dynamics phase leading up to the formation of a quark-gluon plasma (QGP) characterized by deconfined quarks and gluons acting as a nearly perfect fluid \cite{Schlichting:2019abc,Berges:2020fwq}. Experimental measurements of the distribution of final state hadrons are a convolution of the features of all the evolution phases of the medium: initial state geometry, pre-equilibrium dynamics, hydrodynamic evolution, and hadronic transport. As such, the best comparison one can make between simulations and experiment is dependent on the inclusion of all stages of evolution \cite{Heinz:2013th}, especially an accurate description of the initial state.  

Simulations of heavy-ion collisions start with the characterization of the initial energy-momentum tensor $T^{\mu\nu}$ and currents $J^\mu$ of conserved charges, which are far from local thermodynamic equilibrium. This initial state is chosen to occur at some proper time $\tau_0$ and evolved using pre-equilibrium dynamics up to some proper time $\tau_{hydro} > \tau_0$ at which the system is near local thermal equilibrium and a hydrodynamic description is valid. Since the QGP behaves as a nearly perfect fluid \cite{Heinz:2013th}, the initial state geometry, described by $T^{\mu\nu}$ and $J^\mu$, is well translated to the final state hadron distributions \cite{Teaney:2010vd,Gardim:2011xv,Niemi:2012aj,Teaney:2012ke,Qiu:2011iv,Gardim:2014tya,Betz:2016ayq,Hippert:2020kde} (See Sec.~\ref{sec:Eccentricities}). This makes a correct description of the initial condition of heavy-ion collisions essential in the prediction of experimental measurements, while also providing a useful constraint on the initial state physics. The most useful observables, to this purpose, are those that are least sensitive to the hydrodynamic phase. A rather prominent example is the cumulant ratio $v_n\left\{4\right\}/v_n\left\{2\right\}$ measured in central collisions, due to the canceling of linear response coefficients (See Sec.~\ref{sec:Cumulants}), which provides one of the most direct windows into the physics of the initial state \cite{Giacalone:2017uqx,Sievert:2019zjr,Rao:2019vgy}.

As discussed in Chap.~\ref{chap:ICCINGAlgorithm}, initial state models have been focused on the description of the energy/entropy density $\epsilon = T^{00}$, until recently. More initial state variables have been systematically included, such as the initial flow $T^{0i}$ and initial shear $T^{ij}$ \cite{Gardim:2011qn, Gardim:2012yp, Gale:2012rq, Schenke:2019pmk, Liu:2015nwa, Kurkela:2018wud,Plumberg:2021bme,Chiu:2021muk}. Another prominent area of development has been the initialization of conserved charge densities \cite{Werner:1993uh, Itakura:2003jp, Shen:2017bsr, Akamatsu:2018olk, Mohs:2019iee}, $\rho = J^0$, though this has been focused on baryon density, $\rho_B$, because of its role in the search for the QCD critical point. Most recently, the \code{iccing} model (described in Chaps.~\ref{chap:ICCINGAlgorithm} and \ref{chap:ICCINGResults}) developed a method of initializing an arbitrary number of conserved charges, primarily baryon, strange, and electric charge densities. This process is accomplished through a model agnostic method of sampling gluons from an energy density and stocastically allowing them to split into quark/anti-quark ($q\bar{q}$) pairs as determined by some chemistry input. For a full discussion of the physics options available in \code{iccing}, see Chap.~\ref{chap:ICCINGAlgorithm}. 

Constraint of parameters in \code{iccing} must be done through comparison to experimental data, which requires the hydrodynamic evolution of the initial conditions with the new conserved charge densities. To accomplish this, a 2+1D viscous hydrodynamics code that can simultaneously solve all of the hydrodynamic equations of motion for both the energy-momentum tensor and conserved charge currents is needed. While this is a technically challenging task, due to the increase in coupled equations required, and the need for a fully four-dimensional equation of state, progress has been made and will be presented soon in Ref.~\cite{Plumberg:inPrep} as the first full BSQ hydrodynamic code: \code{ccake}.

However, this does not fully address the implementation of conserved charges in heavy-ion simulations since the initial condition is far from equilibrium. Recent work, done by the authors of Ref.~\cite{Plumberg:2021bme}, has elevated concerns about the use of initial conditions directly in hydrodynamics by showing that a large fraction of fluid cells violate nonlinear causality constraints \cite{Bemfica:2020xym}, estimated at around 30\% in \code{trento}, while the remaining cells are deemed uncertain in their causality status. The authors found that by including a pre-equilibrium evolution stage, before hydrodynamics, can reduce the number of acausal fluid cells, but does not fully eliminate them. The proper description of pre-equilibrium dynamics is theoretically desirable and has been shown to have an effect on flow observables across system sizes \cite{Ambrus:2021fej,Kurkela:2018qeb,Ambrus:2022qya}.

One such pre-equilibrium method is from the authors of Refs.~\cite{Kurkela:2018wud} and \cite{Kurkela:2018vqr}, where they developed a non-equilibrium linear response formalism based on a microscopic description in QCD kinetic theory called \kompost. The \kompost code allows the propagation of the energy-momentum tensor, $T^{\mu\nu}$, from early times up to the point of hydrodynamic applicability. In Ref.~\cite{Plumberg:2021bme}, mentioned previously, they used \kompost for the pre-equilibrium stage. Coupling this formalism to \code{iccing} would greatly improve the usefulness of the initial condition generator, though the \kompost code itself does not contain the necessary ingredients to handle conserved charge densities. However, the non-equilibrium Green's functions could be applied to the case of charge perturbations around a vanishing background, which is exactly the conditions generated by \code{iccing}.

The core concept of \kompost involves obtaining the energy-momentum tensor $T^{\mu\nu}(x)$ at a specific time $\tau_{hydro}$ from an initial state model. This is achieved by evolving the system using effective kinetic theory from an initial time $\tau_0$ until it reaches $\tau_{hydro}$. Within this framework, the fluctuations $\delta T^{\mu\nu}(x)$ of the energy-momentum tensor around the background energy-momentum tensor $T_\text{BG}^{\mu\nu}(x)$ are considered. Typically, these perturbations are assumed small and, therefore, able to be linearized, giving rise to linear response theory. By summing $T_\text{BG}^{\mu\nu}(x)$ and a term involving non-equilibrium Green's functions, which captures the evolution of the perturbations, the complete energy-momentum tensor $T^{\mu\nu}(x)$ can be obtained. This approach offers a powerful method for computing $T^{\mu\nu}(x)$ as numerical simulations only need to be performed once to obtain the background evolution and the Green's functions. Over the past few years, various groups have integrated \kompost into their fluid dynamics simulations, establishing direct connections with experimental data \cite{NunesdaSilva:2020bfs,Gale:2021emg,Borghini:2022iym}.

Although \kompost is based on QCD kinetic theory, recent studies have explored computing the same Green's functions in simpler models, such as the Boltzmann equation in the Relaxation Time Approximation (RTA) \cite{Kamata:2020mka,Ke:2022tqf}. The relaxation time approximation drastically simplifies the theoretical description and provides an efficient method for calculating non-equilibrium Green's functions. By assuming the relaxation time approximation, the general formalism of \kompost can be expanded in a much simpler way to include conserved charges and compute Green's functions for the corresponding current. These Green's functions, which describe charge and energy propagation, can be incorporated into \code{iccing} with some careful adjustments, resulting in a meaningful pre-equilibrium evolution for the conserved charge densities.

The 2-particle eccentricity cumulants (see Sec.~\ref{sec:Cumulants}), which characterize the initial state geometry, are used to explore the effect of these new pre-equilibrium charge evolution equations since they are good predictors of the final state flow harmonics \cite{Niemi:2012aj}, except in peripheral collisions \cite{Noronha-Hostler:2015dbi, Sievert:2019zjr, Rao:2019vgy}. This linear mapping makes it possible to remove many medium effects by taking ratios of 4 to 2-particle cumulants, which measure the fluctuations of geometry. While these observables are well understood for energy/entropy density, their use in charge density profiles is quite new and further detailed in Chap.~\ref{chap:ICCINGResults}.

In this chapter, I will first introduce the coordinate space form of the Green's functions, in Sec.~\ref{sec:GreensFunctionsInCoordinateSpace}, and how they are implemented in the \code{iccing} code, in Sec.~\ref{sec:UsingGreensFunctionsInICCING}. Finally, the effect of the different physics introduced by this pre-equilibrium evolution will be investigated in Sec.~\ref{sec:GreensFunctionsImpactOnEccentricities}. This chapter reproduces and refines the work from Ref.~\cite{Carzon:2023zfp}.

\section{Green's Functions in Coordinate Space}\label{sec:GreensFunctionsInCoordinateSpace}

The derivation of the non-equilibrium Green's functions for conserved charge currents is beyond the scope of this work and well detailed in Ref.~\cite{Carzon:2023zfp}. I will, however, introduce the relevant equations and concepts for their implementation in \code{iccing}.

The method used to derive the non-equilibrium Green's functions follows that of Ref.~\cite{Kurkela:2018vqr}, which divides the space-time dynamic into a background evolution and perturbations around this background. During the pre-equilibrium stage, the plasma undergoes rapid longitudinal expansion, along the beam direction, while the transverse evolution starts from rest and grows on a timescale comparable to the system size, allowing it to be neglected. This justifies using the ideal situation of Bjorken flow, which greatly simplifies the problem. In Ref.~\cite{Carzon:2023zfp}, the evolution is derived for charge densities around a vanishing background, which coincides with the \code{iccing} model, but does discuss generalization to non-vanishing densities in appendices. 

The process starts by computing the energy density at late times as a function of the initial energy density, which is dependent on the shear viscosity to entropy density ratio, $\tilde{\eta}/s$, and the constant $C_{\infty}$ quantifying the efficiency of conversion of initial energy into thermal energy. The shear-viscosity to entropy density ratio is taken to be constant, which is reasonable for a conformal system with vanishing net charge density, and $C_{\infty}\approx 0.9$ \cite{Kamata:2020mka,Giacalone:2019ldn}. This is then used to define an attractor curve $\mathcal{E}(\tilde{w})$, that interpolates between free-streaming at early times and viscous hydrodynamics at late times, which is a ratio of the energy density at any given time to that of the asymptotic late time energy density. The energy attractor depends on the dimensionless time variable \cite{Giacalone:2019ldn}
\begin{align}\label{eq:wTilde}
    \tilde{w} = \frac{T(\tau)\tau}{4\pi\tilde{\eta}/s}\, ,
\end{align}
where $T_{eff}(\tau)$ is an effective temperature dependent on the background energy density. Finally, the energy attractor is used to relate the initial energy density to the energy density at some later time with the form:
\begin{align} \label{e:attractor2}
    e_\text{\BG}\pqty{e(\tau_0)} = C_\infty \pqty{\frac{4\pi\tilde{\eta}/s}{T_{eff}(\tau_0)\tau_0^{1/4}}}^{4/9} \frac{(e\tau)_0}{\tau^{4/3}} \mathcal{E}(\tilde{w})   \: ,
\end{align}
where $(e\tau)_0$ is the initial energy density per unit area and rapidity. We can see that the background energy evolution is dynamically dependent on the effective temperature of the initial energy density and the attractor curve.

Having established the behavior of the background evolution, the behavior of perturbations around that background are detailed. In Ref.~\cite{Carzon:2023zfp}, the non-equilibrium Green's functions for the energy-momentum tensor and conserved charge currents are derived in Fourier space and then translated in to coordinate space. The Green's functions describe perturbations of the energy-momentum tensor and charge currents and are decomposed into a basis of Lorentz scalars (s), vectors (v), and tensors (t) (similar to Ref.~\cite{Kurkela:2018vqr}), such that the evolution can be described by the scalar Green's functions. After some normalization, the Green's functions for energy-momentum and charge current perturbations are:
\begin{subequations}
    \begin{align}
        \G{s}{s}\pqty{\kappa,x} &= \frac{\tens{\delta T}{\tau\tau}{\kt}(x)}{e(x)}\, 
        ,\\
        \F{s}{s}\pqty{\kappa,x} &= \tau\tens{\delta N}{\tau}{\kappa}(x)\,
    \end{align}
\end{subequations}
respectively, where $\kappa$ is the propagation phase. For full derivation and analysis of the Fourier space Green's functions see Ref.~\cite{Carzon:2023zfp}.

For the sake of implementing these Green's functions in \code{iccing}, we would like them in coordinate space. This is done by a similar decomposition in coordinate space as was done in Fourier space, with the Green's functions described by a basis of scalars, vectors, and tensors, such that for the two Green's functions we find
\begin{subequations}
    \begin{align}
        \GCS{\tau\tau}{\tau\tau}\pqty{\rt,\tau} &= \GCS{s}{s}\pqty{\vrt,\tau}\, 
        ,\\
        \FCS{\tau}{\tau}\pqty{\rt,\tau} &= \FCS{s}{s}\pqty{\vrt,\tau}\, .
    \end{align}
\end{subequations}
for the energy-momentum tensor and charge current, respectively.
We relate the coordinate space Green's functions to their Fourier space counterparts by the following Fourier-Hankel transforms
\begin{subequations}
    \begin{align}
        \GCS{s}{s}\pqty{\vrt,\tau} &= \frac{1}{2\pi} \int \dd{\vkt} \vkt J_0\pqty{ \vkt \vrt } \G{s}{s} \pqty{\vkt,\tau} \, 
        , \\
        \FCS{s}{s}\pqty{\vrt,\tau} &= \frac{1}{2\pi} \int \dd{\vkt} \vkt J_0\pqty{ \vkt \vrt } \F{s}{s} \pqty{\vkt,\tau} \, ,
    \end{align}
\end{subequations}
where $J_\nu$ are the Bessel functions of the first kind.

\begin{figure}[t!]
    \centering
        \includegraphics[width=0.49\textwidth]{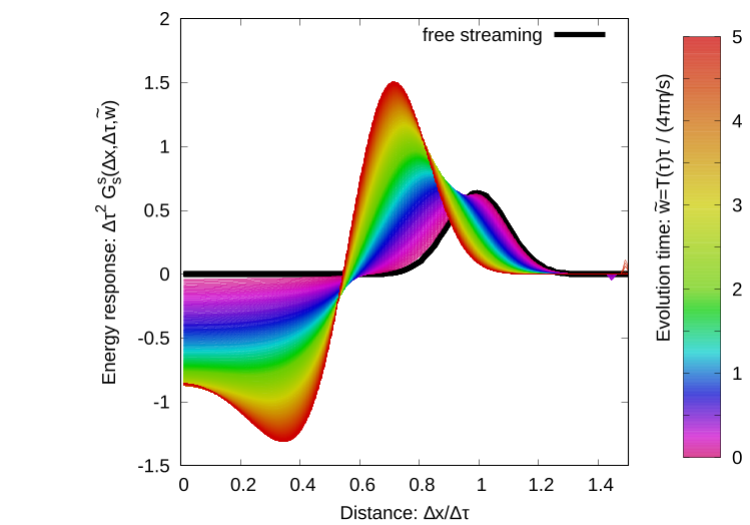}
        \includegraphics[width=0.49\textwidth]{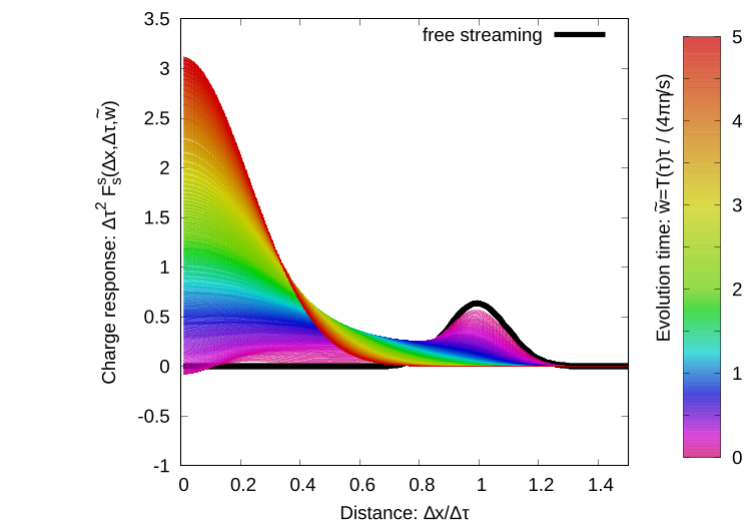}
        \caption{\textsl{Left:} Evolution of the energy Green's function $\G{s}{s}$ in response to initial energy perturbations in coordinate space. \textsl{Right:} Evolution of the charge Green's function $\F{s}{s}$ in response to initial charge perturbations in coordinate space. The different curves in each panel correspond to different times $\tilde{w}$. Figure from Ref.~\cite{Carzon:2023zfp}. \label{fig:ResponseFunctions}}
\end{figure}

The behavior of the Green's functions in coordinate space is presented in Fig. \ref{fig:ResponseFunctions} as a function of the propagation distance in units of elapsed time, $\Delta x/\Delta\tau$, and for a range of evolution times, $\tilde{w}$, which follow a color scale. The free-streaming behavior is shown by a thick black line.

For the energy Green's function, Fig.~\ref{fig:ResponseFunctions} (left), we see that they behave like sound waves propagating at almost the speed of light for the free-streaming case and at early evolution times. As one goes to later times, the peak of the $\GCS{s}{s}$ moves to shorter $\Delta x/\Delta\tau$, approaching the speed of sound $c_{s}=\sqrt{1/3}$, and exhibiting a negative contribution at small $\Delta x/\Delta\tau$ corresponding to a diffusion wake.

The charge response, Fig.~\ref{fig:ResponseFunctions} (right), behaves the same as the energy response at early times since the charge density is carried by free-streaming particles. This behavior, in $\FCS{s}{s}$, persists up to $\tilde{w}\sim 0.5$, where it transitions from the free propagation of charge to a charge diffusion type behavior with a pronounced peak centered around $\Delta x/\Delta\tau=0$.

\section{Implementing Pre-Equilibrium Evolution of Conserved Charges} \label{sec:UsingGreensFunctionsInICCING}

Having obtained both the energy-momentum and charge dependent Green's functions, the coupling of them to the \code{iccing} model is detailed here. This updated version of \code{iccing} is labeled as 2.0 and comes with the code name "Mint"\footnote{Get it! Because it is Green \code{iccing}!}. This version includes many other additions such as more flexible choices for the sampling and depositing profiles.

In this section, I will analyze the effects of integrating linearized pre-equilibrium Green's functions with initial geometries generated by \code{iccing}. We will see that the assumption of linear response has significant implications for the energy and charge perturbations that arise.

As detailed in Chap.~\ref{chap:ICCINGAlgorithm}, \code{iccing} starts with an energy density provided by some initial state constructor, here taken to be \code{trento}. In default \code{iccing}, sampling of $g\rightarrow q\Bar{q}$ is done and energy is redistributed according to the quark production and new charge densities are deposited. This all occurs at a fixed $\tau_0$ and is assumed to happen instantaneously, though this could be relaxed in the future. Adding the Green's functions complicates things. We first start by evolving the incoming event profile to our user provided $\tau_{hydro}$, using the energy attractor $\mathcal{E}(\tilde{w})$ from Eq.~\ref{e:attractor2}. Then the usual sampling is done on the \textit{initial} density profile, while the redistribution is performed on the \textit{evolved} density profile. This evolved profile acts as the background for the whole process. Care is taken in ensuring the sampling is done in a consistent way as to not break energy conservation and redistribution done in a way to avoid negative energy output (more on this later).

The process of an individual gluon splitting produces two types of energy perturbations, relative to the background density. First, there is a negative energy perturbation, or hole, created by the removal of the gluon from the background. Second, there are positive energy perturbations from redistributing the gluon's energy into the quark/anti-quark energy densities. All three of these perturbations (gluon, quark, and anti-quark) are with respect to the background and evolved using the $\GCS{s}{s}$ Green's function until $\tau_{hydro}$. The evolution of the charge densities is done with the $\FCS{s}{s}$ Green's function, which is consistent with the vanishing background charge density generated by \code{iccing}.

The equations, as implemented in the code are expressed as:
\begin{subequations}\label{eq:FullPropagationViaGF}
    \begin{align}
        e\pqty{\thydro, \xt} &= e_\BG\pqty{e_\text{Trento}(\tau_0),\xt} \Big[ 1 + \int_\odot \frac{\dd[2]{x_0}}{\pqty{\Delta\tau}^2} \pqty{\Delta\tau}^2 \GCS{s}{s}\pqty{\frac{\abs{\xt-\xt_0}}{\Delta\tau}, \tilde{w}} \frac{1}{e_\text{Trento}(\tau_0,\xt_0)} \nonumber 
        \\
        &\qquad \times \pqty{\delta e_q(\tau_0,\xt_0) + \delta e_{\overline{q}}(\tau_0,\xt_0) - \delta e_g(\tau_0,\xt_0)}\Big] \, 
        , \\
        n_i\pqty{\thydro, \xt} &= \int_\odot \frac{\dd[2]{x_0}}{\pqty{\Delta\tau}^2} \pqty{\Delta\tau}^2 \FCS{s}{s}\pqty{\frac{\abs{\xt-\xt_0}}{\Delta\tau}, \tilde{w}} \frac{\tau_0}{\tau} \bqty{\delta n_q(\tau_0,\xt_0) - \delta n_{\overline{q}}(\tau_0,\xt_0)} \, .
    \end{align}
\end{subequations}
where we integrate the Green's function evolution over all $\xt_0$ in the past causal light cone $|\xt_0 - \xt| < \Delta \tau$. The last term in these equations refers the source of these perturbations and their sign, i.e. energy perturbation is positive for quark and anti-quark ($\delta e_q$ and $\delta e_{\overline{q}}$, respectively) and negative for the gluon ($\delta e_g$). Furthermore, the reference point $\xt_0$ is distinct for the gluon, quark, and anti-quark and taken to be the center, with respect to each, of the sampled perturbations. The point of interest is referred to as $\xt$ and $\Delta\tau\equiv\thydro-\tau_0$. These equations are implemented in the Event class (see Event.h) with the Green's functions receiving a new class (see GreensFunctions.h). The full implementation is non-trivial, but does not interact with the underlying probability sampling procedure in any way.

\begin{figure}
    \centering
    \includegraphics[width=0.85 \textwidth]{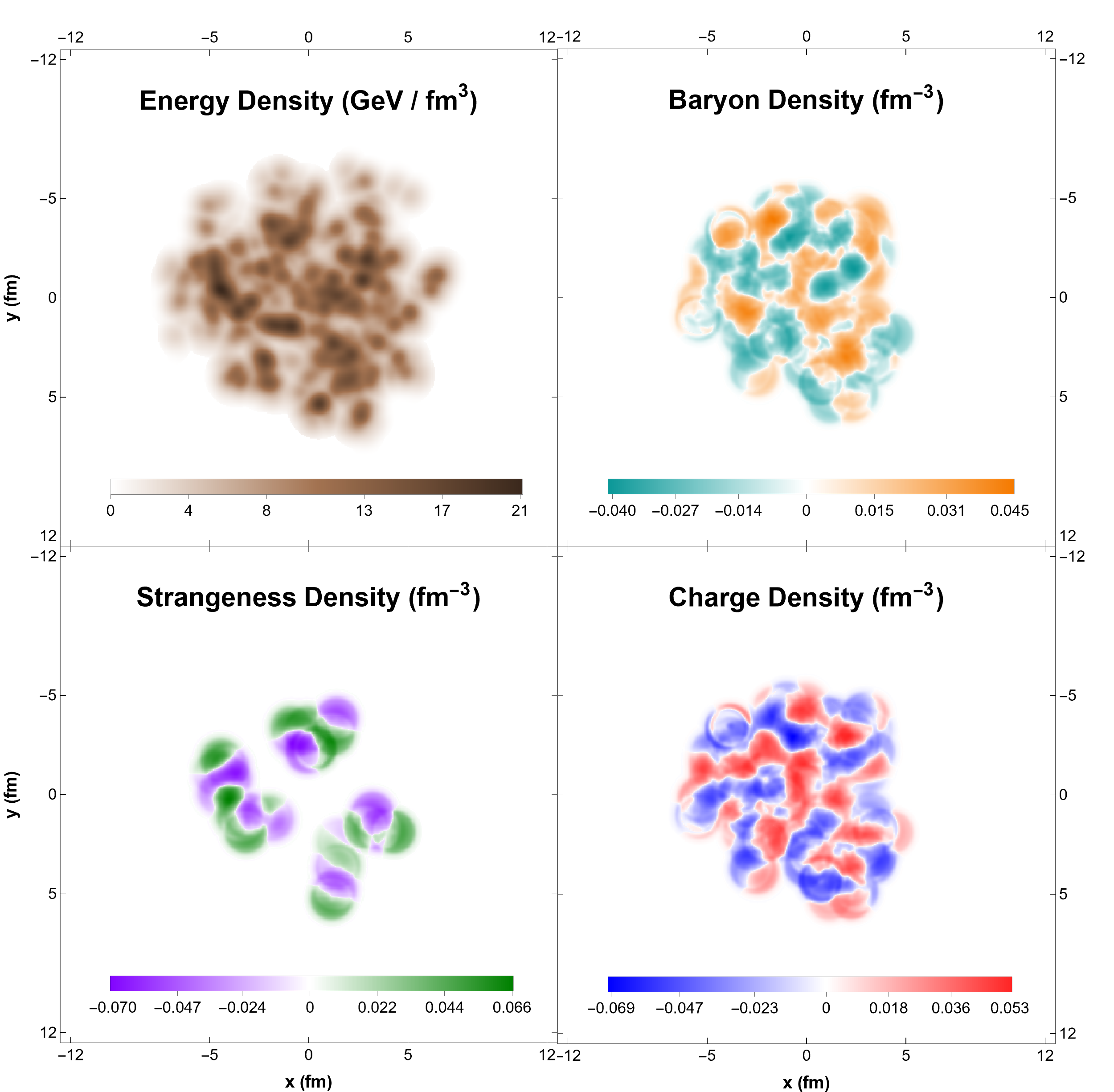}
	\caption{Density distributions for an \code{iccing} event with Green's function evolution of energy and charge perturbations from $g \to q\bar{q}$ splittings after $1 fm/c$ of evolution. Figure from Ref.~\cite{Carzon:2023zfp}.}
	\label{f:FullEventDensities}
\end{figure}

In Fig.~\ref{f:FullEventDensities}, the energy and charge densities of a central \code{iccing}+Green's event is shown using a PbPb collision at $\sqrt{s_{NN}} = 5.02 \, \mathrm{TeV}$ with evolution from $\tau_0=0.1 fm/c$ to $\tau_{hydro}=1.1 fm/c$. All other parameters are the same as the default list from Chap.~\ref{chap:ICCINGAlgorithm}. We see there are similarities in these evolved charge distributions to those from default \code{iccing} (see Fig.~\ref{f:Events}). The baryon and electric charge densities follow the bulk geometry, while the strangeness density retains its more rarefied distribution due to the larger mass threshold. The most striking difference is in the size of the fluctuations, since with the Green's function evolution the radius of the charge perturbations increases with time. We can understand the new features by looking at the background evolution, individual quark splittings, and the radius dependence separately.

%
%__________________________________________________________________________
%
%
%__________________________________________________________________________
%

\begin{figure}
    \centering
	\includegraphics[width=0.85 \textwidth]{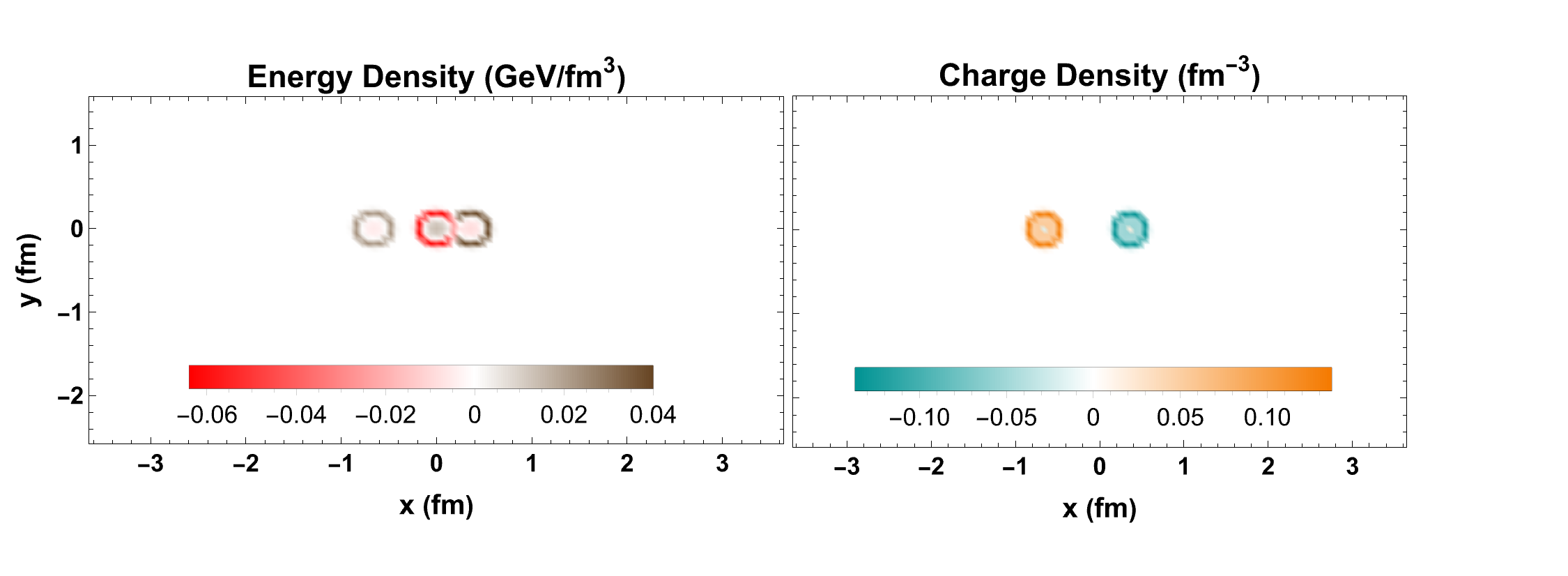} \,
	\includegraphics[width=0.85 \textwidth]{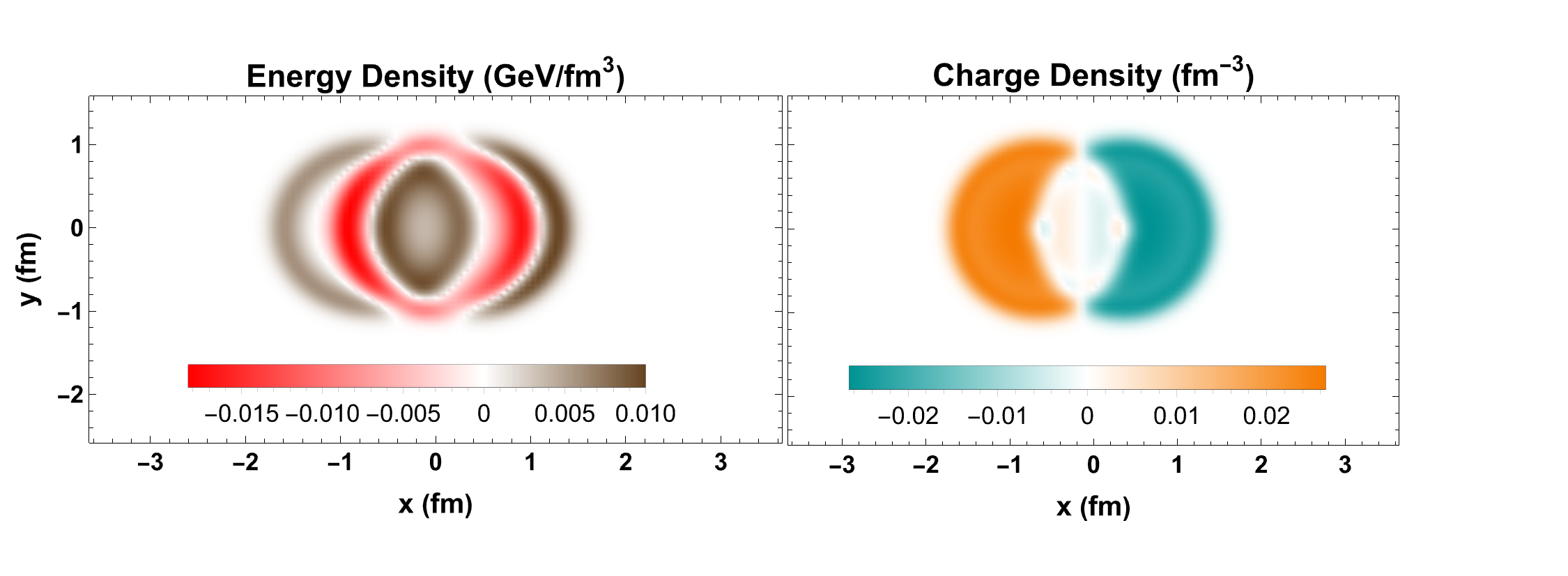}
	\caption{Density distributions for a single strange quark splitting compared for evolution times of $\tau_{hydro} = 0.2 fm$, on the left, and $\tau_{hydro} = 1 fm$, on the right. Figure from Ref.~\cite{Carzon:2023zfp}.}
	\label{f:SingleQuarkEvolution}
\end{figure}

To begin, it is crucial to understand the complete impact that the evolution has on a solitary quark/anti-quark pair. In Fig.~\ref{f:SingleQuarkEvolution}, the energy density and charge density of a quark splitting, that has undergone evolution for $0.2 \: \mathrm{fm/c}$ and $1 \: \mathrm{fm/c}$, is illustrated. The top panel of Fig.~\ref{f:SingleQuarkEvolution} corresponds to the early stage of the quark splitting, while the bottom panel represents the state at the end of the evolution. For a short evolution, the top densities of Fig.~\ref{f:SingleQuarkEvolution} show three distinct types of perturbations: primarily positive-energy perturbations that correspond to the deposition of the $q \bar q$ pair, as well as a mostly negative-energy perturbations that correspond to the subtraction of the parent gluon from the background. The quarks are accompanied by corresponding charge perturbations that contain similar structure as the associated energy. It is difficult to see at early times, but we see a clear wave-like structure to the density perturbations that survives at later times.

%
%__________________________________________________________________________
%
%
%__________________________________________________________________________
%

Comparing the top and bottom of Fig.~\ref{f:SingleQuarkEvolution}, we see that the dominant effect from the evolution time is that the perturbations grow in size over time and that the wave-like structure produces non-trivial features. A simplification is made to the process by fixing the location of the center of each density perturbation, which comes entirely from the sampling algorithm in \code{iccing}. Specifically, the Green's function evolution is essentially an alternative method to the default Gaussian or Kernel profiles and does not interact meaningfully with the sampling algorithm.

\begin{figure}
    \centering
	\includegraphics[width=0.85 \textwidth]{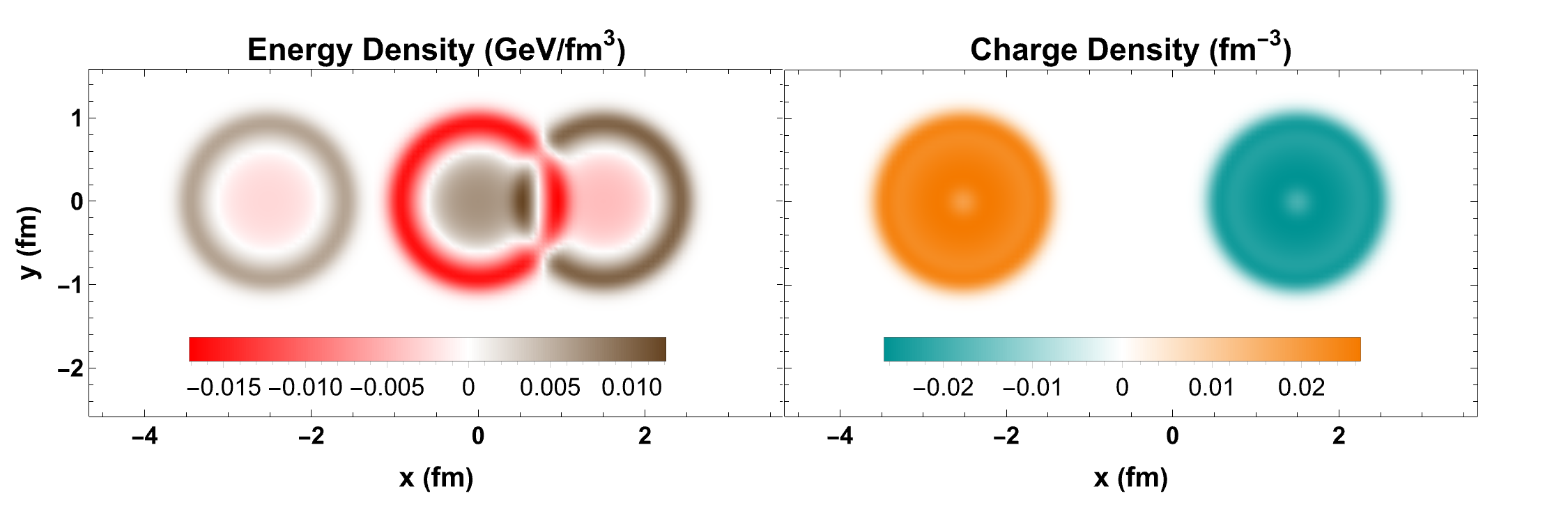} \,
	\includegraphics[width=0.85 \textwidth]{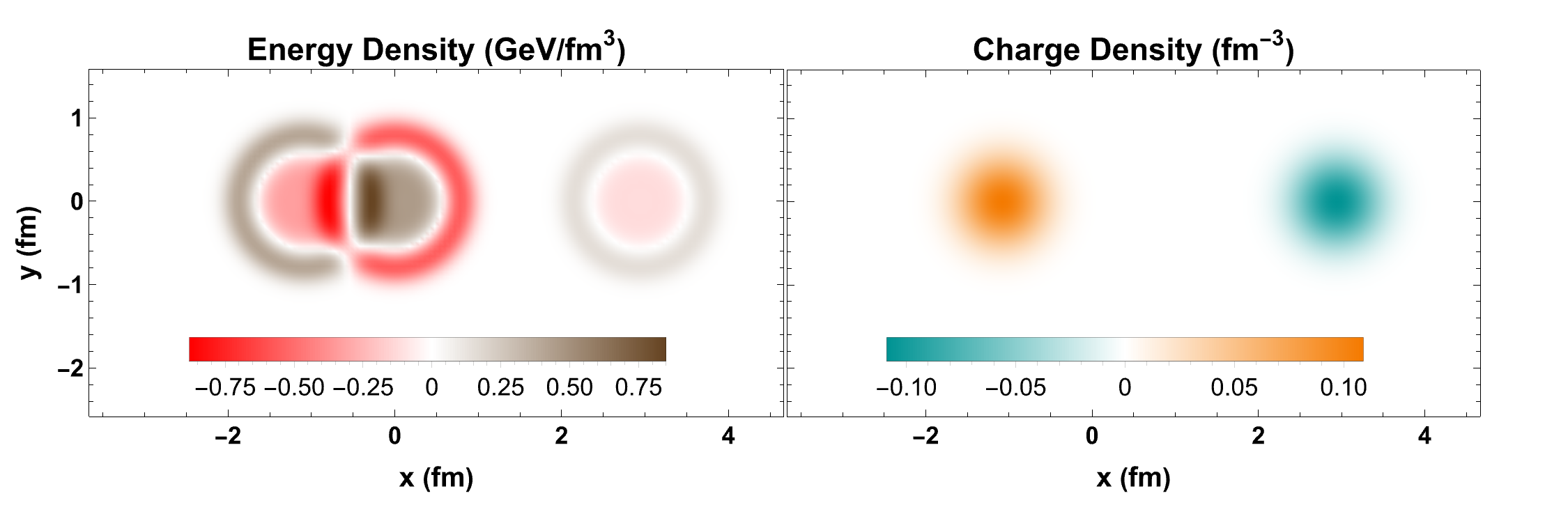}
	
	\caption{Density distributions for a single strange quark splitting from different areas of the event: a cold region on the top, and a hot region on the bottom. Here the separation of the two quarks is artificially increased to better illustrate the behaviour of the Green's functions. Figure from Ref.~\cite{Carzon:2023zfp}. }
	\label{f:TempQuarks}
\end{figure}

The Green's functions depend on the effective temperature of the initial splitting location through the $\tilde{w}$ variable (see Eq.~\ref{eq:wTilde}), and it is reasonable to see if this has an effect on the density perturbations. The spatial profiles of quark pairs produced in low and high temperature regions are illustrated in Fig.~\ref{f:TempQuarks} on the top and bottom, respectively. We see that the charge perturbations in low temperature regimes have a wave like structure, while the high temperature quarks are more concentrated and have smoother profiles. This effect can be seen when revisiting Fig.~\ref{fig:ResponseFunctions} (right), where there is a transition from propagating behavior at small $\tilde{w}$ (low temperature) to diffusive behavior at large $\tilde{w}$ (high temperature). This dependence on the effective temperature of splitting creates an interesting and microscopically noticeable difference based on the location of quark splitting. For long enough evolution times, all of the charge perturbations will look the same regardless of origin since $\tilde{w}$ is dependent on $\tau$ as well.

%
%__________________________________________________________________________
%
%
%__________________________________________________________________________
%

A connection can be made, here, to the Knudsen number, $Kn$, since $Kn=\tau_R/\tau\propto (\tau T)^{-1}$ and so $\tilde{\omega}\propto Kn^{-1}$ \cite{Kamata:2020mka}, such that at late times one expects a smaller $Kn$ number. We can use this relationship to derive a physical intuition for the location dependence of quark/anti-quark production. Hotter spots in the system will have larger $Kn$ and correspond to a Gaussian charge profile, while cold spots have small $Kn$ producing wave-like charge profiles. This difference in behavior may become more important when comparing systems across different energies since the effect will be more consistent on a system scale.

The effects, so far, of including the new physics of the Green's functions are quite interesting, although there is worrisome behavior that comes with it. A core assumption of \code{iccing}, is that all charge must be correlated with some energy. The linearized treatment of the Green's functions generates large negative corrections to the energy density, as seen in Figs.~\ref{f:SingleQuarkEvolution} and \ref{f:TempQuarks}, that are able to win out against the background and consequently produce grid points with net negative energy, which is nonphysical. Additionally, this method may exacerbate issues encountered, in connection to hydrodynamic simulations, when trying to match fluid cells with very low energy density and high charge density to a reasonable equation of state (see Sec.~\ref{SubSec:LocalGlobalImpactDensityProfile}).

The large perturbations, produced by the pre-equilibrium evolution, generate issues that could be solved by returning to the linearized approximation used in the derivation of the Green's functions from Ref.~\cite{Carzon:2023zfp}. This approximation is broken when the local redistributed energy density is greater than or close to the background. We can satisfy this perturbative approximation by either modifying the redistributed densities or restricting quark production. The first case, of modifying the densities, means introducing an artificial damping of the perturbations with respect to the background to ensure the approximation is satisfied. This introduces its own problem, though, when a $q \bar q$ is produced near to the edge of the system and they are suppressed differently leading to a violation of charge conservation. The damping could be done by mirroring the effect on both quark and anti-quark, but this also leads to nonphysical behavior.

The other solution would be to restrict quark production, and is the one implemented in this work. For each $g \rightarrow q\bar{q}$ process, the redistributed energy density is checked against the background and if, at any individual point, the ratio $E_{q} / E_{BG}$ is greater than some perturbative threshold $P$ the splitting is rejected. This solution effectively eliminates quark production near the edge of the system and consequently reduces the contribution of "cold" quarks. An issue with this approach, is that the suppression is dependent on the evolution time since the radius of the quark/anti-quark increases with time. This means that a splitting possible for an evolution of $0.5 \: \mathrm{fm/c}$ may become rejected for a longer evolution time. This introduces an odd acausal effect on quark production. The key issue at play is a mismatch between the evolution of the redistributed energy, which contains transverse expansion, and the background, which is confined to longitudinal expansion. A more consistent application would include transverse expansion in the background evolution, which may fix this issue, but that is beyond the scope of this preliminary study and left for future investigation. The implementation of quark suppression is used here and its effects are reduced as much as possible by choosing a large perturbative regime.

\begin{figure}
    \centering
	\includegraphics[width=0.45 \textwidth]{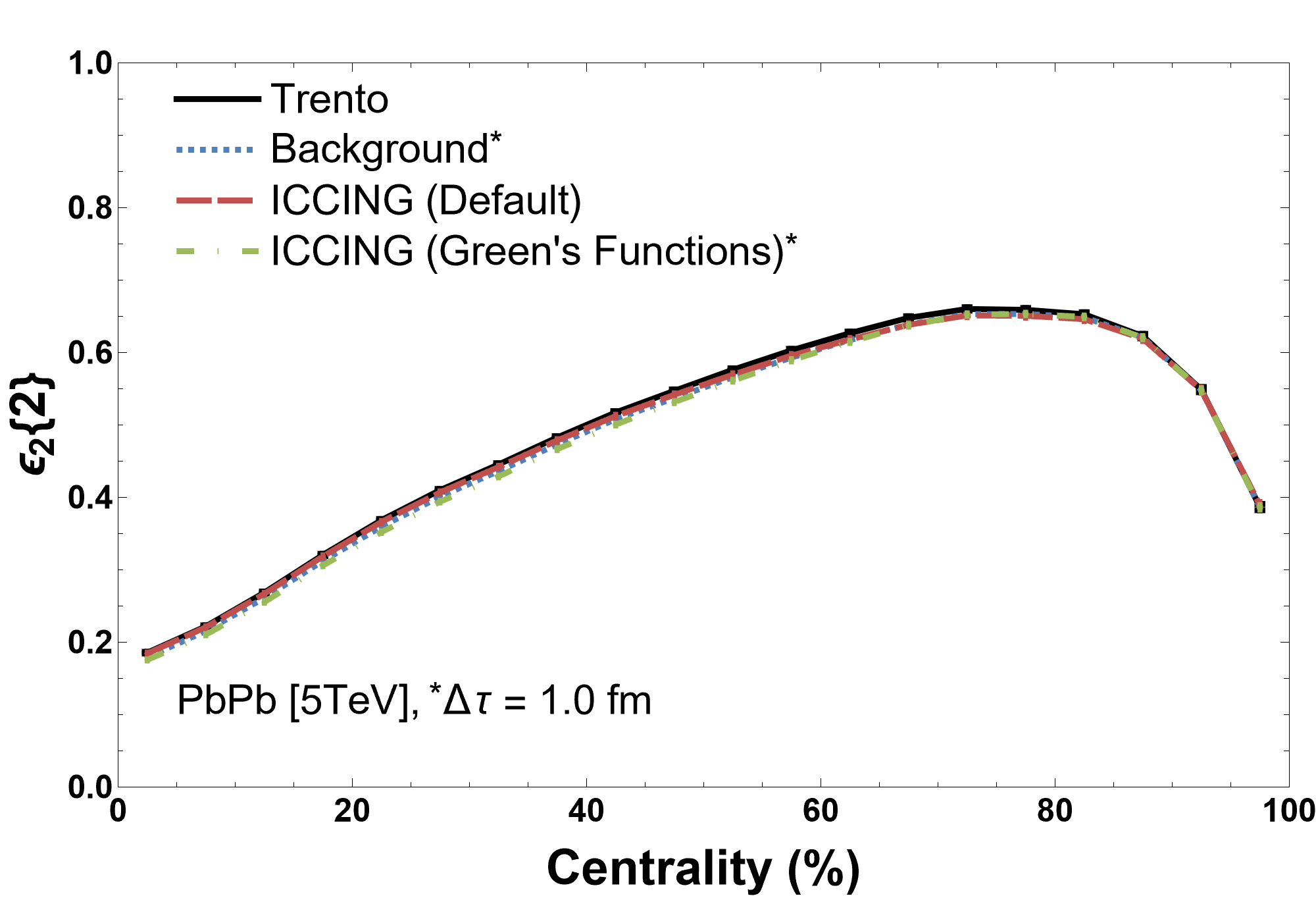} \,
	\includegraphics[width=0.45 \textwidth]{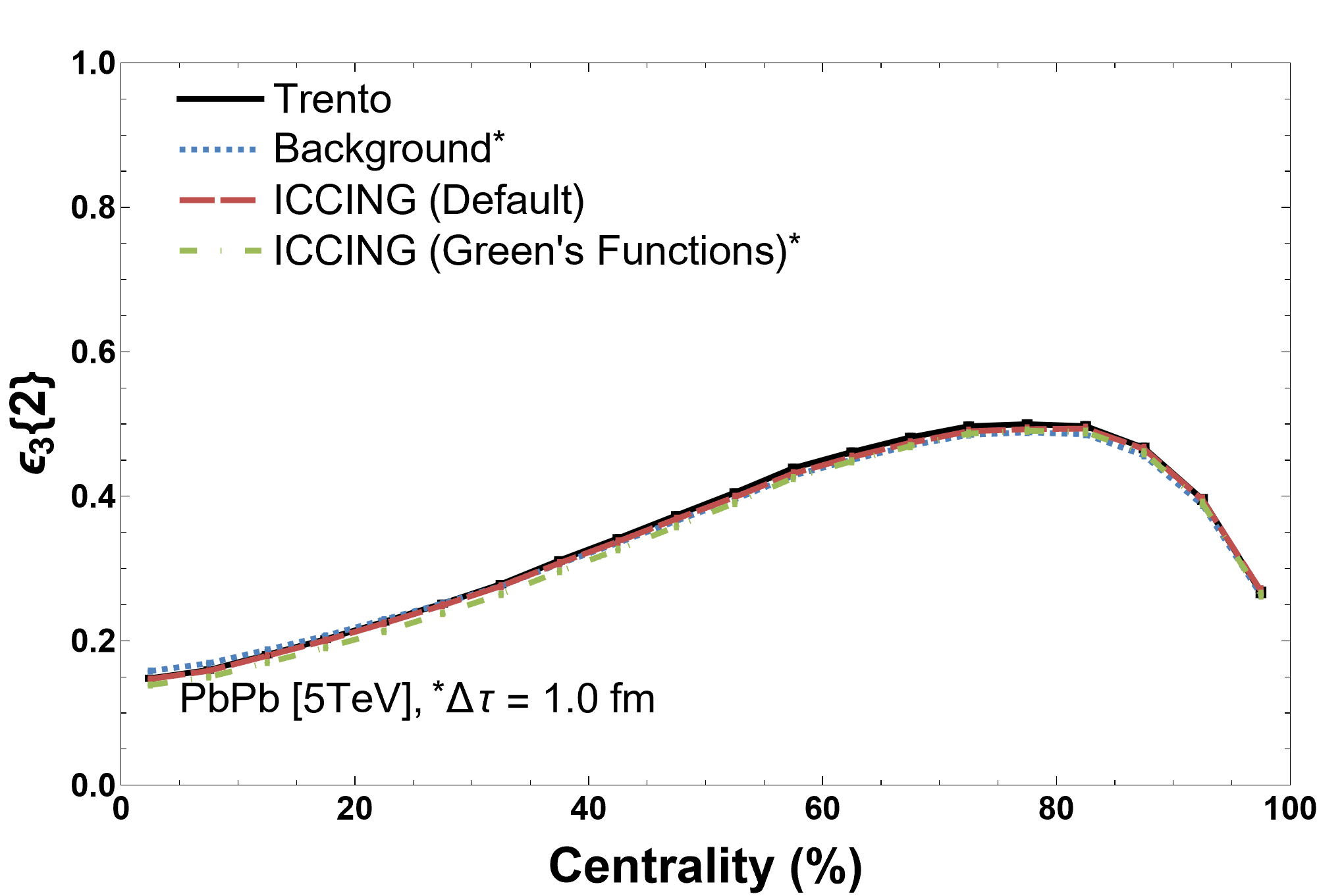}
	
	\caption{Comparison of $\varepsilon_n \{2\}$ across energy and BSQ distributions for different Green's function evolution times. Figure from Ref.~\cite{Carzon:2023zfp}.}
	\label{f:BackgroundGeometry}
\end{figure}

As our procedure is expected to have only a minor impact on the overall energy density, significant alterations of the energy density eccentricities, as defined in Sec.~\ref{sec:Cumulants}, are not anticipated. Therefore, prior to examining the eccentricities of the charge densities, we investigate the impact of the individual processes on the energy density, with the expectation any effect will be negligible. The motivation for a small effect in the energy eccentricities, is that they are already a good predictor of the final state and a large effect may break agreement with experimental data. The energy ellipticity and triangularity is plotted in Fig.~\ref{f:BackgroundGeometry} (left and right, respectively) for the following scenarios: the original \code{trento} profile, evolved background using the longitudinal expansion, \code{trento} modified by default \code{iccing}, and the full \code{iccing} coupled to Green's functions simulation. Comparing the locally evolved background with the initial \code{trento} profile, we notice minor variations at the order of a few percent, which can be attributed to the occurrence of inhomogeneous longitudinal cooling \cite{Ambrus:2021fej,Ambrus:2022qya}. This occurs because thermalization proceeds more quickly in highly energetic regions, quickening a decrease in the energy density of hot regions as compared to cold regions of the QGP. The effect of modifications to the energy density from the default \code{iccing} process are also very small. Adding the effects of Green's function evolution to the \code{iccing} model also has very little effect. This ensures, that for all cases, agreement with experimental data for all charged particles will remain valid.

%
%__________________________________________________________________________
%
%
%__________________________________________________________________________
%

\section{Impact on Eccentricities}\label{sec:GreensFunctionsImpactOnEccentricities}

The next step is to investigate the sensitivity of the pre-equilibrium Green's functions evolution to the parameters it is dependent on. We use the event averaged eccentricities, as defined in Sec.~\ref{sec:Cumulants}, to quantify the effects on the model. Since this is a time dependent process, we define the time evolution for applying the Green's functions as:
\begin{equation}
    \Delta \tau\equiv \tau_{hydro}-\tau_0 ,
\end{equation}
where $\tau_0$ is our initial time, when we begin the Green's function evolution, and $\tau_{hydro}$ is the termination time of pre-equilibrium evolution and signifies a switch to hydrodynamics. The most important ingredient to quantify is the perturbative cuttoff and its effect on the eccentricities. Next, the effect of the microscopic structure of the densities, introduced by the Green's functions, is studied. Finally, the time dependent behavior is quantified, which is the main result of this work.

When defining the eccentricities of charge, we use the same process as for the energy, but define them with respect to the center of mass for the energy density and separate the positive and negative densities. This means that for a given charge density, there are positive and negative eccentricities, which have been found to produce the same event averaged quantity, and so only the eccentricities for the positive density are reported. For a more detailed discussion, see Secs.~\ref{sec:Eccentricities} and \ref{sec:Cumulants}. This is not necessarily the best possible definition for estimators of the charge density and requires more development in future work.

\begin{figure}
    \centering
	\includegraphics[width=0.45 \textwidth]{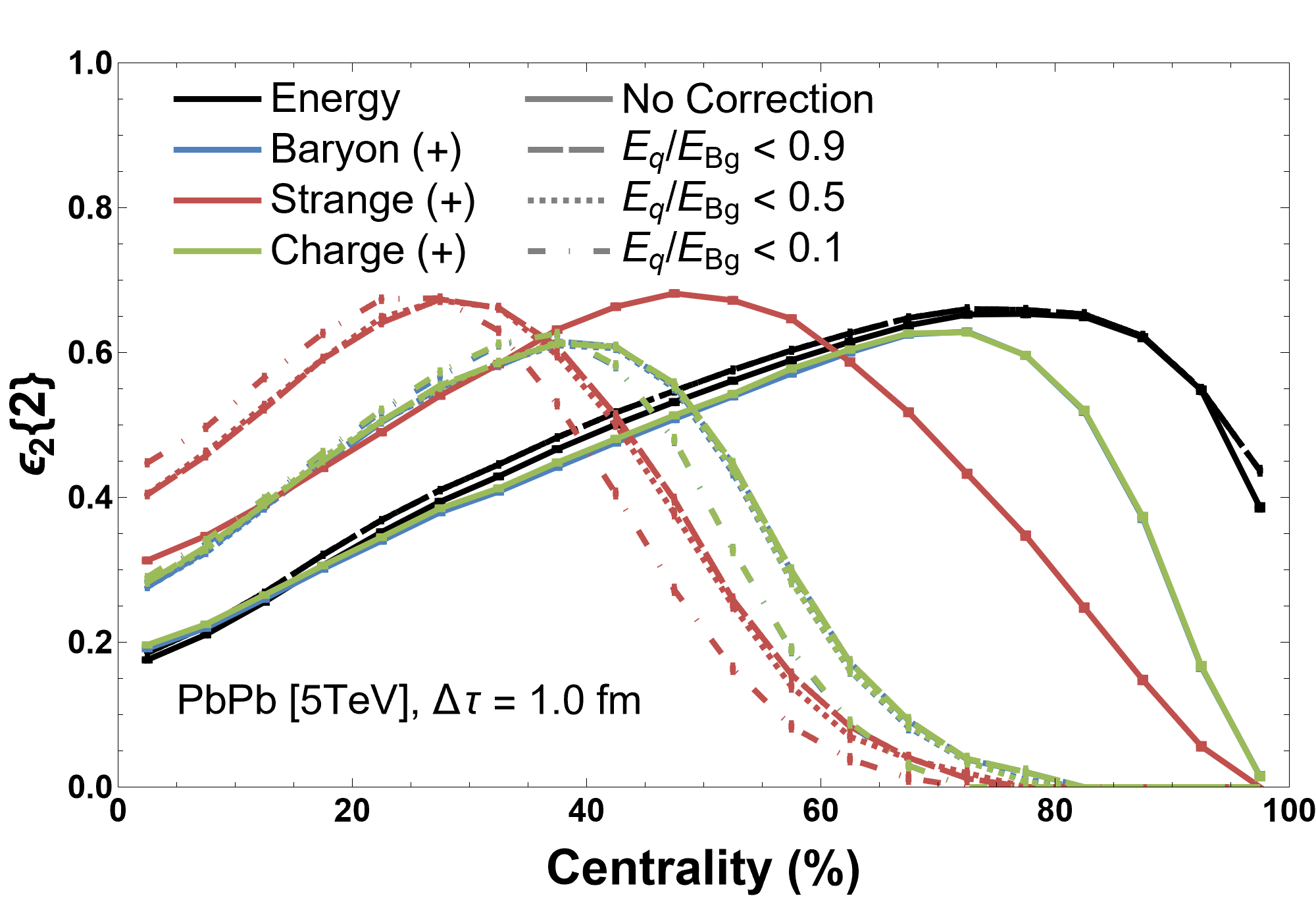} \,
	\includegraphics[width=0.45 \textwidth]{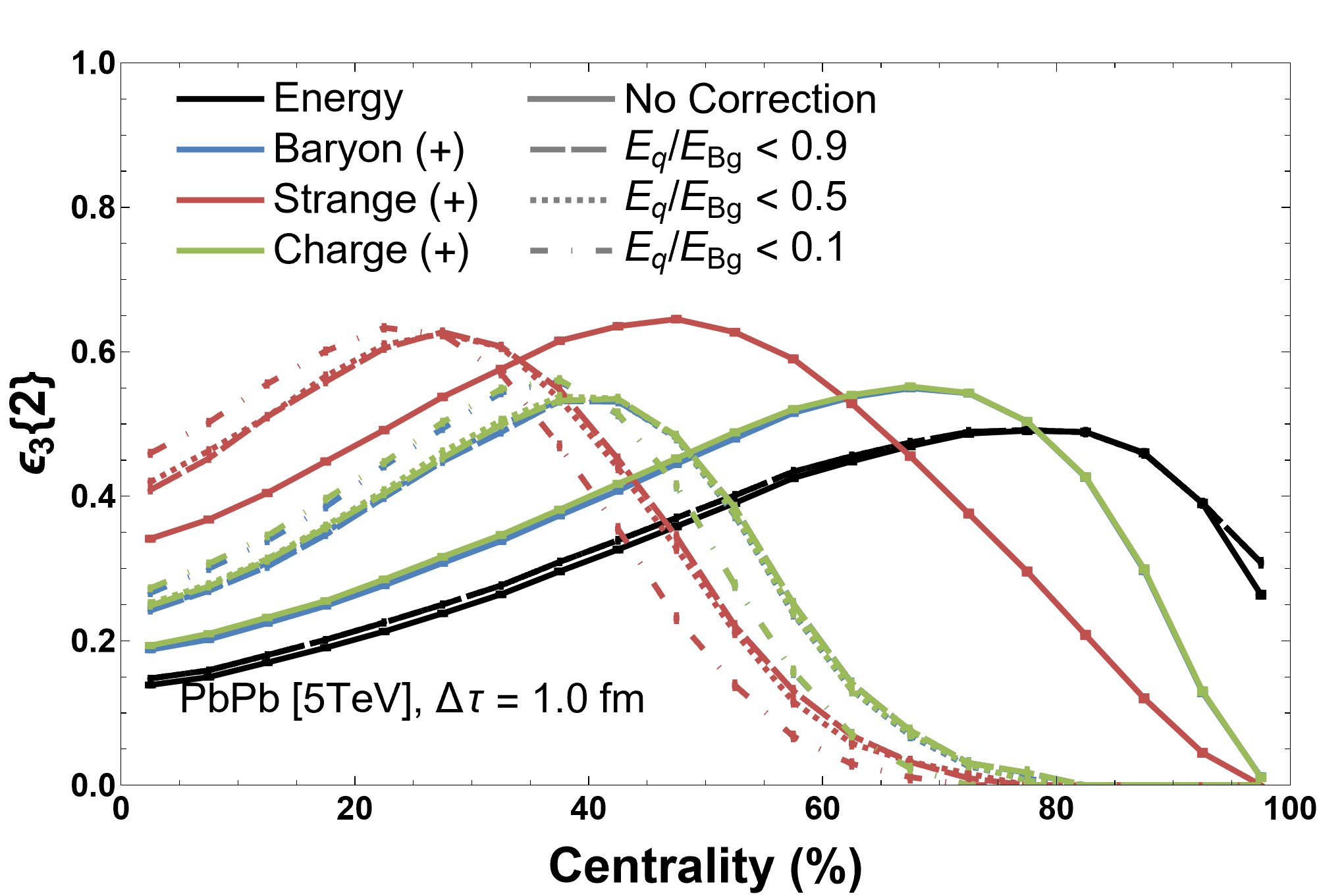}
	
	\caption{Comparison of $\varepsilon_n \{2\}$ for different perturbation cutoff values with a Green's function evolution of $\Delta \tau = 1.0 fm/c$. The solid line is with a Green's function but no cutoff, default \code{iccing} is not shown. Figure from Ref.~\cite{Carzon:2023zfp}.}
	\label{f:PerturbativeCutoff}
\end{figure}

To begin, a quantification of the effect of gluon splitting suppression, required to ensure positive energy densities, is in order. This suppression prohibits quark production in areas where the perturbative approximation is violated and the redistributed densities are larger than the background. A selection criteria is globally defined, and evaluated for each point in the redistributed densities, with the formula:
\begin{equation} \label{eq:PerturbativeCutoff}
        E_q/E_{bg} < P,
\end{equation}
where $P$ is the perturbative cutoff. The dependence of the model, on this cutoff, is shown in Fig.~\ref{f:PerturbativeCutoff} for several values of $P$ at an evolution time of $\Delta\tau=1 fm/c$, with $\varepsilon_2\left\{2\right\}$ and $\varepsilon_3\left\{2\right\}$ on the left and right respectively. The act of enforcing a perturbative cutoff of any magnitude is the dominant effect, with a significant suppression of charge densities, and thus quark/anti-quark production, above 40\% centrality. This means in mid-central to peripheral collisions, a significant proportion of quark production redistributes energy in a way to invalidate perturbative assumptions. In fact, $98.9\%$ of events in the $65-75\%$ centrality class produce no quarks, when $P=0.9$ and $\Delta\tau=1 fm/c$. This extreme response, to the most generous perturbative cutoff, signifies that the model breaks down in  peripheral collisions and/or late times and should not be used for evolution times greater than $1 fm/c$. We see no significant difference between $P=0.9$ and $0.5$ and only a small change for $0.1$, further supporting that the act of imposing a pertubative cutoff is the dominant effect. As such, we will use $P=0.9$ in the rest of the results, unless otherwise specified.

%__________________________________________________________________________
%
%
%__________________________________________________________________________
%

\begin{figure}
    \centering
    \includegraphics[width=0.85 \textwidth]{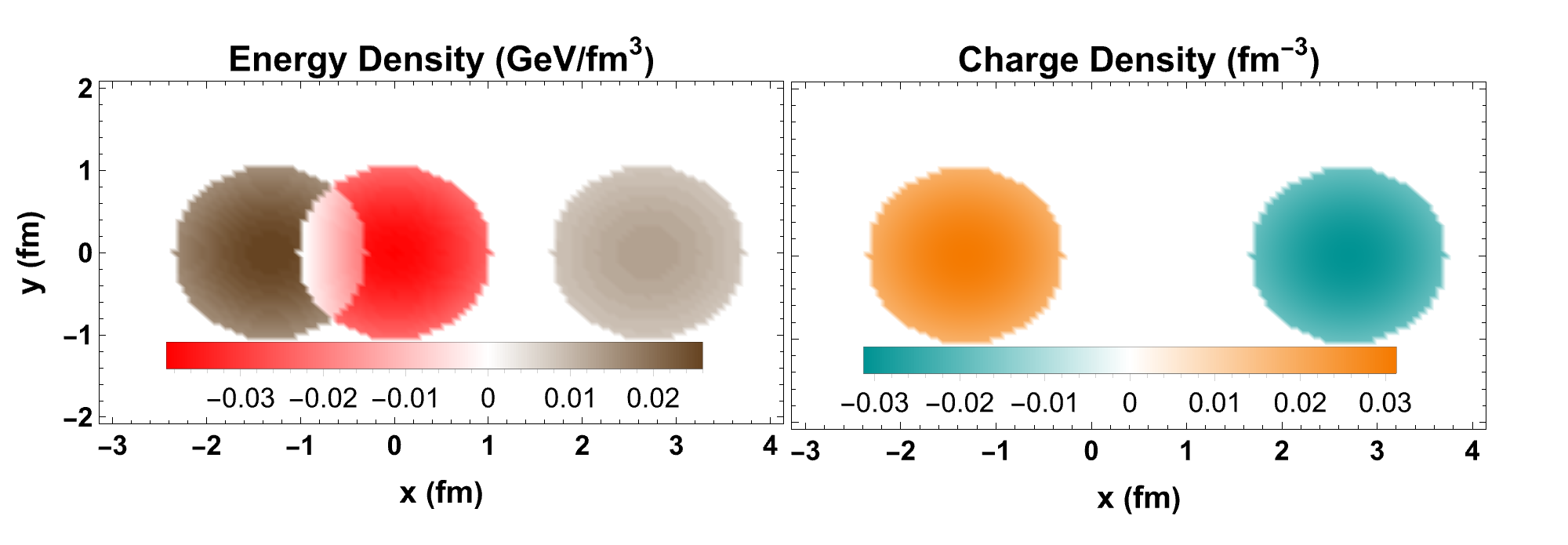}
	\caption{Illustrative density profiles of the Gaussian smearing option at $\Delta \tau = 1.0 fm/c$ which separates structure introduced by the Green's functions from the radial dependence. Figure from Ref.~\cite{Carzon:2023zfp}.}
	\label{f:GaussianSmearing}
\end{figure}

\begin{figure}
    \centering
	\includegraphics[width=0.65 \textwidth]{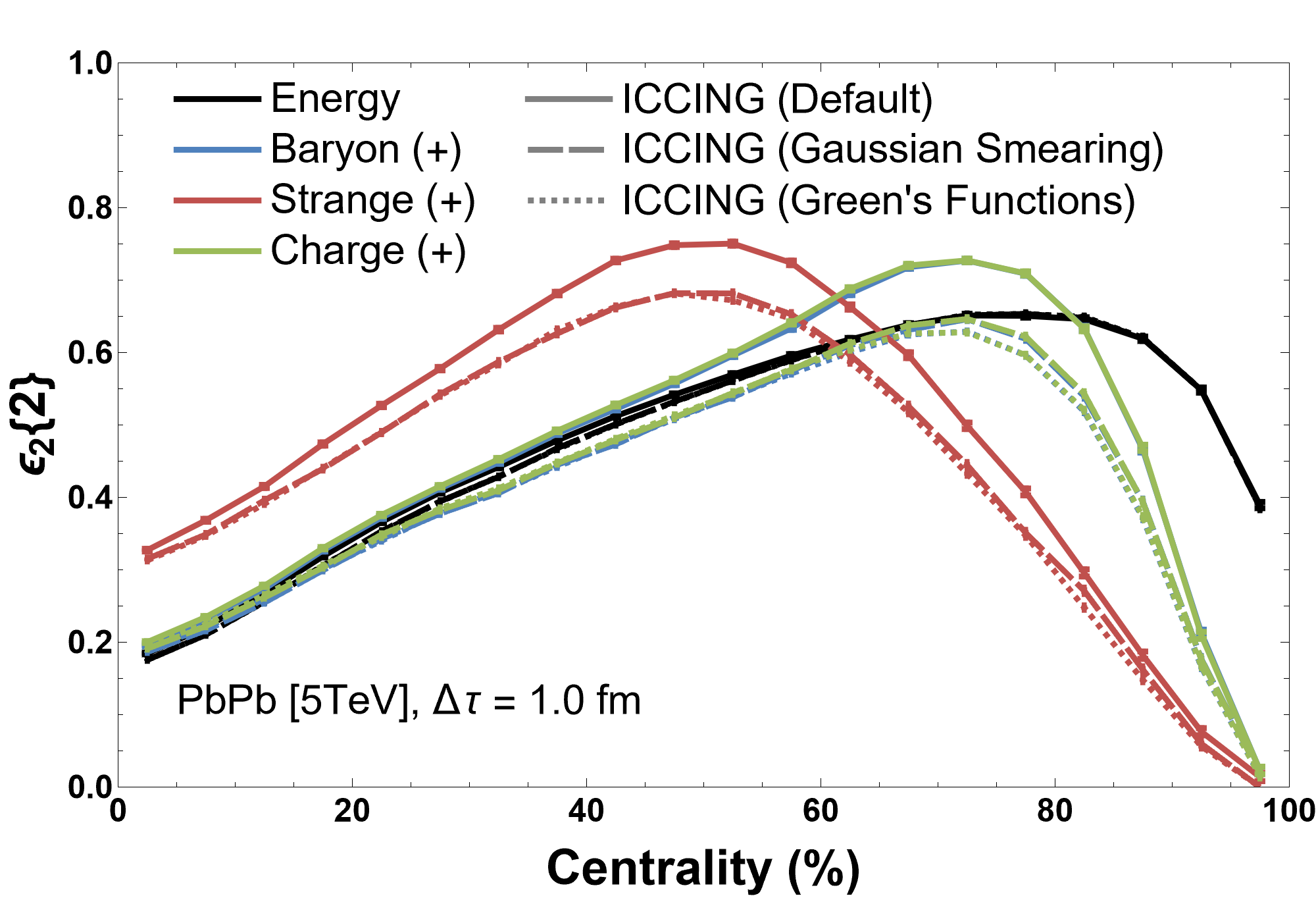}
	\caption{Comparison of $\varepsilon_2 \{2\}$ between default \code{iccing}, Gaussian Smearing, and Green's Functions with an evolution of $\Delta \tau = 1.0 fm/c$. The perturbation cutoff is not used here. Figure from Ref.~\cite{Carzon:2023zfp}.}
	\label{f:GaussianSmearingEccs}
\end{figure}

The key features introduced by the Green's functions are the complicated structure of the redistributed densities and the time dependent size of the perturbations. To disentangle these two effects, we introduce a simplified version of the Green's functions by substituting a simple Gaussian smearing for the transverse structure, defined as: 
\begin{equation}
    \GCS{s}{s}(r,t)=\FCS{s}{s}(r,t)=\frac{exp(-r^2/R(t)^2)}{\pi R^2 (t)}.
\end{equation} 
An illustration of this simplified profile is shown in Fig.~\ref{f:GaussianSmearing}, where we see it removes the shock wave-like structure from the original Green's functions. The effect of this change in structure is presented in Fig.~\ref{f:GaussianSmearingEccs}, where we see a negligible difference between the Gaussian smearing and original Green's functions. This is consistent with the analysis in Sec.~\ref{SubSec:LocalGlobalImpactDensityProfile}, where we saw that the event averaged geometry is insensitive to the fine structure of the redistributed densities.

%
%__________________________________________________________________________
%
%
%__________________________________________________________________________
%

%__________________________________________________________________________
%
%
%__________________________________________________________________________
%

\begin{figure}
    \centering
    \includegraphics[width=0.65 \textwidth]{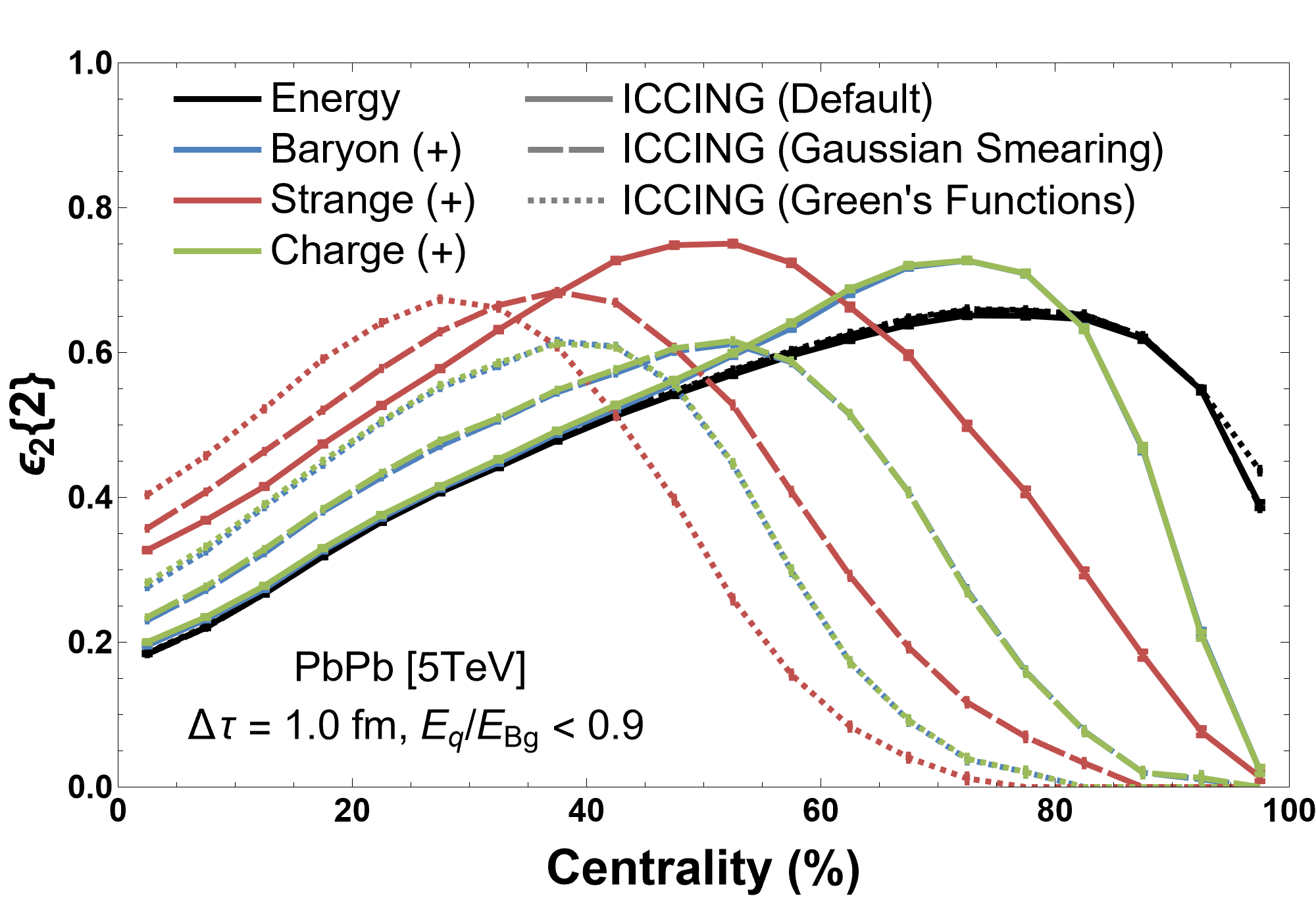}
	\caption{Including perturbation cutoff of $0.9$ for comparison between default \code{iccing}, Gaussian smearing, and Green's functions for $\Delta \tau = 1.0 fm/c$. Figure from Ref.~\cite{Carzon:2023zfp}.}
	\label{f:RadialPerturbative}
\end{figure}

For consistency, we can look at the combined effect of the perturbation cutoff and fine structure of the Green's functions with $\varepsilon_2 \{2\}$, in Fig.~\ref{f:RadialPerturbative}. The solid curves are default \code{iccing}, dashed include evolution using the Gaussian smearing, and the dotted curves include the full Green's functions. There are several effects to be disentangled in Fig.~\ref{f:RadialPerturbative}. First, the simplified profile with a pertubative cutoff pushes the peak of $\varepsilon_2\left\{2\right\}$ to lower centralities and an enhancement in central collisions, when compared to default \code{iccing}. This is due to suppression of quark/anti-quark pair production as the densities expand in size, which leads to a larger convolution of the positive and negative densities and expansion into low energy areas of the system, particularly in peripheral collisions.

%
%__________________________________________________________________________
%
%
%__________________________________________________________________________
%

Now we look at how the structure from the Green's functions interacts with the perturbation cutoff. With respect to the Gaussian smearing, we see a shift in the charge eccentricity peak to lower centralities and a further enhancement of geometry in central events. There are two features that differentiate the Green's functions energy density structure, namely the wave-like behavior, which pushes large densities to the edge of the perturbation, and the presence of negative energy in the redistributed quark densities. Application of the perturbative cutoff, is handled on a point-by-point basis and so suppress quark/anti-quark production more for the Green's function evolved densities because at the edge of the perturbation the density is larger in comparison to the background. For quark/anti-quark production in low energy areas of the system, the Green's function profiles experience a higher suppression because of the microscopic structure. Revisiting Fig.~\ref{f:GaussianSmearingEccs}, there is a negligible difference between the Gaussian smearing and Green's functions, but once the perturbative cutoff is enforced, the behavior diverges. This leads us to conclude that this method of enforcing the perturbative approximation is sensitive to the microscopic structure of the redistributed densities. For future exploration regarding the enforcement of a perturbation approximation, it is important to quantify the sensitivity to microscopic structure. Though microscopic differences may be reduced by alternative perturbative methods, the unique structure of the Green's function density perturbations will remain an important concern when coupling to hydrodynamics, since there would be sensitivity to the non-trivial change in gradients.  

\begin{figure}
    \centering
	\includegraphics[width=0.45 \textwidth]{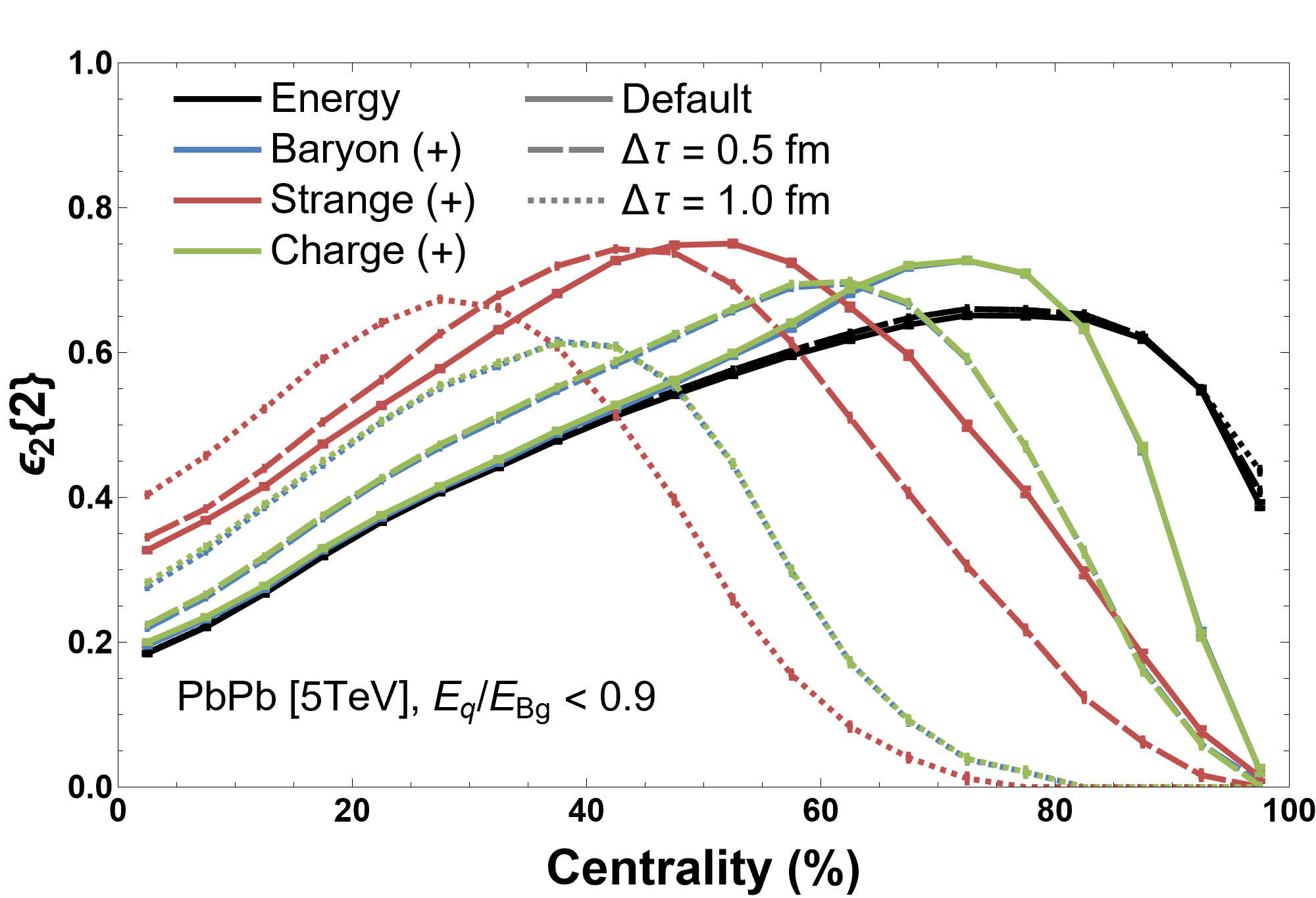} \,
	\includegraphics[width=0.45 \textwidth]{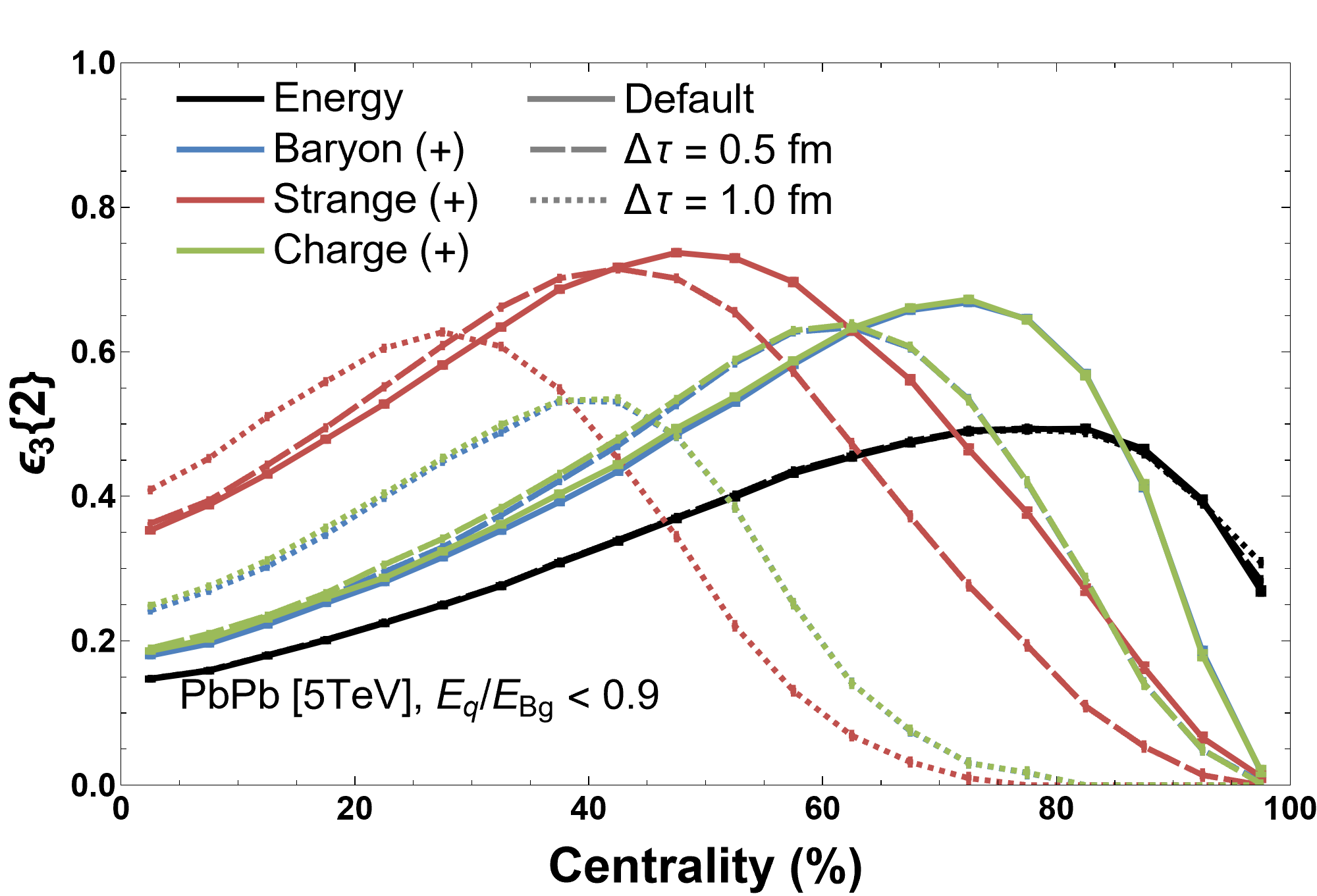}
	
	\caption{Comparison of $\varepsilon_n \{2\}$ across energy and BSQ distributions for different Green's function evolution times using the perturbation cutoff. Figure from Ref.~\cite{Carzon:2023zfp}.}
	\label{f:EccsGreens}
\end{figure}

Having established the effect of the perturbative cutoff, the influence of the microscopic density structure, and the interplay between the densities of a quark/anti-quark pair, we can look at the time dependence of the Green's function evolution. The elliptic (left) and triangular (right) eccentricities for two time steps, $\Delta\tau=0.5 fm$ and $\Delta\tau=1 fm$, is plotted in Fig.~\ref{f:EccsGreens}. As we increase the time of evolution, the peaks of the BSQ eccentricities shift to more central collisions, while the maximum magnitude decreases and an enhancement is seen in central geometry. This behavior suggests that the perturbative assumption becomes worse at later times and the model breaks down. However, there appears to be a limit, up to which, the Green's function evolution can be used in central to mid-central collisions without a significant effect on quark/anti-quark production. The triangularity, $\varepsilon_3$, is effected in a similar way.

%
%__________________________________________________________________________
%
%
%__________________________________________________________________________
%

\begin{figure}
    \centering
	\includegraphics[width=0.45 \textwidth]{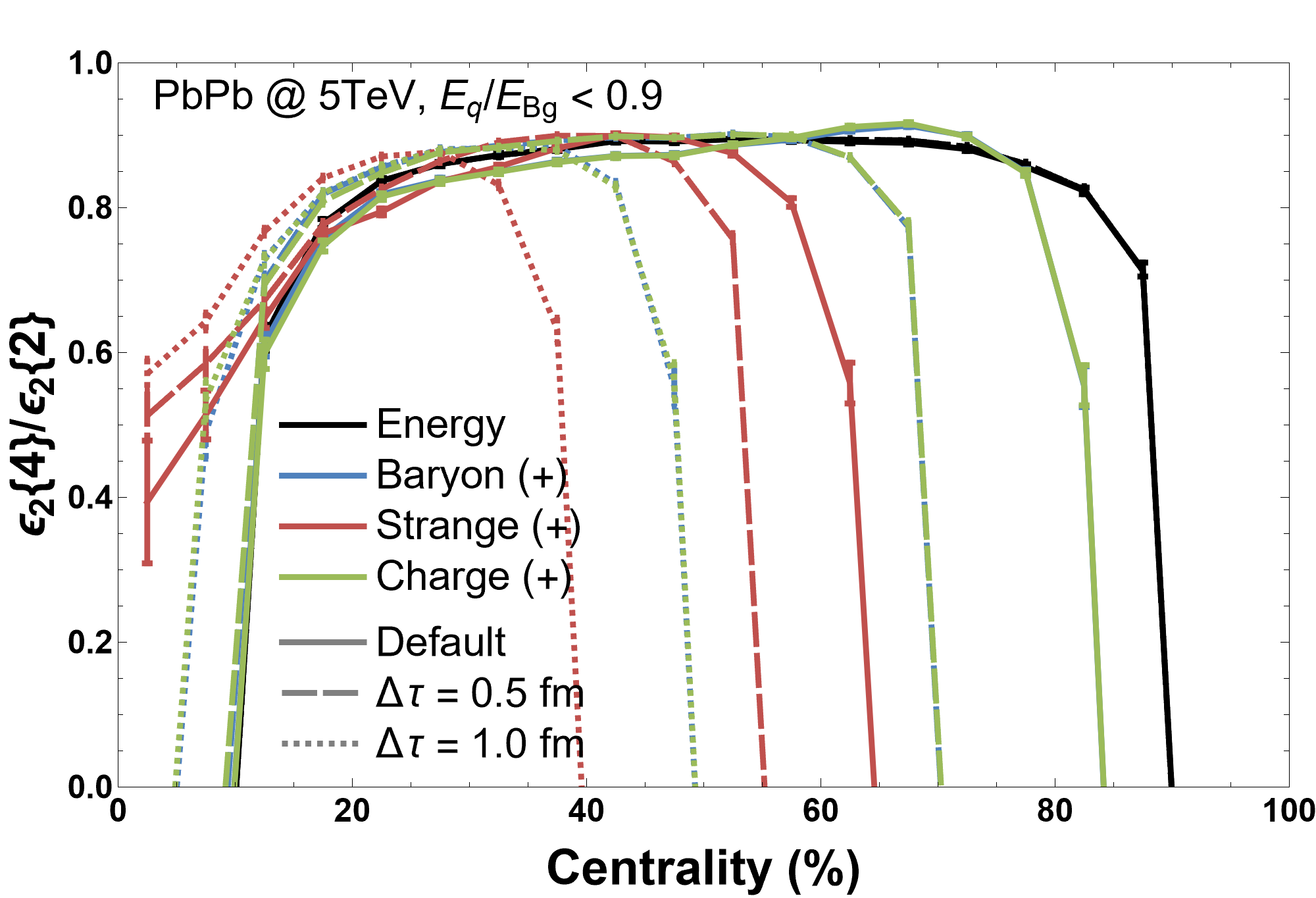} \,
	\includegraphics[width=0.45 \textwidth]{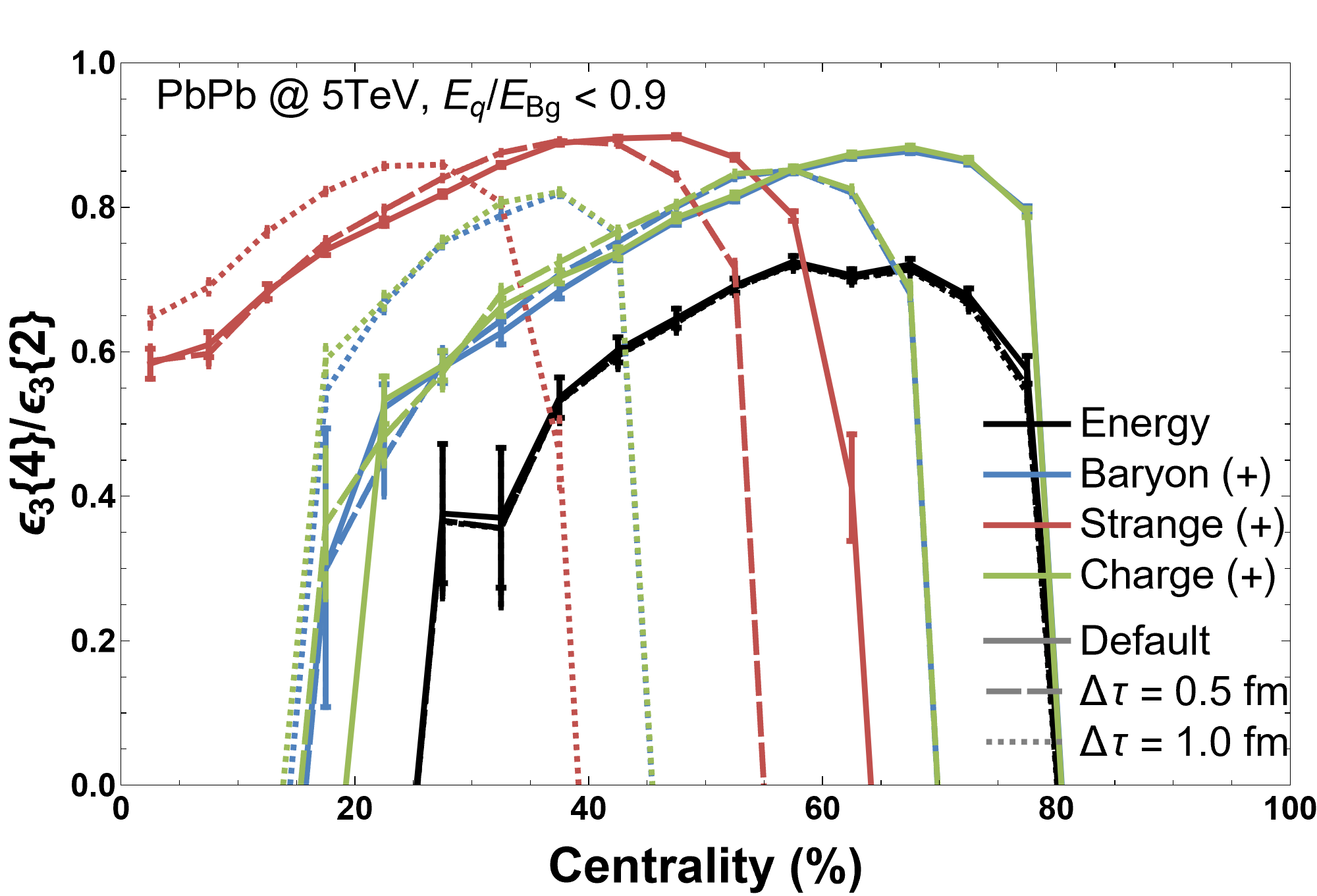}
	
	\caption{Comparison of $\varepsilon_n \{4\} / \varepsilon_n \{2\}$ across energy and BSQ distributions for different Green's function evolution times using the perturbation cutoff. Figure from Ref.~\cite{Carzon:2023zfp}.}
	\label{f:GreensEccentricityRatios}
\end{figure}

An important observable for the constraint of the initial state is the cumulant ratio $\varepsilon_n \{4\} / \varepsilon_n \{2\}$, which cancels out medium effects (see Sec.~\ref{sec:Cumulants}). This ratio is presented in Fig.~\ref{f:GreensEccentricityRatios}, for $n=2$ (left) and $n=3$ (right), at two different evolution times. The $\varepsilon_n \{4\} / \varepsilon_n \{2\}$ observable is a quantification of the fluctuations of the specified geometry, where a value close to $1$ indicates few fluctuations and deviations represent a growth of fluctuations. We see significant differences between the elliptic and triangular cumulant ratios. The BQ elliptic ratios, in central to mid-central collisions, behave similarly to the energy density cumulant ratio with minor differences at ultra-central. There is, however, a large effect at peripheral collisions due to the significant suppression from the perturbative cutoff. While default \code{iccing} BQ ratio observables agree well with the energy, the evolution of those densities creates a divergence at a lower centrality as the evolution time increases. The cumulant ratio of the strangeness density is more sensitive to the evolution since it is more rarefied than baryon and electric charge.

We see a similar response in the triangular cumulant ratio. The fluctuations in BSQ triangular geometry are, generally, smaller than those of of the energy density. The same divergent behavior introduced by the perturbative cutoff, with respect to the evolution time, is seen in peripheral collisions. The only significant difference is seen in the most central $\varepsilon_3 \{4\} / \varepsilon_3 \{2\}$, where differences are seen between the baryon and electric charge densities, although these points have large error bars. 

%
%__________________________________________________________________________
%
%
%__________________________________________________________________________
%

\section{Summary} \label{sec:PreEquilibriumSummary}

The method developed in \cite{Kamata:2020mka}, for computing non-equilibrium Green's functions of the energy-momentum tensor, is extended, here, for conserved charges and the computation of the corresponding Green's functions for charge current perturbations around a vanishing background charge density. These Green's functions were then implemented in the \code{iccing} model, which constructs conserved charge densities in the initial state through $g\rightarrow q \bar{q}$ splittings, producing these fluctuations around a vanishing background. This provides a non-trivial pre-equilibrium evolution phase to the conserved charge initial state. The implementation of the pre-equilibrium evolution is quite complex demonstrates the flexibility of the \code{iccing} algorithm.

Evaluation of the system's response to initial perturbations in terms of Green's functions, requires a description of the background evolution (as explained in Ref.~\cite{Carzon:2023zfp}) used in the energy evolution of the system. It is found that the system dynamics can be quantified in terms of moments and that the Green's functions can be directly obtained from these, providing a powerful tool for obtaining the response functions. Comparison between the energy and charge Green's functions shows a distinct difference in behavior. For the energy Green's functions, we observe the propagation of sound waves, with free-streaming and early times propagating near the speed of light and later times shifting toward the speed of sound. For the charge Green's functions, there is a transition from free-streaming propagation at early times to diffusive behavior as the system continues to thermalize.

To investigate the impact of Green's functions on the initial state charge geometries, we compare the default version of \code{iccing} with \code{iccing} plus Green's functions using $\varepsilon_n \{2\}$. This is supported by a simplifying approximation of the Green's function's structure using a Gaussian smearing profile for the density perturbations. We find that the structural difference, between the Green's functions and Gaussian smearing, has little effect on "macroscale" observables, but becomes important for processes sensitive to "microscopic" features. This implementation contains a mismatch between the background evolution, a local process, and the density perturbations, a non-local process, that opens the possibility of sites with negative energy. We solve this issue by suppressing quark/anti-quark pair production where the assumption, that these densities are perturbative, is broken. This is described by a perturbation cutoff parameter, which has a small effect on the eccentricities, however the act of enforcing this assumption significantly suppresses production in peripheral events and at the edge of collision systems. We find that the primary effect on the initial state geometry from the inclusion of the Green's functions comes from a combination of time dependent expanding perturbations and suppression of non-perturbative quark/anti-quark pairs. This is observed as an enhancement of charge ellipticity and triangularlity in central collisions and a radical suppression in peripheral collisions.

The work presented, in Ref.~\cite{Carzon:2023zfp} and summarized here, is the first step toward including evolution of charge currents in \kompost and an illustration of the effect of pre-equilibrium evolution on conserved charges from the initial state. The issues generated by the mismatch between the evolution of the background and the perturbations, which are local and non-local respectively, would be solved by implementing the pre-equilibrium evolution in \kompost instead of \code{iccing}. As such, this work acts as a proof of concept and supports the importance of such development. The inclusion of charge currents in \kompost is in current development by the authors of Ref.~\cite{Dore:inPrep}. The use of this pre-equilibrium evolution of conserved charges as a bridging stage between the initial state and hydrodynamic evolution is left for future work. Another interesting extension would be the computation of these Green's functions for charges in QCD kinetic theory and how they are different from the approximation used here. A useful generalization of this method would be the formulation of Green's functions around a non-vanishing background, which becomes applicable in systems with baryon stopping.

%%%%%%%%%%%%%%%%%%%%%%%%%%%%%%%%%%%%%%%%%%%%%%%%%%%%%%%%%%%%%%%%%%%%%%%%%%%
%
\chapter{Conclusion and Outlook} \label{chap:Conclusion}
%
%%%%%%%%%%%%%%%%%%%%%%%%%%%%%%%%%%%%%%%%%%%%%%%%%%%%%%%%%%%%%%%%%%%%%%%%%%%

\epigraph{A professor traveled around by car to give the same lecture everywhere. One day the chauffeur said to him, ‘I think I've heard that lecture of yours a thousand times, and I could give it just as well as you do.’ 

‘All right, you give the lecture tonight and I will sit out in the audience in your chauffeur’s uniform.’ 

The chauffeur gave a perfect lecture but at the end someone said, ‘There's a question I would like to ask you. [What is the location of the QCD critical point?]
%When you mix that H2SO4 without any CO2 and compared with the photographic plates of the sun, how do you get the equation that equals M-over-C squared?’ 

The chauffeur responded, ‘That's the most stupid question I ever heard in all my life, and to show you how stupid it is, I'm going to ask my chauffeur to answer.’}{Archbishop Fulton J Sheen}

A correct understanding of the initial state, specifically the event-by-event fluctuations, is important for the extraction of information about the QGP from models of heavy-ion collisions. Much progress has been made, in this regard, with the creation of more complex initial state models that include nucleon fluctuations, both in position and multiplicity, and incorporating nuclear structure from low energy collisions. 

The parameter space of initial state models can be affected by the choice in allowed functional forms for the physical processes. This is well illustrated in Chap.~\ref{chap:ExplorationOfMultiplicityFluctuations}, where the multiplicity fluctuations, in \code{trento}, are sampled from a $\Gamma$ distribution and the functional form for the reduced thickness function is restricted to a generalized mean, Eq.~\ref{e:genmean}. However, the combination of multiplicity fluctuations sampled from a log-normal distribution with a linear functional form for the reduced thickness function is also capable of matching experimental data, indicating that \code{trento} has too restrictive of a parameter space. The log-normal multiplicity distribution and linear scaling for the reduced thickness function were added to \code{trento} and could be used to conduct a new Bayesian analysis to determine if they are preferred.

While the construction of the initial state is generally well understood, there are regimes in which our current models fail, specifically ultra-central collisions where the system is sensitive to small fluctuations. A particular case of this is in ultra-central Pb+Pb collisions, where theoretical models are unable to match the magnitudes of $v_2$ and $v_3$ simultaneously. In Chap.~\ref{chap:V2toV3Puzzle}, an octupole deformation for Pb was included in an attempt to resolve this $v_2$-to-$v_3$ puzzle. Although this was mostly unsuccessful, its failure indicates that there are missing physics in our current modeling of the initial state. The $v_2$-to-$v_3$ puzzle was also discovered to be larger than originally thought, with the need to not only constrain the magnitudes of $v_2$ and $v_3$ but also the fluctuations of $v_3$. 

\code{iccing} is a modular algorithm that initializes conserved charge densities in the initial state of heavy-ion collisions using a Monte Carlo sampling of $g \rightarrow q\bar{q}$ splitting probabilities. The modular approach provides many vectors from which a user can include their own alternative physics. An example, is the ease of implementing alternative splitting functions by creating a new function in the Correlator class and updating parameter handling. Likewise, all physical parameters and theoretical inputs are easily modified by the user. Additionally, the algorithm is well parameterized allowing for various input/output options and specification of all relevant physics parameters through an external configuration file. 

The redistribution of energy through the $g \rightarrow q\bar{q}$ splittings has no discernible effect on the overall energy density geometry, thus retaining experimental agreement for initial conditions that are already well tuned and confirming the perturbative impact of these splittings. We see significant differences between the strange density geometry and the $BQ$ densities, which follow the dominant energy geometry. This indicates that strange quark/anti-quark pairs probe a different geometry of the system. 

The default construction of \code{iccing}, was found to be insufficient for the coupling of the code to hydrodynamic simulations, due to large gradients in the charge density and quark production in non-physical regions of the event. These issues were easily solved by the inclusion of a softer profile for the distribution of the quark/anti-quark densities and a pertubative parameter that can be tuned to suppress production in low energy areas. Additionally, it was found that the inclusion of new degrees of freedom required a retuning of the entropy proportionality constant $a_{entropy}$.

An test of the modular adaptability of \code{iccing}, was done with the inclusion of pre-equilibrium evolution for the quark/anti-quark perturbations. While this pre-equilibrium evolution is an important step in coupling to hydrodynamics, this implementation has limited applicability, only being reasonable for central collisions due to the mismatch in the evolution of the quark/anti-quark perturbations and the background event. A more complete handling of the pre-equilibrium charge evolution requires implementation into \kompost \cite{Dore:inPrep}.

\section{Outlook}

While the dependence of \code{iccing} on its many parameters has been explored in Chap.~\ref{chap:ICCINGResults}, the default values have not been constrained by experimental data. With such a large parameter space, a Bayesian analysis of the \code{iccing} model is warranted. Additionally, the interplay of \code{iccing} and the initial state generator could be explored through both a coupled Bayesian analysis and analysis using initial state models other than \code{trento}. However, charge dependent observables, that are good estimators of the final state charge flow, are needed to constrain the parameter space. Using the standard spatial eccentricity definition to characterize the conserved charge geometry of the initial state is poorly defined and not guaranteed to be a good estimator of the final state flow of charge. As such, there is need for a better estimator of initial state charge geometry and quantification on its effectiveness. This is left to future work but a good direction would be a derivation of charge estimators from the cumulant generating functions.

A key observation from Chap.~\ref{chap:ICCINGResults} is that the strangeness density appears to have a different geometry than the $BQ$ and energy densities. Further investigation into the exact character of the strangeness geometry is needed, particularly as to its connection to the hot spots of the medium. Recent advancements have been made in describing the initial state through a decomposition of the initial density into an average state and a basis of modes \cite{Borghini:2022iym}. Applying this method to the new conserved charge densities produced by \code{iccing} could provide useful information as to the source of the difference between the strange and other event geometries. 

There are many avenues of new physics that can be explored with the \code{iccing} model. While the investigation of this work has been occupied with the conserved $BSQ$ charges, other conserved quantities, such as spin and chirality, could be explored through the \code{iccing} model. Spin and chirality would be particularly interesting for polarized nuclear beam experiments \cite{Bozek:2018xzy, Broniowski:2019kjo}. Alternative sources for the $g \rightarrow q\bar{q}$ splitting probabilities could come from nuclear PDFs. Many initial state and hydrodynamic models are moving toward a description of the rapidity dependence of collision systems, which would require \code{iccing} to be adapted to handle all three dimensions.

The future for the initial state of heavy-ion collisions is very exciting, with emphasis in the field being put on nuclear structure, nucleon sub-structure, ultra-peripheral and ultra-central collisions, and small systems with the Electron Ion Collider (EIC). How the \code{iccing} model will interact with all of these new directions will be exciting to see.

\appendix

%%%%%%%%%%%%%%%%%%%%%%%%%%%%%%%%%%%%%%%%%%%%%%%%%%%%%%%%%%%%%%%%%%%%%%%%%%%
%
\chapter{Tests of \code{iccing} Algorithm} \label{app:TestsOfICCING}
%
%%%%%%%%%%%%%%%%%%%%%%%%%%%%%%%%%%%%%%%%%%%%%%%%%%%%%%%%%%%%%%%%%%%%%%%%%%%

The "test" parameter in \code{iccing} is used to create several of the plots in Chap.~\ref{chap:ICCINGAlgorithm} that illustrate certain key physics options in the algorithm. They are included in the public version of the code for use in reproducing this work and testing any updates the user may make concerning these parts of the model. A list of the included tests is presented in Table~\ref{table:ICCINGTests}, with the location of their implementation in the code and a brief description of their function.

%__________________________________________________________________________
%
\begin{table}[h!]
    \centering
    \includegraphics[width=0.69\textwidth]{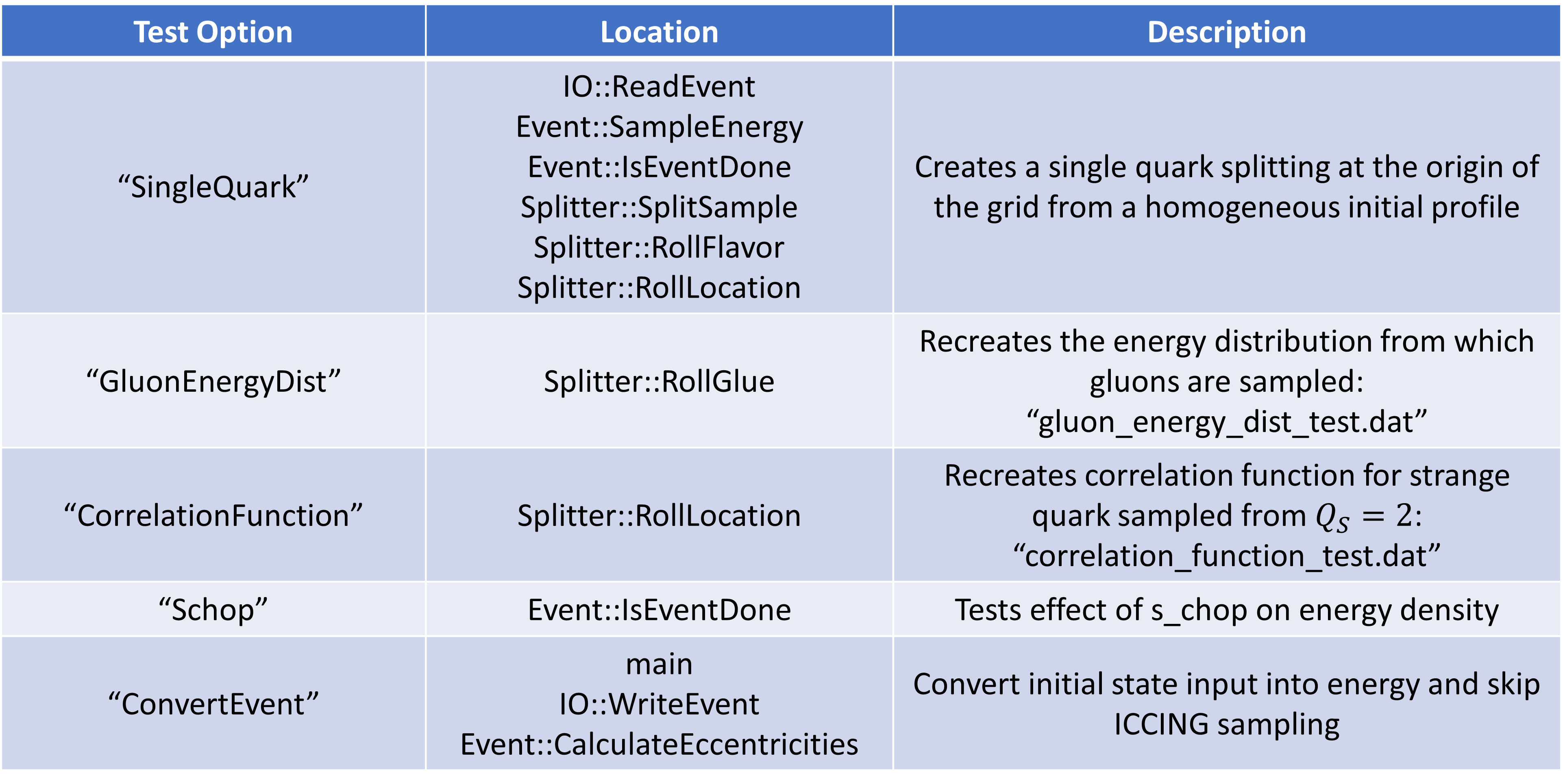}
    
    \caption{A list of the tests included in the \code{iccing} algorithm with a description and location of implementation.}
    \label{table:ICCINGTests}
\end{table}
%
%__________________________________________________________________________

\begin{itemize}
    \item The "SingleQuark" test forces the code to only create one strange $q\bar{q}$ pair centered on the middle of the event with a $Q_s = 2, \tvec{r} = 4, \phi = 0$. This $q\bar{q}$ pair is sampled from a smooth energy density to ensure the method is reproducible. This test can be used to produce plots similar to Figs.~\ref{fig:QuarkRedistribution}, \ref{f:SingleQuarkEvolution}, \ref{f:TempQuarks}, and \ref{f:GaussianSmearing}.

    \item The "GluonEnergyDist" test produces a data file called "gluon\_energy\_dist\_test.dat" which can be used to ensure that the distribution in Fig.~\ref{f:lambdaEffect} (left) is being reproduced as expected.

    \item The "CorrelationFunction" test produces a data file called "correlation\_function\_test.dat" which can be used to reproduce Fig.~\ref{f:MCvalidate} for different quarks and different $Q_s$. The quark flavor and $Q_s$ must be changed in the code and then recompiled and is set to a strange quark with $Q_s = 2$ by default. This will come in handy if the user decides to change the splitting function and wants to confirm that it is behaving correctly.

    \item The "Schop" test is used to visualize the affect of the entropy cutoff on the input energy density and can be used to reproduce Fig.~\ref{f:Echop} (top).

    \item The "ConvertEvent" test bypasses the \code{iccing} sampling algorithm and outputs the energy density that comes from the conversion of the input entropy density using the implemented equation of state.
\end{itemize}
%%%%%%%%%%%%%%%%%%%%%%%%%%%%%%%%%%%%%%%%%%%%%%%%%%%%%%%%%%%%%%%%%%%%%%%%%%%
%
\printbibliography[heading=bibintoc,title={References}]
%
%%%%%%%%%%%%%%%%%%%%%%%%%%%%%%%%%%%%%%%%%%%%%%%%%%%%%%%%%%%%%%%%%%%%%%%%%%%

\end{document}